\documentstyle[epsf,11pt,titlepage]{report4}
\epsfverbosetrue
\renewcommand{\baselinestretch}{1.0}

\def\refname{Notes and References}

\begin{document}
\renewcommand{\baselinestretch}{2.0}
\title{{\Huge \bf The metron model: \\
               elements of a unified
               deterministic  theory
             of  fields  and particles}}
\author{K. Hasselmann}
\renewcommand{\baselinestretch}{1.0}
\pagestyle{empty}
\maketitle
\newpage
~
\newpage
\setcounter{page}{1}
\tableofcontents
\newpage
~
\newpage
\pagestyle{plain}
\setcounter{part}{0}
\setcounter{section}{0}
\typeout{################################}
\typeout{################################}
\typeout{        START OF met4-1.tex}
\typeout{################################}
\typeout{################################}
\newcommand{\ruu}{\rule[-4pt]{0pt}{1pt}}
\newcommand{\ruo}{\rule[0pt]{0pt}{9pt}}
\part{The Metron Concept}
\label{The Metron Concept}
{\em ABSTRACT} \\

\noindent
In the first part of this four-part paper, the framework of
a unified deterministic theory of fields and particles is
presented. The model is based on a single set of field
equations, Einstein's vacuum equations for a
higher-dimensional metric space. The extra space is not
compactified, for example by assuming a spherical topology
with very high
extra-space curvature,
the metric being represented as a perturbation
superimposed on a
flat-space background metric. It is proposed that the
equations contain nonlinear
soliton-type solutions, termed {\it metrons}, which are
strongly
localized in physical space, while carrying far fields which
are independent of or
periodic
with respect
to extra space and time. The solutions are generated through
the mutual interaction between an inhomogeneous mean field
(e.g. a gravitational or electromagnetic field), which acts
as a wave guide, and a wave field, which is periodic in
extra ({\it harmonic}) space and is trapped in the wave
guide. The mode-trapping mechanism is demonstrated for a
simplified Lagrangian which reproduces the basic nonlinear
properties of the gravitational Lagrangian while suppressing
its tensor complexities. The more difficult task of
computing metron solutions for the higher-dimensional
gravitational system is not attempted in this paper.

The model is strictly symmetrical with respect to time
reversal. Thus Bell's basic
theorem on the non-existence of deterministic
hidden-variable theories, which is based on the existence of
an arrow of time, is not applicable. Time-reversal symmetry,
Bell's
theorem and the metron interpretation of the EPR experiment
are discussed in more detail in Part~\ref{Quantum
Phenomena}.

Since the Einstein vacuum equations contain no physical
constants, all particle properties (mass, charge, spin etc.)
and physical constants (the gravitational constant, Planck's
constant, the electroweak and strong coupling coefficients,
the parmeters of the Standard Model, etc.) are inferred from
the  properties of the metron solutions. The paradoxes of
wave-particle duality are explained by the dual nature of
the metron solutions. The localized, strongly nonlinear core
regions of the solutions embody the corpuscular properties,
while the metron far fields, including a periodic
standing-wave {\it de Broglie} field, are responsible for
the
wave-like interference phenomena. The existence of discrete
atomic spectra is explained by resonant interactions between
the eigensolutions of the Maxwell-Dirac field equations and
the orbiting electrons. Thus the metron picture of the
atomic system represents an amalgam of QED (at the tree
level) and Bohr's original orbital theory. The principal
properties of the Standard Model are reproduced assuming a
four-or five-dimensional harmonic-space background metric.
The Standard Model gauge symmetries are explained as a
special case of the diffeomorphic gauge symmetries of the
Einstein equations.  Details are given in Parts \ref{The
Maxwell-Dirac-Einstein System} and \ref{The Standard
Model}.\\

\subsection*{\raggedright Keywords:}
{\small
metron ---
unified theory ---
wave-particle duality  ---
higher-dimensional gravity ---
solitons ---
Maxwell-Dirac-Einstein system ---
Standard Model ---
EPR paradox ---
Bell's theorem ---
arrow of time } \\
%


{\em R\'ESUM\'E} \\

\vspace*{1ex}
Dans la premi\`ere partie de ce travail
est \'elabor\'e le cadre d'une th\'eorie unifi\'ee
d\'eterministe
des champs et particules. Cette th\'eorie s'appuie sur un
ensemble unique d'\'equations de champs: les \'equations
d'Einstein
du vide dans un espace \`a dimensions au-del\`a de quatre.
Les dimensions additionnelles ne sont pas compactifi\'ees
par
cons\'equence d'une tr\`es grande courbure de l'espace
suppl\'ementaire.
La m\'etrique est repr\'esent\'ee par une perturbation
superpos\'ee
sur la m\'etrique de fond de l'espace plat.
Des solutions non-lin\'eaires de type soliton appel\'ee
m\'etrons,
sont propos\'ees pour ces
\'equations. Elles apparaissent de fa\c{c}on
locale dans le domaine spatial et de fa\c{c}on p\'eriodique
dans les
dimensions suppl\'ementaires ainsi que dans le domaine
temporel.
Les solutions r\'esultent d'une interaction mutuelle entre
un
champ moyen inhomog\`ene (par exemple un champ
gravitationnel ou un
champ \'electromagn\'etique) agissant comme guide d'onde et
un
champ ondulatoire p\'eriodique dans l'espace
suppl\'ementaire,
dit espace harmonique captur\'e dans le guide d'onde.
Le m\'ecanisme de capture de modes de champs est d\'ecrit
\`a
l'aide d'un Lagrangien simplifi\'e, qui n\'eammoins
reproduit les
propri\'et\'es fondamentales non-lin\'eaires du Lagrangien
de
gravitation tout en supprimant la complexit\'e des
structures des
tenseurs.
Le but de ce travail se limite aux calculs des solutions
au syst\`eme de gravitation \`a basse dimension, et non aux
calculs plus compliqu\'es des solutions du syst\`eme complet
de gravitation \`a haute dimension

Le mod\`ele est strictement sym\'etrique au sein du domaine
temporel.
Ainsi le th\'eor\`eme fondamental de Bell,
qui s'appuie sur l'existence d'une fl\`eche du temps
et qui \'etablit la non-existence
d'une th\'eorie d\'eterministe \`a variables cach\'ees,
n'est pas valable.
La sym\'etrie d'inversion temporelle, le th\'eor\`eme
de Bell
et l'interpretation de m\'etron de l'exp\'erience
d'Einstein, Podolsky et Rosen (EPR) seront \'etudi\'es
en d\'etail dans la troixi\`eme partie.

Puisque les \'equations d'Einstein du vide ne poss\`edent
pas de
constantes de physique, on d\'eduit
toutes les propri\'et\'es de particules
(masse, charge, spin, etc.) ainsi que les constantes de
physique
(constante de gravitation, constante de Planck, les
coefficients
de couplage des forces fortes et des forces faibles,
les param\`etres du mod\`ele standard, etc.) \`a partir
des propri\'et\'es des solutions de m\'etron.
Les paradoxes de la dualit\'e onde - corpuscule s'expliquent
par la nature duale des solutions de m\'etron.
Les r\'egions localis\'ees, fortement non-lin\'eaires du
noyeau des
solutions, poss\`edent les propri\'et\'es corpusculaires,
tandis que les champs de m\'etron \`a distance, y compris
un champ p\'eriodique d'onde stationnaire de {\it de
Broglie} ,
sont responsables de l'aspect ondulatoire des ph\'enom\`enes
d'interf\'erence.
L'existence de spectres atomiques discrets s'explique
par les interactions r\'esonantes des les solutions propres
des
\'equations de champs de Maxwell-Dirac et des \'electrons
tournoyants.
Ainsi l'aspect de m\'etron d'un syst\`eme atomique
repr\'esente-t-il un
amalgame entre la QED
(si l'on exclut de la s\'erie de perturbation les
diagrammes de Feynman qui contiennent des boucles)
et la th\'eorie quantique originelle des orbites de Bohr.
Les propri\'et\'es principales du mod\`ele standard sont
reproduites, \'etant donn\'ee une m\'etrique de fond de
l'espace
harmonique \`a quatre ou cinq dimensions.
Les sym\'etries de jauge du mod\`ele standard sont ici des
cas
particuliers des sym\'etries de jauge diff\'eomorphiques
des \'equations d'Einstein (cf. parties 2-3-4).\\

\subsection*{\raggedright Mots cl\'es:}
{\small

m\'etron ---
th\'eorie unifi\'ee ---
dualit\'e onde-corpuscule ---
th\'eorie de gravitation \`a haute dimension ---
solitons ---
syst\`eme de Maxwell-Dirac-Einstein ---
mod\`ele standard ---
paradoxe d'EPR ---
th\'eor\`eme de Bell ---
fl\`eche du temps}

\newpage
\section{Introduction}
\label{Introduction 1}

\subsection*{Quantum indeterminacy versus classical
objectivity}

Despite the impressive achievements of quantum theory,
the
discusssion
on the conceptual foundations of the theory has never
completely
abated ever since the theory was first conceived nearly
seventy years
ago \cite{bel3}.  The
debate has
revolved around a number of unusual and somewhat
disturbing
features
of the theory: the limitation to a purely statistical
description of
microphysical systems already at the fundamental level of
the basic
equations; the associated inability of describing
individual
microphysical `events' or assigning physical `objects' to
the
mathematical quantities appearing in the basic equations;
the
imprecise demarkation between the quantum physical system
and the
macrophysical system described by classical physical
observables; and the
concept of a measurement process which induces a sudden
collapse
in the quantum physical state vector.

These difficulties, together with the problem of
divergences, have not
only been a continual source of concern in the
development
of quantum
field theory, but have also presented an obstacle to the
unification
of quantum field theory with the general relativistic
theory
of
gravitation, which, in its elegantly simple foundation on
the
postulate of invariance with respect to coordinate
transformations, is
free of these conceptual intricacies.

Ultimately, the quantum theoretical controversy between
deterministic `realism' and probabilistic `positivism'
reduces to the
pivotal
question: have we no choice but to resort to a
fundamentally
statistical description of nature at the microphysical
level, or is it
conceivable that a deterministic theory can be
constructed
which
provides an objective description of individual
microphysical events?
The alternative viewpoints can be illustrated by the
example
of the
Bragg scattering of a monochromatic beam of particles at
a
periodic
lattice. Quantum theory predicts the mean intensities of
the
scattered
beams in the various discrete Bragg scattering
directions,
but is
unable to `describe' what actually happens when an
individual particle
is scattered.  Yet for a sufficiently low-intensity
particle
beam, it
is perfectly possible to uniquely reconstruct (to
adequate
accuracy
within the Heisenberg uncertainty constraints) the path
of
an
individual particle, which can be registered when it
leaves
the
particle source, must pass through the (small) lattice
target and is detected again at a later time, after
it
has been
scattered into some particular Bragg direction, by
some (also small)
element of a counter array. Quantum theory nevertheless
continues to
describe such an individual event by a scattering wave
function which
contains all the possible Bragg beams up to the instant
when
the
location of the scattered particle is actually measured,
at
which
instant the wave function is suddenly collapsed to a new
state
function describing the localized particle state. The
Copenhagen
interpretation of a sudden collapse can be replaced by
the
more modern representation of a continuous
evolution
of the state function during the measurement process, but
this
does not change the quantum theoretical picture of a
single
particle
scattering into a number of separate beams prior to
the
measurement process. In
contrast, a
deterministic particle theory should be able to
mathematically
describe (although not necessarily predict) such an
individual scattering  event
in terms of the same `objective' particle picture which
an
experimental physicist would normally use to describe the
event. We come back to this example later.

The basic paradigm of quantum theory is that the
experimental finding of both wave-like and corpuscular
phenomena at the microphysical level is fundamentally
irreconcilable within the framework of classical `objective'
physics. The quantum theoretical solution is to
ignore
all corpuscular properties at the basic level of the
dynamical system equations, which describe only
nonlinearly
interacting wave fields. To establish the
connection to the dual nature of observed microphysical
phenomena, a suitable statistical interpretation of the
wave-field computations is then introduced. This enables
the wave-dynamical computations to be
related to either
waves or particles, depending on the experimental situation.
However, the `objective' simultaneous `existence' of both
waves
and
particles is denied.

In the following we question this paradigm. It is argued
that the dual existence of both wave-like and corpuscular
properties does not necessarily contradict `objective'
physics in the classical sense. We develop the basic
elements of a theory of fields and particles which
explicitly incorporates both waves and particles as
objective phenomena in a
conceptually simple manner. The widely held view that such
theories are inherently incompatible with the experimental
evidence, as exemplified by Bell's
theorem, is shown to be invalid for the present model.

\subsection*{The metron approach}

The model contains two basic elements: (i) it is shown
that the apparent  wave-particle
duality conflict can be resolved
in terms of a rather simple soliton-wave picture which
exhibits
both wave-like features, represented by a periodic far field
of the
soliton,  and particle features, associated with the
strongly nonlinear core region of the soliton; and
(ii) a specific soliton model is
developed which unifies
gravity with the other forces of nature.

The model is based on a single simple fundamental equation,
Einstein's
gravitational field equation in a higher-(eight- or nine-
)dimensional
matter-free space, without a cosmological term:
\begin{equation} \label{1.1}
\mbox{
\fbox{
$
R_{LM} = 0
$}}
\end{equation}
where $R_{LM}$ is the Ricci curvature tensor. Apart
from the
trivial flat-space solution, Einstein's vacuum equations
have solutions, such as gravitational waves, for which the
Ricci contraction of the Riemann curvature tensor but not
the full curvature tensor itself vanishes. We postulate
that the higher-dimensional  equations possess also
nonlinear  soliton-type wave solutions,
referred to in the following as {\it{metrons}}.

The determination of the metron solutions themselves is a
major computational task and is not attempted in this
four-part paper.
However, after summarizing the principle concepts and
properties of the metron model in the first three sections
of Part~\ref{The Metron Concept}, we present in
Section~\ref{The
mode-trapping mechanism} some explicit computations of
solitons of the same nonlinear structure as the proposed
metron solution for a simpler
nonlinear system which exhibits the same features
as the
gravitational equations without their tensor complexities.
The
main purpose of this exploratory paper, developed in Parts
\ref{The Maxwell-Dirac-Einstein System}-\ref{The Standard
Model}, is to demonstrate
that the equations (\ref{1.1}) have an extremely rich
nonlinear structure which encompasses all the principal
interactions of quantum field theory and can be used
as the foundation of a unified deterministic theory of
fields and particles. This is shown for electromagnetic
interactions, i.e. for the Maxwell-Dirac-Einstein system, in
Part~\ref{The Maxwell-Dirac-Einstein System} and for all
forces, including weak and strong interactions, in the
discussion of the Standard Model in Part~\ref{The Standard
Model}. Based on the analysis of the Maxwell-Dirac-Einstein
System, Part~\ref{Quantum Phenomena} addresses basic issues
of microphysics and quantum theory, such as irreversibility,
the EPR parodox, Bell's theorem, wave-particle duality and
the origin of discrete atomic spectra.

We note that matter is
not
included
as a separate external source term in (\ref{1.1}). Mass and
other particle properties are
derived instead as
properties of
the solutions of the field equations themselves \cite{ft1}.
The
physical
spacetime components of the Ricci tensor contain not only
the
 Ricci
tensor for the four-dimensional gravitational  field but in
addition (strongly
nonlinear, localized) terms
arising from the contraction of the extra-space
components
of the
Riemann tensor. It will be shown that these  yield the
standard energy-momentum
tensor which appears as external source term in classical
four-dimensional gravitational theory.

Electromagnetic forces and weak and
strong
interactions are represented by the further extra-space or
mixed
physical
spacetime-extra-space components of the Ricci tensor.
These
can
similarly be decomposed into linear or weakly nonlinear
far-field contributions and strongly nonlinear localized
terms, the latter
representing in this case the currents arising from electric
charges and weak and
strong interactions.

Thus all forces
follow, as in classical general relativity, from
the
curvature
of space. However, the curvature is not produced by
prescribed mass fields,
but is a self-generated feature of the nonlinear field
equations
(\ref{1.1}) themselves. Moreover, in contrast to the
standard field theoretical approach, the coupling constants
and symmetries are not postulated in the basic field
equations, but follow from the specific geometrical
properties of the metron solutions. The basic equations are
free of physical constants, and the only postulated symmetry
is the invariance of the field equations with respect to
coordinate transformations. \\

The choice of (\ref{1.1}) as the fundamental set of
equations follows
naturally
from three considerations:
\begin{enumerate}
\item In order to develop a unified description of all
forces we wish to adopt Einstein's successful and elegant
approach of identifying forces with curvature in space.
Since the metric of four-dimensional spacetime is already
fully needed
to
describe classical gravity, the inclusion of additional
forces represented by metric fields
requires an extension of space to higher dimension.
\item We wish ultimately to explain the particle
spectrum,
including the particle masses. This can clearly not be
achieved by a theory in which mass is postulated to exist
from the outset. The mass-dependent source term in the
Einstein equation must therefore be omitted, the
properties
of mass (and other particle properties)
being derived from the solutions of the nonlinear
Einstein
vacuum equations themselves.
\item The equations should be consistent with the
principle
of maximal simplicity.
\end{enumerate}

The present approach clearly lies outside the main stream
of
modern unification schemes. Rather than trying to unify
gravity
and quantum theory by quantizing gravity \cite{mar}, we
attempt to apply
the concepts of
(higher-dimensional) gravity theory to explain quantum
effects.
Thus our
starting point is the Kaluza-Klein approach of the twenties
rather than the super-gravity and super-symmetry theories of
the eighties.

The theory contains a number of common elements with
previous attempts
to develop a classical description of microphysical
phenomena.  As in
Bohm \cite{boh} and de Broglie \cite{debr}, wave-like and
particle-like
properties are not regarded as contradictory and mutually
exclusive
phenomena, but as simultaneously existing `objective'
realities in a
classical, deterministic sense. However, in contrast to
the
de Broglie-Bohm pilot
wave theory, waves and particles are not treated as
separate
entities,
but appear rather as the near and far field expressions
of
the same
physical object: a finite particle, or `metron'.  The
description of
particles as objects of finite extent is clearly
reminiscent
of the
early attempts of Lorentz \cite{lor} to develop a theory of
the
electron
as a finite-sized charge distribution. The present
particle
model differs from that of Lorentz in containing more
space
dimensions
and more fields.

The particles consist of a localized, strongly nonlinear
core and a set of
linear far fields which, in the particle rest frame, are
either time
independent (gravitational, electromagnetic and neutrino
fields) or
periodic in time ({\it de Broglie} waves).  A transitional
weak
interaction region bridges the strongly nonlinear core
and
linear
far field regions.  The core is the origin of the
corpuscular
properties of matter, while the periodic de Broglie far
fields give
rise to the wave-like interference phenomena. The de Broglie
far field represents a trapped standing-wave field, so that
radiative damping does not occur.

\begin{figure}[t] \centering
\begin{minipage}{9cm}
\epsfxsize=250pt
\epsffile{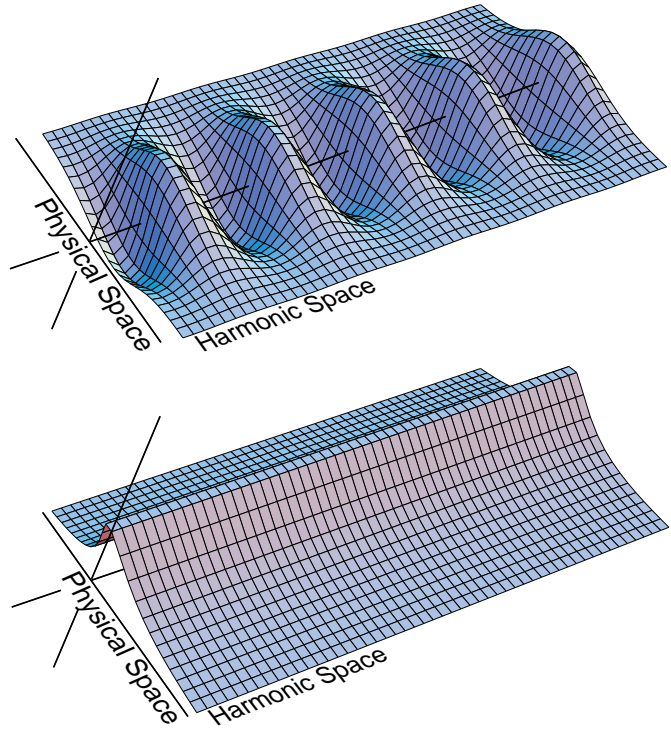}
\caption{Schematic diagram of trapped-mode (upper panel) and
wave-guide (lower panel)
components constituting a metron particle}
\label{Fig.1.1}
\end{minipage}
\end{figure}

These properties apply to physical spacetime. With
respect
to the
extra-space dimensions, the metron fields are
multi-periodic:
they consist of a superposition of a finite number of
fundamental
periodic components and their higher-harmonic interaction
combinations
(including zero-wavenumber fields). Different Fourier
components are
identified with the different constituents (partons) of
elementary
particles, which will be related in Part~\ref{The Standard
Model}
to the partons of the
Standard Model.

All interaction fields are regarded as perturbations
superimposed on an
n-dimensional background metric $\eta_{LM}=
\mbox{diag}\
(1,1,1,-1,\cdots ,\pm 1, \cdots)$. In
contrast
to most
modern Kaluza-Klein theories, the extra space is not
compactified by assuming a spherical topololgy with a
background metric of
very
large
curvature. The concept of periodic fields with respect to
 extra-space dimensions goes back to the original
Kaluza-Klein
papers \cite{kal} and is revived in recent string theories
\cite{gre}. To
emphasize the role of periodicities
in extra-space and the fact that the present extra-space is
more
closely related to the periodic extra-space dimensions of
string
theory than the usual curvature-compactified extra-space
of
higher-dimensional gravity, we shall refer to extra space
in
the
following as {\it{harmonic}} space.

The assumption that all fields can be treated as
perturbations with
respect to a flat background metric implies that the
description is
local with respect to cosmological scales. The manner in
which
 this local description is embedded in a cosmological
model
of
n-dimensional space is not discussed. Similarly, the
origin of
the distinction in the structure of the perturbation
fields
with
respect to physical spacetime and harmonic space - a
fundamental question
of Kaluza-Klein theories which is usually deferred to
cosmology - is
not considered. It is simply noted that in a
higher-dimensional space with the assumed background metric
$\eta_{LM}$, trapped-mode particle-like
solutions
of
(\ref{1.1}) which are locally concentrated in the three
dimensions of
physical space and
periodic
with
respect to the remaining dimensions can be expected to
exist. The geometrical distinction between the locally
concentrated and periodic properties of the metron solution
defines the local physical spacetime and harmonic-space
orientation. The vielbein can in general be a
function of spacetime: different particles at different
locations can have different vielbeins with respect to a
non-local coordinate system. The changes in vielbein
orientation are the origin of forces between particles (see
also item 4 below) \cite{kun}.

These general metron properties are inferred from the
nonlinearity
of the
field equations (\ref{1.1}). It is postulated (and
demonstrated for a simpler prototype system) that the
equations can support nonlinear
soliton-type
solutions in the form of wave modes trapped within a wave
guide. The
wave modes themselves
generate
the
wave guide in which they propagate (cf. Fig.
\ref{Fig.1.1}).  The basic mechanism
is a mutual
interaction between the wave modes and the mean metric
field which governs the wave propagation properties.  The
wave-guide and trapped modes
are
uniform in
harmonic space and time
but inhomogeneous with respect to physical space, the
fields
increasing to large values within the particle
core and falling off exponentially (in accordance with a
trapped mode) or
as $1/r$ (corresponding to a free wave) away from the
core region.
The mean
`radiation
stresses' or currents arising from quadratic and
higher-order wave-wave
interactions are therefore
inhomogeneous in physical space and distort the mean
metric
field in which the modes propagate. This   produces a
wave
guide which refracts
and traps the waves in the neighbourhood of the particle
core. The increase of the trapped-mode amplitudes
towards
the particle core is  in
turn the origin of the
inhomogeneous radiation stresses required to maintain the
wave-guide.

The mechanism is demonstrated for a prototype nonlinear
Lagrangian
obtained by projection of the full gravitational
Lagrangian
onto a
finite set of modes. The reduced Lagrangian captures the
basic
nonlinear properties of the complete gravitational
Lagrangian while
ignoring its tensor complexities. As pointed out, the
computation of
metron
solutions
for the full gravitational equations is considerably more
difficult and
will not be attempted here.

\subsection*{Bell's theorem and time-reversal symmetry}

A deterministic particle model with these properties
clearly falls in the class of hidden-variable models and
must therefore contend with the widely held view that
hidden-variable theories are generically incompatible
with quantum theory and experiment. Although von Neumann's
\cite{von} celebrated proof that all hidden-variable
theories
are necessarily inconsistent with quantum theory has been
shown by Bell \cite{bel2} to rest on invalid assumptions,
Bell's \cite{bel1} own well-known theorem on the Einstein-
Podolsky-Rosen
paradox \cite{ein4} is generally cited as
an irrefutable argument against hidden-variable theories.

Bell showed that any hidden-variable interpretation of
the EPR-Bohm experiment, in which a two-particle state with
zero net angular momentum decays into two separate particles
of opposite spin orientation, must satisfy an inequality
relation regarding the correlations of the spins
of the final particle states, measured in two arbitrary
directions, which is in conflict with the quantum
theoretical
result and experiment. However, an essential assumption
of Bell's theorem, already emphasized by Bell, is forward
causality, or the existence of an arrow of time.  Although
seemingly self-evident in the context of the EPR experiment,
forward causality is in fact incompatible with time
symmetry,
which is a fundamental property of all basic
(deterministic) equations of classical physics.
Time-reversal symmetry is also a basic feature of the
metron model.  Bell's theorem is therefore not applicable
to the metron model, and it will be shown that the EPR
paradox can indeed be readily resolved in the metron
picture without violating the experimental findings.

We shall adopt the classical view that an arrow of time
does not appear at the basic level of microphysical
phenomena
but only at the aggregrated level of macrophysics:
irreversibility arises through the introduction of time
asymmetrical statistical hypotheses, such as the
Boltzmann-Gibbs assumption that fine-grained structure
properties can be neglected when going forwards in time,
but not when reconstructing the past. While this view is
generally accepted for classical wave-wave
interactions or non-relativistic local particle
interactions (collisions), the question of the time-symmetry
of non-local interactions between particles mediated by
fields
which propagate - as required for Lorentz invariance - at
finite speed has been the subject of some debate. The
problem
is to explain the observed irreversible radiative damping
of charged particles under non-uniform acceleration.
Ritz \cite{rit1} believed that this could be
recovered only by introducing the auxilliary axiom that
the electromagnetic field of a charged point particle is
given
by the retarded potential.  Einstein \cite{ein1},
Tetrode \cite{tet1}
and others argued, however, that one should
retain time symmetry by choosing the time-symmetrical Green
function, consisting of half the sum of the retarded and
advanced potentials. Einstein explained the observed
radiative
damping by the time-asymmetrical statistical properties of
other particles with which the radiating particle interacts
\cite{rit3}. According to this view, an
electromagnetically isolated charged particle would not emit
radiation.
The time-symmetrical theory of elecromagnetically
interacting point particles has been developed further by
Tetrode \cite{tet1}, Frenkel \cite{fre},
Fokker \cite{fok1}, Dirac \cite{dir}, Wheeler and
Feynman \cite{whe1}, \cite{whe2} and others, leading to the
prevalent view that the classical electromagnetic
coupling of particles interacting at a distance should in
fact be described by time-symmetrical potentials. Radiation
damping is explained by the time-asymmetrical
statistical properties of a distant perfect absorber
\cite{whe1}.  Alternatively, one can invoke
the time asymmetry of the large-scale cosmological
properties of an absorbing universe \cite{hog}.

We shall similarly describe deterministic interactions
between particles by time-symmetrical potentials,
interpreting irreversibility as a statistical phenomenon
(the
cosmological interpretation is not at our disposal, since we
have
limited ourselves to a locally flat sub-region of the
universe
 -- although it could be argued that local statistical
time-asymmetry can be justified ultimately only by
cosmology \cite{ft2}. However, in contrast to classical
theories
for point particles, we are not forced to take an axiomatic
stance on this question.
The only basic equations of the present theory are the
(time symmetrical) set of equations (\ref{1.1}). As
solutions
of these equations we can in principle admit particle states
which have either time-symmetrical or time-asymmetrical
far fields. It will be shown that the time-symmetrical
particle solutions are closed in the sense that they
conserve
4-momentum within a finite set of interacting particles,
while the time-asymmetrical solutions are open, losing
(or gaining) 4-momentum through radiation to (or from)
space. It will be shown further, following the arguments of
Wheeler and Feynman \cite{whe1}, that the open solutions of
a
finite set of interacting particles correspond to the closed
solutions of an extended system including a distant ensemble
of
perfectly absorbing particles. Thus our option of
describing particle coupling always in terms of
time-symmetrical closed interactions,  introducing an
additional perfect absorber if required, is a matter of
conceptual convenience rather than necessity. Depending
on the system, interactions between particles can be
described
either in closed or open form. It will be argued that for
the EPR
experiment the closed rather than the open interaction
description
is appropriate.

\section{Specific properties of the metron model}
\label{Specific properties of the metron model}

Starting from the basic assumption of the existence of
trapped
modes of the n-dimensional gravitational equations, we
develop
in the following the framework of a unified, deterministic,
time-symmetric theory of fields and particles which is
characterized by the following properties \cite{ft3}.

\begin {enumerate}
\item  All  fields  and   particles   are   derived from
the matter-free full-space gravitational equations
(\ref{1.1}) without introduction of additional fermion,
boson or other mixed fields (in contrast to most modern
higher-dimensional gravity theories \cite{duf}).  Bosons and
fermions are identified with particular components of the
full-space gravitational metric. The theory is thus a pure
higher-dimensional extension of Kaluza-Klein theory
\cite{kal}.
\item The theory is developed by expansion of (\ref{1.1})
about a flat space background metric $\eta_{LM}=
\mbox{diag}\
(1,1,1,-1,\cdots ,\pm 1, \cdots)$.  Thus the
harmonic space is not
compactified. For much of
the analysis concerned with the Maxwell-Dirac-Einstein
system, the space dimension $n\,(>4)$ and the structure of
$\eta_{LM}$ in harmonic space need not be
specified. However, in order to represent fermion fields
in accordance with the standard Dirac Lagrangian, the
dimension of harmonic space must be at least four. The
Maxwell-Dirac-Einstein Lagrangian can then be recovered for
a background harmonic-space metric of  suitable signature.
The principal features of electroweak and
strong interactions, as summarized by the Standard Model,
can also be obtained with a minimal four-dimensional
harmonic-space representation, but a closer
correspondence can be established if an additional dimension
is introduced (see item~\ref{item SM} below).
\item  The  theory  contains  no  universal   physical
constants or particle parameters. The only information on
the structure of the physical world introduced at the
axiomatic level is in the form of the gravitational
equations (which in the matter-free form (\ref{1.1})
contain no physical constants), in the dimension of
space, and in the signs of the normalized background
metric. The normalization of the background metric defines
the length scales of physical space, time and the relative
length scales of harmonic space. Once the reference length
scale
has been specified, for example in terms of the length
scale of some reference particle, all other particle
length and time scales, masses, spins and magnetic moments,
Planck's constant, the elementary charge, the
gravitational constant and other coupling constants
are determined by the intrinsic geometrical properties of
the metron solutions \cite{ft4}. The metron theory must
therefore be able to explain, among other properties,
the force hierarchy, including, in particular, the
extremely small relative magnitude of
gravitational forces.  Although detailed metron solutions
will not be presented in this paper, it will be shown that
gravitational coupling is indeed an exceptionally weak
higher-order nonlinear property of the metron solutions.
\item The only fundamental  symmetry  of   the   theory
is the invariance with respect to regular coordinate
transformations (diffeomorphisms).  All other symmetries,
such as the gauge symmetries of the Standard Model,
follow from this very general gauge symmetry and the
internal geometrical symmetries of the metron solutions.
Thus, in contrast to the standard quantum-theoretical
approach, specific symmetries are not introduced into the
basic field equations, but are derived from the specific
geometrical properties of the solutions of the field
equations.
The geometry of the metron solution for a given particle
defines a canonical local coordinate system in harmonic
space at the location of the particle. This is the
coordinate system for which the harmonic wavenumber vectors
associated with the various periodicities of the metron
solution are oriented in specific harmonic-space directions
assigned to the
individual forces. The gauge symmetries express the
property that these local vielbeins can in general be
functions of physical spacetime. As in the special case of
classical gravitation, the connections describing the
variations of the vielbeins in physical spacetime determine
the forces between particles.
\item  Consistent with the general philosophy of attributing
specific symmetries to the solutions of the field equations
rather than the field equations themselves, parity violation
is
explained as a spatial-reflection asymmetry of certain
sub-components of the metron solution, not as a reflection
asymmetry of the weak-interaction sector of the basic
Lagrangian.
The phenomenon of parity violation is thus removed from
the fundamental level of the field equations and -- just as
circularly polarized light
or left-handed molecules -- is not in conflict with our
intuitive expectation that physics should be invariant with
respect to spatial reflections \cite{ft5}. Although not
discussed, the phenomenon of
CP violation in kaon decay can be similarly interpreted as a
symmetry-breaking property of the metron solutions rather
than of the basic Lagrangian.
\item All metron particles  have finite mass. This is shown
to be
proportional to the metron rest-frame frequency in
accordance
with de Broglie's relation. Finite-mass particles support
periodic (de Broglie) far fields. These are
the origin of the wave-like interference properties of
microphysical phenomena. The classical view that periodic
far fields necessarily lead to irreversible radiative
damping is invalid on the microphysical scale, where time
symmetry prevails, the de Broglie far fields representing
undamped trapped standing waves.
\item All far fields  originate  in  individual  metrons,
or are generated by nonlinear interactions between fields
in the vicinity of metrons.  Free radiation fields without
an associated metron source do not occur.  The fields of
individual particles are time-symmetrical (no net ingoing or
outgoing radiation). Outgoing radiation fields, as mentioned
above, are explained by interactions with a non-time-
symmetrical
statistical ensemble of absorbing particles.
\item Zero-mass particles (photons, neutrinos -
assuming their rest mass is indeed zero)
are not regarded as particles in the metron picture but
as far fields in the classical sense.  They derive their
particle-like properties
from the discrete transitions between discrete particle
states which they mediate \cite{ft6}.
\item  The distinction between Einstein-Bose and
Fermi-Dirac statistics for elementary particles -- which
plays a fundamental role in quantum field theory, where it
is
founded on the different commutation/anti-commutation
relations for bosons  and fermions -- follows in the
metron model simply from the distinction between finite-mass
fermion particles, which as real particles cannot be
superimposed locally and therefore automatically comply
with Pauli's exclusion principle, and massless boson
fields, which, as classical fields, can be superimposed
without restriction. Exceptions from these categories,
however, are the neutrino, which is a fermion but,
according to the metron model, a field, and should
therefore not underly the exclusion principle,
and the finite-mass electroweak bosons, which cannot
be superimposed. The implications of these exceptions
need to be explored further.

\item The theory, if meaningful, should encounter  no
divergence problems: the full set of all nonlinear
interactions should yield finite, singularity-free
particle states.  The historical basis for this anticipation
is not encouraging. Most nonlinear-interaction Lagrangians
which
have been considered in elementary particle physics
have led to divergence problems, many of which could not
be `repaired' by renormalization methods.  The nonlinear
gravitational equations in four-dimensional spacetime
(with mass source terms) are also prone to generate fields
which are not globally regular (e.g. the Schwarzschild
solution) - although it is encouraging that Christodoulou
and Klainerman \cite{chr} have recently shown that Minkowski
space is at least stable to small perturbations.
\item In contrast to quantum  field  theory, which  is
essentially a theory of fields, the metron model has both
a field content, represented by the particle far
fields, and a genuine particle content, represented by the
strongly nonlinear core regions of the fields. This is
the principal  difference between quantum
field theory and the metron model. Quantum field theory
`resolves' the paradox of wave-particle duality by in effect
ignoring corpuscular properties in the basic dynamical
field equations, which are pure wave equations. The
connection to particles is established subsequently
through an appropriate statistical formalism. Since
particles do not appear explicitly in the theory, the
concept of a `particle' is not defined.
In the metron model, on the other hand,  both particles and
fields are well defined `objects' which can be identified
with particular features of the solutions of the basic
field equations. The particle content of the metron model
yields the particle constants, coupling
coefficients and (dimensionless) physical constants. The
field
content is formally equivalent (to lowest interaction
order at the tree level) to the quantum field equations
and therefore reproduces most of the basic results of
quantum field theory. None the less, interactions between
the corpuscular features (contained in the core regions)
and field properties (represented by the far-field regions)
of the metron model can be expected to yield different
results from standard quantum field theory at higher order,
for example in the computation of scattering and interaction
cross-sections and branching ratios.  These could provide a
critical test of the theory (in addition, of course, to the
derivation of the particle properties and dimensionless
physical
constants from the metron solutions).
\item In the metron picture it is meaningful to consider
conceptually the simultaneous position and momentum of an
objectively existing metron particle. Nevertheless,
Heisenberg's uncertainty principle is satisfied in the
sense that metron particles are of finite extent and
support de Broglie fields whose wavenumber widths
and spatial extent are in accordance with the Heisenberg
relation. Moreover, it is in general not possible to devise
an experiment in which an initial statistical distribution
of metron particles with a joint momentum-position
probabilty distribution satisfying the Heisenberg
inequality relation is modified in such a way
that the Heisenberg inequality relation is subsequently
violated. Thus although the Heisenberg uncertainty
principle applies formally only to the field content
of the metron model, the traditional explanation of
the uncertainty principle in classical particle terminology
applies also to the metron model: it is not possible to
accurately measure conjugate particle properties because
of the interaction of the measurement device with the object
being measured.
\item \label{item SM} The principal features  of  the
Standard  Model can
be reproduced assuming a suitable geometrical structure of
the metron solutions and a  four-dimensional  harmonic
space with background metric of suitable signature. A
closer correspondence  can be achieved, however, if an
additional dimension is introduced.  Nevertheless, despite
the close structural similarity of the metron model with the
Standard Model, small differences exist. In particular,
the Standard Model interactions represent only a sub-set of
all possible interactions in the gravitational system.  Thus
from the metron viewpoint the Standard Model appears only
as a first approximation of the fully nonlinear system.
\end{enumerate}

A theory with these properties must clearly rest
ultimately on the demonstration that the n-dimensional
gravitational field equations do indeed support stable,
self-trapping wave-guide type soliton solutions.
The metron solutions must furthermore reproduce all known
elementary particles and their interaction cross-sections
and yield all physical constants.\\

It is also clear that the development of such a complete
theory, involving the numerical solution of the highly
complex nonlinear n-dimensional gravitational equations, is
not
a minor undertaking. Before embarking seriously on this
task,
it therefore appears
appropriate to consider first a number of general
implications of the proposed alternative view of
microphysical
phenomena.  This is the principal purpose of this first
four-part
paper. In short,  we focus here on the feasibility of
developing
a model with the properties listed
above rather than on the detailed structure of the model
itself. Perhaps it would therefore be more appropriate to
speak of
the metron program rather than the metron model.
Nevertheless,
in the process of analyzing the basic concepts of the metron
model, it will be found that most of the basic properties
listed above can indeed be explicitly derived, although some
of the stated features of the model must necessarily remain
speculative at this stage.

\section{Development and implications of the metron
concept}
\label{Development and implications of the metron concept}

The principal differences between the metron and quantum
field
theoretical view of microphysics are summarized
in Table \ref{ta1.1}. The four parts of the paper are
structured in accordance
with the phenomena listed in the table, the last
column of the table indicating the sections  in which the
various concepts are discussed.
\renewcommand{\baselinestretch}{1.2}
\small\normalsize
\begin{table}
\begin{tabular}{|l|l|l|l|}
\hline
{\em Phenomenon} & {\em QFT } & {\em Metron model}&
{\em Sections}  \\
\hline
particles & \parbox{4.2cm}{defined statistically} &
\parbox{4.2cm}{\ruo trapped mode solutions of field
equations \ruu} &
\ref{The mode-trapping mechanism}, \ref{Particle
interactions} \\
\hline
fields & \parbox{4.2cm}{defined statistically in conjunction
with particles by system state} &
\parbox{4.2cm}{\ruo form nonlinear particle core,
experienced as
far-fields \ruu} & \ref{The Maxwell-Dirac-Einstein
Lagrangian}, \ref{The Standard Model}
 \\
\hline
Lagrangians & \parbox{4.2cm}{derived from postulated
gauge symmetries} & \parbox{4.2cm}{\ruo inferred from
n-dimen\-sional
gravitational Lagrangian \ruu} & \ref{The
Maxwell-Dirac-Einstein
Lagrangian}, \ref{The Standard Model}
\\
\hline
\parbox{2.4cm}{physical constants} & postulated &
\parbox{4.2cm}{\ruo derived from metron solutions with
postulated
periodicities \ruu} & \ref{Particle interactions}, \ref{The
Standard Model}  \\
\hline
\parbox{2.4cm}{Bell's theorem} &
\multicolumn{2}{c|}{\parbox{8.4cm}{violates time symmetry
of both theories, not applicable to reversible microphysical
phenomena}} & \ref{The EPR paradox and Bell's theorem}  \\
\hline
\parbox{2.4cm}{wave-particle duality} &
\parbox{4.2cm}{statistical interpretation; non-existence of
`objective' fields and particles} &
\parbox{4.2cm}{\ruo explained by periodic de Broglie far
fields
of `objective'
particles \ruu} & \ref{Bragg scattering}, \ref{Atomic
spectra}
  \\
\hline
\parbox{2.4cm}{atomic spectra} &
\parbox{4.2cm}{eigensolutions~of Maxwell\--Dirac system} &
\parbox{4.2cm}{\ruo same as QED
at
lowest order augmented by Bohr-orbiting electrons \ruu} &
\ref{Atomic spectra}
  \\
\hline
\parbox{2.4cm}{absorption and emission} &
\parbox{4.2cm}{secular~ (resonant) per\-tur\-ba\-tions of
system
state} &
\parbox{4.2cm}{\ruo similar formalism for classical fields
\ruu} &
\ref{Atomic spectra}
\\
\hline
divergences & renormalization & ?
(should not
arise) & --  \\
\hline
\parbox{2.4cm}{Standard Model} &
\parbox{4.2cm}{summarizes~particle spec\-trum, 19 empirical
parameters} &
\parbox{4.2cm}{\ruo general structure reproduced for given
symmetries
of metron solutions; parameters determined by solutions
\ruu} &
\ref{The Standard Model}
\\
\hline
\parbox{2.4cm}{gauge symmetries} & postulated &
\parbox{4.2cm}{\ruo inferred from geometrical symmetries of
metron solutions
and invariance with respect to coordinate transformations
\ruu} &
\ref{The Maxwell-Dirac-Einstein Lagrangian}, \ref{Invariance
properties} \\
\hline
\parbox{2.4cm}{particle interactions} & \parbox{4.2cm}
{S-matrix
formalism} & \parbox{4.2cm}{\ruo not
~discussed,
similarity to S-matrix formalism anticipated from analogy
with
optical
absorption and emission \ruu} & \ref{Atomic spectra}
 \\
\hline
\end{tabular}
\caption[ix] {Relation between metron and quantum field
theoretical picture of microphysical phenomena}
\label{ta1.1}
\end{table}
\renewcommand{\baselinestretch}{1.0}
\small \normalsize

Before commencing with a more detailed analysis of the
metron concept in Parts
 \ref{The Maxwell-Dirac-Einstein System}-\ref{The Standard
Model}, we first investigate in the remaining sections of
Part
\ref{The Metron Concept} whether the basic premise of the
theory, namely that the nonlinear gravitational field
equations
in a higher-dimensional space can support self-trapping
wave-guide modes, appears reasonable. It is demonstrated
that trapped-mode solutions do indeed exist for nonlinear
Lagrangians in  n-dimensional space. Solutions are computed
to lowest interaction order for a prototype
nonlinear Lagrangian whose general structure follows from
the gravitational Lagrangian by projection of the fields
onto a reduced set of modes. Depending on the form of the
coupling, the wave-guide can support trapped wave modes
which fall off exponentially within a short distance
outside the core (representing a model for quark and
gluon fields or weak-interaction bosons) or far fields
which decrease
aymptotically as $1/r$ (gravitational and electromagnetic
fields) or at a very weak exponential rate (de Broglie
fields).

Not resolved is the problem of the discreteness of the
particle spectrum.  The trapped-mode solutions found for
the simplified Lagrangian generally represent a continuum.
Additional considerations, such as stability, need to
be invoked to reduce the
solutions to a discrete set. We regard this as the major
open problem of the metron approach at this point.
Included in the question of discreteness is the problem
of uniqueness. It must be shown that different particle
states at different locations are not only discrete but
also identical. A trivial continuum of solutions always
exists because of the invariance of eq.~(\ref{1.1}) with
respect to an arbitrary common change of the coordinate
scales (without changing the fields -- this  follows from
the homogeneity of the field equations with respect to the
derivatives and is independent of the invariance with
respect to diffeomorphisms). It must therefore be shown that
all solutions exhibit the same
spatial scaling (which can then be used to define a
universal unit of length). This requires some form of
multi-particle interaction leading, presumably, to some
collective  stability criterion.

An alternative philosophy is to simply postulate (in analogy
with string theory) that all solutions of the n-dimensional
gravity equations in our world are periodic, with different
but universal periodicities represented by different
harmonic
wavenumber vectors. The wavenumber components define the
coupling coefficients of the electroweak and strong forces.
The coupling coefficients  can then no longer be regarded as
derived quantities of the metron model, but appear rather
as empirical universal constants (gravitational forces,
however, will still be derived as higher order nonlinear
metron properties). Which of the two views is more
appropriate
must await more detailed stability investigations
(cf.\cite{kun}).

In Part~\ref{The Maxwell-Dirac-Einstein System} the metron
picture of the Maxwell-Dirac-Einstein system is developed.
Assuming that trapped-mode solutions of the
n-dimensioal gravitational equations exist, and that they
are indeed discrete and unique, we address first the
question
whether it is possible to derive the basic boson spin-one
and fermion half-odd-integer-spin fields of standard quantum
field theory from the tensor fields of the gravitational
metric.
In most higher-dimensional gravity theories, boson and
fermion
fields (as well as a large number
of auxiliary mixed fields) are simply introduced as
additional fields. In Sections \ref{Introduction 2},
\ref{Lagrangians}
it is shown that for a metron solution composed
of fields which are periodic in harmonic space, the familiar
free-field equations for bosons and fermions can  be
extracted directly from
the gravitational field equations. Assuming a suitable
background harmonic-space metric with dimension of at least
four and a periodicity of the fermion fields characterized
by a single harmonic wavenumber vector k
$=(k_5,0,0,\cdots)$,
say, the standard fermion-electromagnetic
interaction Lagrangian for these fields is then derived
using simple covariance arguments; an analogous form follows
for the fermion-gravitational interaction Lagrangian.
The $U(1)$ gauge invariance of the Maxwell-Dirac-Einstein
system
is derived from the invariance of the metron
solutions with respect to arbitrary spacetime-dependent
translations
in the $x^5$-direction.

Progressing from the standard interaction Lagrangians for
weak field-field interactions derived in
Sections~\ref{Lagrangians},
\ref{The Maxwell-Dirac-Einstein Lagrangian},
Section~\ref{Particle interactions} considers
the coupling between particles. This is described by the
interactions of the  fields in the nonlinear particle-core
regions with the far fields of other particles.  The
classical
Tetrode-Wheeler-Feynman description of point-particle
interactions at a distance for electromagnetic and, by
extension,
gravitational interactions is recovered. In the process, the
analysis
yields expressions for the particle mass and charge, the
gravitational
constant, Planck's constant and de Broglie's relation.
Similarly, all
particle properties and physical constants are derived
as functions of the metron solution. The exceedingly small
ratio of gravitational to electromagnetic forces is
explained by
the metron geometry: coupling through the gravitational mass
is
found to be a higher-order nonlinear process than the
coupling through
the electromagnetic charge.

The analysis of electromagnetic interactions in
Part~\ref{The Maxwell-Dirac-Einstein System} can be
generalized to weak and strong interactions by considering
fermion fields with periodicities characterized by
wavenumber vectors oriented in other directions than the
electromagnetic direction k $=(k_5, 0,0,\cdots)$. However,
before extending the analysis to the metron
interpretation of the Standard Model in Part~\ref{The
Standard Model}, we address first in Part~\ref{Quantum
Phenomena}
some of the conceptual questions raised by quantum theory,
together with
the basic wave-particle duality paradoxes of
microphysics which originally
lead to the formulation of the theory. These must be
resolved now from the alternative viewpoint of the metron
model. Since the problems involve only atomic-scale
phenomena and are independent of the weak and strong
interactions operating on nuclear scales, they can be
addressed already using only the metron picture of the
Maxwell-Dirac-Einstein system developed in Part~\ref{The
Maxwell-Dirac-Einstein System}.

We first consider the interrelated questions of time-
reversal symmetry (Section \ref{Time-reversal symmetry}),
forward causality, the origin of
the arrow of time (Section~\ref{The radiation condition}),
the Einstein-Podolsky-Rosen paradox and Bell's theorem
(Section \ref{The EPR paradox and Bell's theorem}).
It is shown that conservation of 4-momentum within a
finite set of interacting particles requires
 a time-symmetrical representation of the
particle far fields. Following Einstein \cite{ein1} and
Wheeler and Feynman \cite{whe1}, the empirical finding of
time-asymmetrical outgoing radiation  is explained by
the interaction of the radiating particle with an
infinite distant particle ensemble. This acts as a perfect
absorber for the retarded potential of the particle and
cancels the advanced field of the particle.
The time-asymmetry of the absorber interaction
(which was not explained in detail by Wheeler
and Feynman) is attributed to classical Boltzmann-Gibbs-type
irreversible interactions within a random ensemble of
particles.
Noting that the  distant absorber plays no role in the EPR
experiment and that the forward causality assumption of
Bell's theorem is therefore not satisfied by the time-
symmetrical
metron model, the EPR experiment can then be readily
interpreted
in the metron picture.

The remaining  sections of Part~\ref{Quantum Phenomena}
address the problem of wave-particle duality. The resolution
of the wave-particle duality conflict in the
metron picture is illustrated by two examples: the Bragg
scattering of a particle beam at a periodic lattice
(Section \ref{Bragg scattering}) and atomic spectra
(Section \ref{Atomic spectra}).  In both cases the
corpuscular
phenomena follow from the existence of a particle core,
while interference and other wave-like phenomena are
explained
by the periodic de Broglie far fields of the particles.

The fact that in the case of Bragg scattering the
far-field interference patterns impress their signature
also on the particle
fluxes is explained by resonant interactions between the
scattered far
fields and the oscillating particle cores.
Wave-trajectory
resonance leads to the capture of the scattered particles
in a set of discrete trajectories corresponding to
the Bragg resonance scattering directions.

Resonant interactions between scattered waves and particle
trajectories explain also the existence of discrete atomic
states. The scattered waves are generated in this case by
interactions of the de Broglie far field of the orbiting
electron with the nucleus. The scattered-wave equations are
identical to the standard coupled
Maxwell-Dirac field equations, but contain also a forcing
term representing the interaction of the orbiting electron
with the nucleus.  For a discrete set of orbits
for which the forcing frequency of the orbiting electron
is equal to the frequency of an eigenmode of the
Dirac-electromagnetic equations,
resonance occurs. The resonant interaction between the
orbiting electron and the Dirac eigenmode results in a
trapping of the electron in the resonant orbit. Associated
with the trapping is an interaction current which balances
the radiative damping of the orbiting electron.
For the simplest case of a circular orbit, it can be shown
that the trapping condition is identical to the Bohr orbital
quantum conditions. The metron model thus yields an
interesting amalgam of quantum electrodynamics (at the
tree level) with the original Bohr orbital theory.

In Part~\ref{The Standard Model}, finally, the analysis is
extended to include weak and strong interactions.
In order to recover the $U(1)\times SU(2)\times SU(3)$
symmetry of the Standard Model, specific properties of
the metron solutions and the harmonic-space background
metric must be invoked. The harmonic-space background metric
must be at least four-dimensional, but can have various
signatures.
However,
as pointed out, a closer correspondence between the metron
and Standard Model can be achieved if an additional
dimension is introduced, and we shall accordingly assume as
prototype harmonic metric $\eta_{AB} = \mbox{diag}\
(1,1,1,1,-1)$ or $\mbox{diag}\
(1,1,1,1,1)$.  The first two harmonic-space
dimensions define the electroweak interaction plane,
periodicities with respect to the first and second
dimensions being associated with the electromagnetic forces
and weak interactions, respectively. Periodicities with
respect to the third and fourth dimension (the `color'
plane)
define the strong-interactions, while the fifth
harmonic dimension is needed, together with the other
harmonic dimensions, to establish appropriate polarization
relations between the tensor components of the metric field
and the spinor components of the fermion fields in
accordance with the Maxwell-Dirac-Einstein Lagrangian. For a
suitable wavenumber configuration, the metron
solutions can be shown to reproduce the principal properties
of the Standard Model, although differences remain in the
details of the coupling. The Higgs mechanism is explained as
a
higher-order interaction, but is invoked only to explain the
boson masses, the fermion masses being attributed to the
mode-trapping mechanism. The gauge symmetries of the
Standard Model are explained by the invariance of
the metron model with respect to a  class of
coordinate transformations in which the local
harmonic vielbeins defined by the orientations of the
harmonic wavenumber vectors are varied as functions of
spacetime.

The general correspondence between the metron model and the
Standard Model is established by considering in the metron
model only the boson fields generated by the sub-set of
quadratic difference interactions between pairs of fermion
fields. Quadratic sum interactions and higher-order
interactions are excluded.  From the metron viewpoint, the
Standard Model appears therefore only as a truncated first
approximation of the full nonlinear
n-dimensional gravitational system.

The paper is summarized, finally, in Section \ref{Summary
and conclusions}. We conclude that, although the
existence of a discrete, unique set of metron solutions
of the n-dimensional gravitational equations (\ref{1.1}) has
yet to be demonstrated, the general properties of metron
solutions, if they do indeed exist, appear to capture most
of the salient features of elementary particle
and atomic physics.  The correspondence between quantum
field and metron theory is attributed primarily to the
wave-like properties of the metron solutions. The field
content of the metron model yields naturally the statistical
properties of microphysical phenomena, which are recovered
also by a quantum theoretical description.  The corpuscular
metron features, on the other hand, which are essential for
a
deterministic description of individual particle
interactions, have no counterpart in the quantum field
picture, which for this reason is in principle incapable of
describing individual microphysical events.  The
deterministic
description of the  strongly nonlinear interior core
region of particles in  the metron model also yields all
particle properties, coupling constants and universal
physical
constants as functions of the metron solutions.

A quantitative test of the predictions of the metron
model must await numerical computations of specific metron
solutions.  The purpose of this first analysis was not to
compute numbers, but rather to present an alternative view
of microphysical phenomena which appears able, in principle,
to overcome the conceptual difficulties of standard quantum
field theory while at the same time offering a framework for
a unified theory. It is hoped that the general picture which
has emerged, together with the identification of the
principal
properties of metron solutions needed to explain the
Standard
Model, will motivate attempts to carry out such
computations.
%

\typeout{################################}
\typeout{################################}
\typeout{        START OF met4-2.tex}
\typeout{################################}
\typeout{################################}

\section{The mode-trapping mechanism}
\label{The mode-trapping mechanism}

\subsection*{Metron partons}

The basic premise of the metron model is that the
higher-dimensional matter-free nonlinear gravitational
equations
support trapped wave-guide mode solutions.  Before
preceding further with the implications of the model,
we therefore first investigate this assumption.
Although we shall not attempt to construct explicit
metron solutions of the full gravitational equations in this
paper, the basic nonlinear mode-trapping mechanism can be
illustrated for a simplified nonlinear Lagrangian of the
same structure as the gravitational Lagrangian.
The simplified Lagrangian can be regarded as derived
from the gravitational Lagrangian by projecting the metric
field onto the modes of the metron solution. Anticipating a
few general properties of the metron solutions, one obtains
in this way a Lagrangian which retains the basic nonlinear
interaction structure of the gravitational Lagrangian while
omitting its detailed tensor complexities.

We assume that in a suitably defined coordinate system in a
small region of the universe (e.g. our galaxy), the metric
field $g_{LM}$ of a metron solution can be represented
 as a superposition
\begin{equation} \label{2.1}
g_{LM} = \eta_{LM}   + \sum_{p} g_{L
M}^{(p)}
\end{equation}
of periodic `parton' fields
\begin{equation} \label{2.2}
 g_{LM}^{(p)} := \hat g_{LM}^{(p)}(x) \exp
(iS^p)
+ \mbox{compl. conj.},
\end{equation}
where the phase functions
\begin{equation} \label{2.3}
S^{p} : = k^{(p)}_{A} x^{A}
\end{equation}
have constant harmonic wavenumber vectors $k^{(p)}_{A}$ and
the amplitudes $\hat g_{LM}^{(p)}(x)$ are functions of
physical
spacetime $x$ only. The index and coordinate notation used
here and in the following is defined in Table \ref{ta2.1}.
Non-tensor indices, which are excluded from the
summation convention, are placed in parentheses when
occurring together with tensor indices.

\pagebreak[1]

\begin{table}
\begin{tabular}{|lcl|}
\hline
\multicolumn{1}{|c}{space}  &  components  &
\multicolumn{1}{c|}{vector}\\
\hline
full n-dimensional space     & $x^{L}$        & $X =
(x^1 ,x^2
,\cdots, x^n )$\\
three dimensional physical space & $x^i$    & ${\bf x} =
(x^1 ,x^2 ,x^3)$\\
four dimensional physical spacetime & $x^{\lambda}$& $ x  =
(x^1
,x^2 ,x^3 ,x^4)$\\
$(n - 4)$ dimensional harmonic space & $x^A$ &
x = $(x^5 ,x^6 , \cdots,x^n)$ \\
\hline
\end{tabular}
\caption[ix] {Index and coordinate notation}
\label{ta2.1}
\end{table}

For small perturbations, $|g_{LM}^{(p)}|\ll 1$, the parton
components satisfy the linearized higher-dimensional
gravitational equations \cite{pau}
\begin{equation} \label{2.4}
\partial _{N} \partial ^{N} g_{LM}^{(p)}
= 0,
\end{equation}
or, in terms of the parton amplitudes, the Klein-Gordon
equations
\begin{equation} \label{2.5}
\left( \Box - \hat \omega^{2}_{p} \right) \hat
g^{(p)}_{LM} = 0,
\end{equation}
where
\begin{equation} \label{2.6}
\hat \omega^{2}_{p} : = k_{A}^{(p)} k^{A}_{(p)}.
\end{equation}

Tensor indices are raised or lowered in (\ref{2.4}),
(\ref{2.6}) and in the following using the full metric
$g_{LM}$ and its inverse $g^{LM}$, but to lowest order
the full metric can be replaced by the background metric
$\eta_{LM}$ when applied to perturbation fields (the
non-tensor index $(p)$ is shifted at will for notational
convenience). It should be noted that the perturbations
of the contravariant metric tensor take an opposite sign,
the relation (\ref{2.1}) becoming
\begin{equation} \label{2.9}
g^{LM} = \eta^{LM} - \sum_{p} g^{L
M}_{(p)} + \cdots
\end{equation}
In the following sections, the parton metric fields
$\hat g^{(p)}_{LM} \exp(iS^{p})$ will be identified
with standard boson and fermion fields, the parton
amplitudes $\hat g^{(p)}_{LM}(x)$ being represented
in the general form
\begin{equation} \label{2.10}
\hat g^{(p)}_{LM} = P^{(p)}_{LM}
\varphi_{p},
\end{equation}
where the first factor $P^{(p)}_{LM}$ represents a
constant
polarization
tensor and the second factor $\varphi_{p} =
\varphi_{p} (x)$
a mode amplitude function which satisfies the
Klein-Gordon
equation (\ref{2.5}) to lowest (linear) approximation.

\subsection*{A prototype Lagrangian}

The extension to the general nonlinear case is obtained by
substituting the expression (\ref{2.10}) into the
gravitational Lagrangian 
(cf. Part \ref{The Maxwell-Dirac-Einstein System})
and averaging over harmonic space.
For suitably normalized $\varphi_p$, one obtains then
a Lagrangian of the general form
\begin{eqnarray}
L ( \cdots ,\varphi_{p},  \cdots  )   &  = & - \frac{1}{2}
\left[ \frac{1}{2}  \sum_{p}  \sigma_{p}  \left[
\partial_{\lambda}
\varphi_{p}    \partial^{\lambda}     \varphi_{-p} +
\hat
\omega^{2}_{p}  \varphi_{p}  \varphi_{-p}  \right] \right.
\nonumber \\
 & + & \left. \frac{1}{3}
\sum_{p,q,r}  K_{pqr}  \varphi_{p}   \varphi_{q}
\varphi_{r}   +
\frac{1}{4}  \sum_{p,q,r,s}   K_{pqrs}   \varphi_{p}
\varphi_{q}
\varphi_{r} \varphi_{s} + \cdots \right],
\label{2.11}
\end{eqnarray}
where $\sigma_{p} = \pm 1$ and $K_{pqr},\cdots$ denote
(complex) coupling coefficients. The first term in the sum
represents the Lagrangian associated with the linear
Klein-Gordon equation, while the remaining terms represent
the interactions, expanded in powers of
the mode amplitudes.  Negative indices have been introduced
to denote the complex conjugate terms $\varphi_{-p} =
\varphi_{p}^*$, with $k_{A}^{(-p)} = -
k_{A}^{(p)}$, the summations extending over both index
signs \cite{ft8}.

The coupling coefficients are symmetrical in the indices,
satisfy the reality conditions $K_{pqr} = K^{\ast}_{-p-q-
r},
\cdots$ and (since the gravitational Lagrangian is
homogeneous
of second degree in the derivatives) are quadratic in the
wavenumber components (for simplicity, physical
spacetime derivatives $\partial_{\lambda} \varphi_{p}$ are
neglected in the coupling terms). Note that in contrast to
the standard procedure in quantum
field theory, the coupling coefficients are not postulated
{\em a priori\/}, but follow from the basic gravitational
Lagrangian and the assumed structure (\ref{2.10}) of the
parton
solution.

The diagonal form assumed for the quadratic free-field
Lagrangian implies that different partons have different
wavenumbers. In later specific applications to the
gravitational Lagrangian, the sign $\sigma_{p} $ will
generally be positive, but it
can in principle also be negative for a non-Euclidean
background harmonic-space metric.

The averaging of the Lagrangian over harmonic space implies
that the coupling coefficients vanish unless the sum of the
interacting wavenumbers is zero,
\begin{equation} \label{2.12}
K_{pq \cdots s} = 0 \quad  \mbox{ if} \quad k^{p}_{A}
+ k^{q}_{A}+ \cdots
+ k^{s}_{A}  \neq  0.
\end{equation}

Variation of the Lagrangian with respect to $\varphi_{-p}$
yields the
coupled field equations
\begin{equation} \label{2.13}
\sigma_{p} \left( \Box^{2}  -   \hat   \omega^{2}_{p}
\right)
\varphi_{p}  = \sum_{q,r}  K_{qr-p}  \varphi_{q}
\varphi_{r}   +
\sum_{q,r,s}   K_{qrs-p} \varphi_{q}\varphi_{r}\varphi_{s} +
\cdots.
\end{equation}
The Lagrangian (\ref{2.11}) and field equations (\ref{2.13})
are equivalent to the original gravitational Lagrangian and
field equations (\ref{1.1}) if the set of all parton fields
$p$ is complete, i.e.  if an arbitrary tensor amplitude
function $
\hat g^{(p)}_{LM} $ of an arbitrary periodic metric
field can be represented in the form (\ref{2.10}).
The field equations (\ref{2.13}) represent in this case a
transformation of the
original field equations from the tensor components $\hat
g^{(p)}_{LM}$ to the alternative set of base
functions
$\varphi_{p}$.  In practice, however, the parton fields
 will not form a complete set.  In
fact, an important characterisic of the
metron
solutions considered later in Parts \ref{The
Maxwell-Dirac-Einstein System} and \ref{The Standard Model}
is that the parton constituents consist of only a discrete
set
of Fourier components, and that each individual parton has
special polarization properties involving only a sub-set of
metric tensor components. Thus the field equations
(\ref{2.13})
must be regarded as a strongly truncated version of the
full gravitational field equations: they describe the
interactions only between a particular sub-set of all
possible metric field components, namely those associated
with the partons of the metron solutions.

\subsection*{Special solutions}

The simplest example of mode trapping occurs for the case of
the quadratic interaction between a single mean field
$\varphi_{0}$ and a
single periodic field $\varphi _{1},
\varphi _{-1}$ (with  $\varphi_{-1}   = \varphi_{1}^*$).
The field
equations (\ref{2.13}) reduce in   this   case   to   the
coupled
equations (taking $\sigma_{p}  = 1$)
\begin{equation} \label{2.14}
\left [ \nabla^{2}+ \kappa^{2}\right] \varphi_{1} = 0,
\end{equation}
where
\begin{equation} \label{2.15}
\kappa^{2} : = \omega^{2} - \hat \omega^{2} + \epsilon
\cdot \hat
\omega^{2}\varphi _{0}
\end{equation}
and
\begin{equation} \label{2.16}
\nabla^{2}\varphi _{0} = - \epsilon \hat \omega^{2}  \mid
\varphi_{1}
\mid^{2},
\end{equation}
with the coupling coefficient
\begin{equation} \label{2.17}
\epsilon : = - 2\, K_{10-1}\, \hat \omega^{-2}.
\end{equation}
Equations (\ref{2.14}) - (\ref{2.17}) are seen to have
the right
signature for self-sustained wave trapping, independent of
the sign of
the coupling coefficient $\epsilon$.  For example, for
the case of
spherical symmetry, $\varphi_{0,1} (x) = \varphi_{0,1} (r)$,
with $r =\,
\mid x \mid $, $\varphi_{0}$  has the same sign as
$\epsilon$   for
all $r$ and has a maximum absolute value at $r = 0$. Thus if
$\omega$
is chosen to lie in the interval
\begin{equation} \label{2.18}
\hat  \omega   \left(1   -   \epsilon   \cdot
\varphi_{0}(0)
\right)^{1/2} < \omega < \hat \omega,
\end{equation}
$\kappa^{2}$ will be positive (corresponding to an
oscillatory
behaviour of $\varphi_{1}$ ) in a finite region around $r =
0$ and
negative (corresponding to an exponential fall-off) for
large r,
as required
for a trapped mode.

\begin{figure}[t] \centering
\begin{minipage}{9cm}
\epsfxsize250pt
\epsffile{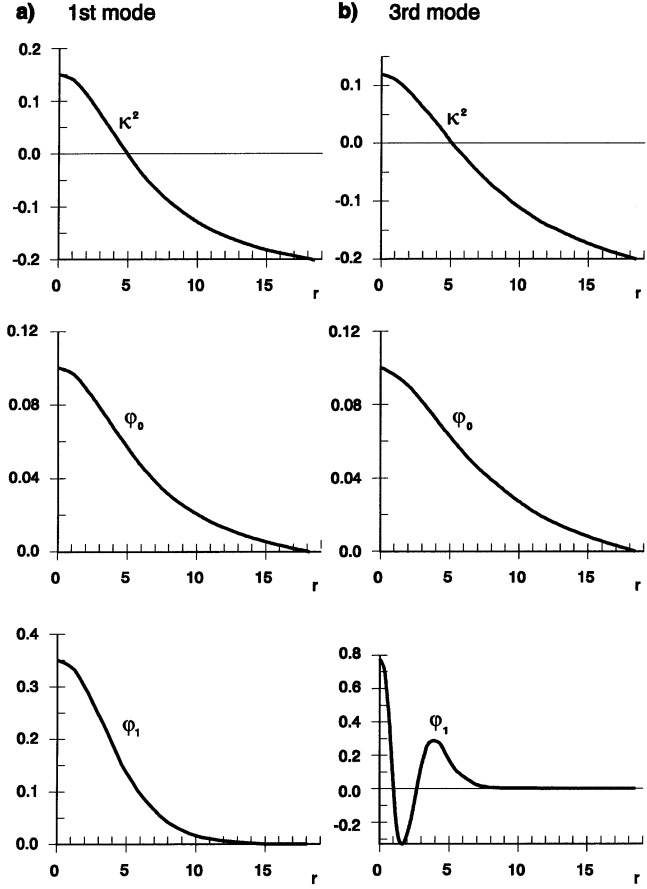}
\caption{\label{Fig.2.1}
Functions $\kappa^2, \varphi_0$ and $\varphi_1$ for
first and third trapped-mode solution of equations
(\protect\ref{2.14})-(\protect\ref{2.17})}
\end{minipage}
\end{figure}

For the spherically symmetric case, mutually consistent
mean-field and
trapped-wave solutions can be constructed for a prescribed
value of
$\hat \omega$ by iteration. Given a mean field
$\varphi^{(n)}_{0}$ at
the n'th iteration level, the associated wave field
$\varphi^{(n)}_{1}$ and eigenvalue $\omega^{(n)}$ for some
given
eigenmode (the lowest, say) is obtained by solving the wave
equation
(\ref{2.14}), (\ref{2.15}).  The mean field
$\varphi^{(n+1)}_{0}$ at
the next iteration level $n + 1$ is then obtained by solving
the
Poisson equation (\ref{2.16}) for given $\varphi^{(n)}_{1}$,
and so
on.  The amplitude of the eigenfunction $\varphi^{(n)}_{1}$,
which is
not determined by the linear equation (\ref{2.14}), can be
fixed by
specifying the physical scale of the wave guide, for example
by
requiring $\kappa^{2}_{(n+1)}$ to cross zero at some given
$r =
r_{0}$. The iteration procedure converges to a unique
solution for a
given eigenmode (cf.~Fig.\ref{Fig.2.1}).

Other solutions with different values $r_{0}' := r_{0}
/\lambda$ of the
zero-crossing point can be obtained by the scale
transformation
\begin{eqnarray}
\label{2.19}
r'                &:=& r/  \lambda \\
\varphi'_{0,1}    &:=&  \lambda^2 \varphi_{0,1}     \\
\omega'^2         &:=&  \hat  \omega^2   -
              \lambda^2  \left (  \hat  \omega^2   -
\omega^2 \right),
\end{eqnarray}
where the values of $\lambda$ are restricted by the
condition
$\omega'^2
\geq 0$ to the interval
\begin{equation} \label{2.20}
0 \leq \lambda \leq \left( 1 - \omega^2 / \hat \omega^2
\right )^{-1/2}.
\end{equation}
Thus for a given eigenmode order there exists a
one-parameter family
of solutions dependent on the nonlinearity scale parameter
$\lambda$.
The upper limit of the interval (\ref{2.20}) corresponds to
the
strongest permissible nonlinearity, yielding a maximum mode
amplitude,
minimum zero-crossing point and zero frequency, while the
linear case,
with $\omega'=\hat \omega$, is recovered in the limit
$\lambda
\rightarrow 0$, or $r'_0 \rightarrow \infty$.

An extension of the analysis to include higher-order
interactions and higher harmonics of the basic field
$\varphi_1$ modifies the solutions and nonlinear dispersion
relation, but does not affect the basic one-parameter
structure of
the
solution.

The model can be readily generalized to an
ensemble of
trapped
modes $\varphi_p, \varphi_q, \varphi _r$, where $p, q,
r,\cdots$ denote
combined indices representing different sub-parton
components and
different trapped-mode orders of a given trapped-mode
branch, and/or a number of different mean fields
$\varphi_a , \varphi_b, \varphi_c, \cdots$.
If it is assumed that there exists no combination of indices
$p,q,r,
\cdots$ for which the coupling conditions
\begin{equation} \label{2.20a}
k^p_{A} +
k^{q}_{A} + k^r_{A} + \cdots = 0
\end{equation}
are satisfied, the modes interact only through the mean
fields,
which they jointly generate.

The coupled set of normal-mode and mean-field equations
(\ref{2.14}) -
(\ref{2.17}) becomes in this case (to lowest quadratic
order, and ignoring again higher harmonics)
\begin{equation} \label{2.21}
\left[ \nabla^2 +  \kappa_p ^2 \right] \; \varphi _p = 0,
\end{equation}
where
\begin{equation} \label{2.22}
 \kappa_p^2 : =  \omega_p ^2 -  \hat  \omega^2_p  +
\sum_a \epsilon_{ap}\, \hat \omega^2_p \, \varphi _a
\end{equation}
and
\begin{equation} \label{2.23}
\nabla^2\varphi_a  =  -  \sum_p  \;  \epsilon_{ap}\,   \hat
\omega^2_p  \left| \varphi_p \right| ^2
\end{equation}
with
\begin{equation} \label{2.24}
\epsilon_{ap} : = - 2 \,\sigma_p K_{pa-p}\, \hat \omega_p^{-
2}.
\end{equation}

The solution can be constructed by iteration in the
same way as in the single-mode case. For $n$ modes, the
solution depends in general on $n$ free parameters
(in addition to the n specified mode
wavenumber vectors $k^p_{A}$ ), which can be related
as before to the mode nonlinearity parameters.

A more appropriate model, however, is one in which
the partons can interact directly, i.e. in which the
resonance condition (\ref{2.20a}) is satisfied for certain
parton sub-sets. This will be discussed in more detail (but
without presenting solutions) in the context of the Standard
Model in Part \ref{The Standard Model}.

\subsection*{Periodic far fields}
\label{Periodic far fields}

The periodic trapped modes in the simplified models
considered above were characterized by  exponentially
decreasing amplitudes for large distances from the particle
kernel.  However, the  metron interpretation of classical
wave-interference effects in particle experiments, discussed
later in Sections \ref{Bragg scattering} and \ref{Atomic
spectra}, depends critically on the assumption that
the metron solution contains also periodic far
fields (de Broglie fields) which extend over distances
large compared with the wavelength of the field and are thus
able to produce resonant interference phenomena. This
requires either that the exponential fall-off is very  weak
or that the fields are asymptotically free, i.e. fall off as
$1/r$ for large~$r$. We discuss both possibilities.
Asymptotically free fields represent a relevant model for
massless fermions (neutrinos), while for finite-mass
particles
a weak exponential fall-off appears to be a more appropriate
description (cf. Parts \ref{The Maxwell-Dirac-Einstein
System} and \ref{The Standard Model}).

\subsubsection*{Asymptotically free fields}

In the above models, asymptotically free wave fields appear
only in the linear limit, in which the amplitudes tend to
zero.  (The mean field, in contrast, always decreases
asymptotically as $1/r$.)  In fact, an asymptotic
finite-amplitude $1/r$ behaviour in the trapped mode would
lead to a divergence in the response of the mean field to
the quadratic current
term in eqs.  (\ref{2.16}), (\ref{2.23}). Although one can
consider a suitable limiting process yielding a finite
mean-field forcing in which the cubic coupling coefficient
approaches zero as the trapped-mode solution approaches the
free-wave limit (cf. Section~\ref{Electroweak
interactions}), one can also obtain finite-amplitude
trapped-mode solutions with asymptotic free-wave properties
more directly by assuming that the lowest-order interaction
term is of higher order than cubic.

Consider, for example, the two-mode, fifth-order Lagrangian
\begin{eqnarray} \label{2.25}
L ( \varphi_0 ,\varphi_1 ,\varphi_2 )& = & -
\frac{1}{4}
 \nabla \varphi_0 \nabla \varphi_0
 \, - \, \frac{1}{2}
 \left\{ \nabla \varphi_{-1}  \nabla
      \varphi_1 +\left( \hat \omega_1^2 - \omega^2_1 \right)
       \varphi_{-1}
      \varphi_1 \right. \\
& &
 +  \left. \nabla \varphi_{-2} \nabla \varphi_2 + (  \hat
      \omega^2_2 - \omega^2_2 )
      \varphi_{-2}\varphi_2
   -  \epsilon_1 \hat \omega^2_1 |\varphi_1|^2 \varphi_0 -
      \eta_2 \hat
      \omega^2_2  |\varphi_2|^4 \varphi_0   \right\}
\nonumber
\end{eqnarray}
with coupling coefficients $\epsilon_1 , \eta_2$.  The
$(\varphi_0,
\varphi_1 )$ interaction sector corresponds to the simplest
model, eqs. (\ref{2.14}) - (\ref{2.16}), discussed above,
while in the $(\varphi_0 ,\varphi_2 )$ interaction sector
the cubic interaction term is replaced now by a
fifth-order term.

The eigenmode equations for $\varphi _1$ are given by eqs.
(\ref{2.14}), (\ref{2.15}), as before, while for $\varphi_2
$ the
corresponding equations become
\begin{equation} \label{2.26}
\left [ \nabla^2 + \kappa^2_2 \right] \varphi_2 = 0,
\end{equation}
where
\begin{equation} \label{2.27}
\kappa^2_2 : = \omega^2_2 - \hat \omega^2_2 + 2 \; \eta_2
\hat  \omega^2_2 \,
\varphi_0 |\varphi_2|^2.
\end{equation}
The mean field generated by the two modes is given again by
a Poisson equation,
\begin{equation} \label{2.28}
\nabla^2 \varphi_0 = - \epsilon_1 \hat \omega^2
|\varphi_1|^2  -  \eta_2 \,
\hat \omega^2_2 |\varphi_2|^4.
\end{equation}
If $ \eta_2$ and $\epsilon_1$ have the same sign, the
coupled system
can support particular solutions in which $\varphi_1$
decreases
exponentially for
large~$r$ while $\varphi_2$ is given by the limiting
trapped-mode
solution $\omega_2 = \hat\omega_2$, which approaches the
free-wave
solution $\varphi_2 \exp(i\omega_2 t)/r $ for large~$r$.

\subsubsection*{Weakly trapped modes}

The parameter $\lambda$ determining the degree of
nonlinearity of the wave-guide mode solutions in the simple
model discussed above could be chosen arbitrarily.  However,
if the model is extended to include higher-order
interactions, the trapping strength can be determined by
stability
arguments.  For a model containing both quadratic and
cubic interactions, for example, the total energy of the
coupled wave mode-wave guide system will generally be some
non-monotonic function $E(\lambda)$. There will therefore
exist some value $\lambda_m$ for which $E(\lambda)$ is a
minimum, which represents the most stable state. The
value $\lambda_m$ depends on the form and relative strengths
of the coupling. Models can be readily constructed for which
the nonlinearity parameter $\lambda_m$ for the minimal
energy solution can be made arbitrarily small.

Solutions with very small but finite $\lambda$ are relevant
for charged finite-mass particles, for which the total
charge of the particle is given by an integral over the
square of the particle field. Extensive far fields must be
postulated for these particles to produce the observed
interference phenomena, but the integrals diverge at
infinity if the fields are assumed to be asymptotically
free in the limit
$\lambda \rightarrow 0$ (cf. Parts \ref{The
Maxwell-Dirac-Einstein System},\ref{The Standard Model})

\subsection*{Open questions}

Although the general analysis outlined above illustrates the
basic
mechanism by which exponentially decreasing or
asymptotically
free-wave
modes can be trapped in a self-generated wave-guide, a
number
of
questions remain.

The first concerns convergence.  Are the higher-order
coupling terms,
beyond the lowest-order interactions considered here,
finite, i.e.  do
the relevant interaction integrals converge? And if this is
the case,
does the resultant interaction series converge?

A second question refers to stability.  Are the trapped-mode
solutions
stable with respect to small perturbations, for example
through far-field interactions with other particles?

A third fundamental problem concerns the discreteness of the
observed
particle spectrum.  The solutions found above represent a
continuum
containing an arbitrary harmonic-space frequency $\hat
\omega$ and a
free nonlinearity parameter for each trapped mode.  To
reduce the
continuum to a discrete spectrum, additional considerations
must be
introduced. The problem is possibly related to the second
question:
stability arguments could lead to the identification of a
discrete
sub-set of most stable (minimal-energy)
solutions, into
which all other solutions of the continuum would then drift
under the
influence of small external perturbations. This argument
could explain also  the existence of the particular
solutions mentioned above containing asymptotically free or
only very weakly
trapped modes.

A fourth, equally fundamental question concerns the
uniqueness of the  solutions. Even if it can be shown that
the solutions are discrete, a given set of solutions is
determined always only up to an arbitrary coordinate-scale
factor. It must be shown that all coordinate-scale factors
of otherwise identical particles are the same. This
problem may again be related to the previous two:
 stable discrete particle states could arise
through a
collective self-organization mechanism. Suppose that the
universe is
indeed composed of only a finite number of particle types
with identical coordinate scales. In this case the
superposition of
the
oscillatory far-fields of all particles will produce a net
oscillatory
background metric field characterized by a finite set of
discrete
wavenumbers in harmonic space and discrete frequencies in
time (the latter would be slightly Doppler broadened by
random motions).  If there existed now for some reason a
particle
from the
available continuum of trapped-mode solutions which did not
correspond
to the assumed discrete spectrum, the state of the particle
(nonlinearity parameters, frequency and harmonic wavenumber
scale) would
in
general be free to drift under the influence of small
external
perturbation forces until it encountered a harmonic-space
wavenumber
and frequency corresponding to one of the discrete
particles. At this
point it would interact in resonance with the background
field. This
could produce a stabilizing force, causing the particle to
lock into
the background field, so that it would stop drifting and be
converted
to a member of the discrete spectrum.  The mechanism is
described in
more detail in the context of particle-field interactions in
Sections~\ref{Bragg scattering} and \ref{Atomic spectra}.

Alternatively, if a satisfactory
explanation of the discreteness and uniqueness of the
particle spectrum cannot be found, one could simply
postulate  -- in analogy
with string theory --
that all solutions of the higher-dimensional gravity
equations in our world exhibit given discrete
periodicities with
respect to the harmonic-space dimensions.

At the present level of analysis, however, these
considerations must
remain speculative. The open questions can be meaningfully
addressed
only within the context of a more detailed analysis of the
trapped-mode solutions of the full n-dimensional
gravitational
equations, which is not attempted in the present paper.

\newpage
~
\newpage
\part{The Maxwell-Dirac-Einstein System}
\label{The Maxwell-Dirac-Einstein System}
\typeout{################################}
\typeout{################################}
\typeout{        START OF met4-3.tex}
\typeout{################################}
\typeout{################################}
{\em ABSTRACT} \\

\noindent

Following the presentation of the general properties of the
metron model and the demonstration of the mode trapping
mechanism responsible for the postulated existence of
discrete soliton-type solutions ({\it metrons}) of
the higher-dimensional Einstein vacuum equations in Part
\ref{The Metron Concept}, we turn in the second part of this
four-part paper to the application of the metron concept to
the Maxwell-Dirac-Einstein system. It is shown that the
standard electromagnetic and fermion fields as well as the
form of their lowest order coupling can be derived from the
 n-dimensional gravitational Lagrangian, assuming a
four-dimensional
extra-({\it harmonic}-)space background metric and an
appropriate geometrical structure of the
metron solutions. Fermion fields are represented by
harmonic-index metric field components which are periodic
with respect to the electromagnetic coordinate $x^5$ of
harmonic space, the wavenumber component $k_5$ determining
the electric charge. The electromagnetic field is described
by a mixed-index metric field.  The $U(1)$ gauge invariance
of the Maxwell-Dirac system is explained by the invariance
of the n-dimensional Einstein equations with respect to
coordinate translations in the $x^5$ direction.

The metron model yields the basic universal physical
constants of the Maxwell-Dirac-Einstein system (the
gravitational constant, Planck's constant, the elementary
charge) and the individual particle constants (mass, charge,
spin) as properties of the metron solutions. The
small ratio of gravitational to electromagnetic forces is
explained by the fact that the gravitational forces
represent a higher-order nonlinear property of the metron
solutions.

The application of the metron picture of the
Maxwell-Dirac-Einstein system for the interpretation of
specific quantum phenomena and paradoxes such as
the EPR experiment,  time-reversal symmetry and Bell's
theorem, Bragg
scattering and atomic spectra is described in Part
\ref{Quantum Phenomena}. A generalization of the present
analysis to
include weak and strong interactions is presented in the
metron interpretation of the Standard Model in Part~\ref{The
Standard Model}.\\

\subsection*{\raggedright Keywords:}
{\small
metron ---
unified theory ---
wave-particle duality  ---
higher-dimensional gravity ---
solitons ---
Maxwell-Dirac-Einstein system ---
physical constants ---
action at a distance ---
force hierarchy}\\



{\em R\'ESUM\'E}

\vspace*{1ex}
Apr\`es avoir pr\'esent\'e dans la premi\`ere partie de ce
travail
les propri\'et\'es g\'en\'erales du
mod\`ele de m\'etron et d\'emontr\'e le m\'ecanisme
de capture de modes  responsable de l'existence
postul\'ee de solutions discr\`etes de type soliton (dites
m\'etrons)
des \'equations d'Einstein du vide \`a haute dimension,
nous consacrons la deuxi\`eme partie de ce travail \`a
l'application du concept de m\'etron
au syst\`eme de Maxwell - Dirac - Einstein.
Nous d\'emontrons, que les champs standards
\'electromagn\'etiques et fermions ainsi que la forme de
leur
couplage \`a l'ordre inf\'erieure peuvent \^etre d\'eriv\'es
du Lagrangien de gravitation \`a $ n $ dimensions
\'etant donn\'ee une m\'etrique de fond de l'espace
harmonique
\`a quatre dimensions et une structure g\'eom\'etrique
appropri\'ee
des solutions de m\'etron.
Les champs de fermions sont
repr\'esent\'es par les composantes de champs m\'etriques
\`a indice
harmonique, ces derniers \'etant p\'eriodiques concernant
la coordonn\'ee \'electromagn\'etique $ x^5 $ de l'espace
harmonique,
la composante $ k_5 $ du vecteur d'onde d\'eterminant la
charge
\'electrique.
Le champ \'electro\-magn\'etique est d\'ecrit par un champ
m\'etrique
\`a indices mixtes. L'invariance de jauge $ U(1) $ du
syst\`eme de
Maxwell - Dirac s'explique par l'invariance des \'equations
d'Einstein
en $ n $ dimensions par translations de coordonn\'ees en
direction $ x^5 $.

\`A partir du mod\`ele de m\'etron on obtient les constantes
de physique universelles du syst\`eme de Maxwell - Dirac -
Einstein
(la constante de gravitation, la constante de Planck, la
charge \'el\'ementaire) ainsi que les constantes de
particules
(masse, charge, spin) en tant que propri\'et\'es des
solutions de
m\'etron.
L'infime rapport entre forces gravitationnelles et forces
\'electromagn\'etiques provient du fait, que les forces
gravitationnelles sont des propri\'et\'es d'ordre
sup\'erieur
non-lin\'eaire des solutions de m\'etron.

La troixi\`eme partie de ce travail va d\'emontrer
l'application du point de vue de m\'etron du
syst\`eme de Maxwell - Dirac - Einstein \`a
l'interpr\'etation
des ph\'enom\`enes quantiques sp\'ecifiques et des paradoxes
tels que
celui de l'exp\'erience d'EPR, celui de la sym\'etrie
d'inversion
temporelle rattach\'e au th\'eor\`eme de Bell et enfin le
paradoxe
de la r\'etrodiffusion de Bragg rattach\'e aux spectres
atomiques.
Une g\'en\'eralisation de cette analyse, ayant pour but
d'inclure les forces fortes et les forces faibles sera
pr\'esent\'ee
dans le contexte de l'interpr\'etation de m\'etron du
mod\`ele
standard dans la quatri\`eme partie.\\

\subsection*{\raggedright Mots cl\'es:}
{\small
m\'etron ---
th\'eorie unifi\'ee ---
dualit\'e onde-corpuscule ---
th\'eorie de gravitation \`a haute dimension ---
solitons ---
syst\`eme de Maxwell-Dirac-Einstein ---
constantes de physique ---
action \`a distance ---
hierarchie des forces}

\newpage
\section{Introduction}
\label{Introduction 2}

In the first Part  of this four-part paper we outlined the
general structure of a unified deterministic theory of
fields and particles based on the postulated existence of
soliton-type trapped-mode {\em metron} solutions of the
n-dimensional vacuum Einstein equations. The trapping
mechanism was illustrated by numerical computations of
metron-type solutions  for a simplified prototype Lagrangian
exhibiting the same nonlinear structure as the gravitational
Lagrangian without its tensor complexities.  Having
established that such nonlinear systems can support
trapped-mode solutions, we turn now in the remaining three
parts of this paper to more specific
implications of
the model.  In Part \ref{The Maxwell-Dirac-Einstein
System} we consider first the mapping of the metron
solutions of
the gravitational equations onto the
Maxwell-Dirac-Einstein system.  The resulting metron model
of electromagnetic interactions sets the stage for a general
discussion of quantum phenomena in Part
\ref{Quantum Phenomena}. In Part \ref{The Standard Model},
finally,
the analysis is extended to include weak and strong
interactions in the development of the metron picture of the
Standard Model.

The analysis in the present part is organized in three
sections. First, the fermion and boson fields of quantum
field theory are identified with specific (to lowest
order, linear wave) solutions of the n-dimensional
gravitational equations (Section \ref{Identification
of fields}). Subsequently, the standard
electromagnetic-Dirac coupling terms are derived from the
gravitational Lagrangian, assuming a suitable harmonic-space
background metric with dimension of at least four (Sections
\ref{Lagrangians}, \ref{The Maxwell-Dirac-Einstein
Lagrangian}). The analysis in these sections is
limited to weakly nonlinear field-field interactions outside
the particle-core regions. As third step we consider then
the
interactions between far fields and the particle-core
regions (Section~\ref{Particle interactions}). This
yields the integral particle properties, i.e. the
gravitational and electromagnetic forces produced by the
particle mass and charge, and the associated basic physical
constants.

In the extension of the analysis to weak and strong
interactions later in Part \ref{The Standard Model}, it will
be shown that the nonlinear gravitational equations
reproduce
the general structure (but with differences in detail) of
all known field-field interactions of quantum field theory,
assuming an appropriate geometrical configuration of the
trapped-mode metron
solutions. Our analysis is nevertheless
incomplete in one essential aspect:  the link between local
field-field interactions and interactions at a distance,
which are governed by the integral particle properties, must
be closed by computations of the postulated trapped-mode
solutions. As
mentioned previously, however, this must await a later
investigation.

\section{Identification of fields}
\label{Identification of fields}

Some general features of metron solutions were summarized
already in Section~\ref{The mode-trapping mechanism}. In
order to establish now the relation between the n-
dimensional
gravitational field equations (\ref{1.1}) and the standard
quantum field equations, further assumptions regarding the
geometrical structure of the metron solutions are needed.
These will be introduced following a general `inverse metron
modelling' approach which will be adopted throughout this
paper. It is assumed that metron solutions
exist. General properties of the solutions are then inferred
from the requirement that the n-dimensional  gravitational
field equations can be mapped into the usual field equations
of quantum field theory.
The existence of the postulated solutions, in accordance
with the mode-trapping mechanism described in Section
\ref{The mode-trapping mechanism}, must, of course,
be subsequently demonstrated.

Parton fields were defined in eq.(\ref{2.1}) as deviations
with respect to the flat background metric $\eta_{LM}$
\cite{ft9}. It will be shown below, and expanded further in
the
discussion of the Standard Model in Part~\ref{The Standard
Model}, that to reproduce  the basic interactions of quantum
field theory, $\eta_{LM}$ must be an at least eight-,
possibly
nine-dimensional metric of the form (in natural coordinates,
with $c=1$)
\begin{equation} \label{3.1}
\eta_{LM} = \mbox{diag} (1,1, 1,-
1,\ldots,\pm1,\ldots),
\end{equation}
where the first four dimensions refer to physical spacetime,
periodicities with respect to the fifth and sixth
dimensions are associated with the electrodynamic and weak
interactions, respectively, the seventh and possibly eighth
dimension represent the
strong-interaction (color) space and the  last (eighth  or
nineth) dimension is needed, together with the other
harmonic-space dimensions, to relate the harmonic-space
components of the metric tensor to the spinor components of
the fermion fields.

To avoid the possible existence of particles or signals
propagating in
physical space-time at speeds greater than the speed of
light,  a background metric with positive harmonic-space
components would be desirable. Also, to ensure that the
signal emitted by an n-dimensional  $\delta$-function pulse
propagates on the surface of an expanding $(n-1)$-
dimensional sphere in
physical-plus-harmonic space, without also filling out the
interior of the sphere, $(n-1)$ should be an odd dimension
\cite{ft10}.
These considerations would favor an
eight-dimensional space with a single time-like coordinate.
However, it is not clear whether the argument that the
general solutions of the n-dimensional
gravitational equations should have analogous properties to
the solutions in physical spacetime are relevant for the
situation which we
envisage:  we assume (for reasons which still have to be
justified) that in practice only the periodic metron
solutions, which have no structure -- apart from their
periodicity -- in harmonic space, prevail in the real world.
We are therefore concerned with signal propagation only in
the four-dimensional sub-space representing physical
spacetime.
At this stage we will accordingly consider all background
metrics as equally acceptable, judging different forms only
on the basis of their ability to reproduce the known
phenomena of particle physics for suitably structured metron
solutions.

In the present and immediately
following sections, only
the first
five (original Kaluza-Klein) dimensions describing
gravitational and
electromagnetic forces will be considered in detail.
However,
the full harmonic space must still be invoked to
relate the fermion fields to the
harmonic-index metric field components.
We summarize in the following some general properties of the
postulated metron
solutions and identify the various classes of fields which
will be encountered, although  details will often not be
needed until we extend the analysis later to the metron
interpretation of the Standard Model in Part \ref{The
Standard Model}.

The set of parton wavenumbers $\mbox{k}^{p}
=(k^{(p)}_{A})$ in harmonic space associated
with a
given metron solution consists of a finite set of
fundamental wavenumbers $\pm\mbox{k}^{1},\ldots,\mbox{$\pm$
k}^{f}$
(negative wavenumbers are associated with the
complex conjugate fields) and their higher harmonics
\begin{equation} \label{3.2}
\mbox{k}^{p}    =     n^{p1}
\mbox{k}^{1}
+    \ldots     +
n^{pf}\mbox{k}^{f},
\end{equation}
where $n^{pj} = 0, \pm1, \pm2, \ldots$.
From the invariance of the gravitational equations
(\ref{1.1}) with
respect to reflections in any coordinate direction, it
follows that
for any
metron solution $m$ a change in sign of all the parton
harmonic wavenumber components, $k^{(p)}_{A}
\rightarrow -\,
k^{(p)}_{A}$,
also yields a solution.  This will be identified with the
anti-particle $\bar m$.  The transformation, corresponding
to the charge conjugation transformation $C$ of quantum
field theory, should be
distinguished
from a sign change $k^{(p)}_{A} \rightarrow - \,
k^{(p)}_{A} $
accompanied by a simultaneous transformation to the complex
conjugate
amplitudes, $\hat g^{(p)}_{LM} \rightarrow \hat
g^{(p)*}_{LM}$, which
yields, of course, the same particle (eq.(\ref{2.2})).
Equivalently,
the anti-particle can be defined by the transformation to
the complex conjugate parton amplitudes, $\hat
g^{(p)}_{LM} \rightarrow \hat g^{(p)*}_{L
M} $, without a
sign
change in
the wavenumber.  It will be assumed that the metron
solutions are
invariant with respect to a change in sign of {\em all}
coordinates,
$g^{(m)}_{LM}(X) = g^{(m)}_{LM}(-X)$
($CPT$ transformation).
In this case a third
equivalent
definition of the anti-particle is the
solution
obtained by changing the signs of only the physical
spacetime
coordinates, $g^{(\bar m)}_{LM}(x,\mbox{x}) =
g^{(m)}_{LM}(-x,\mbox{x})$ ($PT$ transformation).

In the metron restframe, the parton amplitudes,
eq.(\ref{2.2}), are either independent
of time $t = x^4$ or periodic in $t$,
\begin{equation} \label{3.3}
\hat g^{(p)}_{LM}(x) = \tilde g^{(p)}_{L
M}({\bf x}) \exp
\left( - i
\omega^{p}t \right ).
\end{equation}
 The frequency $\omega^{p}$ will be identified in
the next section with the parton mass.

It will be assumed that the
perturbation fields $g^{(p)}_{LM}$ satisfy the
gauge
condition  -- which is always possible through
suitable choice
of
coordinates --
\begin{equation} \label{2.7}
\partial^{L} h_{LM}^{(p)} = 0,
\end{equation}
where
\begin{equation} \label{2.8}
h_{LM}^{(p)} : = g_{LM}^{(p)} -
\frac{1}{2} \eta_{LM} \,
g_{N}^{(p)N}.
\end{equation}
The fields $h_{LM}^{(p)}$ and associated amplitude
functions
$\hat
h_{LM}^{(p)}$ also satisfy the wave and
Klein-Gordon
equations
(\ref{2.4}), (\ref{2.5}), respectively, in the linear
approximation.

In the linear case, the `harmonic' mass
$\hat\omega^{p}
$ appearing in the Klein-Gordon equation (\ref{2.5}) is
equal to the
gravitational mass $\omega^{p}$ defined by the
time-dependent factor
in (\ref{3.3}). However, in the general nonlinear case the
two
masses will
differ (cf. Section~\ref{The mode-trapping mechanism}).

In order to map the n-dimensional gravitational fields
into the
standard quantum-theoretical fields, the
following
field identifications are now made (see also
table~\ref{ta3.1}, which lists
the polarization relations for the various fields discussed
below):\\

\noindent $\bullet $  spacetime metric components
$g_{\lambda
\mu}$:
\hfill {\it classical gravitational field} \\
$\bullet $ mixed spacetime-harmonic space metric components
$g_{\lambda A}$:
\hfill {\it boson fields} \\
$\bullet $ harmonic-space metric components $g_{A
B}$:
\hfill {\it fermion and scalar fields}\\

\noindent The mappings apply only one-way: the standard
classical and quantum theoretical fields on the right hand
side are represented in the metron model by
metric fields
of the indicated index types, but  all components of the
n-dimensional metric field
tensor  cannot in general be related to the
standard quantum theoretical fields. In fact it will be
argued later that the
fields appearing in the  Standard Model represent only an
approximation of the full set of interacting fields in the
metron model.

The fields $g_{\lambda \mu}$ corresponding to
classical
gravitational fields are assumed to be independent of the
harmonic
space coordinates. Ignoring for the present the interactions
of these
fields with other metric fields, one recovers then trivially
for the
gravitational fields not only the linearized equations
(\ref{2.4}),
but also the fully nonlinear classical (matter free)
gravitational
equations.  To avoid confusion in terminology, the term
{\em gravitational
field} will be restricted in the following to the classical
gravitational field in four-dimensional spacetime, while the
corresponding
tensor
field in full space will be referred to as the {\em metric
field}
or,
occasionally, the full-space gravitational field.

\begin{table}
\begin{tabular}{l|c|c}
& physical spacetime & harmonic space \\
~&~&~ \\
 \hline 
~&~&~ \\
physical spacetime & gravity: $g_{\lambda \mu}$ & bosons
$b$:
$\eta_{\lambda A} + B^{(b)}_{\lambda} a^{(b)}_A$ \\
~&~&~ \\
 \hline 
~&~&~ \\
harmonic space & bosons $b$: $\eta_{A \lambda} + a^{(b)}_A
B^{(b)}_{\lambda}$ & fermions $f$: $\eta_{AB} + P^{(f)a}_{AB}
\psi^{(f)}_a $ \\
                 && scalars $s$: $\eta_{AB} + P^{(s)}_{AB}
\phi^{(s)} $
\end{tabular}
\caption[ix] {metric forms for gravitational, vector boson,
fermion and
scalar fields}
\label{ta3.1}
\end{table}

Bosons can be either oscillatory or non-oscillatory, i.e.
have finite
or zero mass.  We will be concerned here in  Part~\ref{The
Maxwell-Dirac-Einstein System}
only with the zero-mass boson field representing the
electromagnetic
four potential $A_{\lambda}$. The associated metric
field will be shown to be of
the form
\begin{equation} \label{3.7}
g^{(a)}_{A  \lambda}  = g^{(a)}_{\lambda  A}  = :
A_{\lambda}
a_{A},
\end{equation}
where $a=(a_{A})$ is a constant vector of length $|a|
:= (a_{A}a^{A})^{^1/2} $ which we take to define the $x^5$
direction.  The
normalization of $A_{\lambda}$, i.e the value of $|a|$, will
be
chosen later in Section \ref{Particle interactions} such
that the metron
Lagrangian  reproduces the classical free-field
electromagnetic Lagrangian.

It will be shown later in Part~\ref{The Standard Model} that
vector bosons
$B^{(b)}_{\lambda}$ can be represented generally in the
metron model in a
form analogous to (\ref{3.7}), in which the mixed-index
metric components
are factorized in the form $g^{(b)}_{A \lambda} =
a^{(b)}_A
B^{(b)}_{\lambda}$ (see table~\ref{ta3.1}).

The Klein-Gordon equation (\ref{2.5}) reduces for
$A_{\lambda}$ to the
wave equation
\begin{equation} \label{3.8}
\Box A_{\lambda} = 0
\end{equation}
and the general metric gauge condition (\ref{2.7}) becomes
the
standard
Lorentz gauge condition
\begin{equation} \label{3.9}
\partial_{\lambda}A^{\lambda} = 0.
\end{equation}
Metric field components associated with complex scalar
fields $\phi^{(s)}$ (required later in Part~\ref{The
Standard Model}
to represent the Higgs field) have the general form
\begin{equation} \label{3.10}
\hat g^{(s)}_{AB} = P^{(s)}_{AB}
\phi^{(s)},
\end{equation}
where, in the linear approximation, $ \phi^{(s)}$ satisfies
the Klein-Gordon equation
\begin{equation} \label{3.11}
\left (\Box - \hat \omega^2_{(s)}\right ) \phi^{(s)} = 0
\end{equation}
and $P^{(s)}_{AB}$ is a constant polarization
tensor.

Similarly, metric fields $g^{(f)}_{AB}$
representing four-spinor
fermion fields \newline $\psi^{(f)} : = \left (
\psi_1^{(f)},
\cdots
\psi^{(f)}_4  \right)$
are given by
\begin{equation} \label{3.12}
\hat g^{(f)}_{AB} = : P^{(f)a}_{AB}
\psi^{(f)}_a,
\end{equation}
where $P^{(f)a}_{AB}$ is again a constant
polarization
tensor and $\psi^{(f)}$ satisfies the Klein-Gordon equation
\begin{equation} \label{3.13}
\left ( \Box - \hat \omega^2_{(f)} \right ) \psi^{(f)} = 0.
\end{equation}

For harmonic-space metric fields associated with scalar or
fermion fields, the physical spacetime components
of the gauge
condition (\ref{2.7}) yield $\partial_{A} h^{A
\lambda}_{(p)} = -
\frac{1}{2} \partial_{A} \eta^{A \lambda}
g^{B}_{(p)B}=
\partial^{\lambda}g^{B}_{(p)B}=0$, or, assuming
that the fields vanish for infinite $\bf x$,
the trace condition
\begin{equation} \label{3.13a}
g^{A}_{(p)A}=0.
\end{equation}
Thus
\begin{equation} \label{3.13aa}
h^{(p)}_{AB}=g^{(p)}_{AB},
\end{equation}
and the remaining harmonic-index components of the gauge
condition become
\begin{equation} \label{3.13b}
k_{A} g^{AB}_{(p)}=0.
\end{equation}

The Klein-Gordon operator acting on a four-spinor $\psi$
(dropping now
the parton index ($f$)) can be factorized into the product
of two
Dirac operators with positive and negative frequencies:
\begin{equation} \label{3.14}
\left ( \Box -  \hat  \omega^2  \right  )  \psi  =
(\partial_{\lambda}
\gamma^{\lambda} + \hat \omega  )
(\partial_{\mu}  \gamma^{\mu}  -
\hat \omega ) \psi = 0,
\end{equation}
where the Dirac matrices $\gamma^{\lambda}$ satisfy the
anti-commutation relations
\begin{equation} \label{3.15}
\gamma^{\lambda} \gamma^{\mu} + \gamma^{\mu}
\gamma^{\lambda}
= 2  \eta^{\lambda
\mu}
\end{equation}
and Hermiticity relations
\begin{equation} \label{3.16}
\gamma^i  =  \left (\gamma^i \right ) ^*, \quad
\gamma^4  =  - \left (\gamma^4 \right )^*.
\end{equation}
The general solution of (\ref{3.14}) may be represented as a
superposition
\begin{equation} \label{3.17}
\psi = \psi^+ + \psi^-
\end{equation}
of the solutions $\psi^+,\psi^-$ of the two Dirac equations
\cite{ft11}

\begin{eqnarray}
\label{3.18}
\left (\partial_{\lambda} \gamma^{\lambda} + \hat \omega
\right ) \psi^+ &=& 0 \\
\label{3.19}
\left (\partial_{\lambda} \gamma^{\lambda} - \hat \omega
\right ) \psi^- &=& 0.
\end{eqnarray}

The frequency $\hat \omega$ will be
defined generally as the positive root of (\ref{2.6}).
However, we
introduce the convention that under a charge conjugation
transformation $C$ (change in sign of the harmonic
wavenumber), $\hat \omega$ also changes sign. From
the definition of an anti-particle given
above
(reflection of the harmonic-space or physical spacetime
coordinates) it follows then
that the anti-particle of a field $\psi^+$ represents a
field
$\psi^-$, and vice versa. It will be assumed that metrons
contain only
one of the two branches (taken in the following to be
$\psi^+$) for
any given parton component.

We point out in conclusion that the Dirac equations
(\ref{3.18}),
(\ref{3.19}), in contrast to the original field equations
(\ref{1.1}),
are not invariant with respect to  diffeomorphisms (if we
wish to maintain the basic $\gamma$-matrix relations
(\ref{3.15}) with invariant $\eta_{\lambda \mu}$).
Rather than generalizing the Dirac equations to an arbitrary
metric \cite{ft12} we will simply restrict the
representation
of the spinor fields and $\gamma$ matrices, when
considering coordinate transformations later in
 n-dimensional space, always to a
local frame with background metric $\eta_{AB}$.

\section{Lagrangians}
\label{Lagrangians}

To describe interactions between the fields identified in
the previous
sub-section, we must consider now the n-dimensional
gravitational
Lagrangian. Details are given here only for interactions
between
fermions and electromagnetic and gravitational fields;
electroweak and
strong interactions are considered later in
Part~\ref{The
Standard Model}.

The gravitational field equations (\ref{1.1}) can be derived
from the
variational principle
\begin{equation} \label{3.21}
\delta \int|g_n| ^{1/2}\, R \, d^nX = 0,
\end{equation}
where $g_n$ is the
determinant
of the
metric tensor in n-dimensional space and $R =
R^{L}_{L}$ is
the scalar curvature,  formed by contraction of the Ricci
curvature tensor
\begin{equation} \label{3.21a}
R_{LM} = \partial_{M} \Gamma^{N}_{L N}
-  \partial_{N}\Gamma^{N}_{LM}
+ P_{LM},
\end{equation}
where
\begin{equation} \label{3.22}
P_{L M} = \Gamma^{N}_{L O}\,
\Gamma^{O}_{M N} -
\Gamma^{N}_{LM}\, \Gamma^{O}_{N O}
\end{equation}
and the connection (Christoffel symbol) is given by
\begin{equation} \label{3.23}
\Gamma^{L}_{M N}  : =  \frac{1}{2} \,  g^{L
O}   \,
\left[
\partial_{M}   g_{O N}   +
\partial_{N} g_{O M} - \partial_{O} g_{M N}
\right].
\end{equation}

In place of the general curvature invariant $R$, it is often
more
convenient to use as Lagrangian density $L$ the equivalent
homogeneous
affine-scalar form
\begin{equation} \label{3.23a}
P = g^{LM}P_{LM},
\end{equation}
 which contains only first
derivatives and differs from $R$ only
through a divergence term.  Multiplying out the
 Christoffel symbols in (\ref{3.22}), this may be written as
\begin{equation} \label{3.24}
L = P = \frac{1}{4} \left[ P_1 - 2P_2 - P_3 + 2P_4 \right ],
\end{equation}
where
\begin{eqnarray} \label{3.25}
\left \{
\begin{array}{c}
P_1 \\ P_2 \\ P_3 \\ P_4
\end{array} \right\} : = g^{L M} g^{N O}
g^{P Q} \left\{
\begin {array}{cc}
\partial_{L}g_{N O} & \partial_{M}
g_{P Q}\\
\partial_{O} g_{L M} &
\partial_{Q}g_{NP}\\
\partial_{P}g_{LN} &
\partial_{Q}g_{M O}\\
\partial_{P}g_{LN} &
\partial_{O}g_{M Q}\\
\end{array}
\right \}.
\end{eqnarray}
Note that, since the products (\ref{3.25}) contain both
covariant and
contravariant fields, the perturbation expansion of the
Lagrangian
(\ref{3.24}) with respect to individual parton fields
$g^{(p)}_{LM}$
(eqs.  (\ref{2.1}),(\ref{2.9})) yields an infinite series of
interaction
terms.

As all fields are assumed to be periodic with
respect to the
harmonic-space coordinates, the Lagrangian density $L$ can
be
replaced in the action integral ({\ref{3.21}) by the
harmonic-space integrated Lagrangian density $\bar L$,
defined
by
\begin{equation} \label{3.26}
\bar L (-g_4)^{1/2} : = \int \left |g_n
\right|^{1/2}Ld^{n-4}\mbox{x},
\end{equation}
which depends on the harmonic-space wavenumber components
but is
otherwise independent of the harmonic coordinates x. The
integration
over full space in the action integral (\ref{3.21}) can then
be
restricted
to the integration over physical spacetime. Unless
otherwise stated, all
Lagrangians
in the following will be regarded as harmonic-space averages
in this
sense, and the overbar will be dropped.

Writing (\ref{3.24}) in the abbreviated form $L = \left [g^
{..}
g^{..} g^{..} \partial.g..\partial.g.. \right ]$, the
free-field
Lagrangian of the linearized gravitational field equations
is given by
\begin{equation} \label{3.27}
L_0 = \left [ \eta^{..}\eta^{..}\eta^{..} \partial. g..
\partial. g.. \right ],
\end{equation}
which yields explicitly \cite{ft13}
\begin{equation} \label{3.28}
L_0  =  \frac{1}{4}  \left  \{-  \partial_{N}  g_{L
M}\,
\partial^{N} g^{LM}   +
\frac{1}{2} \partial_{N}g^{L}_{L}\,
\partial^{N}g^{M}_{M} \right \}.
\end{equation}

For the electromagnetic field, as defined by eq.(\ref{3.7}),
eq.(\ref{3.28}) yields the
free-field Lagrangian
\begin{equation} \label{3.29}
L_0^{em} = - \frac{|a|}{4}^2 F_{\lambda \mu} F^{\lambda
\mu},
\end{equation}
where
\begin{equation} \label{3.30}
F^{\lambda\mu} := \partial^{\lambda}A^{\mu} -
\partial^{\mu}A^{\lambda}.
\end{equation}
For a fermion field, the free-field Lagrangian (\ref{3.28})
reduces, applying the zero-trace condition (\ref{3.13a}),
to
\begin{equation} \label{3.30a}
L_0^f  =  - \frac{1}{2}  \left  \{ \partial_{\lambda} \hat
g_{AB}^{(f)*}\,
\partial^{\lambda} \hat g^{AB}_{(f)}
+ \hat \omega_f^2 \hat g_{AB}^{(f)*}
\hat g^{AB}_{(f)}
\right \}.
\end{equation}
Substit uting  the
general
spinor form (\ref{3.12}), this becomes (dropping the index
$f$)
\begin{equation} \label{3.31}
L^f_0  =  - \frac{1}{2}  \left  (\eta^{\lambda   \mu}
\partial_{\lambda}   \psi^*_a
\partial_{\mu} \psi_b + \hat \omega^2 \psi^*_a \psi_b
\right ) M^{ab},
\end{equation}
where the matrix
\begin{equation} \label{3.31a}
M^{ab} := \left( P^a_{AB}\right)^* P^{b AB}
\end{equation}
will be termed the {\it spinor metric}.

It is shown below that the spinor metric  can be
transformed,
through an appropriate choice of the fields $\psi_a$, to a
matrix
proportional either to $i\gamma^4 $, if the harmonic-space
background metric is non-Euclidean,  or to $ I$, if the
metric
is Euclidean. In either case, the  Lagrangian
(\ref{3.31}), in which the derivatives of the spinor field
appear quadratically, can be reduced  to the standard
Dirac Lagrangian, in which the derivatives  occur only
linearly \cite{ft14}.

A necessary condition, however, is  that the dimension $m$
of
harmonic space must be at least four. The four-component
complex
spinor field must be related, through (\ref{3.12}), to a set
of
complex amplitudes of the periodic
harmonic-space components of the metric field. These
consist at most of $m(m+1)/2$ independent terms. The metric
field components must satisfy also the trace condition
(\ref{3.13a}) and the $m$ divergence conditions
(\ref{3.13b}). This leaves $ l(m) := (m+1)(m-2)/2$
independent
harmonic-index metric components. With $l(3) = 2, l(4)=5$,
it follows that  $m \geq 4$. For the minimal case $m=4$
we shall present
polarization
relations (\ref{3.12}) which satisfy the divergence and
trace
conditions.

\subsection*{Non-Euclidean $\eta_{AB}$}
We consider first the  case that, through a suitable choice
of the polarization tensor, we can set
\begin{equation} \label{A.3}
M = \frac{i\gamma^4}{\hat \omega} =
\frac{1}{\hat\omega} \,\mbox{diag} \,(1,1,-1,-1)
\quad \mbox{(in  the
Dirac representation)}.
\end{equation}
This implies that $\eta_{AB}$ cannot be a
positive or negative
Euclidean metric, since, according to (\ref{3.31a}), this
would
yield
a
positive definite spinor metric $M$.

For the spinor  metric (\ref{A.3}), the
Lagrangian (\ref{3.31}) becomes
\begin{equation} \label{3.32}
L^f_0 = - (2 \hat \omega)^{-1} \left ( \eta^{\lambda  \mu}
\partial_{\lambda}
\bar \psi \partial_{\mu} \psi + \hat \omega^2 \bar \psi \psi
\right ),
\end{equation}
where $ \bar \psi= i \psi^*\gamma^4$ denotes the conjugate
spinor.
Equation (\ref{3.32})  can be factorized, discarding an
irrelevant divergence
term
\[
\partial_{\lambda}
\left\{
\frac{1}{2}\bar\psi \gamma^{\lambda} \psi
+\frac{1}{4 \hat \omega}\bar\psi
\left(\gamma^{\lambda} \gamma^{\mu} -\gamma^{\mu}
\gamma^{\lambda}
\right) \partial_{\mu}\psi
\right\},
\]
in the form
\begin{equation} \label{3.33}
L^f_0  =  -   (2   \hat   \omega)^{-1}
 (\partial_{\lambda}   \bar
\psi\gamma^{\lambda} + \hat \omega \bar \psi  )
 (  \gamma^{\mu}
\partial_{\mu} \psi + \hat \omega \psi  ).
\end{equation}
Considering only the positive-branch solution, $\psi =
\psi^+$,
this  may then be written, noting that
$\bar\psi^+$
satisfies
the conjugate Dirac equation
\begin{equation} \label{3.34}
\partial_{\lambda} \bar \psi^+ \gamma^{\lambda} - \hat
\omega \bar \psi^+ = 0,
\end{equation}
in the standard form
\begin{equation} \label{3.35}
L^f_0 = - \bar \psi^+ \left (\gamma^{\lambda}
\partial_{\lambda} \psi^+ + \hat
\omega \psi^+ \right ).
\end{equation}
Generally, it is not permitted, of course, to substitute
relations valid only for a  particular class of solutions
into a Lagrangian. Doing so implies that we may seek  only
variational solutions of the Lagrangian which belong to that
particular class. In the present case, the substitution of
the positive-branch solution for the conjugate spinor field
into the Lagrangian (\ref{3.33}) has the effect of
automatically
filtering out the negative-branch solution:
the Lagrangian (\ref{3.35}) yields only the
positive-branch free-field Dirac
equation (\ref{3.18}). All variational solutions
of (\ref{3.35}) are automatically variational solutions of
the original fermion sector (\ref{3.33}) of the
gravitational Lagrangian, but the converse  obviously
does not hold. Our approach must be justified ultimately by
the structure of the metron solutions. It is
assumed that the fermion fields occur always as pure
positive- or negative-Dirac-branch components, the
components of the opposite branch appearing in the
corresponding
anti-particles.

In contrast to the original Lagrangian (\ref{3.32}), the
Dirac Lagrangian (\ref{3.35}) is in general
no longer  real (because the divergence term, which
was
discarded in deriving the factorized form
(\ref{3.33}), was not real). Although this is immaterial for
interactions
involving only a single fermion field, for multiple-fermion
interactions considered in the discussion of the Standard
Model later in Part~\ref{The Standard
Model}, we shall
require the real versions of
 eqs.(\ref{3.33}),(\ref{3.35}), which are given here for
future
reference:
\begin{equation} \label{3.33a}
L^f_0  =  - \frac{1}{4   \hat   \omega} \left\{
 (\partial_{\lambda}   \bar
\psi\gamma^{\lambda} + \hat \omega \bar \psi  )
 (  \gamma^{\mu}
\partial_{\mu} \psi + \hat \omega \psi  )
+
 (\partial_{\lambda}   \bar
\psi\gamma^{\lambda} - \hat \omega \bar \psi  )
 (  \gamma^{\mu}
\partial_{\mu} \psi - \hat \omega \psi  )
\right\}
\end{equation}
(general case of both positive- and negative-branch
solutions), or
\begin{equation} \label{3.35a}
L^f_0 = -\frac{1}{2}
\left\{
 \bar \psi^+ \left (\gamma^{\lambda}
\partial_{\lambda} \psi^+ + \hat
\omega \psi^+ \right )
-
\left (
\partial_{\lambda} \bar \psi^+ \gamma^{\lambda}
 - \hat
\omega \bar \psi^+ \right ) \psi^+
\right\}
\end{equation}
(positive-branch solution only).

We turn now to the conditions that the
(non-Euclidean)
harmonic space background metric $\eta_{AB}$ and
polarization matrix $P^a_{AB}$ must satisfy in
order to yield a spinor metric of the form
(\ref{A.3}).
For the minimal dimension $m=4$, a specific solution
can be readily
given for  a   harmonic-space background
metric \cite{ft15}
\begin{equation} \label{A.4}
\eta_{AB} = \mbox{diag} (1,1,1,-1)
\end{equation}
Assuming a finite particle mass,
$k_{A}k^{A} >0$,  so that we can set, through a
suitable harmonic-space
Lorentz
transformation, k$=(k_5,0,0,0)$, the following choice of
polarization matrices
is readily seen to satisfy (\ref{A.3}), together with
the trace and divergence conditions
(\ref{3.13a}) and (\ref{3.13b}):
\begin{equation} \label{A.5}
P^{a}_{AB}\psi_a = \frac{1}{(\sqrt{2 \hat\omega})}
\left(
\begin{array}{cccc}
0& 0& 0& 0 \\
0& \psi_1&  \psi_2 & \psi_3\\
0& \psi_2& -\psi_1 & \psi_4\\
0& \psi_3 & \psi_4 &  0
\end{array}
\right).
\end{equation}
We shall refer to the harmonic metric (\ref{A.4}) with
associated polarization tensor (\ref{A.5}) as the {\it
minimal
model} $(+3,-1)$ \cite{ft16}.

We note that all terms involving the
first harmonic index vanish in the expression (\ref{A.5}).
Thus the sign
of the background
metric for this index is irrelevant, and the solution
(\ref{A.5}) can be applied equally well for the
background harmonic-space metric $\eta_{AB} =
\mbox{diag} (-1,1,1,-1)$.  In this case the square harmonic
mass
$k_{A}k^{A}
<0$. However, a  similar solution can be
given also for the case $k_{A}k^{A} >0$, for
example for the
wavenumber vector k$=(0,k_7,0,0)$ by an interchange $1
\leftrightarrow 2$ and $3 \leftrightarrow 4$ of the harmonic
indices
in (\ref{A.5}).

Noting furthermore that the definition (\ref{3.31a}) for
$M^{ab}$ is
independent of the sign of the background metric
$\eta_{AB}$, it follows then generally that a
spinor representation of the harmonic-space metric field
with a spinor metric
$M^{ab}$ proportional to $i \gamma^4$ can be defined for any
background harmonic metric of mixed sign. However, it will
be shown in Section~\ref{Particle interactions} that to
obtain the right sign for the
electromagnetic forces (like charges repell) we must have
$\eta_{55} = 1$. To simplify the
discussion, we shall consider later as prototype minimal
non-Euclidean model only the model $(+3,-1)$.

The polarization relations (\ref{A.5}) have the shortcoming
that they cannot be applied to
a fermion field with zero harmonic mass: the normalization
factor approaches infinity
as
the mass $\hat\omega$ approaches zero.  This difficulty does
not arise in the following model for a Euclidean background
metric.

\subsection*{Euclidean $\eta_{AB}$}

For a background harmonic metric
$\eta_{AB} = \pm \,\mbox{diag} (1,1, 1,1)$,
an alternative minimal fermion  model $(\pm 4)$ can be
constructed
which
also  yields the standard Dirac Lagrangian. However, it
will be found necessary in this case to restrict
the gravitational solutions
to lie not only on the
positive Dirac branch, but also to have only
positive or negative frequency . We shall apply this model
later  in Part~\ref{The Standard Model}
for the description of leptons in the Standard Model, as, in
contrast to the
non-Euclidean model, it is applicable for particles of both
finite and zero mass.

For the following it is convenient to
change to
the
$\gamma$-matrix representation
\begin{equation} \label{8.54}
\gamma^i =
\left(
\begin{array}{cc}
0& -i\sigma^i \\
i\sigma^i& 0
\end{array}
\right),
\quad
 \gamma^4 = \left(
\begin{array}{cc}
 0 & i\mbox{I}\\
i\mbox{I} & 0
\end{array}
\right),
\quad
\gamma_5 = i \gamma^1 \gamma^2 \gamma^3 \gamma^4 =
\left(
\begin{array}{cc}
 -\mbox{I} & 0\\
0 & \mbox{I}
\end{array}
\right),
\end{equation}
where
$
\sigma_1 = \left(
\begin{array}{cc}
0 & 1\\ 1& 0 \end{array}
\right), \;
 \sigma_2 = \left(
\begin{array}{cc}
0 & -i\\ i& 0 \end{array}
\right),
\;
 \sigma_3 = \left(
\begin{array}{cc}
1 & 0\\ 0& -1 \end{array}
\right)
$
are the Pauli matrices. In this representation
the decomposition of the four-spinor into
right- and left-handed spinors, $\psi =
\psi^R + \psi^L$,  where
\begin{eqnarray} \label{8.52}
\psi^R &:=&(1-\gamma_5) \psi /2 \nonumber \\
\psi^L &:=&(1+\gamma_5) \psi /2
\end{eqnarray}
separates the four spinor into
right- and left-handed two-spinors $\varphi^L,\varphi^R$,
\begin{equation} \label{8.55}
\psi^R = \left(
\begin{array}{c}
\varphi^R\\
0
\end{array}
\right), \quad
\psi ^L= \left(
\begin{array}{c}
0\\
\varphi^L
\end{array}
\right)
\quad
\mbox{and} \quad
\psi = \left(
\begin{array}{c}
\varphi^R\\
\varphi^L
\end{array}
\right).
\end{equation}
The  Lagrangian (\ref{3.35a}) for
(positive Dirac branch) fermions, which may be
expressed in terms of the
right- and left-handed
four-spinor fields in the form
\begin{eqnarray} \label{8.51}
L^f_0 &=&- \frac{1}{2}
\left[
\left(
\bar \psi^L
\gamma^{\lambda}
\partial_{\lambda} \psi^L
-
\partial_{\lambda} \bar \psi^L \gamma^{\lambda}
\psi^L
\right)
+
\left(
\bar \psi^R
\gamma^{\lambda}
\partial_{\lambda} \psi^R
-
\partial_{\lambda} \bar \psi^R \gamma^{\lambda}
\psi^R
\right)
\right]
\nonumber \\
&& + \hat\omega
\left(
\bar \psi^R
\psi^L
+\bar \psi^L
\psi^R
\right)
\end{eqnarray}
then becomes,  written in terms of the left- and
right-handed two-spinors,
\begin{eqnarray} \label{8.52a}
L^{f^L}_0 & = & - \frac{i}{2}
\left[
\varphi^{L*}
\left
(\sigma^i
\partial_i-\partial_t
\right ) \varphi^L
-
(\partial_i \varphi^{L*} \sigma^i - \partial_t \varphi^{L*})
 \varphi^L
\right]
\nonumber \\
&&
+ \frac{i}{2}
\left[
 \varphi^{R*} \left(
\sigma^i
\partial_i+\partial_t \right ) \varphi^R
-
\left(
\partial_i \varphi^{R*} \sigma^i + \partial_t \varphi^{R*}
\right )
\varphi^R
\right]
\nonumber \\
&&
+\hat\omega
\left(
\varphi^{R*}\varphi^{L} +
\varphi^{L*}\varphi^{R}
\right).
\end{eqnarray}

The Dirac equation reduces in this representation to a pair
of left- and
right-handed
Weyl
equations, with a coupling term proportional to the mass:
\begin{eqnarray} \label{8.3.15e}
(\sigma^i \partial_i
 + \partial_t  )
 \varphi^R &=& i \hat\omega \varphi^L
 \\ \label{8.3.15ee}
(\sigma^i \partial_i
 - \partial_t) \varphi^L &=&- i \hat\omega \varphi^R.
\end{eqnarray}

These standard Dirac relations can be recovered from the
gravitational
Lagrangian if the fermion polarization relations yield a
spinor metric
\begin{equation} \label{norm2}
M = I/E,
\end{equation}
where $I$ denotes the unit matrix and the `energy'
$E = k^4 = - \, k_4$. We
assume that $E$ is positive.

In the previous non-Euclidean case, the Dirac Lagrangian
was obtained from the gravitational
Lagrangian by factorizing the Klein-Gordon operator into
positive- and negative-branch Dirac operators. The
negative-branch solutions were then suppressed by replacing
the
negative-branch Dirac operator by a derivative-free term
proportional to the mass,  making use of
the property that the solutions satisfy the positive-branch
Dirac equation. This
approach fails in the
massless limit because the mass factor vanishes. However, a
modification of this technique can be applied in which  the
wave operator rather than the
Klein-Gordon operator is factorized.

Noting that for a two-spinor $\varphi$ the wave operator may
be written
\begin{equation} \label{8.3.14b}
 \partial_{\lambda}\partial^{\lambda}   \varphi  =
(\sigma^i \partial_i
 + \partial_t  )
(\sigma^j \partial_j
 - \partial_t) \varphi = 0,
\end{equation}
the Lagrangian (\ref{3.31}) may be similarly factorized in
the form
\begin{eqnarray} \label{8.3.14dd}
L^f_0  &=& -\frac{1}{4 E}
\left[
\left(\partial_i\varphi^{R*}
\sigma^i + \partial_t \varphi^{R*} \right)
\left(\sigma^j \partial_j
 \varphi^R- \partial_t\varphi^R \right) \right. \nonumber \\
&&
+ \left.
\left(\partial_i\varphi^{L*}
\sigma^i + \partial_t \varphi^{L*} \right)
\left(\sigma^j \partial_j
 \varphi^L- \partial_t\varphi^L \right)
+ c.c.
\right]
\nonumber \\
&& - \frac{1}{2 E} \hat \omega^2
\left( \varphi^{R*}\varphi^{R} + \varphi^{L*} \varphi^L
\right).
\end{eqnarray}

The expression (\ref{8.3.14dd}), which is quadratic in the
first derivatives, can be reduced to the
standard Dirac form (\ref{8.52a}) of the fermion Lagrangian,
in
which the
first derivatives appear only linearly, by suppressing all
solutions except those which  lie on
the positive Dirac branch and furthermore have positive
energy. This can be achieved by invoking
the
relations (\ref {8.3.15e}), (\ref{8.3.15ee}) to replace the
following derivative expressions by their
 equivalent
non-derivative expressions on the right-hand sides:
\begin{eqnarray} \label{8.3.16e}
(\sigma^i \partial_i
 - \partial_t  )
 \varphi^R &=& i \left(  \hat\omega \varphi^L
+ 2 E \varphi^R \right)
 \\ \label{8.3.16ee}
(\sigma^i \partial_i
 + \partial_t) \varphi^L &=&
- i \left(  \hat\omega \varphi^R
+  2 E \varphi^L \right).
\end{eqnarray}

A
polarization tensor which  yields the
desired relation (\ref{norm2}) while satisfying the trace
and
gauge conditions (\ref{3.13a}),
(\ref{3.13b})  for a
fermion field with an harmonic wavenumber vector
$\mbox{k} = (k_5,0,0,0)$ is given, for example, by
\begin{equation} \label{8.3.14h}
\hat P^{a}_{AB}\psi_a = \frac{1}{\sqrt{2E}}
\left(
\begin{array}{cccc}
0& 0& 0& 0\\
0& 0           &\varphi_1^R & \varphi_2^R   \\
0& \varphi_1^R & \varphi_1^L &  \varphi_2^L \\
0& \varphi_2^R & \varphi_2^L &  -\varphi_1^L
\end{array}
\right).
\end{equation}

In contrast to the polarization tensor (\ref{A.5}) for the
non-Euclidean case, the polarization tensor (\ref{8.3.14h})
remains finite in the limit of zero mass.

The above derivation can be applied similarly to the
negative-energy solutions of the Dirac equation,  the
normalization factor $1/\sqrt{2E}$ in
(\ref{8.3.14h}) being replaced in this
case by $1/\sqrt{-2E}$. One
obtains again the standard Dirac Lagrangian (\ref{8.52a}),
but with an opposite sign. The sign change is
immaterial for the Dirac equation itself, but when
interactions
between fermions and bosons are considered, the fermion
currents appearing as source terms in the boson field
equations enter with an opposite sign. For simplicity, we
will restrict the discussion throughout to the
positive-energy solutions.

For the derivation of the Maxwell-Dirac-Einstein equations
considered in this and the following sections, the above
minimal
models are adequate. In the later application of the metron
model
to
weak and strong interactions in Part~\ref{The Standard
Model}, however, it will be found that a closer
correspondence  with
the Standard Model can be achieved (and the analysis can be
simplified) if an additional
dimension is introduced.  For the present
discussion of the Maxwell-Dirac-Einstein system, the
detailed form of the
polarization tensor is in fact immaterial. We need to know
only that
there exists  a representation which yields
the Dirac equation in the linear approximation, and we will
not need to refer to
the expressions (\ref{A.5}) or
(\ref{8.3.14h}) again until we consider the general
structure of the metron solutions in the context of the
Standard Model in Part~\ref{The Standard Model}.

\section{The Maxwell-Dirac-Einstein Lagrangian}
\label{The Maxwell-Dirac-Einstein Lagrangian}

The interaction Lagrangians $L^f_{em}$ and $L^f_g$
describing the
lowest order coupling of fermion fields to electromagnetic
and
gravitational fields, respectively, are obtained by
substituting the forms (\ref{3.7}) and (\ref{3.12}) for
electromagnetic and fermion fields, respectively, into the
gravitational Lagrangian (\ref{3.24}) and collecting the
relevant cubic interaction terms of the required structure
$(\bar
\psi A \psi)$ and $(\bar \psi g^{grav} \psi)$ (with two
appropriately distributed derivatives). Noting that the
derivatives act only on the perturbation fields, this
involves, in effect, replacing in turn each of the three
$\eta^{..}$ factors occurring in the linearized Lagrangian
(\ref{3.27}) by a perturbation field and considering then
all
possible permutations of the perturbation fields in the
resultant cubic interaction expression.

The interaction Lagrangian can be derived, however, in a
simpler and  more illuminating manner by invoking the
invariance of the Lagrangian $L=P$ with respect to affine
coordinate transformations. We present the derivation in the
following only for electromagnetic interactions. The
gravitational
case can be treated in the same way, but will be considered
in a
more general framework in the following section.

It can be seen
first by inspection of (\ref{3.25}), invoking the gauge and
trace conditions,  that interactions containing
derivatives of the electromagnetic field vanish. The
remaining interactions involving the electromagnetic field
itself can then be determined by carrying out an affine
transformation from the original coordinate system $X$ to
a
local
coordinate system $X'$ in which the
electromagnetic field vanishes at some prescribed physical
spacetime point  $x$, say
$x=0$.  (This is the electromagnetic equivalent of removing
the
gravitational forces by transforming to a local inertial
system.) The
interaction Lagrangian then also vanishes at
$x=0$, the Lagrangian reducing to the free-field form in
which
the fermion component is given by (\ref{3.35})
(assuming a pure positive-branch spinor
field $\psi = \psi^+$;  the index $+$ will be dropped in the
following).
The
 fermion-electromagnetic interaction Lagrangian in the
original  coordinate system can then be
recovered by
transforming from the local system back to the
original
system.

The required transformation is
\begin{eqnarray} \label{3.37}
x'^{A}  & =&    x^{A}     +
\xi^{A}_{\lambda} x^{\lambda} \nonumber \\
\xi'^{\lambda} &=& \xi^{\lambda},
\end{eqnarray}
where (cf.eq.(\ref{3.7}))
\begin{equation} \label{3.38}
\xi^{A}_{\lambda} := \left\{ g^{(a) A }
_{\lambda}\right\}_{x=0}
= a^{A} \left\{ A_{\lambda}\right\}_{x=0}.
\end{equation}

In transforming back to the original coordinates $X$, the
free-field Lagrangian for the electromagnetic field is, of
course,  recovered
unchanged. However, the
affine back-transformation (\ref{3.37}),(\ref{3.38}) is not
applied to the free-fermion Lagrangian, since this would
involve transforming not only the tensor
components
$g'^{(f)}_{AB}$ of the fermion field, but also
the $\gamma$
matrices
and the square harmonic mass term $\hat\omega'^2 =
k'_{A}k'^{A}$.  As pointed out at the end of the
previous section, this would destroy
the basic
structure
of the Dirac equation, which is
invariant
only with respect to Lorentz transformations, not with
respect to the
present affine
transformation.

Thus we define the spinor fields
$\psi$
 at each point in space, in accordance with the
form
(\ref{3.12}), in terms of the fermion components
$g'^{(f)}_{AB}$ of the metric tensor in the
local
frame with vanishing electromagnetic field.
The resultant fields $\psi$ are then
regarded
as functions of the original physical spacetime coordinates
$x$.

Adopting this definition, the fermion
Lagrangian in the
presence of
an electromagnetic field is obtained by simply replacing
the
local-frame derivative $\partial'_{\lambda}$ in the local
free-fermion
Lagrangian (\ref{3.35}) by the derivative with respect to
the original
coordinates,
$\partial_{\lambda}$,
\begin{equation} \label{3.42}
\partial'_{\lambda}  =  \partial  _{\lambda}  -
\xi^{A}_{\;\lambda}
\partial_{A},
\end{equation}
or, applying (\ref{3.38}) and introducing the usual
covariant derivative  notation
$D_{\lambda}$ for $ \partial'_{\lambda}$,
\begin{equation} \label{3.45}
D_{\lambda} : = \partial_{\lambda} - ie' A_{\lambda},
\end{equation}
where the electromagnetic coupling constant is given by
\begin{equation} \label{3.46}
\mbox{
\fbox{
$
e' : = k_{A}a^{A} = k_5 |a|
$}}
\end{equation}
and $k_5$ is the electromagnetic component of the harmonic
wavenumber
vector k of the fermion field.

The covariant derivative $D_{\lambda}$ is seen to be
identical in form
to the covariant derivative of the
QED $U(1)$ gauge group, yielding the standard
fermion-electromagnetic interaction Lagrangian
\begin{equation} \label{3.47}
L^f_{em} = j^{\mu}\;A_{\mu},
\end{equation}
with
\begin{equation} \label{3.48}
j^{\mu} : = i\;e' \left(\bar \psi \; \gamma^{\mu}\;\psi
\right).
\end{equation}

To within a calibration factor, the coupling constant $e'$
can be identified with the
elementary
charge $e$. The calibration factor depends
on the
normalization of the Dirac fields, which is different in the
metron
model and in standard QED.  It will be
determined in the following section.

The gauge invariance of the QED Lagrangian can now be
readily
recognized as the invariance of the gravitational Lagrangian
with
respect to a particular class of coordinate transformations.
Consider an infinitesimal
translation in the harmonic-space direction $a^A$ of
the electromagnetic field,
\begin{eqnarray} \label{3.49}
\check x^{A} & = & x^{A} - \xi(x) \,
a^{A} \nonumber \\
\check x^{\lambda} & = & x^{\lambda},
\end{eqnarray}
where the infinitesimal displacement factor $\xi(x)$ depends
on
physical
spacetime only.
The associated transformation of the metric tensor
\begin{equation} \label{3.50}
g^{L M}   \rightarrow   \check   g^{L
M}(\check   X)   =
\partial_{N}\check x^{L}   \,
\partial_{O} \check x^{M} \, g^{N O}(X)
\end{equation}
yields for the periodic harmonic-space tensor components of
the fermion fields,
for which
$\partial_{A} \check x ^{B} =
\delta^{B}_{A}$,
simply a phase shift,
\begin{equation} \label{3.51}
\check g^{AB}_{(f)}(\check X) =  g^{A
B}_{(f)}(X)  =
\tilde g^{AB}_{(f)}(x) \exp \left \{ \,i\,
k_{A}
\check x ^{A}
+ i\, k_{A}a^{A} \xi \right \},
\end{equation}
so that the spinor field transforms as
\begin{equation} \label{3.52}
\check \psi(x) = \psi(x) \; \exp (i \xi\,  e')
\end{equation}
.

The transformation for the mixed tensor components
representing the
electromagnetic field yields, to linear order in the
deviations from
the reference metric $\eta^{LM}$ (cf.(\ref{2.9})),
\begin{equation} \label{3.53}
\check g^{\lambda A}_{(a)}  (x) = g^{\lambda
A}_{(a)}  (x)  +
\eta^{\lambda \mu}a^{A} \partial_{\mu} \xi.
\end{equation}
Thus the electromagnetic field (eq.(\ref{3.7}))
transforms as
\begin{equation} \label{3.54}
\check A^{\lambda} = A^{\lambda} + \eta^{\lambda
\mu}\partial_{\mu}\xi.
\end{equation}
Equations  (\ref{3.52}), (\ref{3.54}) represent the standard
spinor and
electromagnetic $U(1)$ gauge transformation relations.

As already mentioned, the above treatment of
 fermion-electromagnetic interactions
can be
applied in principle also to fermion-gravitational
interactions. The
deviation of the gravitational field from the flat
background metric
would be regarded again as a perturbation which can be
removed
locally by an appropriate coordinate transformation.
However, it is
more convenient in this case to simply apply the
fully
nonlinear general invariant theory of gravitation in
physical
spacetime, as
indicated in the following section.

\section {Particle interactions}
\label{Particle interactions}
\typeout{################################}
\typeout{################################}
\typeout{        START OF met4-4.tex}
\typeout{################################}
\typeout{################################}

Having demonstrated that we can recover from the
n-dimensional nonlinear gravitational equations the basic
quantum-theoretical free-field equations and, to lowest
interaction order, the coupling between
fermion, electromagnetic and gravitational fields, as
described by the Maxwell-Dirac-Einstein Lagrangian, we turn
now to the next question: can we derive from the matter-free
Einstein equations also particle properties and the classical
picture of point-particle coupling through particle far-field
interactions? To this end we must clearly consider the sources
of the fields, which reside in the localized regions of
strongly nonlinear interactions in the particle kernels.
It will be shown in the following section that the classical
gravitational and electromagnetic source terms
corresponding to quasi-point particles can indeed be derived
from the full-space nonlinear gravitational Lagrangian,
the derivation yielding not only the structure of the source
terms but also the associated physical
constants (mass, gravitational constant and charge) as
functions of the metron solutions. In addition, we shall
explain the extremely weak strength of the gravitational
forces by the higher-order nonlinearity of gravitational
coupling and recover, in the process,  de Broglie's relation
and Planck's constant.

We recall first the classical point-particle
interaction relations which  the metron model must
reproduce.

\subsection*{Classical  point-particle interactions}

The classical equations describing the interactions of point
particles
coupled through gravitational and electromagnetic fields are
given by:
the particle propagation equations (in natural coordinates, with $c=1$)
\begin{equation} \label{4.1}
\frac{du}{ds}^{\lambda}   +    \Gamma^{\lambda}_{\mu
\nu}u^{\mu}u^{\nu}    =
\frac{q}{m}\,F^{\lambda}_{\mu}\,u^{\mu},
\end{equation}
where $m$ and $q$ denote the charge and mass of a particle
and
$u^{\lambda}$ is the particle 4-velocity, with the usual
normalization
$u^{\lambda}u_{\lambda} = -1$; the (linearized) field
equations
\begin{equation} \label{4.2}
\Box\;h_{\lambda \mu} = - 2\,G\,T_{\lambda \mu}
\end{equation}
for the divergence-free gravitational field $h_{\lambda\mu}
=
g'_{\lambda\mu} - \frac{1}{2} \,\eta_{\lambda \mu}
g'^{\nu}_{\nu}$ ,
where $g'_{\lambda\mu} = g_{\lambda \mu} - \eta_{\lambda
\mu}$ is the
perturbation of the gravitational field about the reference
background
metric $\eta_{\lambda\mu}$; and the field equation for the
electromagnetic potential $A^{\lambda}$,
\begin{equation} \label{4.3}
\Box \, A^{\lambda} = j^{\lambda}.
\end{equation}
Here $G$ is the gravitational constant and the source terms
are given
by the energy-momentum tensor
\begin{equation} \label{4.4}
T_{\lambda  \mu}  : =  \sum_i  \,m^{(i)}\,
\int_{T^{(i)}}\;ds
\,u_{\lambda}\,
u_{\mu} \delta^{(4)} (x- \xi (s))
\end{equation}
and the electric current
\begin{equation} \label{4.5}
j^{\lambda} : = \sum_i \; \int_{T^{(i)}}\;ds \,q^{(i)}
u^{\lambda} \delta^{(4)}
(x- \xi (s)),
\end{equation}
where $T^{(i)}$ is the path $x = \xi(s)$ of the particle $i$
(the
index $i$ is dropped, e.g. in eq.(\ref{4.1}) and in the
trajectory
$\xi(s)$, where the reference is clear). In computing the
fields
acting on a given particle $j$, the self-interaction terms
are
excluded, i.e. the source terms are summed over all
particles
$i$ excluding the particle $j$.

Since we shall not be concerned with classical nonlinear
gravity-field interactions, the gravity field equations
(\ref{4.2}) have
been
linearized on the left hand side, although the gravity field
nonlinearities have been retained in the familiar form in
the
gravity
connection term in (\ref{4.1}).

The coupled field-particle equations  (\ref{4.1}) -
(\ref{4.5}) can be derived from the classical action
principle $\delta W_{cl} = 0$, where  $W_{cl} = \int L d^4 x$ and,
specifically \cite{ft18},
\begin{equation} \label{4.6}
W_{cl} : = W_g + W_A + W_{pg} + W_{pA},
\end{equation}
with
\begin{eqnarray} \label{4.7}
W_g     &: =& -  \frac{1}{4}\,   \int
              \left\{
                \partial_{\lambda} g^{\mu\nu}
                \partial^{\lambda}  g_{\mu  \nu}  -
\frac{1}{2}  \,
                \partial_{\lambda} g^{\mu}_{\mu}
                \partial^{\lambda} g^{\nu}_{\nu}
              \right \}\; d^4x \\
\label{4.8}
W_A    &: =& - \frac{G}{2}\, \int \, F_{\lambda
\mu}F^{\lambda\mu}\; d^4x \\
\label{4.9}
W_{pg} &: =&  - 2 G\,  \sum_i \, \int_{T^{(i)}}\; m_{(i)}
             \left\{
                - g_{\lambda\mu} u^{\lambda}_{(i)} \,
u^{\mu}_{(i)}
             \right \}^{1/2} \; ds \\
\label{4.10}
W_{pA}  &: =&  2G   \,      \sum_i   \int_{T^{(i)}}
q_{(i)}A_{\lambda}
             \, u^{\lambda}_{(i)} \; ds.
\end{eqnarray}

The variations are carried out with respect to the particle
paths \cite{ft19} (yielding (\ref{4.1})) and the fields
$g_{\lambda \mu}$  and $A_{\lambda}$ (yielding (\ref{4.2})
and (\ref{4.3}), respectively).

We remark that from the viewpoint of classical
gravitational and electromagnetic interactions it would be
more natural to choose a different
normalization of the Lagrangians in which
all expressions (\ref{4.7}) - (\ref{4.10}) are divided by
$G$. This removes the gravitational
constant  from the electromagnetic Lagrangians
(\ref{4.8}), (\ref{4.10}) and the inertial Lagrangian
(\ref{4.9}), the gravitational constant appearing only in
the free-field gravitational Lagrangian (\ref{4.7}).
However, we have retained here the
same normalization for the 4-dimensional gravitational
Lagrangian as used for the n-dimensional  gravitational
Lagrangian (\ref{3.26}) in the previous section, in which no
physical constants appear,  in order to directly relate
the metron and classical
descriptions of gravitational and electromagnetic particle
interactions.

The system of equations (\ref{4.1}) - (\ref{4.5}) does not
yet
uniquely determine the particle coupling: the field
equations must be
augmented by boundary conditions at infinity. Normally, the
outgoing
radiation condition is invoked. However, this introduces a
time
asymmetry into the problem.  As already mentioned and
discussed in more detail in Sections~\ref{Time-reversal
symmetry}, \ref{The radiation condition}, this is
justified
in macrophysical applications, in which one is normally
concerned with
time asymmetrical solutions in an irreversible macrophysical
world, but
is not appropriate for the microphysical approach adopted
here, which
is founded on the basic postulate of time-reversal symmetry.
Accordingly, the Tetrode-Wheeler-Feynman condition of no net
ingoing or outgoing
radiation will be invoked.  This is required not only for
time-reversal
symmetry, but also, in a relativistic theory, in order to
describe the interactions between a finite set of particles
as a
closed system, in which total energy and momentum are
conserved without
losses to infinity.

Applying this boundary condition to solve for the fields,
the action
integrals $W_g$, $W_A$ for the fields can be expressed in
terms of the
line-integral source terms, and one obtains the action
expression
(extending the distant-interaction result of Wheeler and
Feynman for
electromagnetic coupling, using (\ref{4.2}), (\ref{4.4}), to include
linearized gravity-field
coupling)

\begin{eqnarray} \label{4.11}
\lefteqn{W_{cl}} & & =
    - 2\,  \sum_p \int_{T^{(i)}} \, m_{(i)}\,
             \left \{ -\eta
               _{\lambda \mu} u^{\lambda}_{(i)} u^{\mu}
_{(i)}
             \right \}^{1/2}\, ds
 \\
    & & + \frac{1}{2\pi} \sum_{i,j \atop (i \not= j)}
        \int_{T^{(i)}} \int_{T^{(j)}}
          \left\{
               q_{(i)} q_{(j)} u^{\lambda}_{(i)}
                  u^{(j)}_{\lambda}  +  2\,
                 G^2 u^{\lambda}_{(i)} u^{\mu}_{(i)}
u^{(j)}_{\lambda}
                  u^{(j)}_{\mu}
          \right\}
           \delta \left (\xi^2_{[ij]} \right
)ds_{(i)}ds_{(j)}\nonumber,
\end{eqnarray}
where
\begin{equation} \label{4.12}
\xi^{\lambda}_{[ij]} : = \xi^{\lambda}_{(i)} -
\xi^{\lambda}_{(j)}.
\end{equation}
The form (\ref{4.11}) depends only on the particle paths and
demonstrates explicitly the particle-interaction symmetry
resulting
from the time-symmetrical inclusion of interactions on both
forward
and backward light cones. The result will be generalized in
Section~\ref{Time-reversal symmetry} to
actions-at-a-distance
governed by the dispersive
Klein-Gordon equation.

\subsection*{Point-particle interactions in the metron
picture}

We show now that the basic structure of the classical relations for
interacting point particles listed above are reproduced by the metron
model. Specifically, we show that the Ricci tensor of the n-dimensional
Einstein vacuum equations, integrated over the nonlinear particle-core
regions, yields the energy-momentum tensor and electric current
representing the point-particle source terms of the classical gravitational
and electromagnetic field equations. In the process, we shall determine the
basic physical constants characterizing the interactions and the Planck
constant in terms of the metron solutions.

To derive the classical point-particle action expressions
(\ref{4.6})-(\ref{4.10}) from the n-dimensional
gravitational
action
\begin{equation} \label{4.13}
W_n : = \int \left|g_n \right| ^{1/2} L \, d^nX,
\end{equation}
we divide now the gravitational action integral into
near-
and far-field contributions $W_F$ and $W_{T^{(i)}}$. We
ignore the integral over
harmonic space,
since all fields are assumed to be homogeneous in harmonic
space and
the Lagrangians can therefore be regarded as harmonic-space
averages.
The near-field integrals are defined to extend over the set
of
quasi-line (tube) regions $T^{(i)}$ in the vicinity of the
metron-kernel trajectories in which the local metron field
dominates
over the far field of the other particles, while the
far-field
integral covers the remaining spacetime region $F$:
\begin{equation} \label{4.14}
W_n = \left [ \int_F + \sum_i \int_{T^{(i)}} \right ]
\sqrt{-g_4}\, L \, d^4x  = :
W_F + \sum_i  W_{T^{(i)}}.
\end{equation}

In accordance with the linearization of the gravitational field
assumed in (\ref{4.7}), the Lagrangian need be considered in
the far field region $F$ only to quadratic order, and in the
near field regions $T_{(i)}$ only to linear order with respect
to the far fields. It is assumed that the near-field integrals
of the interaction Lagrangian converge sufficiently rapidly
that the near-field region can be regarded formally as
extending to infinity in the evaluation of the action integrals
over the metron `tubes'.  Conversely, the near-field 'holes'
in the integration region F for the free field
Lagrangian are assumed to be sufficiently small that the far
fields can be smoothly interpolated across the metron trajectories
(the interpolated fields formally define the coupling fields
in the non-self-interaction point-particle limit).

In accordance with the classical point-particle interaction
theory, the nonlinear self-interaction terms in the metron
core regions will be ignored.  Although these are essential
in determining the internal structure of the metron solutions,
they do not affect the far-field coupling between different
particles and need therefore not be
considered in the point-particle interaction limit with
which we are concerned here.

Assuming that the metric field $g_{AB}$ consists only of the
four-dimensional spacetime metric $g_{\lambda \mu}$ and the
electromagnetic mixed-index tensor field $g_{A
\lambda}$, as
defined by
eq.(\ref{3.7}) together with the gauge condition
(\ref{3.9}), the
general free-field gravitational Lagrangian (\ref{3.28})
yields for
the far-field action term $W_F$ in (\ref{4.14}), to lowest
quadratic
order
\begin{equation} \label{4.15}
W_F = W'_g + W'_A,
\end{equation}
where $W'_g$ represents the gravitational action for
four-dimensional spacetime, which is trivially identical to
the classical gravitational action $W_g$ \cite{ft20},
\begin{equation} \label{4.16}
W'_g = W_g,
\end{equation}
while the metron equivalent of the
action for
the electromagnetic free field is given by
\begin{equation} \label{4.17}
W'_A =  -  \frac{1}{4}  \,  a_{A}  a^{A}  \int  \,
F_{\lambda  \mu}
F^{\lambda \mu} d^4x.
\end{equation}
Comparing this result with the classical electromagnetic
action expression (\ref{4.8}) and the definition (\ref{3.7})
for the electromgnetic field, it follows that the normalization of the
constant constant metron electromagnetic vector
$a_{A}$ in (\ref{4.8}) must be defined as
\begin{equation} \label{4.18}
\mbox{
\fbox{
$
a_{A} a^{A} =  |a|^2 =2 G
$}}
\end{equation}
This implies, in particular, that the sign of the
electromagnetic part of the background
metric must be positive,
$\eta_{55} = 1$, in order to reproduce the correct sign of
the electromagnetic forces, as expressed by the negative
sign in the definition of the electromagnetic action $W_A$
in (\ref{4.8}).

It appears at first sight curious that the gravitational
constant
should be related to a property of the metron's
electromagnetic field. However, as discussed above, this
follows from the
appearance of $G$ in the classical electromagnetic action in
a unified gravitational-electromagnetic description of
field-particle interactions, in which the normalization of
the Lagrangians is derived from the original n-dimensional
gravitational Lagrangian.

The metron expressions for the remaining physical constants
characterizing gravitational and electromagnetic particle
interactions, namely the particle mass $m$ and charge
$q$, follow from the line-integral action expressions. The
determination of the charge will yield also the calibration
of the  electromagnetic coupling constant $e'$, which was
defined through eq.(\ref{3.46}) only to within an
undetermined normalization factor.

To match the line integrals of the classical action with
the corresponding tube integrals of the metron
action expression, we first reduce the tube integrals in
the metron action
(\ref{4.14}) to line integrals by integrating across the
tube cross-sections.  Introducing for this purpose local 4d
coordinates
$\check x$ defined with respect to the restframe with
local spacetime
metric $\eta_{\lambda
\mu}$, in which the
volume element in the action integral is given by
$\sqrt{-g_4} d^4x=d^3
\check {\bf x} d \check x^4 =  d^3  \check  {\bf  x}   \left
(-g_{\lambda \mu} u^{\lambda} u^{\mu} \right )^{1/2} ds$,
the
action integral over a trajectory may be written
\begin{equation} \label{4.19}
W_{T^{(i)}}  = \int_{T^{(i)}} \sqrt{-g_4} \, L\,d^4x =
\int_{T^{(i)}} <\!L\!>  \left
( -g_{\lambda \mu} u^{\lambda} u^{\mu} \right ) ^{1/2}ds,
\end{equation}
where $<\!L\!>$ denotes the integration across the
three-dimensional tube
cross-section in the coordinate system $\check
{\bf x}$.

We assume that to first order, in accordance with the
lowest-order interaction analysis of Sections~\ref{Lagrangians},
\ref{The Maxwell-Dirac-Einstein Lagrangian}, the principal
contribution to the integral across the cross-section comes
from the extended `weak coupling' region, in which the metron
field of particle~$i$ can be
represented by the metron free-wave Dirac field, as
determined
by eq.(\ref{3.18}), modified only by the far field
$g^{(F)}_{AB}$ of the
remaining particles. In the local particle restframe with
metric
$\eta_{\lambda
\mu}$, the only external mean field is $A_{\lambda}$ \cite{ft21}.
Retaining only the electromagnetic far field, the integral of
the Lagrangian across the tube cross-section is given,
according to  eqs. (\ref{3.45})-(\ref{3.48})), by
\begin{equation} \label{4.20}
<\!L\!>  =  -  \left  \langle  \bar  \psi  \left  (
\gamma^{\lambda}   \left   [
\partial_{\lambda} - ie'A_{\lambda} \right ]  +  \hat
\omega  \right  )  \psi
\right \rangle.
\end{equation}

For a solution of the n-dimensional gravitational equations
in full space, the
variation of the gravitational action must vanish for
arbitrary
variations of the fields.  In establishing the equivalence
with
the
classical action for interacting point particles, however,
the
variations can be restricted to: a) variations in the far
fields
$g_{\lambda \mu}$ and $A_{\lambda}$ without changes in the
metron
trajectories and the associated metron near fields $\psi$;
and b)
variations in the metron trajectories without changes in the
metron
near fields $j$ relative to the local inertial frame of the
changed
path, and without modification of the far fields. Variations
in the
metron near fields $\psi$ need not be considered, since they
yield the
nonlinear field equations determining the internal structure
of the
metron solutions, which are assumed for the present
discussion to be
given. It is assumed also that the far fields have
negligible impact on the structure of the near fields in the
nonlinear metron core region.

Variation of (\ref{4.19}) with respect to the
electromagnetic field
$A_{\lambda}$ yields the electric current
appearing  as
source term in the electromagnetic field equations. The
relevant
component in (\ref{4.20}) which determines this source term
is the tube-averaged interaction Lagrangian
\begin{equation} \label{4.21}
\left \langle L_{int} \right \rangle = i \, e' \left \langle
\bar \psi  \gamma
^{\lambda} \psi \right \rangle A_{\lambda}.
\end{equation}

We assume that the particles are isotropic, so that the
vector
$v^{\lambda}=
i<\!\bar\psi \gamma^{\lambda} \psi\!>$ in (\ref{4.21})
reduces
in the
metron restframe to the
fourth component
\begin{equation} \label{4.22}
v^4 = i < \bar \psi \gamma^4 \psi> = <\psi^* \psi> =: \beta
=
\mbox{const}.
\end{equation}
The isotropy assumption is consistent with the neglect of
the far fields on the structure of the metron fields in the
nonlinear particle core region. It implies, in particular,
that we ignore the coupling of the particle spin to the
magnetic far field \cite{ft22}.
In the global frame, the mean interaction Lagrangian then
takes the
form (noting that $v^{\lambda}$ transforms in the same way
as
$u^{\lambda}$ and must therefore be parallel to
$u^{\lambda}$)
\begin{equation} \label{4.23}
<L_{int}> = \beta e'  A_{\lambda}  u^{\lambda}  \left  (-
g_{\mu  \nu}  u^{\mu}
u^{\nu} \right )^{-1/2}.
\end{equation}
The metron line integral representing the influence of the
particle
charge on the electromagnetic field -- and also the
complementary influence of the electromagnetic field on the
particle
trajectory -- thus becomes
\begin{equation} \label{4.24}
W'_{pA}   =   \sum_i   \int_{T^{(i)}}   \beta_{(i)}
e'_{(i)}    A_{\lambda}
u^{\lambda}_{(i)} \, ds.
\end{equation}
This is seen to be identical to the corresponding classical
expression (\ref{4.10})
if the metron and classical physical constants are related
through
\begin{equation} \label{4.25}
\mbox{
\fbox{
$
e'_{(i)}\, \beta_{(i)} = 2 \, G \, q_{(i)}
$}}
\end{equation}
For the special
case that the
particle is an electron with elementary  charge $e$,  eq.
(\ref{4.25}) yields the calibration factor relating the
non-normalized coupling coefficient $e'$ of  eq.(\ref{3.46})
to the elementary charge:
\begin{equation} \label{4.29a}
\mbox{
\fbox{
$
e' = \frac{2G}{\beta_e} \, e
$}}
\end{equation}
where $\beta_e = <\psi^*
\psi>_e $.

Expressed in terms of the electromagnetic wavenumber component $k^{(i)}_5$,
eq.(\ref{4.25}) may be written, using (\ref{3.46}),
\begin{equation} \label{4.29b}
k^{(i)}_5 = \frac{(2G)^{1/2}}{\beta_{(i)}} q_{(i)}
\end{equation}
Thus the wavenumber component $k^{(i)}_5$ determines the electric charge.

 For a reference particle, for example the electron, eq.(\ref{4.29b}) may
be regarded as the defining equation for the (so far undetermined)
reference length scale of the metron solutions (the scaling of the other
metron property $\beta_{(i)}$ in (\ref{4.29b}) is not free, but is fixed
by the definitions of the background metric and fermion metric). Once the
reference scale has been determined through the elementary charge of the
electron, the metron solutions should predict the charges of all other
particles (provided the harmonic wavenumbers of other metron solutions are
computed and not prescribed, cf. discussion in Part~\ref{The Metron
Concept}).

In order that the coefficient $\beta$ in eq.(\ref{4.25}) is finite, the
integrals in eqs.(\ref{4.21}),
(\ref{4.22}) must converge. This implies that the spinor
field $\psi$  must be a genuine trapped field which falls
off exponentially rather than as $1/r$, in accordance with a free wave,
for large distances
$r$ from the particle core (cf. discussion in
Section~\ref{The mode-trapping mechanism}). However, the
e-folding
length scale $l_e$ of the field $\psi$ must be large
compared with typical atomic scales in order that the
periodic (de Broglie) far field can produce resonant
interference phenomena
(cf. Sections~\ref{Bragg scattering}-\ref{Atomic spectra}).
It will be shown in the
following
section that this condition can indeed be satisfied, and that the length
scale $l_e$ can be inferred from the ratio of gravitational
to electromagnetic forces.

Turning now to variations with respect to the
gravitational field
$g_{\lambda \mu}$, we may replace the expression
(\ref{4.19}) for $W_{T^{(i)}}$ by the action expression
\begin{equation} \label{4.26}
W'_{pg}  =  -  \sum_i  \int_{T^{(i)}}  M_{(i)}  \left  \{
-g_{\lambda  \mu}
u^{\lambda} _{(i)} u^{\mu}_{(i)} \right \}^{1/2} \, ds,
\end{equation}
in which
\begin{equation} \label{4.27}
M_{(i)} : = - <\!L\!>_{(i)} = \left <\bar  \psi  \left
(\gamma^{\lambda}  \left  [
\partial_{\lambda} - ie' A^0_{\lambda} \right ] + \hat
\omega  \right  )  \psi
\right >_{(i)}
\end{equation}
is treated as constant, i.e $A_{\lambda}$ is not varied but
is
regarded as a given function $A^0_{\lambda}$ along the
trajectory.

We note that the second factor on the right hand side of
eq.(\ref{4.27}) corresponds to the field equation for a
Dirac field $ \psi$ interacting with an electromagnetic
field $A^0_{\lambda}$ and is thus zero to lowest order. The
mass $M_{(i)}$ must therefore represent a higher-order
property of the nonlinear metron solution. The apparent
dependence of
$M_{(i)}$ on $A_{\lambda}$ in eq.  (\ref{4.27}) also vanishes to
first
order, so that $M_{(i)}$ is essentially a constant property of the
metron
solution (apart from a possible
higher-order
dependence of the nonlinear core region on the external
electromagnetic far field, which we have neglected).
The fact that the mass vanishes to lowest interaction order
is the origin of the weakness of gravitational forces. We
return to this question in the next section.

The form (\ref{4.26}) is identical to the
corresponding
classical form (\ref{4.9}) for $W_{pg}$,  where the metron
and classical
physical
constants are related in this case through
\begin{equation} \label{4.28}
\mbox{
\fbox{
$
M_{(i)} = 2 \, G \, m_{(i)}.
$}}
\end{equation}
Since the free length scale has already been fixed by the elementary
charge,
we have no free scaling factors left in the metron solution. Thus
eq.(\ref{4.28} implies that the particle masses -- or, equivalently, the
dimensionless gravitational/electromagnetic coupling ratios $G
m_{(i)}^2/e^2$ -- are determined by the metron solutions.

Finally, it remains to be confirmed that the variation of
the metron line-integral
action $W_{T^{(i)}}$ with
respect to the particle trajectories $\xi_{\lambda}(s)$ is
equivalent
to the variation with respect to $\xi_{\lambda}(s)$ of the
two action expressions $W'_{pA}$,
$W'_{pg}$, which were inferred from the field
variations, i.e. that
$W'_{pA}$ and $W'_{pg}$ yield not only the source terms in
the field
equations but also the particle trajectory equations. For
the action $W'_{pA}$ this is obvious from the derivation. In
the case of the action $W'_{pg}$, however, this must still
be demonstrated, since the replacement of  $W_{T^{(i)}}$ by
the action expression $W'_{pg}$ was justified only for
variations with respect to $g_{\lambda \mu}$.

To distinguish between the contributions which arise from the
variations induced in the electromagnetic field by changes
in the trajectory from the variations of
the trajectory itself, the tube-averaged Lagrangian may be
written in the form (dropping now the index $i$)
\begin{equation} \label{4.29}
<\!L\!>\, =\,<\!\!L>^0+ <\!L\!>',
\end{equation}
where $<\!L\!>^0$ is defined in
eqs.(\ref{4.26}), (\ref{4.27})
and $<\!L\!>'$ represents
the variation of $<\!L\!>$ arising from variations in the
electromagnetic
field.  Thus $<\!L\!>'$ is zero along the trajectory itself,
but
is
non-zero in the neighbourhood of the trajectory. Variation
of $W_T$
with respect to $\xi _{\lambda}(s)$ yields
\begin{eqnarray} \label{4.28a}
\delta W_T &=& \int_T \left\{ -M  \left(  \partial_{\lambda}
-   \frac{d}{ds}
\frac {\partial}{\partial  u^{\lambda}}  \right)  \left(
g_{\mu  \nu}  u^{\mu}
u^{\nu} \right)^{1/2} \right. \nonumber \\
&+ & \left. \left( \partial_{\lambda} - \frac{d}{ds} \frac
{\partial}{\partial u^{ \lambda}} \right) \left\{ <\!L\!>'
\left(  g_{\mu  \nu}
u^{\mu}  u^{\nu}  \right
)^{1/2} \right \} \right \} \delta \xi^{\lambda} (s) \, ds.
\end{eqnarray}
The first term in (\ref{4.28a}) is identical to the result
obtained by
varying the action $W_{pg}'$ with respect to the trajectory
and yields
the geodetic contribution to the trajectory equations. The
second term
can be recognized as the expression obtained by varying the
action
$W'_{pA}$ with respect to the trajectory; it yields the
Lorentz force
term.

Thus the metron model reproduces the classical action
describing the
coupling of point particles through gravitational and
electromagnetic
fields and determines the gravitational constant $G$ and
particle mass $m$ and charge
$q$ through
the three equations (\ref{4.18}), (\ref{4.25}) and
(\ref{4.28}).

\subsection*{The ratio of gravitational to electromagnetic
forces and Planck's constant}

One of the fundamental properties of nature which the metron
model
must explain is the extremely small ratio $\epsilon :=
G(m/q)^2$ of
gravitational to electromagnetic forces.  For the electron,
the ratio is
\begin{eqnarray} \label{4.30aa}
\epsilon_e = G (m_e/e)^2 &=& (6.67\; 10^{-8} \mbox{cm}^3
g^{-1} s^{-2})
 \left
\{(9.12\; 10^{-28}g) / (4.80\; 10^{-10} \mbox{esu}) \right
\}^2 \nonumber \\
&=& 2.4\;
10^{-43}. \nonumber
\end{eqnarray}
According to the metron model (cf.
eqs.(\ref{3.46}),
(\ref{4.18}), (\ref{4.25}) and (\ref{4.28})),
\begin{equation} \label{4.30}
\epsilon = \frac{1}{2} \left \{ \frac{M}{\beta \, k_5}
\right \}^2.
\end{equation}

As has already been pointed out, the definitions of M,
eq.(\ref{4.27}), and $\beta$, eq.(\ref{4.22}), yield a
very interesting result. Whereas $\beta$ is a finite
quantity to lowest linear order in the
metron field $\psi$, the expression for $M$ vanishes to this
order. The mass $M$ must therefore be determined by
higher-order nonlinearities of the metron solution.  Thus
while the source terms for the electromagnetic far fields are
governed to first order by the extensive weak interaction
region outside the strongly nonlinear core region, to compute
the source terms for the gravitational far fields we must
consider the higher-order nonlinear coupling within the metron
core.

This requires an extension of the electromagnetic interaction
analysis of Section~\ref{The Maxwell-Dirac-Einstein Lagrangian}.
Additional interactions involving the coupling between two
fermion fields $f_1$ and $f_2$ through electroweak boson or
strong-interaction gluon fields $b_{1 \bar
2}$ ,
in accordance with the Standard Model picture, will be considered
later in Part~\ref{The Standard Model}. However, to
lowest interaction order these again
represent cubic interactions of the same form ($\bar f_1
b_{1 \bar
2} f_2$) as the electromagnetic interactions, except that
two
different fermion fields $f_1$,$f_2$  are now involved. They
therefore also yield no
contributions to M at lowest interaction order: generally,
$<\!L\!>$ vanishes for all
Lagrangians which are linear in the individual fermion fields
(adjoint fields being regarded formally as independent fields).
The lowest-order interaction Lagrangian which yields a
gravitational mass term is of the form
$L^g_{int} = (\bar f_1 \bar f_1 b_{11 \bar 2} f_2)$ or
$(\bar
f_1 \bar f_1 f_{11 \bar 2} f_2)$,
where $b_{11 \bar 2}$ or $f_{11 \bar 2}$ represent
higher-order boson or
fermion coupling fields, respectively.

Assuming that the coupling is
mediated, as in the Standard Model, by bosons rather than
fermions, it
can readily be seen, by inspection of the general form of
the
gravitational Lagrangian (\ref{3.24}), (\ref{3.25}) (noting
that boson
fields are defined as mixed-index fields and invoking the
gauge
condition (\ref{2.7})), that in the metron restframe
$L^g_{int} \sim
k_4^{(2)}$, i.e.  the gravitational interaction Lagrangian
is
proportional to the frequency $k_4$ of the fermion field
$f_2$.
This was identified in Section~ \ref{Identification of
fields}
with
the (de Broglie) particle mass.
The structure of these higher-order interactions will not be
investigated here.  However, it appears reasonable to assume
that the
dominant interactions will involve a single fermion field
$f_1 =
f_2 = \psi$, so that one can simply write
\begin{equation} \label{4.31}
<L^g_{int}> = - 2\, G' \, k_4,
\end{equation}
where the constant $G'$ depends on the geometrical structure
of the
metron
solution (or the relevant parton components of the
metron solution which determine the particle mass).  If it
is
assumed, finally, that the relevant partons all
have the
same basic structure with respect to this higher-order
gravitational
interaction Lagrangian, so that $G'$ is a universal
constant, the
gravitational mass can be identified with the de Broglie
mass, defined
by $m = \hbar \, k_4$, and the constant $G'$ yields the
Planck constant
\begin{equation} \label{4.32}
\mbox{
\fbox{
$
\hbar = G'/G
$}}
\end{equation}

For a quartic interaction of the assumed form $(\bar
f_1 \bar f_1 b_{11 \bar 2} f_2)$, the
ratio $\epsilon$ of gravitational to electromagnetic forces,
eq.(\ref{4.30}), can be readily estimated. Noting that
the polarization factor relating the
metric field to the the fermion field has the
dimension $k^{-1/2}$ (eq.(\ref{A.5})), the
mass term $M = - <L^g_{int}>$ is of order
$(g_m^6 l_m^3 k_m^2)$, where $l_m$ is the spatial
extent of the strongly nonlinear core region and $k_m,
g_m$ denote the orders of magnitude of the harmonic
wavenumber and  amplitude, respectively,  of the parton
field which generates the mass term in the core region.
We take  $l_m$ to be of the same order as $k_m^{-1}$.

The term $\beta$, on the other hand, is of order $(g_e^2
l_e^3 k_e)$, where $k_e, g_e$ denote the orders of
magnitude of the relevant wavenumber and amplitude,
respectively, of the
charge-generating parton field in the nonlinear core region
and $l_e$ is the e-folding scale of the (weak)
exponential fall-off of the field. It was pointed out in
Section~\ref{The mode-trapping mechanism} that
$l_e$ must be
 large compared with the parton wave length, and, in fact,
large compared with the atomic scale $10^{-8}\, \mbox{cm}$,
in
order that there exists a far field of sufficient extent to
produce the interference phenomena discussed later in
Sections~\ref{Bragg scattering} and \ref{Atomic spectra}.
In Part~\ref{The
Standard Model} it will be shown that in the metron picture
of the Standard Model  the
wavenumbers $k_e \approx k_m$ and $k_5$ of the electron are
of comparable
magnitude.
This is
generally the case if the `harmonic mass' $\hat \omega =
(k_{A}k^{A})^{1/2} = O(k_m)$ (eq.(\ref{2.6})) is
determined mainly by
the wavenumber component $k_5$.  This
holds for the electron, but not for nucleons, for which
$\hat \omega
= O(10^3) k_5$. For the electron, the metron ratio of
gravitational to
electromagnetic forces is thus given by (for nucleons the
ratio is accordingly of
order $10^6$ larger)
\begin{equation} \label{4.37}
\epsilon_e = 0 \,
\left(
[g_m^6/ g_e^2]^2
[l_m^3/l_e^3]^2
\right).
\end{equation}

Setting, for lack of other information, $g_m^6/
g_e^2 = O\,(1)$ (for strongly nonlinear fields,
one would expect typically $O(g^{(m))} =O(g^e) = O(1)$),
the
experimental value $\epsilon_e \approx 10^{-43}$ yields
\begin{equation} \label{4.38}
l_m/l_e = O(10^{-43/6}) \approx 10^{-7}.
\end{equation}
Taking $l_m$ to be of the order of the nucleus scale $10^{-
13}\, \mbox{cm}$, it follows that $l_e = O\, (10^2) \times$
atomic scale ($ 10^{-8}\,$cm), i.e. $l_e$ is large compared
with the atomic scale, as required for effective
interference phenomena. However, the scale separation factor
$10^2$ is not exceedingly large, implying a
limitation on the resonant sharpness of, for example, Bragg
scattering
phenomena (cf. Section~\ref{Bragg scattering}). This
could conceivably be detected by experiments.

As a side comment we note that the proportionality of
the coupling
constants $q_{(i)}$ and $m_{(i)}$ for electromagnetic
and
gravitational forces to the respective wavenumber components
$k_5$ and
$k_4$ (eqs.  (\ref{3.46}), (\ref{4.25}), (\ref{4.28}) and
(\ref{4.31})) is
consistent with the particle and anti-particle definition
given in
Section~\ref{Identification of fields}.  For an
anti-particle, the
electric charge, being
proportional to a harmonic-space wavenumber component, is of
opposite
sign to that of a particle, while the mass, which is
proportional to a
physical-spacetime wavenumber component, is the same for
both particle
and anti-particle.

In summary, the metron picture of
the Maxwell-Dirac-Einstein system is able to  explain, in
terms of properties of the trapped-mode metron solution, the
origin
of
gravitational and electromagnetic forces, the magnitudes of
the
masses, charges and coupling constants which characterize
these
forces, and the de Broglie relation, including Planck's
constant. Distinguishing between dimensional physical constants which
follow from the normalization of the metron solutions and dimensionless
physical constants which represent genuine predictions of the metron model,
the metron model yields the ratio $G m^2/e^2$ of gravitational to
electromagnetic forces, the fine-structure constant $e^2/\hbar$ and the
charge and mass ratios $q_{(i)}/e,\, m_{(i)}/m_e$ for all particles.
Whereas forces arise already at
lowest order in the interaction between the particle core
region and an external electromagnetic far field, the
corresponding gravitational forces vanish to lowest
interaction order and must therefore be described by
higher-order nonlinearities. The large disparity in the
strengths of the
gravitational and electromagnetic forces is accordingly
attributed to the disparity in the spatial scale of the
strongly nonlinear interaction region in the metron core (of
the order of the nucleus scale), which determines the
particle mass, and the weakly nonlinear far-field region (of
the order of $10^2$ times the atomic scale), which defines
the electric charge of the particle.

\part{Quantum Phenomena}
\label{Quantum Phenomena}
\typeout{################################}
\typeout{################################}
\typeout{ START OF met4-5.tex}
\typeout{################################}
\typeout{################################}
%
{\em ABSTRACT} \\

\noindent
In the third part of this four-part paper we apply the
unified, deterministic model of particles and fields based
on the postulated existence of soliton-type ({\it metron})
solutions of the higher-dimensional gravitational equations,
which was summarized in Part \ref{The Metron Concept} and
developed in more detail for  the Maxwell-Dirac-Einstein
system in Part \ref{The Maxwell-Dirac-Einstein System}, to
explain various quantum phenomena. The  wave-particle
duality paradoxes, which motivated
the formulation of quantum theory, are resolved in terms of
the
deterministic metron picture. The widely held view, based
on Bell's theorem for the EPR experiment,  that
deterministic
hidden-variable theories are inherently incapable of
explaining microphysical
phenomena, is shown to be invalid for the metron model.
Essential for Bell's theorem is the existence of an arrow of
time, which contradicts the time symmetry of
the metron model.  Following a general discussion of time
symmetry, the metron interpretation of the
EPR experiment  is presented.

The wave-like interference phenomena of microphysics  are
explained by the periodic ({\it de Broglie}) far fields of
the
metron particles. The appearance of interference patterns in
particle scattering distributions is attributed to resonant
interactions between the particles and the
scattered wave fields.  The mechanism is illustrated for
Bragg scattering and atomic spectra. In the latter case, the
existence of discrete atomic spectra results from the
resonant interaction between an eigenmode of the metron
 Maxwell-Dirac system (which is identical to QED at the
lowest-order tree
level) and the orbiting electron. For circular orbits the
resonant condition reproduces the Bohr condition. Thus the
metron picture of atomic spectra represents an interesting
amalgam of QED and the original Bohr orbital theory.
The metron formalism for computing
radiation-induced or spontaneous transitions between
discrete atomic states is shown to be essentially
identical to the QED computations at the tree level. It is
anticipated, but not demonstrated, that higher-order metron
computations will not encounter divergence problems. It
remains also to be investigated whether higher-order
computations with the metron model will
reproduce observed atomic spectra to the same
accuracy as QED.\\

\subsection*{\raggedright Keywords:}
{\small
metron ---
unified theory ---
wave-particle duality  ---
higher-dimensional gravity ---
solitons ---
EPR paradox ---
Bell's theorem ---
arrow of time ---
Bragg scattering ---
atomic spectra}\\

%
{\em R\'ESUM\'E} \\

\vspace*{1ex}
Dans la troixi\`eme partie de ce
travail, le mod\`ele unifi\'e d\'eterministe
des champs et particules, qui, comme nous l'avions
r\'esum\'e
dans la premi\`ere partie et d\'evelopp\'e en d\'etail
pour le syst\`eme de Maxwell - Dirac - Einstein dans la
deuxi\`eme
partie,
est fond\'e sur l'existence postul\'ee
de solutions de type soliton (dite m\'etrons) des
\'equations
gravitationnelles \`a haute dimension. Ce mod\'ele est
utilis\'e
afin d'expliquer les differents ph\'enom\`enes quantiques.
Les paradoxes provenant de la dualit\'e onde - corpuscule et
qui avaient abouti \`a la formulation de la th\'eorie
quantique,
sont r\'esolus gr\^ace \`a l'introduction du point de vue de
m\'etron d\'eterministe.
Nous d\'emontrons ici que l'id\'ee
g\'en\'eralement admise
qui s'appuie sur le th\'eor\`eme de Bell concernant
l'exp\'erience d'EPR, et qui consid\`ere que toute th\'eorie
d\'eterministe \`a variables cach\'ees est incapable
par inh\'erence d'expliquer les ph\'enom\`enes
microphysiques,
n'est pas valable dans le cas du mod\`ele de m\'etron.
Dans le th\'eor\`eme de Bell, l'essentiel est l'existence
d'une
fl\`eche du temps: ce qui contredit la sym\'etrie
d'inversion
temporelle du mod\`ele de m\'etron.
Tout en poursuivant une discussion g\'en\'erale sur
la sym\'etrie d'inversion temporelle,
nous pr\'esentons ici l'interpr\'etation de m\'etron de
l'exp\'erience d'EPR.

En microphysique les ph\'enom\`enes d'interf\'erence
ayant un aspect ondulatoire, sont compris gr\^ace aux
champs p\'eriodiques de distance (champs de {\it de
Broglie})
des particules de m\'etron.
L'apparition de figures d'interf\'erence dans la
distribution
des particules diffus\'ees est attribu\'ee aux
interactions r\'esonantes entre particules et champs
ondulatoires
diffus\'es. Ce m\'ecanisme trouve son illustration dans le
cas
de la r\'etrodiffusion de Bragg rattach\'ee aux spectres
atomiques.
Dans le dernier cas, l'existence de spectres atomiques
discrets
r\'esulte d'interactions r\'esonantes entre un mode propre
du m\'etron du syst\`eme de Maxwell - Dirac (qui est
identique
\`a la QED
si l'on exclut de la s\'erie de perturbation les
diagrammes de Feynman qui contiennent des boucles)
et l'\'electron tournoyant.
Dans le cas des orbites circulaires, la condition de
r\'esonance reproduit
les r\`egles de quantification de Bohr.
Ainsi le point de vue de m\'etron des spectres atomiques
repr\'esente-t-il un amalgame int\'eressant entre QED et
la th\'eorie quantique originelle des orbites de Bohr.
Nous d\'emontrons que le formalisme de m\'etron servant \'a
calculer
les transitions spontan\'ees induites par radiation,
parmi les \'etats atomiques discrets,
est essentiellement identique \`a celui
de la QED
si l'on exclut les
diagrammes de Feynman ayant des boucles.
Bien que ce ne soit pas d\'emontr\'e, nous supposons que le
calcul des ordres sup\'erieurs ne conna\^itra pas de
probl\`emes
de divergences.
De m\^eme il reste \`a examiner si le
calcul d'ordres sup\'erieurs
reproduisera les spectres atomiques observ\'es avec une
pr\'ecision \'egale \`a celle atteind par la QED.\\

\subsection*{\raggedright Mots cl\'es:}
{\small
m\'etron ---
th\'eorie unifi\'ee ---
dualit\'e onde-corpuscule ---
th\'eorie de gravitation \`a haute dimension ---
solitons ---
paradoxe d'EPR ---
th\'eor\`eme de Bell ---
fl\`eche du temps ---
r\'etrodiffusion de Bragg ---
spectres atomiques}
\newpage
\section{Introduction}
\label{Introduction 3}
The metron representation of gravitational and
electromagnetic
interactions developed in Part
\ref{The Maxwell-Dirac-Einstein System} of this paper
can be
readily
extended to weak and strong interactions. However, before
pursuing the interaction
analysis further in Part \ref{The Standard Model}, we return
first
to
some of
the more fundamental questions raised in the overview of the
metron model in Part \ref{The Metron Concept}.
These can
be addressed now within the framework of the
metron
picture of the Maxwell-Dirac-Einstein system
which has already emerged.

In the first two sections we consider the
question
of time-symmetry and the origin of irreversibility.  In
Section~\ref{The EPR paradox and Bell's theorem} it
is
then
shown
that the EPR paradox can be readily resolved by the
metron
model, the
conflict with Bell's fundamental theorem on the inherent
incompatibility of the EPR experiment with any causal
(in the sense of directed-time) hidden-variable
interpretation of the experiment
being avoided by the
time-symmetry of the metron model.

The remaining Sections~\ref{Bragg scattering} -
\ref{Atomic spectra}
address
general questions of wave-particle duality. Bragg scattering
is chosen in Section \ref{Bragg scattering} as a simple
example illustrating the dual nature of the metron particle
model. The interference patterns of Bragg-scattered particle
beams are explained by resonant interactions between the
scattered de Broglie far fields of the particles  and the
periodic fields within the particle core from which the de
Broglie fields emanate. Similar resonant interactions
between particles and scattered de Broglie waves are invoked
in Section~\ref{Atomic spectra}  to explain the
existence of discrete atomic spectra. Here the resonance
occurs between the orbiting electron (in accordance with
Bohr's original picture) and the eigensolutions of the
standard Maxwell-Dirac equations for a fermion field in a
Coulomb potential. In the last sub-section of
Section~\ref{Atomic spectra} it is shown that the
computation of spontaneous
or forced emissions in the metron model is closely analagous
to
the standard QED computations.

The examples chosen represent only a small selection of the
many quantum
phenomena which the metron model must be able to explain.
However, they
capture the salient features of the metron approach to the
resolution of
the wave-particle duality problem, and the application of
these concepts
to other phenomena is basically straightforward.  After
addressing these
basic questions
we return in the last Part \ref{The Standard Model} of this
paper to the
general
interaction analysis. Through the introduction of more than
one fermion field, in the form of leptons and quarks with
different flavors and colors, together  with appropriate
weak and
strong coupling bosons, the metron picture of the
Maxwell-Dirac-Einstein system is extended in a natural
manner to
an interpretation
of the Standard Model.

\section{Time-reversal symmetry}
\label{Time-reversal symmetry}

The Tetrode-Wheeler-Feynman representation of the
time\--sym\-met\-rical elec\-tro\-magnetic distant
interaction between point particles, which was generalized
already in Section \ref{Particle interactions} to
gravitational forces, can be readily extended to
arbitrary interactions, including periodic coupling fields.
We consider, as in Section \ref{Particle interactions},
interactions between the near fields within a
particle `tube' $(i)$ and the far field
\begin{equation} \label{5.a1}
g_{L M}^{(j)}=\hat{g}_{L
M}^{(j)}\exp (ik^{(j)}_{A}
x^{A}) + c.c.,
\end{equation}
defined again as the deviation from the n-dimensional
background metric $\eta_{L M}$, of a distant particle $(j)$
\cite{ft23}.
The field is now allowed to be periodic with respect to the
harmonic coordinates, but can represent also, as before, an
electromagnetic, gravitational or, possibly, neutrino field
with
$k_{A}^{(j)} =0$.

From Section \ref{Particle
interactions} it follows that the action integral describing
the coupling of the far
field
of particle $(j)$ to
particle $(i)$ is of the form (generalizing
eqs.(\ref{4.19} -- \ref{4.21}))
\begin{equation} \label{5.a3}
W_{(ji)}   = \int_{T^{(i)}}g_{L M}^{(j)} I^{L
M}_{(i)}
\left
( -g_{\lambda \mu} u^{\lambda} u^{\mu} \right ) ^{1/2}ds,
\end{equation}
where $I^{L M}_{(i)} $ denotes the integral over the
tube cross-section of all (in
general periodic) expressions involving  the fields of
particle
$(i)$
which
interact linearly with the far field $g_{L M}^{(j)}
$. The
expression
$g_{L M}^{(j)} I^{L M}_{(i)} $ is obtained
by collecting
all
terms in
the general gravitational Lagrangian
(\ref{3.24}), (\ref{3.25}), with
$g_{L M} = \eta_{L M}+g_{L
M}^{(j)}+g_{L M}^{(i)} $, which
are
linear in
$g_{L M}^{(j)} $, and then integrating over the tube
cross-section.

Variation with respect to the far field $g_{L
M}^{(j)}$ of
the
free-field action (\ref{3.28}) for
$g_{L M}^{(j)}$
together with the
interaction integral (\ref{5.a3}) yields the field equations
\begin{equation} \label{5.a4}
\frac{1}{2}
({\partial}_{\lambda}{\partial}^{\lambda}-
\hat{\omega}^2)
(g_{L M} - \frac{1}{2} \eta_{L
M}g_{N}^{N})_{(j)} =  -
\int_{T^{(i)}} I_{L M}^{(i)}  \delta^{(4)}(x-
\xi^{(i)}(s))ds,
\end{equation}
where
\begin{equation} \label{5.a5}
\hat{\omega}^2 : = (k_{A}^{(j)}k^{A}_{(j)}).
\end{equation}
The harmonic masses $\hat{\omega}$ of all
interacting particles (or partons) are assumed to be the
same.  This is
necessary for interactions involving periodic far fields in
order that
the product of  the far fields $g_{L M}^{(j)}$ and
the
core-region fields $I^{L M}_{(i)}$ yield a resonant
mean
force.

The solution of (\ref{5.a4}) is given by
\begin{equation} \label{5.a5a}
g_{L M}^{(j)} = - 2 \int_{T^{(i)}} \{ I_{L
M}^{(i)} -
\frac{1}{n-2} \eta_{L M} I_{N}^{N(i)} \}  G (\xi
_{[ij]})ds,
\end{equation}
where the
Green function
$G(x)$ is defined by
\begin{equation} \label{5.a6}
({\partial}_{\lambda}{\partial}^{\lambda}
- \hat{\omega}^2)G : = \delta^4(x),
\end{equation}
\begin{equation} \label{5.a8}
\xi_{[ij]} : = x_{(i)} - x_{(j)}
\end{equation}
and $n$ is the dimension of full space.

Substituting the solution (\ref{5.a5a}) into
(\ref{5.a3}), one obtains
the action
integral
describing particle coupling in the
distant-interaction form (considering now an ensemble of
particles
rather than a single pair)
\begin{equation} \label{5.a7}
W_{int} = \sum_{i,j \atop (i \not= j)}\int_{T^{(i)}}
\int_{T^{(j)}}
ds_{(i)}ds_{(j)}\left
( -g_{\lambda \mu} u^{\lambda}_{(i)} u^{\mu}_{(i)}
\right ) ^{1/2}  I_{(ij)} G (\xi _{[ij]}),
\end{equation}
where
\begin{equation} \label{5.a9}
I_{(ij)} : = \frac{2}{n-2}I_L^{L(i)}
I_{M}^{M(j)} -
2I_{L M}^{(i)}I_{(j)}^{L M}.
\end{equation}

Equation (\ref{5.a7}) represents the generalization to
other
forces, including, in particular, periodic fields, of
the Wheeler-Feynman \cite{whe21} distant-interaction form
(\ref{4.11})
of
the
action integral describing the electromagnetic (and
linearized
gravitational) coupling between quasi-point-particles. Self
interactions, $i=j$, are again
excluded. These
determine the internal structures of the particles, which
are regarded
here as given.

In contrast to the coupling between non-periodic
electromagnetic and
gravitational fields, the coupling through periodic (de
Broglie) far
fields is effective only for special resonant
trajectories
for which
the frequencies of the far field $g_{L M}^{(j)}$ and
the
particle's
near-field form $I_{L M}^{(i)}$ are matched.

 Not included in
(\ref{5.a7}) are self-interactions of a particle with its
own field
which has been scattered at other particles. These
processes
can be
important, but in the present context of direct
particle-particle interactions are of higher order. They
are
discussed in
sections \ref{Bragg scattering} and \ref{Atomic spectra}.

The Green function $G$ is determined by the definition
(\ref{5.a6}) only to within an arbitrary solution of the
homogeneous
Klein-Gordon equation. For initial value
problems
associated with an arrow of time the appropriate Green
function is normally the retarded Green function $G^R$,
which can be represented
in Fourier integral form as
\begin{equation} \label{5.a10}
G^R := (2\pi)^{-4}\int d{\bf k }\int_{C^R} d\omega
\frac{e^{i({\bf k.x} -\omega t)}}
     {\omega ^2 - \omega_k^2},
\end{equation}
where
\begin{equation} \label{5.a11}
\omega_k^2 : = \hat{\omega}^2 +k_ik^i
\end{equation}
and the integration over $\omega$ is carried out along a
curve
$C^R$ in
the complex plane which follows the $\omega$ axis except
for
indentations passing above the poles at $\omega =
\pm \omega_k$. Closing the integral in the upper or lower
half plane for $t<0$ or $t>0$, respectively, one obtains
\begin{eqnarray} \label{5.a11a}
G^R &=&(2\pi)^{-3} \Theta(t) \int \sin ({\bf k.x}-
\omega_k t)  \omega_{{\bf k}}^{-1} d{\bf k }  \nonumber
\\
&=& - (2\pi)^{-2}  \frac{\Theta (t)}{r}  \int \{\cos(kr-
\omega_k t) - \cos(kr+\omega_k t)\} \frac{k}{ \omega_k}\,
dk,
\end{eqnarray}
where $r:=|{\bf x}|$ and
$$ \Theta(t) = \left\{ \begin{array}{c}  1 \; \mbox{for}
\; t > 0\\ 0 \; \mbox{for} \; t < 0
\end{array} \right \}. $$

The associated advanced Green function $G^A$ is obtained
by
replacing the
curve $C^R$ by the curve $C^A$ passing below the poles,
yielding
\begin{eqnarray} \label{5.a11b}
G^A &=& - (2\pi)^{-3}\Theta(-t)  \int  \sin ({\bf
k.x}-\omega t) \omega_k^{-1} d{\bf k }  \nonumber \\
&=& (2\pi)^{-2} \frac{\Theta (-t)}{r}  \int \{\cos(kr-
\omega_kt) - \cos(kr+\omega_kt)\} \frac{k}{\omega_k}\, dk.
\end{eqnarray}

For the non-dispersive case $\hat\omega = 0,\; \omega_k =
k$,
the expressions (\ref{5.a11a}), (\ref{5.a11b}) yield
\begin{eqnarray} \label{5.a11c}
\left\{ \begin{array}{l}G^R \\ G^A \end{array} \right\}
&=&
- \frac{1}{(2\pi)^{2}r}
\int  \left\{ \begin{array}{l} \Theta(t) \cos (kr-
\omega_k t) \\ \Theta(-t) \cos (kr+\omega_k t)
\end{array} \right\}dk \\
&=&  - \frac{1}{2 \pi r} \left\{ \begin{array}{c}
\Theta(t)
\delta
(r-t) \\ \Theta(-t) \delta (r+ t)  \end{array} \right\},
\end{eqnarray}
while in the dispersive case, $\hat\omega > 0$, one
obtains
 in the stationary-phase approximation
\begin{equation} \label{5.a11d}
\left\{\begin{array}{l} G^R \\ G^A \end{array}\right\}
\approx
-(2\pi)^{-2} \frac{1}{r}  \left(
\frac{\pi}{(\omega''_k)_0}\right)^{1/2} \left\{
\begin{array}{c} \Theta (t) \cos (k_0 r - \omega_0 t) \\
\Theta (-t)\cos (k_0 r + \omega_0 t) \end{array} \right\},
\end{equation}
where $k_0,\,\omega_0 := (\omega_k)_0$ correspond to the
stationary-phase wavenumber and frequency for which
\begin{equation} \label{5.a11e}
 \frac{d \omega_k}{dk} = \omega'_k =  r/t = v \; (<1).
\end{equation}
Thus in the dispersive case, for any given propagation
cone
$r/t = v < 1$ , the
retarded (advanced) Green function  consists in the
stationary phase approximation of
a single outgoing  (ingoing)
spherical wave component with group velocity equal to
$v$.

In the following we shall need only the result that in both
cases,
dispersive and non-dispersive, the retarded and advanced
Green functions can be represented as  a superposition of
outgoing and ingoing spherical waves, respectively,  of the
form
\begin{equation} \label{5.a11f}
\left\{ \begin{array}{c}
W^R \\ W^A \end{array}\right\} = \frac{A}{r}  \left\{
\begin{array}{c}
\Theta (t) \exp \left[ i(kr - \omega_k t) \right] \\ \Theta
(-t) \exp \left[-i(kr +
\omega_k t)\right] \end{array} \right\} \;\; (A =
\mbox{const}).
\end{equation}

In contrast to macrophysical applications, in which the
typical task of
predicting
 the future
state of a system given the present state leads naturally to
the choice of
the time-asymmetrical retarded
Green function $G^R$, the initial value problem is
irrelevant
for discrete
particle interactions.  One is concerned here
simply
with a physically consistent description of the
interactions between a
finite set of particles, and with the
further problem of embedding the particular finite system
being studied within the
universe of all other particles with which the system can
interact. It is argued below that in this case a
time-symmetric description of particle interactions is
appropriate, for which the
relevant potential is the time-symmetric Green
function
\begin{equation} \label{5.a12}
G^S : = (G^R +G^A)/2
\end{equation}
consisting of a superposition of equally large outgoing and
ingoing
spherical wave
components.

We distinguish between internal
interactions within the finite system considered and
external interactions
with other particles. For a single non-interacting
particle, it was assumed in Sections~\ref{The mode-trapping
mechanism}  and
\ref{Particle interactions} that the internal structure of
the particle was determined
entirely
by the local
nonlinear mode-trapping mechanism. The dependence of the
internal dynamics of the particle on the interactions of the
particle's
far fields with other particles was ignored. Similarly, in
describing the
interactions between a finite set of interacting
particles,
we shall ignore the
coupling of the set of particles to the rest of the
universe
in a first step. We require that in
such a closed particle set,  total momentum and
energy are conserved, i.e. that
the distant interactions between a finite set of
particles
can lead
only to an
exchange of 4-momentum between the particles, without
loss
(or gain) of
4-momentum by radiation to (or from) infinity.

It is readily verified that the conservation of 4-momentum
for a finite set of interacting
particles requires the
choice of the time-symmetric Green
function
$G^S$. Consider a finite set of particles which are far
separated
and therefore no longer interact at the beginning and end
of
their
paths, for $s \rightarrow \pm \, \infty $.  From the
trajectory
equations
for the particle $(i)$, obtained by varying the
zero'th
order action integral
 (cf. equations (\ref{4.19}), (\ref{4.26}),
(\ref{4.27})) and the
distant-interaction integral (\ref{5.a7}) with respect to
the
trajectory $\xi_{(i)}$,
\begin{equation} \label{5.a14}
M_{(i)}\frac{du^{\lambda}_{(i)}}{ds} = \sum_j \int_{T^{(j)}}
ds_{(j)}
\left( \frac{\partial}{\partial x_{\lambda}}
 -\frac{d}{ds}\frac{\partial}{\partial u_{\lambda}}
\right)
\{I_{(ij)}
G (\xi_{[ij]})\},
\end{equation}
we obtain for the change in the
total
\mbox{4-momentum}
between the beginning and end of the interaction:
\begin{equation} \label{5.a15}
\sum_{i}M_{(i)}[u^\lambda_{(i)}]^{s=\infty}_{s=-
\infty}= \sum_{i,j \atop (i \not= j)}\int_{T^{(i)}}
\int_{T^{(j)}} ds_{(i)}ds_{(j)}
 I_{(ij)}
\frac{\partial
G (\xi _{[ij]})}
{\partial x_{\lambda}^{(j)}}.
\end{equation}
Noting the symmetries $\xi_{[ij]} =
- \,\xi_{[ji]},\,I_{(ij)}=I_{(ji)}$ (cf. eqs. (\ref{5.a8}),
 (\ref{5.a9})), the
change in total \mbox{4-momentum} is seen to vanish for
arbitrary
trajectories if and only if
\begin{equation} \label{5.a16}
G (\xi _{[ij]})=G (-\xi _{[ij]}).
\end{equation}
This symmetry condition is satisfied only by the Green
function
$G^S$.

The symmetry property (\ref{5.a16}) ensures that in
the path-integral expression (\ref{5.a15}) for
the net \mbox{4-momentum} exchange,
4-momentum
is
conserved already at the
elementary interaction level: for any pair of interacting
infinitesmal line
elements $ds_{(i)},ds_{(j)}$ of the particle paths, the
momentum gained or lost
by
particle
$(i)$ is exactly balanced by the momentum lost or gained
by
particle
${(j)}$.

In contrast to the  closed-system description of particle
interactions in terms of the time-symmetrical Green
function, the
 open-system description
using
the retarded Green function fails to conserve 4-momentum
within the  system, since 4-momentum is  radiated to
infinity. The question of the proper
choice
of the
Green function has been the subject of some debate in
classical theories of the electromagnetic interaction of
charged
point particles. The
closed-system description has the formal advantage of
preserving
time-reversal symmetry, but must then explain the origin of
the
observed irreversible radiative damping of acclerated
charges. Wheeler and Feynmann \cite{whe21} resolved the
paradox
of radiative damping
for time-symmetrical electromagnetic interactions by showing
that the open-system description for a finite set of
particles can be derived from the closed-system description
if the finite particle set is extended to include an
infinite
statistical ensemble of distant
particles which completely absorbs
all outgoing radiation. The time-reversal asymmetry of the
outgoing radiation condition follows then from  the assumed
time-asymmetrical property of complete absorption by the
distant particle ensemble.

We shall adopt this interpretation also for the general
interaction case. Thus  we shall extend
our
finite physical system to include interactions with an
infinite external particle ensemble, assuming still that the
interactions
between individual particle pairs can be described by
time-symmetrical Green functions. The outgoing radiation
condition
will then be shown to follow from assumed
time-asymmetrical
statistical
properties of the external ensemble of particles with
which
the finite system interacts.

It should be pointed out, however, that there is a subtle
but fundamental difference between  the development of a
classical theory of electromagnetic interactions between
point particles and the metron model. In classical theory,
point particles are simply postulated to exist, and the type
of Green function must therefore also be postulated
axiomatically in defining the electromagnetic coupling
between
particles. In contrast, the only basic equations of the
metron
model are the n-dimensional Einstein vacuum equations
(\ref{1.1}). All particle properties and the details of
their coupling must follow from these equations.  From
the metron viewpoint, the closed-system and open-system
descriptions are both  permissible solutions of
eqs.(\ref{1.1})
(provided trapped-mode particle solutions exist, as
assumed). Both representations provide legitimate
descriptions
of particle interactions for suitably defined particle
ensembles.  Which of the two descriptions is more
appropriate
depends on the experimental situation.  It will be argued in
Section~\ref{The EPR paradox and Bell's theorem} that in
the case of the EPR experiment, the time-symmetrical
closed-system description is the relevant picture. On the
other hand, in Section~\ref{Atomic spectra} it will be found
more convenient to treat radiation mediating the coupling
between discrete atomic states in the traditional
sense as an independent external field interacting in an
open  particle system.

\section{The radiation condition}
\label{The radiation condition}

To derive the (electromagnetic) radiation condition,
Wheeler and Feynmann \cite{whe21} considered a charged test
particle, moving under the influence of an external force,
which was
 assumed to be coupled through
time-symmetrical electromagnetic interactions with a
distant ensemble of charged particles.
Under suitable assumptions regarding the absorbing
properties of the distant particle ensemble, they
showed that the back-interaction of the particle ensemble
on the test particle produces the Dirac \cite{dir2}
damping force and an associated net
field in the neighbourhood of the test particle in
accordance with the usual
picture of outgoing
radiation. Four different
derivations of the radiative damping were presented. The
first three were based on explicit electrodynamical
interaction properties and can be
generalized to the case considered here only through
a more detailed analysis of the perturbations induced in
the
coupling function $I_{L M}^{(i)}$ in
eq.(\ref{5.a3}). But
the
last, particularly simple derivation
 assumed
only that the interactions between the test particle and
the
absorber lead to complete absorption of all  fields in
the
absorber. This derivation can be readily generalized to
the
present case and will therefore be presented first.
However,
it fails to explain the origin
and nature of the absorption mechanism and the resultant
time asymmetry. We shall accordingly
present subsequently also a more detailed derivation of
the
radiation condition based on an explicit description of
the
interactions
between the test particle and the absorbing medium.
\subsection*{Simple  derivation}
Consider a test particle $e$ which is
perturbed by some external force, producing a
perturbation
in the particle's local coupling form $I_{L
M}^{(e)}$ in
eq.(\ref{5.a3}).
This
will cause a perturbation of the test particle's far
field
$g_{L M}^{(e)}$
 as given by (\ref{5.a5a}). For the following, the
tensor structure of the field $g_{L M}$ is an
irrelevant
complication, and we shall accordingly consider simply a
scalar emitted field  $\phi_e$ consisting of half the sum
of the
retarded and
advanced
potentials,
\begin{equation} \label{5.a17}
\phi_e = (\phi_e^A + \phi_e^R)/2.
\end{equation}

The perturbed field $\phi_e$ will in turn generate
perturbations in the particles~$j$ of the absorber,
producing response fields
\begin{equation} \label{5.a18}
\phi_j = (\phi_j^A + \phi_j^R)/2.
\end{equation}

Assume now that outside the absorber, beyond some large but
finite
sphere of radius $R$, the total field
\begin{equation} \label{5.a19}
\phi_{tot} = \phi_e +  \phi_r,
\end{equation}
consisting of the sum of the emitted field and the net
response field
\begin{equation} \label{5.a19a}
\phi_r= \sum_j \phi_j,
\end{equation}
effectively vanishes \cite{ft24}. The statistical properties
of the
particle
ensemble required to produce the assumed absorption will
be discussed later. If the total field vanishes, so must the
total retarded and advanced fields individually, since
the two fields have different propagation signatures and
therefore cannot be superimposed to yield a zero field.
Thus
\begin{eqnarray} \label{5.a20}
\phi_{tot}^R &=& \phi_e^R + \phi_r^R = 0 \;\mbox{
for
}r>R\\
\phi_{tot}^A &=& \phi_e^A +  \phi_r^A = 0 \;\mbox{
for
}r>R.
\end{eqnarray}
The difference field
\begin{equation} \label{5.a21}
\phi_{tot}^D = (\phi_{tot}^R -\phi_{tot}^A)/2 = (\phi_e^R
-
\phi_e^A)/2  +  (\phi_r^R -
\phi_r^A)/2 = 0
\end{equation}
therefore also vanishes outside the absorbing sphere. But
the field $\phi_{tot}^D$ has no sources. Thus it must
vanish identically in all space.

The response field $\phi_r$ may then be expressed in the
form
\begin{eqnarray} \label{5.a22}
\phi_r &=& (\phi_r^R + \phi_r^A)/2 = -
(\phi_r^R - \phi_r^A)/2 +  \phi_r^R \nonumber
\\
&=& (\phi_e^R -\phi_e^A)/2 + \phi_r^R.
\end{eqnarray}

Applied at the test particle, the term $(\phi_e^R -
\phi_e^A)/2$ of the last expression represents the
(generalized) Dirac  radiative damping force $\phi_e^D$
acting on a particle radiating energy to space. It was shown
by Dirac \cite{dir2} for the electromagnetic case that the
energy
extracted from the emitting particle by this force
corresponds exactly to the energy flux radiated away to
infinity. The second term represents the retarded response
field
of the absorber.

Adding to the response field the field emitted by the test
particle itself one obtains finally for the total field
\begin{eqnarray} \label{5.a23}
\phi_{tot} & =&  (\phi_e^R + \phi_e^A)/2 + \phi_r
\nonumber
\\
&=& \phi_e^R +  \phi_r^R.
\end{eqnarray}
Thus the total field consists of the retarded potential
of the test particle and the retarded response field of
the absorber.

Equations (\ref{5.a22}) and (\ref{5.a23}) are in
accordance with the classical time-asymmetrical picture
of a test particle emitting radiation into space and an
absorber emitting a retarded field in response to the
test
particle field -- although
the relations were derived using time-symmetrical
interaction potentials only. The result is rather
curious, as the only assumption introduced was that of
complete absorption of both advanced and retarded fields
by the absorbing particle ensemble, which in itself is
not a time-asymmetrical hypothesis. In fact, it can readily
be verified that the fields $\phi^R$ and $\phi^A$ can be
interchanged in the above derivation, yielding (as
pointed out by Wheeler and Feynamn) the equally valid
result
\begin{equation} \label{5.a23a}
\phi_{tot}  = \phi_e^A +   \phi_r^A.
\end{equation}

The resolution of this paradox is that, although not
explicitly stated, the assumed absorption is, in fact, a
time-asymmetrical process. This results in basically
different structures of the
advanced and retarded fields
for
later and earlier times relative to the perturbation
time of the test particle.  Thus although both
eqs.(\ref{5.a23}) and (\ref{5.a23a}) are formally
correct,
the retarded and advanced fields appearing in the
equations have quite different time-symmetry
properties and
different physical interpretations.

The origin of the assumed absorption and time asymmetry was
not discussed by Wheeler and Feynman. It was simply
observed that a certain phase integral occurring in their
first radiative damping derivation, which would otherwise
have been indeterminate, converged if a weak damping term
(breaking time symmetry) was introduced.   Since the basic
time asymmetrical statistical
hypothesis underlying the phenomenon of absorption is
fundamental for the understanding of radiative damping
and irreversibility in general -- including the question
of whether time symmetry applies for the EPR
experiment -- we discuss this point in more detail in the
following two sub-sections.

\subsection*{Detailed derivation}

It may be assumed without loss of generality that the
perturbation applied to the test particle is a
$\delta $-function in space and time, the
general case following by superposition. For a
$\delta $-function input the  retarded and advanced
fields emitted by the test particle appear in separate half
spaces $t>0$ and $t<0$. In accordance with
eqs.(\ref{5.a11c}),
(\ref{5.a11d}), the fields may be furthermore decomposed
into spherical-wave Fourier components $\sim \Theta(\pm
t)
\exp (kr \mp \omega_k t)$.  We
consider first only the retarded wave component
\begin{equation} \label{5.a23b}
 (\phi^R_e/2)_{\omega} =: W^R = \Theta (t)
\frac{A}{r}    \exp i(kr - \omega t).
\end{equation}

In response to the forcing of the retarded field $\phi_e^R$,
the absorber particles emit  a net response
field $ \phi_r$, yielding
a total field $\phi_{tot} =   \phi_e^R + \phi_r $. The
response field can be represented again as
half the sum of the advanced and retarded response
fields, $ \phi_r =(\phi^R_r + \phi^A_r)/2$. Regarding the
absorber as a statistical distribution of particles,
the response field
$\phi_r$ can be furthermore  decomposed into the sum
$\phi_r
= \; <\!\phi_r \!> + \; \phi'$ of the statistically
averaged
field $<\!\phi_r \!>$ and a residual scattered field
$\phi'$ whose ensemble mean value vanishes. In this
sub-section we shall be concerned only with the
ensemble mean field (the coherent component).  The
evolution (anticipating in the use of this term already the
later appearance of an arrow of time)  of the incoherent
 scattered field  will be considered in the following
sub-section.

The determination of the back-interaction of the absorber
involves two steps: First, we note that the coherent
component of the response field is (i)
spherically symmetric (we assume a spherically symmetric
absorber), (ii) proportional to $\exp (-i\omega t)$ and
(iii) regular at $r = 0$. It follows that in the
neighbourhood of $r = 0$ the coherent component of the
response field must have the form
\begin{equation} \label{5.a24}
<\!\phi_r\!> = \frac{B}{2ir} ( e^{ikr} - e^{-ikr}) e^{-
i\omega t}.
\end{equation}

In the second step we  determine the constant $B$ by
evaluating the field
at $r=0$:
\begin{equation} \label{5.a25}
<\!\phi_r\!>_{r=0}\; =  B k e^{-i\omega t}.
\end{equation}
To compute $<\!\phi_r\!>_{r=0}$ we determine the total
coherent field $<\!\phi_{tot}\!>$ propagating through the
absorbing medium, evaluate the perturbations induced
in the individual absorber particles by
$<\!\phi_{tot}\!>$ and then sum over the far fields
generated by
these perturbations at the location of the test particle.

The total coherent field will be shown to have the
general
form
\begin{equation} \label{5.a26}
<\!\phi_{tot}\!> = \frac{C}{r} \exp[i(kr-\omega t)
+i\theta(r) - \mu(r)],
\end{equation}
where $C=$ const, $\theta(r)$ is the phase shift induced
by
the dispersion of the absorbing medium and $\mu(r)$ is
the
damping (anticipating here the result of the next
sub-section) due to scattering into the incoherent field.

The amplitude $C$ is related to the amplitudes $A$ and
$B$
by the condition that for $r \rightarrow 0$ the retarded
spherical wave $<\!\phi_{tot}\!>$ is given by the sum of
the
retarded emitted wave and the outgoing component of the
coherent response wave (\ref{5.a24}).  Setting $\mu(r)$
and
$\theta(r) =
0$
for $r = 0$ (the variables are defined only up to
arbitrary
additive constants, which can be absorbed in the
definition
of $C$), we have
\begin{equation} \label{5.a27}
C = A + \frac{B}{2i}.
\end{equation}

The dispersive phase shift and absorption arising from
interactions with the particles of the absorbing medium
follow from
eq.(\ref{5.a4}). In the present
non-tensorial notation, and generalized to an
ensemble of particles~$j$ generating a single net field,
the equation may
be written

\begin{equation} \label{5.a28}
 ({\partial}_{\lambda}{\partial}^{\lambda}-
\hat{\omega}^2)\phi_{tot}
 = - \sum_j \int_{T^{(j}} \delta I^{(j)} \delta^{(4)}(x-
\xi^{(j)}(s))ds,
\end{equation}
where $\delta I^{(j)}$ represents the perturbation of the
coupling term $I^{(j)}$ induced by the field $\phi_{tot}$.
It can be expressed generally in the form
\begin{equation} \label{5.a29}
\delta I^{(j)} =  R^{(j)} \phi_{tot},
\end{equation}
where $R^{(j)}$ is a response function. It will be
assumed
that $R^{(j)}$ is real. This will be seen to be
equivalent to the assumption that there is no absorption
of radiation within the particles themselves.

Substituting the response relation (\ref{5.a29}) into
(\ref{5.a28}), the latter may be written
\begin{equation} \label{5.a30}
 ({\partial}_{\lambda}{\partial}^{\lambda}-
\hat{\omega}^2)\phi_{tot} = - R \, \phi_{tot},
\end{equation}
where
\begin{equation} \label{5.a31}
 R(x) := \sum_j \int_{T^{(j}} R^{(j)}  \delta^{(4)}(x-
\xi^{(j)}(s))ds.
\end{equation}

The evolution equation for the coherent field component
follows by taking the ensemble mean of eq.(\ref{5.a30}):
\begin{equation} \label{5.a31a}
 ({\partial}_{\lambda}{\partial}^{\lambda}-
\hat{\omega}^2)<\!\phi_{tot}\!> =
 - <\!R\!> <\!\phi_{tot}\!> - <\!R'\phi'\!>.
\end{equation}
We ignore in this sub-section the second term on the right
hand side of (\ref{5.a31a}). It will be shown in the
following sub-section that the correlation between the
incoherent fields
$R'$ and $\phi'$ is of second-order and yields the
damping term in eq.(\ref{5.a26}).

Regarding the statistical particle ensemble as locally
stationary and homogeneous, the ensemble mean response
factor
$<\!R\!>$ represents
a slowly varying function of space
and time which is proportional to the local particle
density.
Retaining then only the first term on the right hand
side,
equation (\ref{5.a31a}) yields, for real $<\!R\!>$,
the modified local
dispersion relation
\begin{equation} \label{5.a32}
 \omega^2  =  \hat{\omega}^2 + k_i k^i - <\!R\!> =:
\omega'^2_k.
\end{equation}
For small $<\!R\!>$, the perturbation $\delta k =
<\!R\!>/2k$ induced in the local wavenumber for given
frequency $\omega$ can be
represented, as in (\ref{5.a26}), as a phase shift
$\theta$
in a wave with unperturbed wavenumber, where the local
rate of change of the phase is given by
\begin{equation} \label{5.a33}
\frac{d\theta}{dr} = \frac{<\!R\!>}{2k}.
\end{equation}

The mean interaction term $-<\!R\!> <\!\phi_{tot}\!>$ on
the right hand side of (\ref{5.a31a}) not only modifies
the local propagation characteristics of the coherent field,
but also represents a source term generating retarded and
advanced far fields. Our interest is in the advanced
far field at the location $r=0$ of the test particle,
which is synchronous with the emitted
retarded field of the test particle. Noting that the
time-symmetric response at $r=0$ to a time-periodic,
spatial $\delta $-function source term  $ \delta ({\bf x -
x}_0)
\exp(-i\omega t)$ on the right hand side of (\ref{5.a31a})
is
\[-\frac{1}{8\pi r}(e^{ik\rho}+e^{-ik\rho})e^{-i\omega
t}, \]
where  $\rho := |{\bf x}_0| $, the net coherent
advanced-field
response at the location of
the test particle is given by
\begin{equation} \label{5.a35}
<\!(\phi_r)\!>_{r=0}\; = \frac{C}{2} e^{-i\omega t} \int
<\!R\!>
e^{i\theta(r)-\mu(r)} dr.
\end{equation}
Transforming from the integration variable $r$ to
$\theta' :=
\theta (r) + i\mu (r)$, applying (\ref{5.a33}) and
assuming
that the damping rate is small, $d \mu /dr \ll d \theta
/dr$, so that $d \theta' /dr \approx d \theta /dr$, but
nevertheless finite, so that $\mu(r) \rightarrow \infty$
for
$r \rightarrow \infty$, the
integration can be carried out explicitly. The response
factor $<\!R\!>$ cancels and one obtains
\begin{equation} \label{5.a36}
<\!\phi\!>_{r=0}\; = i k C e^{-i\omega t}.
\end{equation}
From eqs. (\ref{5.a36}) and (\ref{5.a25}) we find then
\begin{equation} \label{5.a37}
B=iC,
\end{equation}
so that, from eq.(\ref{5.a27}), finally
\begin{equation} \label{5.a38}
B=2iA.
\end{equation}

\begin{figure}[t] \centering
\begin{minipage}{10.5cm}
\epsfxsize300pt
\epsffile{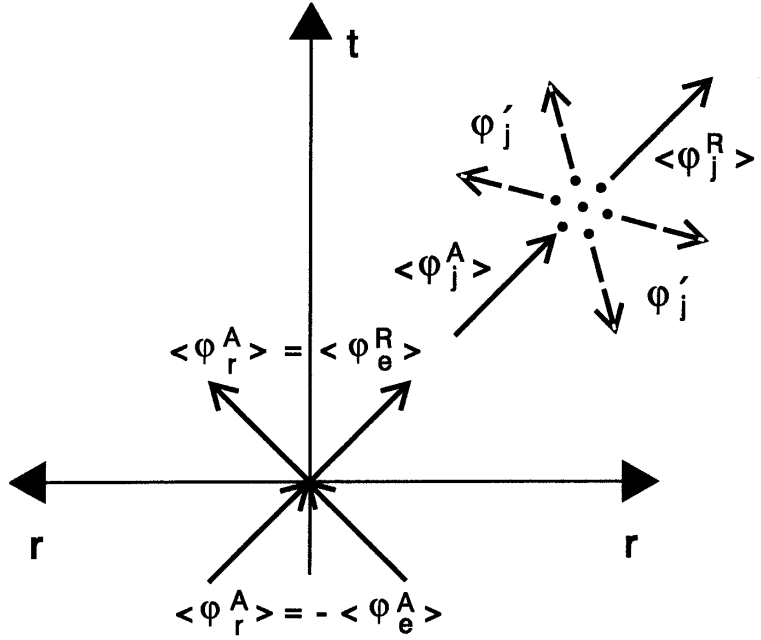}
\caption{\label{Fig50} Emitted, coherent response and
scattered fields in
an absorbing medium. For a perfect absorber, the net
advanced coherent response field reinforces the emitted
retarded field by a factor of two and exactly cancels the
emitted  advanced field. The absorption of the fields is due
to incoherent scattering by individual particles --  an
irreversible process responsible for the time-asymmetry of
the net radiation.}
\end{minipage}
\end{figure}

Comparing the original expression (\ref{5.a23b})
for the emitted retarded field with the expression
(\ref{5.a24}) for the
coherent response field  in the neighbourhood
of the test particle, the  outgoing coherent response
component
(relative to $r=0$) is seen to be exactly
equal to the emitted retarded wave $
(\phi^R_e/2)_{\omega} =
W^R$. Thus the net coherent outgoing field, consisting of
the emitted retarded field and the outgoing component
(relative to $r=0$) of the advanced response field,
is $ \phi^R_e$.
The emitted advanced wave, on the other hand,  is exactly
cancelled by the ingoing component  of the coherent advanced
response field.
Thus there exists no net advanced field. Although we have
considered so far only the interactions of the
absorber with the retarded wave of the test particle,
our analysis is therefore completed: there exists no net
advanced field which can interact with the absorber --
provided the assumed time asymmetrical description of
the absorption mechanism is valid.

These results are in accordance with the classical radiation
picture as derived in the previous sub-section. However, in
contrast to the previous, apparently  time-symmetrical
derivation, the present detailed derivation demonstrates
that
the ingoing and outgoing response fields
are not time symmetrical. In fact, they are exactly time
anti-symmetrical,   thereby yielding the desired
time-asymmetry of the net fields (cf.Fig.
\ref{Fig50}).

Another aspect clarified by the present derivation is that
the relations deduced in the previous sub-section from the
property of complete absorption apply only for the coherent
fields. The incoherent wave fields are not absorbed. In
fact, it will be shown in the next sub-section that  the
energy of the coherent wave field  is in fact also not
absorbed, but is converted rather to incoherent wave energy,
which -- assuming no net wave energy loss in the system --
is
then radiated to infinity.

\subsection*{The damping mechanism}

We turn now to the origin of the absorption assumed in eq.
(\ref{5.a26}). For simplicity, we use the WKB approximation:
the coherent wave field is represented locally as a plane
wave
\begin{equation} \label{5.a39}
<\!\phi_{tot}\!> = a e^{i({\bf k}_0.{\bf x}-\omega_0 t)}
+c.c,
\end{equation}
where the amplitude $a({\bf x},t)$ and wavenumber ${\bf
k}_0$ are slowly varying functions of space and time. An
index $0$ has been introduced to distinguish the
discrete wavenumber and frequency of the coherent wave
from
the continuous wavenumber-frequency spectrum of the
incoherent scattered field, and we have now included
explicitly the complex conjugate terms, as required for
the following nonlinear
analysis. We regard the statistical properties of the
particle distribution through
which the coherent field propagates and  the incoherent
scattered field generated by the interaction of the
coherent
wave with the particle distribution  as slowly
varying in space and time. Thus, in accordance with the
usual two-scale description, the fields can be characterized
by variance spectra $F({\bf k},\omega)
 = F({\bf
k},\omega;{\bf x},t)$ representing a locally
statistically stationary and homogeneous process  which
varies slowly with
${\bf x}$ and $t$.

We shall  attribute the damping of the coherent wave
in the following to scattering. We could alternatively
simply invoke absorption
processes within the particles of the absorber
themselves.
These can be modelled by imaginary components in the
response functions $R^{(j)}$, which must then be assumed to
have different signs for the
retarded and advanced wave components (thereby introducing
an irreversibility hypothesis).  However, we
prefer the scattering explanation. It applies generally
for conservative interactions, is an inherent property of
wave propagation in a microscopically heterogeneous
medium, and illustrates more clearly the statistical
origin of irreversibility.

The damping arising from scattering is represented formally
by the second-order correlation term in eq.(\ref{5.a31a}),
which was neglected in the first-order treatment of the
coherent field in the previous sub-section.
For  statistically stationary and homogeneous fields and
particle distributions, the random components appearing
in
this term may be represented by
Fourier (strictly, Fourier-Stieltjes) integrals
\begin{equation} \label{5.a42}
\left\{ \begin{array}{c}
\phi' \\ R' \end{array}\right\}
 = \int \left\{ \begin{array}{c}
\phi_{{\bf k}\, \omega} \\ R_{{\bf k}\,\omega}
\end{array}\right\}
 \exp[i({\bf
k.x} -\omega t)]d{\bf k}d\omega,
\end{equation}
where, for real fields $\phi'$, $R'$,
\begin{equation} \label{5.a42a}
\phi_{{\bf k}\,\omega} = \phi_{-{\bf k}-\omega}^*
\end{equation}
\begin{equation}  \label{5.a42b}
R_{{\bf k}\,\omega} =R_{{-\bf k}-\omega}^*.
\end{equation}
The expectation values of the Fourier components vanish,
and
the second moments are given by
\begin{eqnarray} \label{5.a43}
\left\{ \begin{array}{c}
<\phi^*_{{\bf k'}\omega'}\phi_{{\bf k}\,\omega}>  \\
<R^*_{{\bf k}'\omega'}R_{{\bf k}\,\omega}>
\end{array}\right\}
&=& \left\{ \begin{array}{c}
F^{\phi}({\bf k},\omega) \\ F^R({\bf
k},\omega) \end{array}\right\}
 \;\delta({\bf k}-{\bf
k}')\delta(\omega - \omega') \\
<\phi_{{\bf k'}\omega'}\phi_{{\bf k}\,\omega}> & = &
<R_{{\bf
k}'\omega'}R_{{\bf k}\,\omega}> = 0 \nonumber,
\end{eqnarray}
where
$F^{\phi}({\bf k},\omega)$, $F^R({\bf k},\omega)$ are the
variance spectra of the fields $\phi'$, $R'$,
respectively.

The  Fourier component $\phi_{{\bf k}\,\omega}$ can be
determined from the Fourier transform of eq.(\ref{5.a30}).
Invoking (\ref{5.a39}) and (\ref{5.a42}) this yields, to
lowest interaction order \cite{ft25},
\begin{equation} \label{5.a44}
\phi_{{\bf k}\,\omega} = - (aR_{{\bf k-k}_0,\omega-
                        \omega_0} + a^* R_{{\bf
k+k}_0,\omega+\omega_0})(\omega^2 - \omega^2_k)^{-1}
\end{equation}

Applying (\ref{5.a42b}),(\ref{5.a43}), we obtain then for
the second-order
correlation term in
(\ref{5.a31a})
\begin{equation} \label{5.a45}
<\phi' R'> =
-a\exp[i({\bf k}_0.{\bf x} -\omega_0 t)]
\int \frac{F^R({\bf k - k}_0, \omega -
\omega_0)}{\omega^2
-\omega^2_k}d{\bf k} d\omega \;
 + \; \mbox{c.c}.
\end{equation}
Apart from the contributions from the singularities at the
resonance frequencies $\omega = \pm \omega_k$, which at this
point are indeterminate, the integral is real. Thus the
contributions from the non-resonance forced waves represent
an
additional higher order modification of the dispersion
relation rather than a damping term, which should be
imaginary.

To determine the damping we must investigate the
contributions
from the resonance singularities. In scattering
computations,
it is usually assumed that the eigenfrequency $\omega_k$
contains
a small negative imaginary component, i.e. that the waves
are
weakly damped. The integration path then passes close by but
not
through the singularity, and one obtains automatically the
desired result that the expression (\ref{5.a45}) represents
a damping term. However, for the present discussion this is
clearly a circular argument. To understand the origin of the
irreversible damping we need to investigate more closely
the nature of the resonant interactions.

Physically, the resonance singularities are associated with
a
transfer of energy from the coherent wave field to the
free-wave components of the incoherent field.  To describe
this process we must admit a non-local, non-stationary
response in the neighbourhood of the resonance frequencies.
Accordingly, we represent the spectrum of the incoherent
scattered field more generally as the sum
\begin{equation} \label{5.a46}
F^{\phi}({\bf k}, \omega)= \hat F^{\phi}({\bf k}, \omega)
+
F^{\phi}({\bf k}) \delta (\omega - \omega_k)
\end{equation}
of a four-dimensional local, stationary forced-wave
contribution $\hat F^{\phi}({\bf k},\omega)$, for
wavenumbers and frequencies which lie off the dispersion
surface, and a secularly changing three-dimensional spectrum
$F^{\phi}_{\bf k} $ for free waves, whose frequencies lie
on the dispersion surface itself.

To determine the secular change in the spectrum
$F^{\phi}({\bf k}) $ (and the impact of the secular
change
on the damping expression (\ref{5.a45})) we replace the
representation (\ref{5.a42}) for a statistically
stationary,
homogeneous field $\phi'$ by the  representation
\begin{equation} \label{5.a47}
\phi' = \int \phi'_{{\bf k}} e^{i({\bf k.x}-\omega_k
t)}d{\bf k} +
c.c.,
\end{equation}
where $\phi'_{{\bf k}}= \phi'_{{\bf k}}({\bf x},t)$ consists
generally of a
superposition of
stationary components at off-resonance frequencies and
secular components at the resonance frequencies. The
notation deserves a comment. In contrast to
eq.(\ref{5.a42}), the integral in (\ref{5.a47})  no
longer extends
over positive and negative wavenumbers and
frequencies, but only over positive and negative
wavenumbers, the
frequency $\omega_k$ being prescribed on the positive
branch of the dispersion surface. Thus the conditions
(\ref{5.a42a}) no longer apply, the secular contributions to
the amplitudes  $\phi'_{{\bf
k}}$ and
$\phi'_{-{\bf k}}$ representing independent free waves
propagating in opposite directions. In
contrast
to the normal variance spectra $ F^R({\bf k},\omega)$,
$F^{\phi}({\bf k},\omega)$ and $\hat F^{\phi}({\bf
k},\omega)$, the free-wave spectrum $F^{\phi}_{\bf k}$ is
therefore not, in general, an even function, the spectrum
representing the variance density of waves propagating in
the $+{\bf k}$ direction.

The evolution equation for the slowly varying amplitude
$\phi'_{{\bf k}}$ is obtained by substituting the form
(\ref{5.a47}) into (\ref{5.a30}),
including now also the secular first derivative terms of the
amplitude:
\begin{equation} \label{5.a48}
2i\omega_k \frac{d}{ds}\phi'_{{\bf k}}({\bf x},t) = -
\int
 (aR_{{\bf k-k}_0,\omega- \omega_0}
+ a^* R_{{\bf k+k}_0,\omega+\omega_0})
e^{i(\omega_k -\omega)t} \, d \omega,
\end{equation}
where
\begin{equation} \label{5.a49}
\frac{d}{ds} := \frac{\partial}{\partial t} +
\frac{k_i}{\omega_k}
\frac{\partial}{\partial x_i} = \frac{\partial}{\partial t}
+ v_i
\frac{\partial}{\partial x_i}
\end{equation}
and $v_i = v_i({\bf k})$ is the group velocity of the
free
wave of wavenumber ${\bf k}$.
The integration of
(\ref{5.a48}) along a characteristic $x^i = x^i_0 + s
v^i$,
$t= t_0 + s$, with initial amplitude
$\phi'_{{\bf k}}({\bf x}_0,t_0) = \phi'^0_{{\bf k}}$ at
$s=0$, yields
\begin{equation} \label{5.a50}
\phi'_{{\bf k}}({\bf x},s)=\phi'^0_{{\bf k}} +
\frac{i}{2\omega_k}
\int
 (aR_{{\bf k-k}_0,\omega- \omega_0}
+ a^* R_{{\bf k+k}_0,\omega+\omega_0})
\Delta(\omega -\omega_k,s)\, d\omega,
\end{equation}
where
\begin{equation} \label{5.a51}
\Delta(\omega -\omega_k,s) := \frac{1-e^{-i(\omega -
\omega_k)s}}{i(\omega -\omega_k)}.
\end{equation}

In contrast to the stationary solution (\ref{5.a44}), the
solution (\ref{5.a50}) has a finite response $\sim s$ at
the resonance frequency $\omega =\omega_k$. However, the
assumption
that the amplitude $\phi'_{{\bf k}}({\bf x},t)$ is slowly
varying, so that only the secular first derivatives needed
to be retained in eq.(\ref{5.a48}), is clearly valid
only for frequencies in the neighbourhood of the
resonance
frequency. Equation (\ref{5.a50}) therefore applies
only for near-resonance frequencies, the response at
off-resonance frequencies being described as before by the
stationary solution
(\ref{5.a44}). The integral over the frequency
in
(\ref{5.a45}) can be divided accordingly into a resonance
contribution from a narrow  frequency band $\omega_k -
\epsilon < \omega < \omega_k + \epsilon$, where $\epsilon$
is small,   and the remaining
non-resonance integral. The non-resonance contribution
represents the principal value of the integral
(\ref{5.a45}) and is real, yielding
a second order  perturbation of the
dispersion relation. The resonant contribution will be shown
to yield the imaginary part responsible for the damping.

To evaluate this contribution, we note that in the
integration
across the resonance frequency, the response relation
(\ref{5.a51}) can be replaced  for large positive $s$  by
the asymptotic relation
\begin{equation} \label{5.a52}
\Delta(\omega,s) = \pi \delta(\omega) \quad \mbox{for}
\quad s \rightarrow \infty.
\end{equation}
In  deriving the response (\ref{5.a50}), it was assumed
that the slowly varying amplitudes $a$ and $R_{{\bf
k},\omega}$ could be regarded as constant. Thus the
relation is valid only for a finite $s$ interval. We
assume nevertheless that the amplitudes change so
slowly that  $s$ can still be chosen sufficiently
large that the asymptotic response relation (\ref{5.a52})
can be applied. Computing the correlation $<\phi' R'>$ under
this two-timing approximation, and
 assuming that the
initial free-wave  and scattering amplitudes
$R({{\bf k},\omega})$ and $\phi'^0_{{\bf k}}$,
respectively, are uncorrelated, we obtain then as the
resonant-interaction contribution:
\begin{eqnarray}
\label{5.a53}
\lefteqn{\hspace*{-2em} <\phi' R'>_{res} \:=} \nonumber
\\
&&  \hspace*{-2em}
    \frac{i\pi}{2\omega_k}
   \left\{
           a\exp[i ({\bf k}_0.{\bf x} -\omega_0 t) ]
       \int F^R({\bf k - k}_0, \omega -\omega_0)
       \delta (\omega - \omega_k) d{\bf k} d\omega
   \right.
- \nonumber \\
&& \hspace*{-2em}
    \left.
       \hspace*{1.5em}
       - a^* \exp[ -i ({\bf k}_0.{\bf x} -\omega_0 t) ]
       \int F^R({\bf k + k}_0, \omega +\omega_0)
       \delta (\omega - \omega_k) d{\bf k} d\omega
   \right\}.
\end{eqnarray}
Comparing eqs.(\ref{5.a53}), (\ref{5.a26}) and
(\ref{5.a31a}),
this  is seen to correspond to a positive differential
damping
coefficient
\begin{equation} \label{5.a54}
\frac{d\mu(r)}{dr} =
\frac{\pi}{4\omega_k^2} \int F^R({\bf k - k}_0, \omega -
\omega_0) \delta (\omega - \omega_k) d{\bf k} d\omega.
\end{equation}

How has the arrow of time entered in this derivation?
Clearly, in the assumption that the scattering amplitude
$R_{{\bf k}\,\omega}$ and the initial scattered free wave
amplitude $\phi'^0_{{\bf k}}$ are uncorrelated, and that
it was appropriate to take the asymptotic form (\ref{5.a52})
with $s \rightarrow + \infty$ for the resonant response.
If we had taken the opposite limit  $s \rightarrow -
\infty$, the right hand side of (\ref{5.a52}) would have
taken
an opposite sign, and we would have obtained, under the same
hypothesis of no correlation, a negative damping
coefficient. If our results are assumed to be valid for an
arbitrary initial eigentime $s_0$, so that the slow-
eigentime
derivative cannot have a cusp at $s_0$, this is a
contradiction. It follows that our basic hypothesis that the
amplitudes $R_{{\bf k}\,\omega}$ and $\phi'^0_{{\bf k}}$ are
uncorrelated must be wrong.  If we wish to maintain the
result (\ref{5.a54}), we must invoke the basic Boltzmann-
Gibbs
time-asymmetrical hypothesis that it is permissible to
regard
the amplitudes as uncorrelated when computing the evolution
of
the field forwards in time, but  not when attempting to
reconstruct the past.

 Within the framework of a local two-timing derivation,
as presented here, the Boltzmann-Gibbs hypothesis
cannot be justified beyond the intuitive argument that it
appears reasonable to assume that the incoherent free wave
components entering a local scattering region are initially
uncorrelated with the scattering field, a correlation
developing only in the course of the interaction. But this
is, of course, a circular argument, as it presupposes
intuitively an arrow of time. However,
by extending the analysis from a local to a global time
frame,
it has been shown by Prigogine \cite{pri}, through
integration of the
interaction equations to arbitrary order in slow time,
that the Boltzmann-Gibbs hypothesis can be derived from the
assumption that at some distant
time in the past the interacting fields were genuinely
uncorrelated. The existence of an the arrow of time
for all later  times is thus a consequence of a specially
`prepared' initial state.

The damping of the coherent wave is accompanied by a
corresponding growth of the free-wave component
$F^{\phi}({\bf k})$ of the incoherent wave spectrum.
Applying (\ref{5.a43}), (\ref{5.a48}) and (\ref{5.a50}),
we find
\begin{equation} \label{5.a55}
\frac{ dF^{\phi}({\bf k})}{ds} = \frac{\pi}{2\omega_k^2}
|a|^2 \int
[F^R({\bf k - k}_0, \omega -\omega_0) + F^R({\bf k +
k}_0,
\omega +\omega_0)]
\delta(\omega -\omega_k) d\omega.
\end{equation}

Integration of the transport equation (\ref{5.a55}) for
the free-wave spectrum of the incoherent scattered field,
in conjunction with the coupled propagation equation for
the damped coherent wave field, yields a spherically
symmetrical scattered-wave spectrum which initially grows
with
distance $r$ from the emitting test particle and then, when
the
coherent wave amplitude which generates the scattered
field has been sufficiently attenuated, decreases again,
decaying asymptotically as $r^{-2}$. The total outward
radiated energy of the coherent and incoherent field
remains constant, the energy flux being slowly transferred
from the coherent wave to the incoherent wave field. The
inclusion of higher order scattering processes
(interactions between incoherent waves) modifies the
incoherent wave spectrum  towards a more isotropic
distribution,
with a resultant slower decrease of the energy level of the
scattered field, but has no impact to lowest order on the
distribution of the net outward radiation between the
coherent
and incoherent fields.

\subsection*{Homogenization}

From the point of view of time-symmetrical particle
interactions, it appears more appropriate to speak of
`homogenization' rather than radiative damping: the
interactions of any given test particle with an ensemble of
other particles leads in general -- in accordance with the
second law of thermodynamics -- to a redistribution of
energy between all particles towards a statistically uniform
distribution. The radiative damping of the test particle and
the associated `heating' of the particle ensemble considered
in the previous sections represent only one side of this
homogenization process. The complementary side is the
heating of the test particle by the radiation from the
particle ensemble. This is ignored when focussing on the
radiative damping mechanism. However, in the asymptotic
homogeneous thermodynamic equilibrium state, both transfer
processes balance: there is no net `radiative damping'.

It is of interest to speculate on the relevance of this
homogenization process for the open questions regarding the
origin of the discreetness and uniqueness of the particle
spectrum discussed briefly in the last sub-section of
Section~\ref{The mode-trapping mechanism}. It is conceivable
that the coupling between similar metron particles with
initially slightly different nonlinearity levels (and
therefore de Broglie frequencies) through their de Broglie
far fields results in the equalization of the particle
energies and frequencies. However, the homogenization
process is complicated by the fact that an effective
coupling occurs only when the particles are close to
resonance, i.e. when the de Broglie frequencies lie within a
narrow frequency band whose width is determined by the
Doppler broadening associated with the statistical particle
motions.

Nevertheless. regardless of the details, it appears
reasonable
to assume that for the high-frequency de Broglie far fields,
statistical homogenization is achieved very rapidly, so that
radiative damping through the de Broglie far field does not
in fact arise \cite{ft26}.

\section{The EPR paradox and Bell's theorem}
\label{The EPR paradox and Bell's theorem}

We turn now to the application of these results to the
Einstein-Podolsky-Rosen experiment. The EPR gedanken-
experiment
was originally proposed \cite{ein24} to highlight a general
concern regarding the  quantum theoretical measurement
concept,
in which a spatially distributed state function is suddenly
collapsed at the instant of measurement. This appears
inconsistent with the finite speed of propagation of
information. The EPR experiment is perhaps the best known
example of a number of gedanken-experiments which have
been proposed to illustrate the paradoxes which this can
lead to.
In the usual Bohm version of the EPR experiment, a zero-
angular-momentum
state decays into two spin 1/2 particles with opposite but
unknown
spin orientations. A measurement of the spin of one particle
will
then immediately produce a change in the state not only of
that particle, but also of the other particle, since its
spin is
now also known, even though the two particles have space-
like
separations and are therefore not causally connected.

It is nevertheless just this experiment which is normally
cited, in the context of Bell's well known theorem, as proof
that
it is impossible to construct a deterministic microphysical
theory which is consistent with experiment. Bell
\cite{bel22} has shown,
under very general conditions, that for any deterministic
(hidden-variable) model of the EPR experiment, in which the
outcome of the two spin measurements is predetermined at the
time of emission of the particles by some unknown
(hidden) parameter $\lambda$, the covariance function
$C({\bf a,b)} =\,
<\!s_1s_2\!>$ of the values $s_1, s_2$ of the spins of
the two particles measured (in units of $\hbar/2$) by
Stern-Gerlach magnets pointing in
directions {\bf a} and {\bf b}, respectively, must
satisfy the inequality
\begin{equation} \label{5.1}
|C({\bf a.b}) - C({\bf a,c})| \leq 1 + C({\bf b,c}).
\end{equation}
This contradicts the  quantum  theoretical  result
\cite{ft27}
\begin{equation} \label{5.2}
<\!s_1s_2\!> \; =  - {\bf a.b}
\end{equation}

The quantum theoretical prediction has been verified for an
alternative version of the EPR experiment \cite{cla}, in
which the
spin-1/2 particles are replaced by a pair of photons emitted
in an
atomic cascade process and the Stern-Gerlach magnets by
polarization filters.

It was already stressed by Bell, however, that an essential
although seemingly self evident assumption of his theorem is
(forwards) causality (or `locality', in the terminology of
Bell).
It is assumed that for each particle the measured spin
depends
on the common hidden variable $\lambda$ and the orientation
of the Stern-Gerlach magnet used for the spin measurement of
that particle, but is independent of the orientation of the
other Stern-Gerlach magnet (cf.~Fig. \ref{figepr}a).  This
assumption is incompatible with time-symmetry, a basic
property
of the metron model. The possibility of circumventing Bell's
theorem with general time-symmetrical models has also been
discussed in some detail by de Beauregard \cite{debe} and
Dorling
\cite{dor}.
\begin{figure}[t] \centering
\begin{minipage}{8.5cm}
\epsfysize300pt
\epsffile{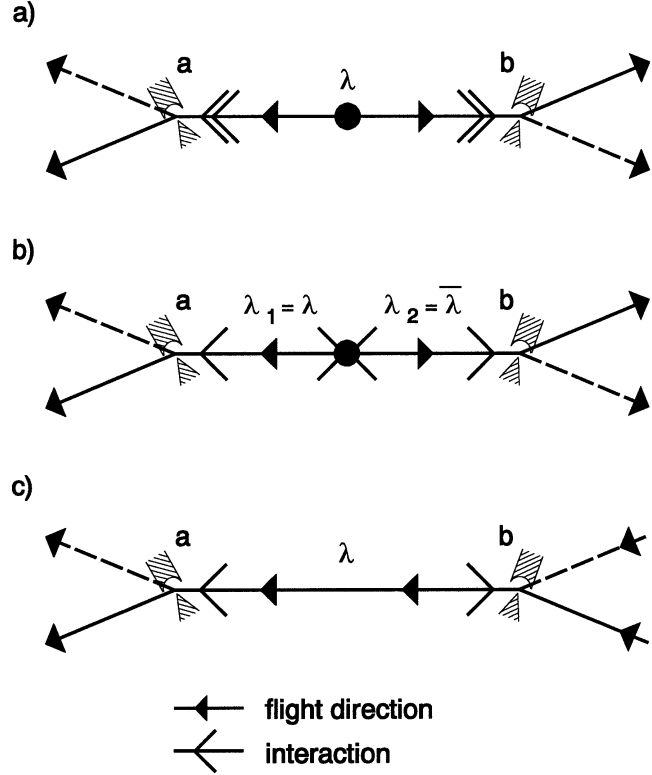}
\caption{\label{figepr}
Relation between (a) causal and (b) time-symmetrical
picture
of EPR experiment and corresponding one-particle
experiment (c)}
\end{minipage}
\end{figure}

Time-reversal symmetry requires in the case of the EPR
experiment that the interactions involved in the emission
and measurement processes of an
individual event must have the symmetrical structure
indicated in Fig. \ref{figepr}b. This is clearly
incompatible with the decoupling of the two measurement
processes assumed by Bell (Fig \ref{figepr}a).
For the two-photon version of the EPR experiment, Dorling
\cite{dor} has
pointed out that a time-symmetrical interpretation of the
EPR experiment is readily available in the Wheeler and
Feynmann
\cite{whe21}
distant-interaction theory of electromagnetism (cf.
Sections~\ref{Time-reversal symmetry}, \ref{The radiation
condition}). By replacing the standard retarded potentials
by
time-symmetrical advanced and retarded potentials, the
distinction
between particles which emit and absorb photons is lost
except in
the geometrical, non-causal sense of indicating the relative
locations of electromagnetically interacting particles on
the two
separate light cone branches.  This yields naturally the
time-symmetrical interaction picture of Fig.\ref{figepr}b.

Essentially the same picture applies also for the more
general interactions involved in the metron model. For
the metron interpretation of the EPR experiment, the
details of the model are not important. Relevant is only
that the particle trajectories, spin orientations and
interaction
fields are determined by the variation of an action integral
over space and time which contains in addition to the
particle
trajectories the time-symmetrical particle far fields
\cite{ft28}.
 The symmetrical occurrence of forward and backward
potentials
implies that the solution of the variational problem cannot
be
determined in the standard manner by forwards integration in
time, but only through a non-separable space-time integral
analysis. The basic assumption in our interpretation of the
EPR experiment is that the measurement process does
not involve irreversible statistical interactions with
a distant absorber, which, as discussed in the previous
section, would give rise to an arrow of time, but depends
only on the direct, time-symmetrical interaction between the
emitting system and the measurement apparatus.

To recover the  experimentally verified  quantum theoretical
result for the EPR spin correlations, we need in addition
two
further assumptions. The first states that if a single
particle
travels from a region with a magnetic field in the
direction {\bf b}, with its spin aligned parallel or
anti-parallel to {\bf b}, into a region with a magnetic
field in the direction {\bf a}, in which the spin is
realigned
parallel or anti-parallel to {\bf a}, the correlation of the
spins is given by the quantum theoretical
relation $<s_as_b> = \, {\bf a.b}$ (Fig \ref{figepr}c).
The relation is symmetrical in {\bf a} and {\bf b} and
clearly satisfies the condition of time symmetry. It implies
a
statistical assumption regarding the probability
distribution
of the particle's hidden variable $\lambda$, which governs
which of the different possible spin polarizations is
actually realized in a given particle trajectory (in the
metron model, $\lambda$ has many components representing the
polarizations and phases of the internal particle fields).
The
hidden variable may be associated with the particle state at
the beginning or end of the trajectory (or at some point in
between), but necessarily characterizes the trajectory as a
whole: the $\lambda$ sub-set associated with a particular
spin-sign combination changes if either {\bf a} or {\bf b}
is
changed.

The second assumption concerns the internal hidden
variables
$\lambda_1$, $\lambda_2$ of the two EPR particles at the
time $t_0$ of
emission.  It is assumed that the hidden variables are
`conjugate',
$\lambda_2 = \bar\lambda_1$, in the sense that the
trajectory of
particle 2, as defined by $\lambda_2$, is identical to
the
backwards
extension $(t< t_0)$ of the trajectory of particle 1, as
determined by
$\lambda_1$, in all respects (including internal particle
properties)
except for a sign reversal in time and spin. This is
clearly
the
simplest internal symmetry hypothesis which conserves
angular momentum
and ensures that the spins are exactly anti-correlated,
as
observed,
when both magnets are aligned, ${\bf a} = {\bf b}$.

Comparing the single-particle case (Fig \ref{figepr}c)
with the
EPR two-particle geometry (Fig \ref{figepr}b), the only
difference is
seen to lie in the replacement of the backwards
trajectory
$(t < t_0)$
of particle 1 in the single-particle case by the
conjugate
trajectory
$(t > t_0)$ for particle 2 in the EPR case. If the
internal
variable
$\lambda_1$ defines a spin value $s_b(1)$ for particle 1
at
the
Stern-Gerlach magnet $b$ in the single-particle case, the
conjugate
internal variable $\lambda_2 = \bar \lambda_1$ for
particle
2 in the
EPR case must define a spin $s_b(2) = - s_b(1)$.  It
follows
that the
EPR two-particle experiment must yield exactly the same
correlation
for the spins measured at the Stern-Gerlach magnets a and
b
as the
one-particle experiment, except for a sign change -- in
agreement with
the quantum theoretical result.

This straightforward time-symmetrical hidden-variable
interpretation
of the EPR spin correlation result is clearly very close
in
spirit to
the equally simple standard quantum theoretical
derivation,
in which
the two-particle problem is also effectively reduced to
the
one-particle case. It suggests that despite considerable
differences
in the basic concepts, there exists in practice a rather
close
correspondence between metron-model and
quantum-theoretical
computations. This will be demonstrated further in the
dicussion of
wave-particle duality in the following sections.

\section{Bragg scattering}
\label{Bragg scattering}
\typeout{################################}
\typeout{################################}
\typeout{        START OF met4-6.tex}
\typeout{################################}
\typeout{################################}

\subsection*{Wave-particle duality}

Although it has been shown in the previous section that the
time-reversal symmetry of the metron model circumvents the
fundamental conflict of deterministic hidden-variable
theories with quantum theory expressed by Bell's theorem, 
it remains to be demonstrated that the metron model is in 
fact able to resolve the basic wave-particle
duality dilemma which was the motivation for the
creation of quantum theory in the first place.

As outlined in Part \ref{The Metron Concept}, the
simultaneous occurrence of wave-like and
corpuscular phenomena is explained in the metron model by
the existence of periodic `de Broglie' far fields for all
finite-mass particles.  Although the forces exerted directly
by the de Broglie fields vanish in the mean and therefore 
have no impact on the mean particle motions, mean forces 
can arise at higher order through interactions involving 
scattered de Broglie waves. These are generated whenever a 
particle interacts with another object. The interaction 
between a scattered de Broglie wave and the internal
periodic field of the particle kernel, which represents the
source of the de Broglie far field,  can result in mean forces 
which affect the particle trajectory. The mean forces are 
modulated by the interference patterns of the scattered wave 
fields, giving rise to similar interference patterns in the 
particle distributions. The mechanism is illustrated in this 
section  for the case of Bragg scattering. In the following 
sections it is shown that interactions with scattered 
de Broglie waves explain also the existence of discrete 
atomic states.

\subsection*{Wave-particle resonance in Bragg scattering}

Consider a particle $(i)$ with constant
velocity $u^{\lambda}_{(i)}$ impinging on a periodic
lattice. Assume that the particle has a
periodic de Broglie far field $\sim \tilde g^{(i)}_{L M} ~
\exp(ik_{\lambda}^{(i)}\,x^{\lambda})$ whose wavenumber
four-vector $k_{\lambda}^{(i)} = \omega_0 u_{\lambda}^{(i)}$ 
satisfies the de Broglie free-wave dispersion relation
\begin{equation} \label{6.1}
k^{(i)}_{\lambda}k^{\lambda}_{(i)} = -\, \omega_0^2,
\end{equation}
where $\omega_0$ is the particle mass \cite{ft29}. 
The interaction of the incident
field $\tilde g^{(i)}_{L M}$ with the individual elements of
the periodic lattice generate a
scattered de Broglie field $\tilde g^{(s)}_{L M}$ whose
wavenumber
$k^{(s)}_{\lambda}$ satisfies the Bragg scattering condition
for constructive interference,
\begin{equation} \label{6.2}
k^{(s)}_{\lambda} = k^{(i)}_{\lambda} + k^{(l)}_{\lambda},
\end{equation}
where $k^{(l)}_{\lambda}$ is one of the periodicity
wavenumbers of the lattice (i.e. an integer linear 
combination of the fundamental wavenumbers which define 
the lattice structure). To represent a propagating de Broglie 
field, $k^{(s)}_{\lambda}$must satisfy the free-wave 
dispersion relation
\begin{equation} \label{6.3}
k^{(s)}_{\lambda} k^{\lambda}_{(s)} = -\, \omega^2_0.
\end{equation}
For a three dimensional lattice, the conditions (\ref{6.2}),
(\ref{6.3}) can be satisfied simultaneously only for
particular
`glanz' incidence and scattering wavenumbers, while for two
dimensional surface scattering lattices (where the
lattice wavenumber components in the direction orthogonal to
the lattice plane represent a continuum), a set of discrete
Bragg
scattering directions exists for any incident wavenumber.

The interaction of the metron trajectory with its scattered
field can
be treated using the same formalism as developed in Section
\ref{Particle interactions}.  The
relevant action integral describing the coupling is the line
(tube)
integral (\ref{4.19}).
The only difference relative to the analysis of
Section \ref{Particle interactions} is
that $<\!L\!>$, the integral of the Lagrangian density
across the three-dimensional tube cross-section of the
particle in the particle's restframe,
is regarded now as modified not by the mean far fields of
other particles, but by the
scattered de
Broglie far field of the particle itself.  The relevant de
Broglie contribution $L_{dB}$ to
$<\!L\!>$ is governed by the interaction
between the
scattered far field $\tilde g^{(s)}_{L M}$ and the
near-field
component
of the particle's de Broglie field.

In the particle rest frame, the frequency of the de Broglie
near-field
is $\omega_0$, while the frequency of the scattered far
field $\tilde
g^{(s)}_{L M}$, measured at the position of the particle in
the particle's
restframe, is given by the `frequency of encounter'
\begin{equation} \label{6.5}
\omega_e : = - k^{(s)}_{\lambda} u^{\lambda}_{(0)},
\end{equation}
where $u^{\lambda}_{(0)}$ is the local velocity of the
outgoing
particle after the scattering event.  If $\omega_e$ differs
from the
intrinsic particle frequency $\omega_0$, the interaction of
the
scattered de Broglie field with the de Broglie field of the
particle
kernel will yield an oscillatory contribution to $<\!L\!>$
which has no
impact on the mean particle trajectory.  However, in the
case of
resonance, $\omega_0 = \omega_e$, mean forcing terms result
which can
affect the particle trajectory.

The resonant interaction condition $\omega_e = \omega_0$
yields a
condition on the direction of the scattered velocity
$u^{(s)}_{\lambda}$.  Denoting the direction of the
scattered
wavenumber by $v_{\lambda}^{(s)}$, so that, from
(\ref{6.3}),
\begin{equation} \label{6.6}
k_{\lambda}^{(s)} = : \omega_0 v_{\lambda}^{(s)},
\end{equation}
the condition $\omega_0 = \omega_e$ implies, from
(\ref{6.5}),
\begin{equation} \label{6.7}
v_{\lambda}^{(s)} u^{\lambda}_{(0)} = -1.
\end{equation}
This can be satisfied for two normalized subluminal vectors
$v$, $u$ with $v_{\lambda} v^{\lambda} = u_{\lambda}
u^{\lambda} = -
1$ only for $v = u$. Thus the scattered particle is in
resonance with
its scattered de Broglie wave if and only if it propagates
in the same
direction as its scattered wave, \cite{ft30}:
\begin{equation} \label{6.8}
u_{\lambda}^{(0)} = k_{\lambda}^{(s)} \omega^{-1}_{0}.
\end{equation}

Consider now the dependence of $L_{dB}$ on the particle
trajectory.
Regarding the incident section of the trajectory as fixed,
the
dependence on the scattered section of the trajectory takes
the form
of a sequence of $\delta$-functions: $L_{dB}$ effectively
vanishes
except for a discrete set of values of the particle velocity
which
satisfy the wave-particle resonant interaction condition
(\ref{6.8}).
The sharp resonant extrema in the Bragg directions act as
potential
energy canyons which will tend to trap the particles in
these preferred
directions after they have been scattered.

\subsection*{A Bragg scattering model}
\label{A Bragg scattering model}

To investigate the trapping mechanism in more detail, some
assumptions
must be made regarding the de Broglie interaction
Lagrangian. If the
coupling between the scattered de Broglie (fermion) far
field $\psi^{(s)}$
and
the
intrinsic de Broglie field $\psi
^{(o)}$ of the scattered particle is
mediated by a
bosonic field $V^{\lambda}$ of the scattered particle, the
de Broglie
interaction Lagrangian can be assumed to be given, in
accordance with the
Maxwell-Dirac-Einstein interaction Lagrangian, cf.
Section~\ref{The Maxwell-Dirac-Einstein Lagrangian},
and the more general fermion-boson interactions considered
later
in Part~\ref{The Standard Model}, by an expression of the
general form
\begin{equation} \label{6.8a}
L_{dB}  =  \mbox{const}  \;  i  \bar  \psi  ^{(s)}
\gamma_{\lambda}    <\! \psi
^{(o)} V^{\lambda}\!> + \;\; c.c.,
\end{equation}
where the cornered parentheses $<\ldots>$ denote the
integral over the particle core in the particle's rest frame
(cf. Section~\ref{Particle interactions}) and the adjoint
scattered wave is given by
\begin{equation} \label{6.8b}
 \bar \psi^{(s)} =  \bar \psi^{(s)}_0 \,
\exp[i(k_{\lambda}^{(s)} x^{\lambda})],
\end{equation}
with constant (or slowly varying) amplitude $ \bar
\psi_0^{(s)}$. For an
isotropic particle,
$V^{\lambda}$ is parallel to $u^{\lambda}$, so that

\begin{equation} \label{6.8c}
<\psi^{(0)}V^{\lambda}>  \,=  \psi^{(0)}_0  u^{\lambda}  \,
(-g_{\lambda   \mu}
u^{\lambda} u^{\mu})^{-1/2} \, \exp[ i \omega_0s],
\end{equation}
where $\psi_0^{(0)}$ is again a constant (or slowly varying)
amplitude factor. Thus
\begin{equation} \label{6.8d}
W_{dB} = \int_{T^{(i)}} \; L_{dB}(-g_{\lambda
\mu}    u^{\lambda}
u^{\mu})^{1/2}ds = \int_{T^{(i)}} \alpha' \, W_{\lambda}
u^{\lambda}e^{iS}ds + \;\; c.c.,
\end{equation}
where
\begin{equation} \label{6.8e}
W_{\lambda} : = i \; \bar \psi^{(s)}_0 \gamma_{\lambda} \,
\psi^{(0)}_0,
\end{equation}
\begin{equation} \label{6.8f}
S : = k_{\lambda}^{(s)} x^{\lambda} + \omega_0s,
\end{equation}
and $\alpha'$ is a complex coefficient. The slowly varying
amplitude
of the scattered field can be regarded as included in the
definition
of $\alpha'$ (the dependence of $\alpha'$ and
other
slowly varying factors on $x$
will be neglected anyway in the following compared with the
derivatives of the more rapidly varying exponential factor).
For
simplicity, the dependence of the coupling vector
$W_{\lambda}$ on the
velocity $u^{\lambda}_{(0)}$, which could affect
the relative spin
orientations
of $\psi_0^{(s)}$ and $\psi_0^{(0)}$, will also
be ignored.

Variation of the action integral (\ref{6.8d}) with respect
to the
particle trajectory for a given scattered field (noting that
the
particle phase function $\omega_0s$ must be replaced by the
normalization-free form $\omega_0 \int(-g_{\lambda
\mu}
u^{\lambda} u^{\mu})^{1/2} \, ds$ when carrying out the
variation)
yields the trajectory equation
\begin{equation} \label{6.8g}
du^{\lambda}_{(0)}/ds = (F^{\lambda}_{\;\mu} +
G^{\lambda}_{\;\mu}) u^{\mu}_{(0)},
\end{equation}
where $F^{\lambda}_{\;\mu}$ represents the
mean field producing the particle scattering at the
lattice (for
example, an
electromagnetic field) and $G^{\lambda}_{\;\mu}$ is the de
Broglie
interaction field, given by
\begin{equation} \label{6.8h}
G^{\lambda}_{\;\mu} : = i \,  \alpha  \,
\{W_{\mu}(k^{\lambda}_{(s)}  -  \omega_0
u^{\lambda}_{(0)}) - W^{\lambda}(k_{\mu}^{(s)} -  \omega_0
u_{\mu}^{(0)})  \}
\,e^{iS} + \;\; c.c.,
\end{equation}
with a complex constant $\alpha$.  The de Broglie
interaction
field can
be represented in terms of the de Broglie interaction
potential
$B_{\lambda}$,
\begin{equation} \label{6.8i}
G^{\lambda}_{\;\mu} : = \partial^{\lambda}B_{\mu} -
\partial_{\mu}B^{\lambda},
\end{equation}
where
\begin{equation} \label{6.8j}
B_{\lambda} : = \alpha \, W_{\lambda} \,e^{iS} + \;\;c.c.
\end{equation}
and the phase (\ref{6.8f}) is given in the neighbourhood of
the scattered particle by \cite{ft31}
\begin{equation} \label{6.8k}
S = k_{\lambda}^{(s)}x^{\lambda} - \omega_0
u_{\lambda}^{(0)}
     x^{\lambda}  + \mbox{const}
\end{equation}

Resonance of the trajectory of the outgoing particle with
its
scattered field occurs if $dS/ds = 0$, i.e.  if
\begin{equation} \label{6.8ka}
u^{\lambda}_{(0)} =
v^{\lambda}_{(s)} = k^{\lambda}_{(s)}/\omega_0.
\end{equation}
 For $dS/ds
\not= 0$,
the field $B_{\lambda}$ is oscillatory and does not
significantly
affect the mean particle trajectory.

Substituting (\ref{6.8ka}) into (\ref{6.8h}), the de Broglie
force (acceleration)
\begin{equation} \label{6.8l}
A^{\lambda} : = G^{\lambda}_{\;\mu} u^{\mu}_{(0)}
\end{equation}
is seen to vanish in the resonance direction itself. For a
small
velocity
perturbation $\delta u^{\lambda}$ about the resonance
direction,
\begin{equation} \label{6.9a}
u^{\lambda}_{(0)} = v^{\lambda}_{(s)} + \delta u^{\lambda},
\end{equation}
the perturbation force is given, to lowest order in $\delta
u^{\lambda}$, by
\begin{equation} \label{6.9b}
\delta A^{\lambda} = - \beta \delta u^{\lambda} ,
\end{equation}
where
\begin{equation} \label{6.9c}
\beta : = 2\, W_{\mu} u^{\mu}_{(0)} \mbox{Re} \{i \alpha
\exp
i (S_0+ \delta S) \},
\end{equation}
$S_0$ is the initial phase at $s = 0$, and use has been
made
of the
relation
\begin{equation} \label{6.9d}
\delta u^{\lambda} \, v^{(s)}_{\lambda} = - \frac{1}{2}
\delta
u^{\lambda} \delta u_{\lambda}
\cong 0 ,
\end{equation}
which follows from the normalization of $u^{\lambda}_{(0)}$
and
$v^{\lambda}_{(s)}$. The phase perturbation is given by
\begin{equation} \label{6.9e}
\delta S = - \frac{\omega_0}{2} \, \int^s \, \delta
u^{\lambda}_{(0)} \delta
u^{(0)}_{\lambda} \, ds.
\end{equation}
Although of second order, this is retained in (\ref{6.9c})
as a
potentially rapidly oscillating term which determines the
resonance width of $\beta$.

Since the perturbation force is parallel to $\delta
u^{\lambda}$, it
will act initially as a pure restoring or amplifying force,
depending
on the sign of $\beta_0 := \beta(s=0)$.

For positive $\beta_0$ (restoring force), the velocity will
relax back
exponentially to its resonance value $u^{\lambda}_{(0)}$. If
the
deviation from the resonance trajectory occuring during this
relaxation process is
small, the
change occuring in $\beta$ during the relaxation process can
be
neglected.
If $\beta_0$ is negative, however, $\delta u^{\lambda}$
grows
exponentially. In this case $\beta$ cannot be regarded as
constant. As
$\delta u^{\lambda}$ grows, the phase perturbation $\delta
S$ grows,
leading after some time to a change in sign of $\beta$. When
this
occurs,
the perturbation begins to decay again exponentially. If the
change of
phase during the decay phase is not large enough to cause
another
change in sign of $\beta$ back to the unstable state, the
velocity
becomes trapped at its resonant value. Thus the net effect
of
the initial
unstable state is simply to cause a small displacement of
the particle
from its initial position on a potential ridge to the
neighbouring
potential valley.  If the initial perturbation $\delta
u^{\lambda}$ is sufficiently
large,
however, the particle does not become trapped but retains
its original
finite velocity perturbation, with superimposed fluctuations
as the
particle passes through successive potential valleys and
ridges.

The trapping condition can be readily determined by
integrating the
coupled equations for $\delta u^{\lambda}$ and $\delta S$.
Setting $E
= \delta u^{\lambda} \delta u_{\lambda}/2$ (which for small
$\delta
u^{\lambda}$ is always positive), and ignoring the non-de
Broglie
forces, equations (\ref{6.9b}), (\ref{6.9e}) yield the
coupled
equations
\begin{equation} \label{6.9f}
dE/ds = - \,\gamma \, E \, \cos(\delta S + \varphi),
\end{equation}
\begin{equation} \label{6.9g}
d \delta S/ds = -\, \omega _0E,
\end{equation}
with initial conditions
\begin{equation} \label{6.9h}
E = E_0, \quad \delta S = 0 \quad \mbox{for} \; s = 0,
\end{equation}
where the (real) constants $\gamma$ and $\varphi$ are
defined by
\begin{equation} \label{6.9i}
\gamma \, \exp (i \varphi) : =  2  \;  i  \alpha  W_{\mu}
\,
u^{\mu}_{(0)}  \,
\exp(iS_0).
\end{equation}
From equs. (\ref{6.9f}), (\ref{6.9g}) one can immediately
derive the
first integral
\begin{equation} \label{6.9j}
E - \frac{\gamma}{\omega_0} \, \{\sin (\delta S + \varphi) -
\sin
\varphi \} = E_0,
\end{equation}
which can be used to eliminate $E$ in the phase equation,
yielding
\begin{equation} \label{6.9k}
d \delta S /ds = - \,\omega_0\, E_0 + \gamma \sin \varphi -
\gamma \sin  (\delta
S + \varphi ).
\end{equation}
Equation (\ref{6.9k}) can be integrated in closed form.
However,
without writing down the result explicitly, it can be seen
that the
solutions are trapped, with $E \rightarrow 0$ for $t
\rightarrow \infty$, if
\begin{equation} \label{6.9l}
B : =  \omega_0 \, E_0/ \gamma -\sin \varphi  \leq 1
\end{equation}
and indefinitely oscillatory otherwise. For if
inequality (\ref{6.9l}) is not satisfied,
eq.(\ref{6.9k})
implies that $d \delta S/ds \leq \gamma (1 - B) < 0$ for all
values of
$\delta S$. Thus $\delta S$ decreases monotonically with
$s$ and $E$, according to (\ref{6.9j}), will oscillate
indefinitely.
On the other hand, if (\ref{6.9l}) holds, $\delta S$
approaches an
equilibrium solution $\delta S_{\infty}$ defined by
\begin{equation} \label{6.9m}
\sin (\delta S_{\infty} + \varphi ) = -\, B.
\end{equation}

Equation (\ref{6.9m}) has two solutions in the range $0 \leq
\delta
S_{\infty} \leq 2 \pi$. One of these is unstable,
corresponding to a
position on the top of a potential energy ridge, while the
other is stable,
representing a position at the bottom of a potential energy
valley.

If we combine now the resonant interactions with the non-de
Broglie forces producing the scattering of the particle at the lattice, we
may
expect the latter
to produce a continual deflection of the
particle as it passes by a lattice element until
the direction of the particle's velocity happens to be
sufficiently close to a Bragg scattering
direction
for the inequality (\ref{6.9l}) to apply. At this point the
particle will
become trapped in the Bragg scattering direction by the
resonant de Broglie forces.

Into which of the possible discrete Bragg scattering
directions any given incident particle
is actually scattered depends not only on the
resonant field-trajectory interaction, but also on
the forces exerted on the particle when it passes
close to an element of the lattice. This will depend on the
sub-lattice-scale details of the particle trajectory. In
practice,
these details  cannot be known well enough in advance to
predict the outcome of any single particle scattering event.
Thus although the basic microphysical equations are
deterministic, scattering experiments can in fact be
predicted
only statistically. This is consistent with the quantum
theoretical
result, but the origin of the indeterminacy is explained now
in the standard terms of classical statistical mechanics.

A quantitative analysis of the statistical distribution of
the scattered particles resulting from the resonant
trajectory-trapping
mechanism
requires a more detailed specification of the metron model
than is possible in the present paper. However, it is
qualitatively clear
that the scattered particle distribution will
correspond in general appearance to the scattered wave
distribution.  Nevertheless, it can be anticipated that
computations of the scattered trajectories of an
ensemble of incident particles within the framework of the
metron model will not
map one-to-one onto the corresponding quantum theoretical
wave
scattering computations, but will
yield relative intensities for the different Bragg
scattering
beams which differ in detail from the standard quantum
theoretical
results. Bragg scattering experiments should therefore
provide a good test of the metron model. This may
not be entirely straightforward, however. Quantitative
verifications of quantum theory using particle diffraction
data - beyond the verification of the essentially
kinematical Bragg scattering conditions - have proved 
notoriously difficult \cite{van}. Normally, measured
diffraction intensities are used in the inverse modelling
mode to reconstruct the unknown lattice scattering 
potentials. A conclusive discrimination between the two 
theories will require independent information on the 
lattice scattering properties.


\section{Atomic spectra}
\label{Atomic spectra}
\typeout{################################}
\typeout{################################}
\typeout{        START OF met4-7.tex}
\typeout{################################}
\typeout{################################}

\subsection*{The metron approach}
The most impressive quantitative success of quantum
theory is
undoubtedly the explanation of atomic spectra by quantum
electrodynamics, in particular the highly accurate
prediction of the
hydrogen spectrum.  Can the metron model reproduce these
results?

At first sight it appears unlikely that a discrete particle
theory
should be able to yield the same results as a continuous
field theory.
However, as in the case of Bragg scattering, a close
correspondence
between the metron model and quantum theory can be established also in
the case
of atomic spectra, since the metron model describes not only
discrete particles and field-particle coupling, but also
nonlinear interactions between fields alone. It was shown in
Section~\ref{The Maxwell-Dirac-Einstein Lagrangian} that the
fermion-electromagnetic sector of the metron field-field
interaction equations are
identical to lowest order to the standard coupled
Maxwell-Dirac
field
equations of quantum electrodynamics. The field-trajectory
interactions of the metron model, on the other hand, have no
counterpart in QED.
However,
they exhibit an interesting correspondence to Bohr's
original orbital
theory: the conditions for resonant field-orbit coupling
will be found
to be essentially the same as the quantum orbital conditions
of Bohr.
The resonant field-orbit interactions give rise to an
additional force
(current) which balances the radiative damping of the
orbiting
electron, thereby also resolving the classical dilemma that
an
orbiting electron does not represent a stable
steady state. It is of interest in this context that the
existence of particle-like atomic states corresponding to
electrons travelling on Kepler orbits has recently been
demonstrated using picosecond pulse technology \cite{gae}, although
the results can be explained also in
the standard quantum theoretical picture \cite{mad}.

The solutions of the metron field-field interaction
equations for the scattered de Broglie field of an electron
in the Coulomb field of a nucleus are just
the standard Maxwell-Dirac eigenmodes, which are forced in
the present case, however, by
additional
field-particle interaction terms. The eigenmodes in turn
determine the
conditions under which an electron can become trapped in a
stable
orbit of the coupled electron-nucleus system. Once the
eigenmodes of
the field-field interaction equations have been determined,
the
computation of the associated electron orbits from the
field-orbit
coupling can be treated as a second independent problem.
Thus at the lowest interaction order, the metron
computation of the stable states of the interacting
electron-nucleus
system reduces  essentially to the
standard
eigenmode problem of QED at the tree level. It remains to be
investigated whether the metron computations reproduce the
observed atomic spectra also at higher order, where the
details of the metron and QED computations differ.

\subsection*{Atomic field interactions}

To derive the field-interaction relations, let the total
field of
the interacting nucleus-electron system be represented
generally as
the superposition (suppressing tensor indices)
\begin{equation} \label{6.9}
g_{tot} = g_n +g_e +g_{int}
\end{equation}
of the fields $g_n$, $g_e$ of the nucleus and orbiting
electron,
respectively, and the interaction field $g_{int}$. The
fields $g_n$
and $g_e$ are defined as the fields which would be
associated
with each particle for a given position of the nucleus and
given
electron orbit if there were no interactions with the other
particle.
From Section~\ref{The Maxwell-Dirac-Einstein Lagrangian} it
follows that the interaction field
$g_{int}$
is determined generally by an equation of the form
\begin{equation} \label{6.10}
D \, (g_{int}, g_n) = F_1 (g_e, g_n) + F_2(g_n, g_e) + F_3
(g_e, g_n, g_{int}),
\end{equation}
where $D$ denotes a linear differential operator acting on
$g_{int}$
that describes the propagation of the field $g_{int}$ in the
presence
of the distortions of the background metric caused by the
nucleus
field $g_n$, and
$F_1$, $F_2$ and $F_3$ represent forcing functions
describing,
respectively, the first order scattering of the field $g_e$
at the
nucleus, the first order scattering of the field $g_n$ at
the orbiting
electron, and higher-order field-field interactions.

It was shown in Section~\ref{The Maxwell-Dirac-Einstein
Lagrangian} that if the
nucleus field $g_n$ consists of
an electromagnetic mean field, described by a mixed-index
tensor
$g^{(n)}_{ \lambda\alpha}$, the propagation of the fermion
component of
the interaction field $g_{int}$, represented by a periodic
harmonic-space tensor $\tilde g^{(int)}_{\alpha \beta}$, is
given to
lowest order by the Dirac equation in the presence
of an
electromagnetic field. Thus the homogeneous equation
\begin{equation} \label{6.11}
D \, (g_{int}, g_n) = 0
\end{equation}
reduces to the QED eigenmode equation for an electron in a
Coulomb
field (regarded here as a classical field equation rather
than an
operator equation).

Solutions of the coupled field and orbit equations can be
constructed
by standard iteration methods.  Starting from the
 first-order Kepler
orbit of the electron in the presence of the first-order
(Coulomb)
nucleon field $g_n^{(1)}$, the
lowest-order interaction field $g_{int}$ can be computed by solving the
inhomogeneous eq.
(\ref{6.10}) with forcing terms $F_1$, $F_2$ determined from
the
quadratic interaction of the first order fields $g_n^{(1)}$
and
$g_e^{(1)}$ (where $g_e^{(1)}$ is taken as the
time-symmetric
non-radiating field). The procedure can then be iterated
using higher-order orbit and field approximations and
higher-order
coupling terms
in the forcing functions $F_1$, $F_2$ and $F_3$.

Apart from possible divergences in the expansion procedure
--
which
will not be investigated here -- this straightforward
approach breaks down when the
inhomogeneous eq.(\ref{6.10}) is forced in resonance. If
the
forcing frequency is equal to the eigenfrequency of one of
the normal
modes of the homogeneous eq.(\ref{6.11}), a stationary
solution
does not exist and the expansion procedure must be modified.
It will be shown, in analogy with the wave-trajectory
resonance
phenomena in
the case of Bragg scattering, that these resonant solutions
represent
stable states into which all solutions will slowly drift if
exposed to
external perturbations.

\subsection*{Field-orbit interactions for a circular orbit}

The behaviour of the solution in the neighbourhood of a
resonance can
be investigated by expanding the field $g_{int}$ with
respect to the
eigenfunctions $\psi_p$ of~(\ref{6.11}),
\begin{equation} \label{6.12}
g_{int} = \sum \, a_p (t) \, \psi_p ({\bf x}),
\end{equation}
where the coordinates ${\bf x}$, $t = x^4$ refer to the
restframe of
the nucleus, and for the present qualitative discussion
tensor and spinor indices
and the polarization relations between the Dirac field
$\psi_p$
and the metric tensor have been suppressed.

The evolution of the amplitudes $a_p$ of the individual
modes can be
determined by projecting the inhomogeneous equation
(\ref{6.10}) onto
the eigenfunction $ \psi _p$. Noting that the interactions
between the
emitted de Broglie field of the electron and the mean field
of the
nucleus at either of the two particle kernels yields a
forcing
function $F_1$ or $F_2$ which exhibits the periodicity of
the de
Broglie wave, modulated by the more slowly varying
dependence on the
position of the electron along its orbit, one obtains then
an equation
of the general form
\begin{equation} \label{6.13}
(d/ \, dt - i \, \omega_p) a_p = \gamma_p(t) \exp[iS(t)] = :
f_p(t),
\end{equation}
where $\omega_p$ is the eigenfrequency of the mode $p, S(t)$
denotes
the phase function of the (quasi-periodic) de Broglie field
of the
orbiting electron and $\gamma_p(t)$ is a modulation factor
which
depends on the position ${\bf x} = \mbox{\boldmath
$\xi$}(t)$ of the electron on its
(given) orbit and thus exhibits the
periodicity of the
electron orbit. The local frequency of the de Broglie wave
in the
restframe of the nucleus is given by
\begin{equation} \label{6.14}
\omega = dS/dt = dS/ds \, {(u^4)}^{-1} = \omega_e {(u^4)}^{-
1},
\end{equation}
where $u^{\lambda} = dx^{\lambda}/ds$ is the electron
velocity with
respect to the electron eigentime $s$ and $ \omega_e$ is the
electron
frequency (rest mass).

The net forcing function $f_p(t)$ can be represented as the
product of
a forcing term with the central frequency
\begin{eqnarray} \label{6.15}
\bar \omega & : = & T^{-1} \{S(t+T) - S (t) \} \nonumber \\
& = & \omega_e\, T^{-1}  \int^T_0 [u^4 (t)] ^{-1} \, dt
\end{eqnarray}
and a slowly varying modulation factor which is periodic
with the
orbital period~$T$:
\begin{equation} \label{6.16}
f_p(t) = \gamma_p(t) \, \exp(i \delta S) \exp(i \bar
\omega t),
\end{equation}
where
\begin{equation} \label{6.17}
\delta S (t) : = \{S(t) - S(0) \} - (t/T) \{S(T) - S(0) \}.
\end{equation}
The frequency spectrum
\begin{equation} \label{6.18}
f_p(t) = \sum_n \gamma_{pn} \exp(i \omega_n t)
\end{equation}
of the forcing $f_p$ consists then of a central line $n = 0$
at the
mean de Broglie frequency $\omega_0 = \bar \omega (\cong
\omega_e)$
and a sequence of split lines at the frequencies
\begin{equation} \label{6.18a}
\omega_n = \bar \omega + n\, \Omega \qquad (n = \pm 1, \pm
2,
\ldots )
\end{equation}
where $\Omega = 2 \pi/T$ is the fundamental orbital
frequency. The
amplitudes $\gamma_{pn}$ depend on the modulation of $u^4$
(which is
determined by the ellipticity of the orbit) and on the
spatial form
of the
eigenfunction $\psi_p$ in relation to the electron orbit
(which
determines the modulation factor $\gamma_p(t)$ in eq.
(\ref{6.13})).

The solution of (\ref{6.13}), with $f_p$ given by
(\ref{6.16}), is
\begin{equation} \label{6.19}
a_p(t) = \sum_n \tilde \Delta( \omega_n - \omega_p)
\gamma_{pn}
\exp(i \omega_nt),
\end{equation}
where
\begin{equation} \label{6.20}
\tilde \Delta( \omega_n - \omega_p) : = -i (\omega_n -
\omega_p)^{-
1}.
\end{equation}
To avoid the singularities at the resonance frequencies
$\omega_n =
\omega_p$, the stationary response function
$\tilde \Delta(\omega)$, eq.
(\ref{6.20}), should be replaced by the more general non-
stationary
response function (cf. eq.(\ref{5.a51}))
\begin{equation} \label{6.21}
 \Delta (\omega) : = - i \, (1 - e^{-i \omega t}) \,
\omega^{-1}
\end{equation}
representing the solution to the initial-value problem $a_p
= 0$ for
$t = 0$. This exhibits secular growth, $ \Delta(0) = t$,
instead of a singularity  at  the  resonance
frequencies.  In applications involving the integration of
the response function $\Delta (\omega)$ across resonances,
it
 can be represented by the asymptotic
$\delta$-function relation, valid for large $t$,
\begin{equation} \label{6.22}
\Delta (\omega) \cong  \pi \, \delta(\omega).
\end{equation}
Alternatively, a small damping term $\mu a_p$
(parametrising, for
example, the higher-order effects of the back-interaction,
discussed
below, of the scattered field on the electron motion) can be
added to
the left hand side of (\ref{6.13}), yielding the response
function
\begin{equation} \label{6.23}
\hat \Delta (\omega) = (i \omega + \mu)^{-1}.
\end{equation}
This can also be approximated by a $\delta$-function for
small
$\mu$.  In
the neighbourhood of a resonance both functions (\ref{6.21})
and
(\ref{6.23}) exhibit the same behaviour for $t = O( \mu^{-
1})$.

As long as the electron orbit is not near a resonance, small
external
disturbances and the radiative damping of the oscillating
electron
will produce a drift from one Kepler orbit to another,
normally from
higher energies to lower. However, when the orbit drifts
into a
resonance, energy is transferred from the orbit to the
resonant
eigenmode field; the quadratic interaction between the
electron's own
de Broglie field and the Dirac eigenmode field which is
generated by
the electron-nucleus interaction results then in a mean
force which is
able to counterbalance the external drift forces. Thus the
electron
orbit will generally drift freely under the influence of
small
external disturbances until it encounters a resonance, when it
becomes trapped in the effective $\delta$-function
potential-energy
canyon arising from the orbit-eigenmode resonant
interaction.

We illustrate the orbit-trapping mechanism  for the case of
a
circular
orbit. Applying the secular perturbation techniques of
Keplerian
mechanics, the orbit drift can be described generally by an
equation
of the form
\begin{equation} \label{6.24}
dr/dt = -\, d + \mbox{Re}(\alpha A_p),
\end{equation}
where $d$ represents the rate of change of the orbit radius
$r$ due to
external forces and radiative damping and Re$(\alpha A_p)$
represents
the net drift due to quadratic interactions between the scatttered de
Broglie
field and the periodic kernel field of the electron. The latter term can be
represented by
the real part of the product of a  complex constant
$\alpha$, which defines a reference phase for the forcing,
and the amplitude $A_p $ of
the eigenmode $a_p = A_p \, \exp(i \bar \omega
t)$. The form of this forcing term follows from the
observation
that, despite the local Doppler frequency shift induced by
the electron
motion, the de Broglie field of the electron within the
electron core is always in
resonance
with the scattered field with which it interacts when
considered over an orbital
period: the
number of waves emitted by the electron during an orbital
period is
the same as the number of scattered waves the electron
encounters, so
that in the reference frame of the nucleus the central
frequency $\bar \omega$  (and in fact also the line
splitting
through the orbital motion) are the same for both the
periodic field in the electron core and its
scattered field. For simplicity, only a single eigenmode
$p$ and the
central forcing frequency $\bar \omega = \bar \omega(r)$
will be
considered.

In the neighbourhood of the resonant frequency $\omega_p$,
the forcing
frequency can be represented as
\begin{equation} \label{6.25}
\bar \omega (r) = : \omega_p + \beta \, \delta r,
\end{equation}
where $\delta r : = r - r_p$ is the deviation from the
resonant radius
$r_p$ and $\beta$ is a constant. In the neighbourhoood of a
resonance
$d$ and $\alpha$ can similarly be regarded as constant.

Substituting the expansion (\ref{6.25}), together with the
stationary
response relation (\ref{6.23}) for $\tilde\Delta$, into the
single-mode
version of the response equation (\ref{6.19}), the amplitude
of the
eigenmode is given by
\begin{equation} \label{6.26}
A_p = \gamma \, (i \beta \, \delta r + \mu)^{-1},
\end{equation}
where $\gamma = \gamma_{po}$.  This yields for the
drift equation
(\ref{6.24})
\begin{equation} \label{6.27}
d \delta r/ dt = d \{ -1 + (C_1 \, \delta r + 2 C_2) /
(\delta r^2
+ C_3) \},
\end{equation}
where
\begin{eqnarray} \label{6.28}
C_1 & : = & \mbox{Im}(\alpha \gamma) / (2 \beta d) \nonumber
\\
C_2 & : = & \mu\, \mbox{Re} (\alpha \gamma) / (\beta^2 d) \\
C_3 & : = & \mu^2 / \beta^2 \nonumber.
\end{eqnarray}
Equation (\ref{6.27}) has two equilibrium solutions (cf.
Fig.\ref{figtrap}):
\begin{figure}[t] \centering
\begin{minipage}{9cm}
\epsfxsize220pt
\epsffile{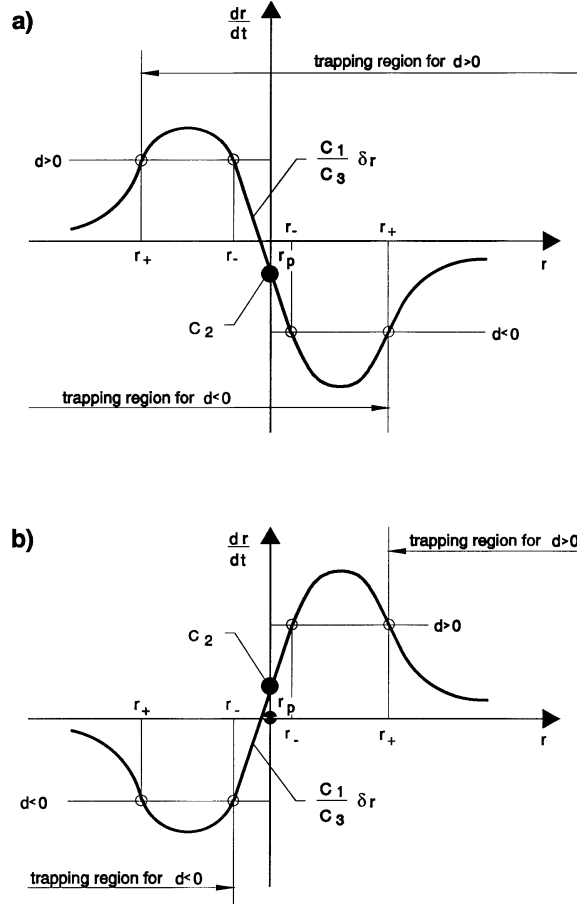}
\caption{\label{figtrap}
Orbit trapping in resonant potential energy canyon
(panel a) and at potential energy barrier (panel~b)}
\end{minipage}
\end{figure}

\begin{equation} \label{6.29}
\delta r_{\pm} = C_1 \pm \left[ C_1^2 + C_2 - C_3 \right
]^{1/2}.
\end{equation}
For sufficiently small $\mu$, the solutions are always real.

If $C_1$ is negative, the solution $\delta r_-$ closest to
the
resonant point is stable, independent of the sign of the
drift term.
In this case the resonant mode-orbit interaction potential
corresponds
to a potential energy `canyon' (Fig.{\ref{figtrap}a).  The
second solution $\delta r_+$ is
unstable; it represents the boundary of the orbit-trapping
region. If
$d$ is positive (radiative damping), $\delta r_+ < \delta
r_-
$; in this
case, the orbit can escape from the attractor and drift to
smaller
values of $r$ if initially $\delta r < \delta r_+$, while
for $\delta
r > \delta r_+$, the orbit falls always into the stable
solution
$\delta r_-$.  Conversely, for negative $d$ (radiative
heating),
$\delta r_+ > \delta r_-$, and the orbit can escape from the
attractor if
initially $ \delta r > \delta r_+$, drifting otherwise into
the stable
solution $\delta r_-$.

Positive $C_1$ corresponds to repulsive resonant mode-orbit
interactions (potential energy barrier rather than canyon,
Fig.{\ref{figtrap}b). In this
case the solution $\delta r_+$ furthest away from the
resonant point
is stable and prevents orbits which are drifting towards the
resonant
radius from reaching the resonant point. Between $r_+$ and
$r_-$, the
orbits are attracted to the stable solution $r_+$, while
beyond $r_-$ (on the side oppposite to $r_+$)
the orbits follow the direction of the drift away from the
resonance point.

Trapping at a potential energy barrier is less stable with
respect to
changes in the external drift forces than trapping in a
potential
energy canyon. If the drift rate $d$ is randomly varying, a
change in
sign of $d$ releases the electron from a positive-energy
barrier and
allows it to drift in the opposite direction away from the
resonance
point, whereas an electron trapped in a potential energy
canyon can be
freed only through a very large external drift force
(cf.~Fig.\ref{figtrap}), or through interactions with
external
fields which destroy the trapping field $A_p$, as discussed
in the
following sub-section.

The determination of the sign of $C_1$ involves the
determination of
the scattering source terms in eq.(\ref{6.10}), which
requires
computing
the quadratic interaction between the scattered de Broglie
fields
and the periodic source fields of the electron core (cf.
also eq.(\ref{6.24})).  This is
beyond the
scope of the present analysis.  It will be assumed in the
following
that $C_1$ is negative and the mode-orbit interactions yield
trapping
potential energy canyons.

If the drift rate $d$ represents radiative damping, it
follows
that in
the stable trapped-orbit state, the quadratic interaction
between the
scattered, resonant eigenmode field $\psi_p$ and the Dirac
(de
Broglie) field within the electron core must give rise to a
force
which,
averaged over an orbit, exactly balances the radiative
damping. In Section~\ref{The Maxwell-Dirac-Einstein
Lagrangian}, it was shown that the periodicity of the Dirac
fields is
associated with the electric charge. Thus  the force
generated
by the quadratic interactions must represent  an
electromagnetic force,
with an associated electromagnetic current. The balancing of
the
classical radiative damping force and the internally
generated
electromagnetic interaction force implies that the
inherent electromagnetic current of
the orbiting electron is balanced by the current
generated by the
interactions with the scattered field. But if there exists
no net electromagnetic current there can exist no
net radiative damping, thereby resolving the classical
dilemma of the radiative collapse of the Bohr orbits.

\subsection*{Relation to Bohr's orbital theory}

In the case of a circular orbit, the orbit-eigenmode
resonance
condition can be shown to reduce to the familiar Bohr
orbital
quantum
condition. Consider the dominant interaction involving the
central
orbital frequency $\bar \omega$. In the non-relativistic
limit, the
eigenfrequency of the eigenmode can be written
\begin{equation} \label{6.30}
\omega_p = \omega_0 + \omega'_p,
\end{equation}
where $\omega'_p$ is the eigenfrequency of the solution of
the
Schr\"odinger equation.  To first non-relativistic order, the
central
forcing frequency $ \bar \omega$, eq.(\ref{6.15}), for a
circular
orbit is given by
\begin{equation} \label{6.31}
\bar \omega = \omega_0
\left[
1 - T^{-1}c^{-2} \int \frac{{\bf v}^2_p}{2} dt
\right]  =
\omega_0 (1 + E_p / (mc^2)),
\end{equation}
where ${\bf v}_p : = d{\bf x}/dt, E_p$ is the total (kinetic
plus
potential) energy of the orbiting electron and m is the
electron rest
mass (using here dimensional units). Thus the resonance
condition
$ \bar \omega = \omega_p$ yields
\begin{equation} \label{6.32}
\omega'_p = E_p \, \omega_0 / (mc^2)
\end{equation}
or, with $mc^2 = \hbar \, \omega_0$,
\begin{equation} \label{6.33}
E_p = \hbar  \, \omega'_p.
\end{equation}
This is identical to Bohr's result. Bohr determined discrete
values of
$E_p$ from his orbital quantum condition, and then defined the
associated
frequencies through (\ref{6.33}). We have followed the
reverse path of
determining $\omega'_p$ from the eigenmode of the
Schr\"odinger
equation and have
found then that the energy of the orbiting electron
satisfies (\ref{6.33}),
in agreement with Bohr's relation.

\subsection*{Generalization to elliptical orbits}

The above analysis can be extended to arbitrary elliptical
orbits and
the general Bohr-Sommerfeld quantum orbit conditions.  For
this
purpose it is convenient to transform from standard (e.g.
spherical)
canonical variables $q_k$, $p_k$ to the Delaunay canonical
elements
$\alpha_k$, $J_k$. These consist of three action variables
$J_k$,
where $J_1$ is related to the total energy $E, J_2$ is the
total
angular momentum $P$ and $J_k$ the angular momentum $P_z$
in the
$z$-direction, and their three associated cyclic coordinates
$\alpha_k$.  For small perturbations about a spherically
symmetrical
field, the particle motion is multi-periodic with
periodicity 2$\pi$
with respect to the angle variables $\alpha_k$, which grow
linearly in time,
\begin{equation} \label{6.34}
\alpha_k = \omega_k  t + \mbox{const}.
\end{equation}
The three frequencies $\omega_k$ are identical in the
degenerate case
of a Coulomb field, but differ in the presence of
symmetry-breaking
perturbations.

The transformed Hamiltonian $\tilde H ({\bf J}) = H({\bf
q},{\bf p})$
is independent of {\boldmath $\alpha $}.  The transformed
Lagrangian is
accordingly given by
\begin{equation} \label{6.35}
\tilde L (\mbox{\boldmath $\alpha $},{\bf J})  =  \sum_k
J_k\,  d\alpha
_k/dt  -  \tilde
H({\bf J}) = \sum_k \, J_k  \,  \omega_k  -  \tilde  H
({\bf
J})  =  \tilde  L
(\mbox{\boldmath $ \omega$}, {\bf J}),
\end{equation}
so that the action variables are related to the Lagrangian
through
\begin{equation} \label{6.36}
J_k = \partial \tilde L/ \partial \omega_k.
\end{equation}
Perturbations of the system by additional time-dependent
interactions
give rise to drifts of the action variables $J_k$.  These
can be
treated by regarding the variables $\omega_k, J_k$ in the
Lagrangian
$L(\mbox{\boldmath $\omega $}, {\bf J})$, eq.(\ref{6.35})
(where the tilde has now
been dropped), as slowly varying with time.  Variation of
$L$ with
respect to the orbital elements $J_k$ yields again the usual
Kepler
orbital relations $\omega_k = \omega_k({\bf J})$, modified
by additional perturbation terms. Variations with respect
to
the angle variables or phase
functions $\alpha_k$, with $\omega_k = d\alpha_k/dt$, yield
the orbit drift equations
\begin{equation} \label{6.37}
\frac{d}{dt}  \left(  \frac  {\partial  L}{\partial
\omega_k}  \right   )   =
\frac{d}{dt} J_k = 0.
\end{equation}

Assuming now, as before, that the Kepler orbit is perturbed
by
interactions with external fields (radiative damping or
heating) and
by the de Broglie field-particle interactions, the
Lagrangian takes the form
\begin{equation} \label{6.38}
L({\bf J}, \mbox{\boldmath $\omega $}) = L^n + L^e + L^{dB},
\end{equation}
where $L^n$ describes the interaction with the
time-independent field
of the nucleus, $L^e$  the
interactions with
external time-dependent fields and $L^{dB}$ represents the
interaction between the electron and
scattered
de Broglie fields.  Equation (\ref{6.37}) can
thus be written
\begin{equation} \label{6.39}
\frac{d}{dt} J^n_k = - \frac{d}{dt} J^e_k - \frac{d}{dt}
J^{dB}_k,
\end{equation}
where $J^n_k$ denotes the action variables as defined by the
unperturbed Lagrangian, $\frac{d}{dt} J^e_k = : d^e_k$ is
the
drift of
$J^n_k$ induced by the external interactions and
$\frac{d}{dt}
J^{dB}_k = \frac{d}{dt} ( \partial L ^{dB}/ \partial
\omega_k) = : d^{dB}_k$ is the drift due to the de Broglie
field-particle interactions.

The Lagrangian $L^{dB}_k$ is given by the interaction of the
electron's de Broglie field $\psi_e$ with the scattered de
Broglie field $g_{int}$ (eq.(\ref{6.12})). This is
regarded
as a given external field in the variation of the Lagrangian
with respect to the orbit. The de Broglie field of the
orbiting electron is composed in general of a component at
the  central
frequency $ \bar \omega$ (eq.(\ref{6.15})) and a spectrum
of
split lines at the frequencies
\begin{equation} \label{6.40}
\omega_{kn} : = \bar \omega + n \, \omega_k \; \quad (n = \pm1,
\pm2, \ldots )
\end{equation}
In place of the expression (\ref{6.15}) appropriate for a single
periodicity, we define now more
generally the
mean
frequency $\bar \omega$ as the mean rate of change of the
phase of the
de Broglie field of the electron averaged over a
sufficiently long
time interval to effectively include all of the orbit
periodicities.  Similarly,
the line
splitting is characterized now not just by a single orbital
frequency,
but by the three cyclic frequencies $\omega_k$. The
time-averaged
Lagrangian $L^{dB}$ thus has the general form
\begin{equation} \label{6.41}
L_{dB} = \sum_p  \,  \bar  \alpha_p  <\!a_p  (t)  \exp  (i
\bar  \omega  t)\!>  +
\sum_{k,n,p} \, \alpha_{knp} <\!a_p  (t)  \exp  (i
\omega_{kn}  t)\!> + \; c.c.,
\end{equation}
where $<\ldots >$ denotes a time average and
$\bar\alpha_p,\alpha_{knp}$ are constant complex
coefficients
characterizing the quadratic coupling of the eigenmode
$\psi_p$ in
the expansion (\ref{6.12}) of the scattered field $g_{int}$
with the frequency
components $ \bar
\omega, \omega_{kn}$, respectively, of the de
Broglie field of the electron.

In computing now the action variable $J^{dB}_k$
from (\ref{6.41}), applying
 (\ref{6.36}), secular terms arise through the time
derivative of the exponential factors, yielding for the de
Broglie
drift
\begin{eqnarray} \label{6.42}
d^{dB}_k &=& i \, \partial \bar  \omega  /\partial  \omega_k
\sum_p \bar \alpha_p <\! a_p (t) \exp (i \bar \omega t) \!>
\nonumber\\
&& \quad\quad\quad + \sum_{n,p} \, i
\, n \, \alpha_{knp} <\!a_p (t) \exp (i \omega_{kn} t)\!> +
\; c.c.
\end{eqnarray}
Retaining, as before, only the interaction with the dominant
central frequency $\bar \omega$, represented by the first
term on the
right hand side of (\ref{6.42}), we obtain,
substituting the solution
(\ref{6.19}), (\ref{6.20}) for $a_n$ and using the
asymptotic form
(\ref{6.22}) for the response function $ \Delta$,
\begin{equation} \label{6.43}
d^{dB}_k =  \pi i \, \partial \bar \omega/ \partial
\omega_k \, \sum_p  \bar
\alpha _p \, \gamma_{po} \delta( \bar \omega - \omega_p) +
c.c.
\end{equation}
The de Broglie drift term is thus limited to the field-orbit
resonance
surfaces $ \bar \omega = \omega_n$, where the effectively
infinite
$\delta$-function factor ensures that the particle becomes
trapped. On
the resonance surface, the orbit can continue to drift with
respect to
the remaining two degrees of freedom until it reaches a
stable point where the remaining de Broglie drift terms become zero.
Thus it can be expected that for each eigenmode with
eigenfrequency $\omega_p$ there will exist in general
(at least) one stable attractor, an elliptical orbit which is in trapped
resonant interaction with the Dirac eigenmode.

\subsection*{Interaction with radiation}

Having outlined the general metron picture of the origin and
nature of discrete atomic states, there remains the question
of the interaction of these discrete states with
electromagnetic radiation, and  the mechanism of the
transition
from one stable atomic state to another.
In quantum theory, transitions between atomic states are
associated with interactions with electromagnetic
radiation for
which the frequencies of the three fields involved (atomic
states 1
and 2, electromagnetic radiation $1 \bar 2$) satisfy the
resonant
interaction conditions
\begin{equation} \label{6.44}
\omega_1 - \omega_2 = \omega_{1 \bar 2}.
\end{equation}
The same interaction principles apply also for the metron
model.

The transition mechanism can be illustrated by a simple
generalization
of the model eqs. (\ref{6.24}) - (\ref{6.29}) for a
circular orbit. We
consider again only the dominant interactions with the
central orbital
frequency $\bar \omega$, which is assumed to be close to
resonance
with the eigenmode 1, $\bar\omega = \omega_1 + \beta \delta
r$, where
$\delta r = r - r_1$.  Writing $a_p = A_p \exp(i\omega_p t)$
for the
modes $p = 1,2$ , where the amplitudes $A_p= A_p(t)$
(referred now to
the frequencies $\omega_p$ rather than $\bar \omega$) are
slowly
varying
with time, and setting similarly the electromagnetic field
proportional to $A_{1\bar 2}(t) \exp(i \omega_{1 \bar 2}
t)$, the
interactions between the three fields have the general
structure
\begin{equation} \label{6.45}
dA_1/dt + \mu_1A_1 = i \, K \, A_{1  \bar  2}  A_2  +
\gamma  \,  e^{i  \beta
\delta rt},
\end{equation}
\begin{equation} \label{6.46}
dA_2/dt + \mu_2A_2 = i \, K^* A^*_{1 \bar 2}A_1,
\end{equation}
\begin{equation} \label{6.47}
dA_{1 \bar 2}/dt = i \, K^* A_1 \, A_2^*,
\end{equation}
where the forcing $\gamma e^{i \beta \delta rt}$ by the
orbiting
electron is included only for the near-resonant first mode
and $K$ is
a coupling coefficient appearing in the cubic interaction
Lagrangian $
\sim K A^*_1 A_{1 \bar 2}A_2$.  The third equation
(\ref{6.47}) is
needed only in the case of emitted electromagnetic
radiation. If
radiation is absorbed, the field $A_{1 \bar 2}$ is regarded
as a
specified external field.

The mode interaction equations must be augmented by the
orbit equation
(\ref{6.24}), which in the present case takes the form
\begin{equation} \label{6.48}
dr/dt = - d + \mbox{Re} (\alpha_1 e ^{-i \beta \delta rt}
A_1) + \mbox{Re} (\alpha_2 e ^{-i \beta \delta rt} A_2).
\end{equation}

The evolution of the coupled system $A_1, A_2, A_{1 \bar
2}$, $r$
depends in detail on whether the absorption or emission of
radiation is being considered. However, in both cases the
cross-coupling through the field $A_{1 \bar 2}$ has the
effect that the resonant forcing of the eigenmode 1 is no
longer confined to  mode~1 but is communicated also to
mode~2.

Consider first the case of the interaction with a
prescribed, time
independent coupling field $A_{1 \bar 2}$. The coupled
homogeneous
equs (\ref{6.45}), (\ref{6.46}), without the forcing terms,
then have
(weakly damped) coupled harmonic oscillator solutions of
frequency
$\omega_c = |K A_{1 \bar 2}|^{1/2}$, in which the amplitudes
$A_p$ of
both eigenmodes oscillate with constant relative phase and
with
the same
oscillation amplitude.  Suppose now that the external
radiation field
$A_{1 \bar 2}$ is suddenly turned on at a time when the
electron is
trapped in
the resonant orbit $r = r_1$, so that initially $A_1 \not=
0$, $A_2 =
0$. Since this no longer represents the equilibrium solution
of the
coupled system, a free oscillation will be excited. The
alternating
signs of the amplitudes $A_1$ and $A_2$ of the free
oscillation
produce oscillating forcing terms in the orbit drift
equation
(\ref{6.48}), instead of the restoring forces of
the previous
decoupled single-mode system. This results in a breakdown of
the `potential energy
canyon', and the electron can escape from the resonant
orbit.

The details of the escape mechanism are a little more
complicated than
outlined here, since the (nonlinear) feedback of the
changes in the
orbit radius through the orbital forcing term in eq.
(\ref{6.45})
must be taken into consideration. However,  the basic
mechanism of the breakdown of the potential energy
canyon in the presence of
sufficiently strong cross-mode coupling should remain valid,
independent of these details.

An alternative, probably more realistic model of the
interaction with an externally
prescribed radiation field is to represent $A_{1 \bar 2}$ as
a
stochastic process characterized by a continuous variance
spectrum
$F_{1 \bar 2} (\omega)$. In the absence of orbital forcing,
the
evolution of the variances $N_p = <\!|A_p|^2\!>$ can be
shown to
be
governed in this case by the coupled equations \cite{has1}

\begin{equation} \label{6.49}
\frac{d}{dt} \, N_1 - 2 \, \mu_1 \, N_1 = K'(N_2 - N_1)
\end{equation}
\begin{equation} \label{6.50}
\frac{d}{dt} \, N_2 - 2 \, \mu_2 \, N_2 = K'(N_1 - N_2),
\end{equation}
where
\begin{equation} \label{6.51}
K' : = 2 \pi \, F_{1 \bar 2} (0) |K^2|
\end{equation}
(note that the spectral density $F_{1 \bar 2} (0)$ of
the amplitude $A_{1 \bar 2}$ at zero frequency
corresponds to the spectral density  of
the electromagnetic radiation at the resonant coupling
frequency $
\omega_{1 \bar 2}$). Ignoring  the  damping,  the  solutions
tend
again to an equilibrium in which both eigenmodes have equal
variances
and energy is continually exchanged between the two
modes. The mode-orbit interaction terms in the orbit
equation
(\ref{6.48})
will exhibit in this case random fluctuations similar to the
oscillations
in the
case of a constant field $A_{1 \bar 2}$, resulting again in
a
breakdown
of the potential energy canyon and a release of the trapped
electron.

Similar considerations apply for the case of emitted
radiation,
except that
here the field $A_{1 \bar 2}$ is not prescribed, but is
generated
spontaneously through an instability of the coupled set of
modes $A_1,
A_2, A_{1 \bar 2}$: for given finite $A_1$, a pair of
initially
infinitesmal perturbations $A_2, A_{1 \bar 2}$ will grow, according to
equations (\ref{6.46}), (\ref{6.47}), as $e^{\nu t}$, where
$\nu : = -
\mu/2 + \{(\mu /2)^2 + |KA_1|^2\}^{1/2}$ \cite{ft32}. As the fields $A_1$
and $A_{1
\bar
2}$ grow,
the field $A_1$ decreases, and the field-orbit interaction
terms in
the orbit drift equation begin to oscillate, leading again
to a
release of the trapped electron.

Not considered in this discussion is the further fate of the
electron
after it has been freed from its trapped-orbit state 1.  The
subsequent capture of the electron in the trapped-orbit
state 2 involves  a higher-order analysis of the
interactions
between the
scattered fields and the electron's Dirac and
electromagnetic fields
for non-resonant orbits, which will not be attempted here.
While conceptually straightforward - since all fields and
particles are well
defined 'objects' with well-defined interactions - the
analysis of the
three-way interactions between an orbiting electron, its
associated
scattered fields and additional electromagnetic radiation
fields,
whether internally or externally generated, is clearly a
non-trivial
task if carried out in quantitative detail.  The purpose of
this
rather cursory preliminary analysis is only to demonstrate
that the known
general interrelations between atomic eigenstates and
atomic
radiation appear to be basically consistent with the metron
picture.

We note in conclusion that we have treated
electromagnetic radiation here in the traditional manner
as prescribed incident radiation or as emitted radiation
into space, although we specifically made the point in
Section~\ref{The radiation condition} that in the metron
model
electromagnetic fields should not be viewed
 as independent radiation fields, but
rather as interaction fields describing the
coupling between pairs of charged particles. The traditional
radiation
picture is obtained in the present case by dividing the
complete system of
interacting particle pairs into the orbiting electron, as
the object of immediate interest, and all remaining
particles,
which are regarded as external to the system under study.
While this description is convenient for the present
analysis, it is
important to keep the particle-interaction picture in mind
when
considering the statistical properties of radiation. In
general, either picture can be applied.
Thus the Stefan-Boltzmann spectrum, and the
corpuscular properties of electromagnetic radiation with
which this is normally associated, can be readily interpreted
in the metron picture, following Einstein \cite{ein22},  in terms of
electromagnetic interactions mediating the transitions
between discrete atomic states for an ensemble of atoms in
thermodynamic equilibrium.

\subsection*{Open questions}

A number of basic questions have clearly not been addressed
in this brief outline of a possible metron theory of atomic
spectra. The orbital parameters and stability properties
of the set of resonant electron orbits associated with the
set of eigenmodes have not been determined for the general
elliptical-orbit case.  It has also not been demonstrated -
although it appears intrinsically plausible - that  for a
multi-electron system there exists only one
stable electron orbit per eigenmode, so that Pauli's
exclusion principle can be derived, rather than having to be
postulated.

Other questions relate to the higher-order quantitative
equivalence of the metron Dirac-electromagnetic interaction
equations and the standard QED formalism. The equivalence
of the field equations shown in Section~\ref{The
Maxwell-Dirac-Einstein Lagrangian}  applies only
at the tree level, ignoring closed loop contributions,
and to lowest interaction order. Barut \cite{bar} has claimed that
higher-order first-quantization  computations yield atomic
spectra at least to the same accuracy as QED computations.
It remains to
be investigated whether this holds also with the inclusion
of the higher-order interaction
terms of the gravitational Lagrangian (the higher-order
terms of the infinite interaction series of the
gravitational
Lagrangian have no counterpart in the cubic
Dirac-electromagnetic interaction Lagrangian of QED). Of
particular interest are the
higher-order interactions within the metron near-field
regions, which we anticipate will be
needed to  ensure a divergence-free
interaction
expansion.

\newpage
~
\newpage

\newcommand{\psivec}{\mbox{\boldmath$\psi$}}
\part{The Standard Model}
\label{The Standard Model}
\typeout{################################}
\typeout{################################}
\typeout{        START OF met4-8.tex}
\typeout{################################}
\typeout{################################}
%
{\em ABSTRACT} \\

\noindent
In the first three parts of this  paper we
developed a unified, deterministic model of fields and
particles based on the postulated existence of soliton-type
({\it metron}) solutions of the higher-dimensional vacuum
gravitational equations. Following the demonstration in
Part~\ref{The Metron Concept}
that
such solutions exist for a simplified, scalar prototype of
the gravitational Lagrangian, the metron model was
investigated in more detail for the Maxwell-Dirac-Einstein
system in Part~\ref{The Maxwell-Dirac-Einstein System} and
then applied in Part~\ref{Quantum Phenomena}
to explain basic quantum phenomena
such as the EPR paradox, interference effects in scattering
experiments and atomic spectra.

In the final part of this paper we generalize the
interaction analysis of the Maxwell-Dirac-Einstein system to
include weak and strong interactions. It is shown that the
principal properties of the Standard Model can be recovered
by a four-dimensional, or (with closer agreement)
five-dimensional non-Euclidean or Euclidean
harmonic-space background metric, assuming a suitable
geometric structure of the trapped-mode metron solution. The
solution is assumed to
be composed of the basic fermion fields, representing
leptons and quarks of different color and flavor, and the
associated boson fields, which are generated by quadratic
difference interactions between the fermions. The fermions
are described by harmonic-space metron components which are
periodic with respect to extra ({\it harmonic}) space, the
wavenumber components $k_5$ and $k_6$ defining the coupling
constants for the electromagnetic and weak interactions,
respectively, while the components   $k_7$ and (in the case
of a
five-dimensional harmonic space) $k_8$ determine the
strong-interaction coupling. The last harmonic dimension is
needed, in combination with the other harmonic dimensions,
to
define an appropriate polarization tensor relating  the
metric-tensor components to the Dirac-field components such
that the  standard Dirac Lagrangian is recovered from the
gravitational Lagrangian.

 A higher-order interaction corresponding to the Higgs
mechanism is invoked to explain the electroweak boson
masses. A quartic interaction in which the  neutrino field
appears quadratically exhibits the desired properties.
Fermion masses are attributed to the $SU(2)$-breaking
mode-trapping mechanism.

The analysis is restricted to a single family; it is
suggested that the second and third families can be
described by higher-order trapped modes. The Standard Model
gauge symmetries are explained as a special case of the
general gauge invariance of the gravitational equations with
respect to diffeomorphisms, applied to a particular class of
coordinate
transformations reflecting the geometrical symmetries
of the metron solutions.

The purpose of the inverse modelling approach pursued in
this paper is twofold: it is demonstrated generally that
soliton-type
solutions of the higher-dimensional vacuum  gravitational
equations exhibit a sufficiently rich structure to reproduce
the principal results or quantum field theory, as summarized
in the Standard Model, while at the same time specific
geometrical features of the anticipated metron solutions are
identified, which one can then seek to confirm with exact
numerical computations. Such computations should yield not
only the symmetries of the Standard Model but also all
universal physical constants and
particle
parameters.\\

\subsection*{\raggedright Keywords:}
{\small
metron ---
unified theory ---
higher-dimensional gravity ---
solitons ---
Standard Model ---
physical constants ---
gauge symmetry}\\

{\em R\'ESUM\'E} \\

\vspace*{1ex}
Dans les troix premi\`eres parties de ce travail nous avons
d\'evelopp\'e un mod\`ele unifi\'e d\'eterministe des champs
et particules s'appuyant sur l'existence postul\'ee de
solutions de type soliton (dites m\'etrons)
des \'equations d'Einstein du vide \`a haute dimension.
Apr\`es avoir d\'emontr\'e dans la premi\`ere partie que de
telles
solutions existent dans le cas d'un Lagrangien de
gravitation prototype
scalaire de forme simplifi\'e, le mod\`ele de m\'etron a
\'et\'e
examin\'e plus en d\'etail dans la deuxi\`eme partie
dans le cas du syst\`eme de
Maxwell - Dirac - Einstein. Il a \'et\'e
utilis\'e dans la troixi\`eme partie pour expliquer les
ph\'enom\`enes quantiques fondamentaux tels que celui du
paradoxe
de l'exp\'erience d'EPR, celui des effets d'interf\'erence
lors
des exp\'eriences de diffusion et celui des spectres
atomiques
discrets.

Dans la derni\`ere partie de ce travail nous
g\'en\'eralisons
l'analyse des interactions dans le syst\`eme de
Maxwell - Dirac - Einstein  en incluant les
forces faibles et les forces fortes.
Nous montrons que les propri\'et\'ees principales du
mod\`ele standard peuvent \^etre retrouv\'ees
\`a l'aide d'une m\'etrique de fond Euclidienne ou non-
Euclidienne
de l'espace harmonique \`a quatre ou
(en augmentant l'accord) cinq dimensions
s'il existe une structure g\'eom\'etrique appropri\'ee des
solutions de m\'etron \`a  modes captur\'es.
On suppose que la solution est compos\'ee de champs de
fermions \'el\'ementaires repr\'esentant les leptons et les
quarks aux couleurs et go\^uts diff\'erents, ainsi que les
champs
de bosons associ\'es, provenant des interactions
de diff\'erences quadratiques entre les fermions.
Les fermions sont d\'ecrits par les composantes de m\'etron
d'espace harmonique qui eux, sont des fonctions
p\'eriodiques
dans cet espace. Les vecteurs d'onde
$ k_5 $ et $ k_6 $ repr\'esentent les constantes de couplage
r\'espectivement des forces \'electromagn\'etiques et des
forces faibles;
les composantes $ k_7 $ et (dans le cas d'un espace
harmonique \`a
cinq dimensions) $ k_8 $ repr\'esentent les constantes de
couplage fort.
La derni\`ere dimension harmonique, suppl\'ementaire
aux autres dimensions harmoniques, est n\'ecessaire, afin de
pouvoir d\'efinir
un tenseur de polarisation appropri\'e qui doit relier les
composantes du tenseur m\'etrique \`a ceux du champ de
Dirac.

Une interaction d'ordre sup\'erieur correspondant au
m\'ecanisme
d'Higgs est \'evoqu\'ee afin de pouvoir expliquer l'origine
des masses
des bosons \'electro-faibles.
Une interaction quartique dans laquelle le champ de neutrino
appara\^it
de fa\c{c}on quadratique montre les propri\'et\'es
d\'esir\'ees.
Les masses des fermions sont attribu\'ees au m\'ecanisme
de capture de mode qui brise la sym\'etrie de $ SU(2) $.

L'analyse se restreint \`a une seule famille de leptons et
de quarks;
on sugg\`ere
que la seconde et la troixi\`eme famille peuvent \^etre
d\'ecrites par des modes d'ordres sup\'erieurs captur\'es .
Les sym\'etries de jauge du mod\`ele standard sont
attribu\'ees \`a
des cas particuliers de l'invariance de jauge g\'en\'erale
des \'equations gravitationnelles sous des
diff\'eomorphismes,
quand ceux-ci sont appliqu\'es \`a une classe particuliaire
de transformation des coordonn\'ees refl\`etant
les sym\'etries g\'eom\'etriques des solutions de m\'etron.

L'approche de mod\'elisation inverse poursuivie dans ce
travail
tend \`a identifier les structures
anticip\'ees des solutions exactes de type soliton des
\'equations
gravitationnelles du vide \`a haute dimension
et ainsi \`a montrer
les cons\'equences importantes du concept de m\'etron dans
le
domaine de la physique th\'eorique des particules.
On suppose que les calculs num\'eriques des solutions de
m\'etron
donneront toutes les constantes universelles de physique
ainsi que
les param\`etres des particules.
Les conceptions fondamentales de la th\'eorie sont encore
une fois
r\'esum\'ees dans le dernier paragraphe d'un point de vue
plut\^ot
constructif s'ajoutant ainsi \`a l'approche d\'eductive,
laquelle
a \'et\'e poursuivie dans la premi\`ere partie.\\

\subsection*{\raggedright Mots cl\'es:}
{\small
m\'etron ---
th\'eorie unifi\'ee ---
th\'eorie de gravitation \`a haute dimension ---
solitons ---
mod\`ele standard ---
constantes de physique ---
sym\'etrie de jauge}

\newpage
\section{Introduction}
\label{Introduction 4}

After the digression in Part \ref{Quantum Phenomena} of this
paper to basic
quantum-theoretical
questions concerning the metron interpretation of the EPR
paradox, Bell's theorem, time-reversal symmetry and the
problems
of wave-particle duality, we continue now with the detailed
description of field interactions and particle properties in
the metron model which we had begun in Part~\ref
{The Maxwell-Dirac-Einstein System} with the analysis of the
Maxwell-Dirac-Einstein system. This has two motivations.
First, we need
still to demonstrate that the metron solutions of the
n-dimensional vacuum
Einstein equations yield not only the
Maxwell-Dirac-Einstein theory, including the
gravitational constant
and all par\-ticle properties and physical constants
relevant
for the description of
microphysical phenomena at the atomic level, but
also a
description of
high-energy phenomena at nuclear and sub-nuclear scales.
Secondly, by estabishing a correspondence between the metron
model and the Standard Model, following as before  the
inverse modelling approach adopted in
Part~\ref{The Maxwell-Dirac-Einstein System}, we shall
identify the relevant features of the metron solutions which
one can then later attempt to reproduce in numerical
computations of specific trapped-mode solutions of the
n-dimensional Einstein equations.

To recover
the
Maxwell-Dirac-Einstein equations and the
Wheeler-Feynman point-particle interaction formalism, we
assumed  in Part~\ref{The Maxwell-Dirac-Einstein System}
that charged fields were
periodic with
respect to
some direction in harmonic space (the electromagnetic
coup\-ling direction $x^5$) and  that the metron
solutions support (de Broglie) fermion far fields which are
periodic in physical spacetime. The relevant interactions
occurred in the far-field regions outside the nonlinear
metron
core regions,
in which the field-field coupling was weak.

In
extending the metron model now to weak and strong
interactions, we will be concerned with interactions
 within the strongly nonlinear
particle core regions and will need to invoke periodicities
with respect to the
remaining coordinates of harmonic space. Although a general
structural correspondence between the metron model and the
Standard
Model can be established already with the minimal models
introduced
in Section \ref{Lagrangians} in the derivation of the
Maxwell-Dirac-Einstein
equations, a closer interrelationship can be established
(and the analysis simplified) if
the dimension of harmonic space is increased from four to
five.  Periodicities
with respect to the second
harmonic-space coordinate
$x^6$ will be associated with weak
interactions, the two coordinates $x^5,x^6$ together
defining the  electroweak interaction plane, while strong
interactions
are represented by periodicities in the chromodynamic
$(x^7,x^8)$-plane. The last harmonic-space
coordinate $x^9$ is needed for all interactions to construct
the polarization tensor relating the metric fields to the
fermion
fields such that the relevant sector of the gravitational
Lagrangian is mapped into the Dirac Lagrangian. The
harmonic-space background metric is assumed to be given by
$\eta_{AB} = \mbox{diag}(1,1,1,1,\pm 1)$
(Euclidean  or non-Euclidean model).

In accordance with our inverse modelling
approach, we
assume that trapped-mode solutions of the n-dimensional
gravitational field equations exist and attempt then
to identify the
structures that these solutions must exhibit in order to
reproduce
the observed properties of elementary particles.  Although a
close similarity between the metron
and Standard Model will be found, the similarity should not
be over-emphasized. The Lagrangians play a fundamentally
different role in the two models. In quantum field theory,
the Lagrangian is used to determine the evolution of field
operators defining
expectation values of predefined particle states, while in
the metron model the Lagrangian provides only the starting
point for the
computation of deterministic trapped-mode particle
solutions; once these have been determined,  the transition
probabilities between different particle states
must still be computed in a further interaction analysis.
None the less,
a general agreement in the structure of the interaction
Lagrangians is encouraging, since it may be anticipated, as
shown for the analagous problem of atomic
state transitions in Section~\ref{Atomic spectra},  that the
$S$-matrix
computations for quantum field
theory and the metron model will exhibit certain formal
similarities.

Only a
single family of solutions is considered. It is speculated
that the second and third Standard Model families correspond
to higher modes of the trapped-wave solutions (cf.
Section~\ref{The mode-trapping mechanism}). If this is
indeed the
case, there is no reason that the number of families
should be restricted to three, although higher modes
presumably become increasingly unstable.

It is not claimed that the models developed in the following
are in any way
unique. But they do
represent a rather simple and natural way of relating the
Standard
Model to the metron picture. A test of these concepts must
await again the construction of specific trapped-mode
solutions of the n-dimensional gravitational equations.

\section{Strong interactions}
\label{Strong interactions}

We consider first strong interactions, as these exhibit
somewhat
simpler symmetry properties than electroweak interactions.
We adopt essentially the same approach is  in
Part~\ref{The Maxwell-Dirac-Einstein System} for
electromagnetic-fermion
interactions, but consider now instead of a single
fermion field $\psi$ three quark fermions $\psi^{(q)}$ of
different color  $q=1,2$ or $3$. The lowest order coupling
between
fermions can then no longer be mediated by a single boson
field $A_{\lambda}$ with an interaction Lagrangian of the
form (suppressing indices) $\bar \psi A \psi$, but requires
a
set of boson fields $B^{(p \bar q)}_{\lambda}$, with an
associated coupling Lagrangian
$\bar
\psi^{(p)}B^{(p \bar q)}\psi^{(q)}$. The metron boson fields
$B^{(p \bar q)}$ will be
related to the gluons  of the chromodynamic $SU(3)$  gauge
group of the Standard Model. The detailed form of
the interaction Lagrangian will be derived, as in the
analysis of the Maxwell-Dirac-Einstein system in
Section~\ref{The Maxwell-Dirac-Einstein Lagrangian}, by
invoking
the invariance of the  gravitational Lagrangian
with respect to  coordinate
transformations. The gauge
symmetries of the Standard Model
are similarly explained by the invariance of the metron
model with respect to  a particular
class of diffeomorphisms.

\subsection*{Metron representation of strong-interaction
fields}

We assume that in the strongly nonlinear core region the
metron solutions contain trapped-mode quark constituents
$(q)$ of different color $q=1,2,3$
represented by fermion fields (defined as
usual as deviations from the background metric $\eta_{AB}$)
\begin{equation} \label{8.1}
g_{AB}^{(q)} = P_{AB}^a
\psi^{(q)}_a e^{iS^q} + c.c.,
\end{equation}
where
\begin{equation} \label{8.2}
S^q :=  k^{(q)}_{A} x^{A}.
\end{equation}
We identify, as in Part~\ref{The Maxwell-Dirac-Einstein
System}, fermions with harmonic-space components of the
metric tensor,
while bosons will be represented by mixed harmonic space-
physical
spacetime metric components \cite{ft33}.

The fermion polarization tensor $P_{AB}^a$ is
assumed to be independent of the color index
$q$. This is the reason, as will be seen below, that the
harmonic-space dimension needs to be extended from four to
five. The analysis can be carried through without this
assumption, but becomes more cumbersome, and the
correspondence between the metron model and the Standard
Model is not so close.

The wavenumber vectors
$\mbox{k}^{(q)}$ of the three
quarks are assumed to lie in a symmetrical star
configuration in the color plane $(x^7,x^8)$ (cf. Fig
\ref{Fig.8.1}a), with
\begin{eqnarray} \label{8.2a}
k_{A}^{(1)} + k_{A}^{(2)} + k_{A}^{(3)}&
= & 0, \\
\label{8.2ab}
 k^{(q)}_5 = k^{(q)}_6=k^{(q)}_9 &=& 0.
\end{eqnarray}

\begin{figure}[t] \centering
\begin{minipage}{10.5cm}
\epsfxsize300pt
\epsffile{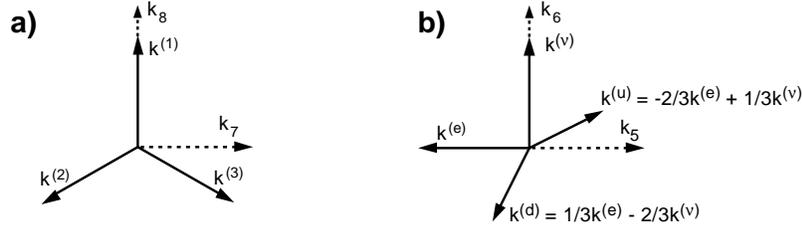}
\caption{\label{Fig.8.1} Fermion harmonic wavenumber
configurations for {\it (a)} the three colored quarks in the
chromodyamic plane $k_7,k_8$ and {\it (b)} the leptons
$e,\nu$ and quarks $u,d$ in the electroweak plane $k_5,k_6$}
\end{minipage}
\end{figure}

We ignore in this section the coupling with leptons and
electroweak
bosons, which involve non-zero wavenumber components
$ k^{(q)}_5 ,k^{(q)}_6,k^{(q)}_9$ (cf. Fig \ref{Fig.8.1}b).
This  will be
discussed later in Section~\ref{Electroweak interactions}.

 The motivation for these
symmetry assumptions is to recover the
strong-interaction $SU(3)$
symmetry of the Standard Model.
The assumed symmetry implies that all quarks
have the same harmonic mass
\begin{equation} \label{8.5a}
\hat \omega_f := \left(
k_{A}^{(q)}k^{A}_{(q)}
\right)
^{1/2}
\end{equation}
and  gravitational (de Broglie) mass
\begin{equation} \label{8.5b}
\omega_f := - k_4^{(q)}
\end{equation}
(we assume a stationary solution with $\psi^{(q)}_a \sim
\exp i k_4^{(q)}
x^4$,
where $\omega_f =  k^4 = -k_4 > 0$). Since the system is
strongly
nonlinear, the two masses will be different. We do not
inquire into the origin of the masses. We ascribe
finite quark masses to the mode-trapping mechanism (rather
than
the Higgs mechanism), which is not
investigated further here.

A fermion polarization tensor $P_{AB}^a$ which is
independent of  color can be obtained by generalizing the
minimal-model
tensors (\ref{A.5}) or (\ref{8.3.14h}) from
four to five
dimensions. In the minimal-model,
the single non-zero wavenumber component, which was taken as
the first harmonic wavenumber component $k_5$, induced,
through the gauge condition, a zero first column and first
row in the  polarization tensor. In the
present case we wish to define a polarization tensor which
satisfies the trace and divergence gauge conditions
(\ref{3.13a}), (\ref{3.13b})
for two non-zero
wavenumber components $k_7,k_8$ in the chromodynamic plane.
If the polarization tensor is to be independent of the
wavenumber vector in the
chromodynamic plane, we must introduce then
two zero columns and rows. Thus the generalization of the
polarization tensor (\ref{A.5}) of
the minimal non-Euclidean model $(+3,-1)$ becomes for the
five-dimensional
model $(+4,-1)$, with a non-zero wavenumber  vector confined
to the color plane,
\begin{equation} \label{8.2aa}
P^{a}_{AB}\psi^{(q)}_a =
\frac{1}{(\sqrt{2 \hat\omega^{(q)}})}
\left(
\begin{array}{ccccc}
\psi^{(q)}_1 & \psi^{(q)}_2 & 0 & 0 & \psi^{(q)}_3 \\
\psi^{(q)}_2 &-\psi^{(q)}_1 & 0 & 0 & \psi^{(q)}_4 \\
0& 0& 0& 0& 0\\
0& 0& 0& 0& 0\\
\psi^{(q)}_3 &\psi^{(q)}_4 &  0 & 0 & 0
\end{array}
\right),
\end{equation}
while the corresponding generalization of the polarization
tensor (\ref{8.3.14h}) for the minimal Euclidean model
$(+4)$ to the five-dimensional model
$(+5)$ is given by
\begin{equation} \label{8.2aab}
P^{a}_{AB}\psi^{(q)}_a =
\frac{1}{(\sqrt{2 E})}
\left(
\begin{array}{ccccc}
0 & \varphi^R_1 &  0 & 0 & \varphi^R_2  \\
\varphi^R_1 &\varphi^L_1  & 0 & 0 & \varphi^L_2  \\
0& 0& 0& 0& 0\\
0& 0& 0& 0& 0\\
\varphi^R_2 &\varphi^L_2  & 0 & 0 & -\varphi^L_2  \\
\end{array}
\right).
\end{equation}
However, as in the investigation of the
 Maxwell-Dirac-Einstein system in Section~\ref{The
Maxwell-Dirac-Einstein Lagrangian},
the detailed form
of the fermion polarization tensor is irrelevant at the
present level of analysis, provided only that it yields the
standard Dirac
Lagrangian. A discrimination between competing
models must await the computation of specific trapped-mode
metron solutions.

The minimal models must then also be considered
as serious contenders.  In this case the basic forms
(\ref{8.2aa}), (\ref{8.2aab}) are retained without the
addition of a
second color dimension, the color plane  being replaced by a
single dimension. The polarization
tensors contain only a single row and column of zeros,
and to
satisfy the gauge condition, the polarization tensors must
be defined in a specific coordinate system in which the zero
rows and columns correspond to the direction of
the
quark wavenumber vector. Thus for a quark of different
color, the polarization tensor must be rotated. The
dependence of the polarization tensor on color has no
impact on interactions involving a single quark, i.e. on the
quark coupling through diagonal bosons, but
modifies the interactions involving non-diagonal bosons. The
basic
interaction structure of the metron model  is nevertheless
not significantly affected.

We have excluded the minimal models in the following
primarily
to simplify the analysis and obtain a somewhat closer
correspondence to the
Standard Model, rather than because of fundamental
shortcomings of the minimal  models. In fact, as pointed out
in Section~\ref{Identification of fields}, the minimal
Euclidean model $(+4)$ has some attractive intrinsic
features (avoidance of tachyons, signal propagation confined
to the surface of the seven-dimensional sphere) -- although
it is not clear whether these general properties are
relevant for the special case of periodic-homogeneous metron
solutions in harmonic space with which we are concerned
here.

Since the  metron solutions are real, there exists for each
complex
quark field $(q)$  an associated complex conjugate quark
$(\bar q)$ with negative wavenumbers $(k^{(\bar q)}_4,
k^{(\bar q)}_{A})= (-k^{(q)}_4 , -k^{(q)}_{A})$.
There exists also for each quark $(q)$ an independent
anti-quark $(q')$ (although presumably not as a constituent
of the same metron solution)  with wavenumbers $(k'^{(q)}_4
,
k'^{(q)}_{A}) = (k^{(q)}_4 , -k^{(q)}_{A})$ (cf.
Section~\ref{Identification of fields}). We shall not
distinguish between quarks and anti-quarks in the following,
denoting either constituent simply as $(q)$.

The empirical finding that the net color of  hadrons
is white translates in the metron picture into the
property that the wavenumber sum  of all the  quarks in a
hadron vanishes. It will be shown below that this implies
that the net mean fields generated by the set of all
quark  interactions of a  hadron vanish: strong interactions
-- as opposed to electroweak and gravitational interactions
-- generate no mean far fields.

A perturbation  expansion of the nonlinear coupling between
quarks yields quadratic  sum and difference interactions
at lowest order, and further cubic and higher-order
interactions.
We will consider in the following only the
mixed-index (boson) quadratic difference-interaction  fields
\begin{equation} \label{8.3}
g_{\lambda A}^{(p\bar q)} = \left[ g_{\lambda
A}^{(q\bar p)}\right ]^* = B_{\lambda A}^{(p\bar
q)} e^{iS^{p\bar q}},
\end{equation}
where
\begin{equation} \label{8.4}
S^{p\bar q} = - S^{q\bar p} :=
k^{(p\bar q)}_{A} x^{A},
\end{equation}
with
\begin{equation} \label{8.5}
k^{(p\bar q)}_a := k^{(p)}_a - k^{(q)}_a.
\end{equation}
We focus on this sub-set of interaction fields because it
will be found to provide  a close analogy to the gluons of
the Standard Model. However, from the metron viewpoint,
the need to restrict the analysis to the quadratic
difference interactions implies that  the Standard
Model  describes only a single sector of
the full set of metron interactions and represents therefore
only an approximation of the complete nonlinear system.

From the symmetry of the quark configuration it follows that
all non-diagonal bosons have the same harmonic mass, which,
according to (\ref{8.5a}), (\ref{8.5}), is given by
\begin{equation} \label{8.5c}
\hat \omega_b :=
\left(
k_{A}^{(p\bar q)}k^{A}_{(p\bar q)}
\right)
^{1/2} = \sqrt{3} \hat \omega_f
 \quad \mbox{for} \quad p \neq q
\end{equation}
The harmonic masses for the diagonal bosons $p=q$ vanish.
For all
bosons, also,  the gravitational masses $\omega_b = -
k_4^{(p\bar
q)}$ are zero \cite{ft35}. In contrast to the fermion-
electromagnetic
interactions
considered in Part~\ref{The Maxwell-Dirac-Einstein System},
the present fields can no
longer be regarded as approximately linear, so that the
harmonic and gravitational boson masses will in general not
be
equal: $\hat \omega_b \neq \omega_b
(=0) $. The same holds for the quark masses.

We  note that while different non-diagonal fermion
interactions
$(p \bar q)$, $p \neq q$, generate bosons
distinguished by different harmonic wavenumbers
$\mbox{k}^{(p\bar
q)}$, all diagonal fermion interactions $(q \bar
q)$
excite the same zero-wavenumber Fourier component. Since
different
fields are  distinguished in the variation of
the
Lagrangian only by their
different
wavenumbers, all
diagonal interaction fields must be collected together
in
a single net diagonal
boson field :
\begin{equation} \label{8.5d}
g_{\lambda A}^{(d)} :=
\sum_q B_{\lambda A}^{(q \bar q)}
=:  B^{(d)}_{\lambda A} \quad \mbox{(real)}.
\end{equation}
The net diagonal field will be decomposed again later,
however,
on the basis of the different source terms in the field
equation, or, alternatively,  with respect to the different
orientations of the resultant metric tensor in harmonic
space.

\subsection*{Metron representation of the
strong-interaction
Lagrangian}

We consider in the
following only
the cubic fermion-boson-fermion products $(\bar p)(p\bar
q)(q)$
in the interaction Lagrangian which describe the lowest
order
coupling between fermions and bosons.
These
correspond to the quark-gluon-quark coupling products in
chromodynamics. Both the metron model  and the chromodynamic
model contain also further interactions, in
the metron model
in the form of an infinite series, in  the chromodynamic
model as
cubic and
quartic  boson-boson coupling terms. We anticipate a similar
general agreement in the structure of the boson-boson
interactions in the two models, but restrict the analysis
here for illustration to fermion-boson interactions. An
equivalence of the
two models cannot be expected at higher
interaction order. Indeed,  already
at the lowest  cubic fermion-boson coupling level considered
here, it will be found that, although the general
structures agree, the models are not
completely
identical. This is not particularly disturbing, however,
since, as pointed out, the Lagrangians play a basically
different role in the two models
in the computation
of particle states and transition probabilities.

To derive the interaction
Langrangian, we
apply the same technique as in
Section~\ref{The Maxwell-Dirac-Einstein Lagrangian}:  we
carry out a local
coordinate transformation  $X\rightarrow X'$ such that the
transformed boson fields vanish at some prescribed (four
dimensional) world point $x$, which we take to be the
origin.
At this point the affine-invariant gravitational Lagrangian
$L=P$ reduces to the free-field
Lagrangian (\ref{3.35}). Subsequently, we transform back
again to
the global
coordinate system, retaining, however, the local definition
of the fermion fields.

As before, the method requires that there exist no
fermion-boson interactions involving  derivatives of the
boson fields, so that the interaction Lagrangian does indeed
vanish at the  location where the boson fields (but not
necessarily their derivatives) are zero. It can be seen by
direct inspection of the Lagrangian (\ref{3.24}),
(\ref{3.25}) that this is the case not only for the diagonal
boson fields $B_{\lambda A}^{(q \bar q)}$ -- which have
essentially the same properties as the electromagnetic field
discussed in Part~\ref{The Maxwell-Dirac-Einstein System} --
but also
for the
non-diagonal bosons $B_{\lambda A}^{(p \bar q)}$, with
$p\neq q$ -- provided the fermion polarization relations
(\ref{8.2aa}) are independent of color and the quarks have
the same mass, both of
which we have assumed.

The required transformation, generalizing (\ref{3.37}) to
the case of periodic boson fields, is given by
\begin{eqnarray} \label{8.13}
x'^{A}  & =&    x^{A}     +
\xi^{A}_{\lambda} x^{\lambda} \nonumber \\
x'^{\lambda} &=& x^{\lambda},
\end{eqnarray}
where (cf.eqs.(\ref{8.3}),(\ref{8.5d}))
\begin{equation} \label{8.14}
\xi^{A}_{\lambda}= \sum_{p \neq q}
\left\{B^{ (p\bar
q) A}_{\lambda}
\right\}_{x=0}
 \exp(iS^{p\bar q}) +
\left\{B^{(d) A }_{\lambda}
\right\}_{x=0}.
\end{equation}

In contrast to the transformation (\ref{3.37}) involving
only diagonal
bosons, the
transformation (\ref{8.13}),(\ref{8.14}) including periodic
non-diagonal bosons is no longer an
affine transformation because of the presence of the
non-constant
factor $\exp(iS^{p\bar q})$.
However, it can be readily shown
that the derivative of this factor induces an interaction
which is of second order in the boson fields and can
therefore be ignored in the present context.

In the local coordinate system, the resulting free-fermion
Lagrangian    is obtained by generalizing the form
(\ref{3.35}) to a superposition of fermions with a common
polarization tensor. Anticipating interactions between
different fermions,  we write this now in the symmetrized
real form corresponding to (\ref{3.35a}):
\begin{eqnarray} \label{8.6}
L_f^0 &=& -\frac{1}{2} \sum_{p,q}
\left\{
  \bar \psi^{(p)}
  \left(
  \gamma^{\lambda}
  \partial'_{\lambda} \psi^{(q)} + \hat
  \omega_f \psi^{(q)}
  \right )
  \exp(-iS^{p\bar q})
\right.
\nopagebreak
  \nonumber \\
&&
\left.
  -
\left (
   \partial'_{\lambda} \bar
   \psi^{(q)}\gamma^{\lambda}
   - \hat
   \omega_f \bar \psi^{(q)}
\right )
   \psi^{(p)}
   \exp(-iS^{q\bar p})
\right\}.
\end{eqnarray}
Note that contrary to the Lagrangian in the global
coordinate system, the oscillatory terms have been retained
in the local free-fermion Lagrangian (\ref{8.6}). This is
necessary, since the transformation (\ref{8.13}) from
global to local coordinates introduces oscillatory terms in
the transformation Jacobian, resulting in a contribution
from the oscillatory terms
to the
 action integral over harmonic space
in the coordinate system $X'$. However, we are concerned
here not with the action integral, but only with the
structure of the Lagrangian at $x=0$.

On transforming back to global coordinates, we retain as
before the
definition of the
fermion fields $\psi^{(q)}$ as given  in
Section~\ref{Identification of fields}
with respect to the harmonic-index metric
field components in the local coordinate system, but regard
the fields $\psi^{(q)}$ now as functions of the global
coordinates. Thus we express the local derivative
$\partial'_{\lambda}$ in (\ref{8.6}) in terms of the global
derivative,
\begin{equation} \label{8.6a}
\partial'_{\lambda}= \partial_{\lambda} -
\xi^{A}_{\lambda}\partial_{A},
\end{equation}
or, applying (\ref{8.14}),
\begin{equation} \label{8.25a}
\partial'_{\lambda} \psi^{(r)}= \partial_{\lambda}\psi^{(r)}
- i k_{A}^{(r)} \psi^{(r)}
\left\{
\sum_{p \neq q}
B^{ (p\bar
q) A}_{\lambda}
 \exp(iS^{p\bar q}) +
B^{(d) A }_{\lambda}
\right\}.
\end{equation}

In the previous treatment of electromagnetic interactions,
we had simply identified the local derivative
$\partial'_{\lambda}$ with the covariant derivative
$D_{\lambda}$. However, in the present case, in order to
remove the
phase-function factors in (\ref{8.6}) and (\ref{8.25a}), it
is
more convenient to define the covariant derivative as
\begin{equation} \label{8.25}
D_{\lambda}\psi^{(p)} = \partial_{\lambda}\psi^{(p)} -
i \sum_{q\neq p} k_{A}^{(q)}
B^{(p\bar q) A}_{\lambda}\psi^{(q)}
-ik_{A}^{(p)} B^{(d) A}_{ \lambda } \psi^{(p)}.
\end{equation}
Comparing the derivative forms (\ref{8.25a}) and
(\ref{8.25}), we see that the effect of the
non-diagonal boson interaction on  a given `input' fermion
$(p)$ is represented in the local derivative
$\partial'_{\lambda}\psi^{(p)}$  as a scatter operation
affecting the `output'
fields at different wavenumbers, whereas the covariant
derivative
$D_{\lambda}\psi^{(p)}$ is defined as a gather operation, in
which all
boson-`input' fermion  interactions  which affect a given
`output'
fermion field $(p)$ are collected together.

Applying the definition (\ref{8.25}), the complete fermion
Lagrangian
becomes (noting that oscillatory terms can be dropped
again after
returning to the global coordinate system)
\begin{eqnarray} \label{8.6b}
L_f &=& -\frac{1}{2} \sum_p
\left\{
  \bar \psi^{(p)}
  \left(
  \gamma^{\lambda}
  D_{\lambda} \psi^{(p)} + \hat
  \omega_f \psi^{(p)}
  \right )
\right.
  \nonumber \\
&&
\left.
  -
\left (
   D_{\lambda} \bar
   \psi^{(p)}\gamma^{\lambda}
   - \hat
   \omega_f \bar \psi^{(p)}
\right )
   \psi^{(p)}
\right\},
\end{eqnarray}
which yields the
quark-boson interaction
Lagrangian
\begin{equation} \label{8.27}
L_{qb} = \frac{1}{2} \sum_{q\neq p}
 \left(k_{A}^{(p)} + k_{A}^{(q)}
\right)
J_{\lambda}^{(\bar p q)}
 B^{\lambda A}_{(p\bar q)}
 +\,  \sum_p
k_{A}^{(p)}
J_{\lambda}^{(\bar p p)}
B^{\lambda A}_{(d)},
\end{equation}
where the currents are defined as
\begin{equation} \label{8.27a}
J_{\lambda}^{(\bar p q)} := i \left(\bar
\psi^{(p)}\gamma_{\lambda}\psi^{(q)}
\right).
\end{equation}

To complete the Lagrangian for the quark-boson system, we
need also the boson
free-field  Lagrangian. Here we
must distinguish
between the
non-diagonal fields $B_{\lambda
A}^{(p\bar q)}$, $p \neq q$, with non-zero harmonic
wavenumber, and the
zero-wavenumber diagonal field $B^{(d)}_{\lambda A}$.
For
the
former we obtain from (\ref{3.28}) and (\ref{8.3}) -
(\ref{8.5})
\begin{equation} \label{8.7}
L_b^{(n
d)} = - \frac{1}{4}\sum_{p \neq q}
\left[
F_{\lambda \mu A}^{(p \bar q)*} F^{\lambda \mu
A}_{(p \bar q)} + H_{A \lambda B }^{(p \bar
q)*} H^{A \lambda B }_{(p \bar q)}
\right],
\end{equation}
where
\begin{equation} \label{8.8}
F^{\lambda \mu A}_{(p \bar q)} = \partial^{\lambda}
B^{ \mu A}_{(p \bar q)} - \partial^{\mu}
B^{ \lambda A}_{(p \bar q)}
\end{equation}
and
\begin{equation} \label{8.9}
H^{A \lambda B }_{(p \bar q)} = i \left[
k^{A }_{(p \bar q)} B^{\lambda B}_{(p \bar q)}
- k^{B }_{(p \bar q)} B^{\lambda A}_{(p \bar q)}
\right].
\end{equation}

For the zero-wavenumber diagonal boson, the
free-field Lagrangian reduces to the simpler form of the
electromagnetic Lagrangian (\ref{3.29}),
\begin{equation} \label{8.10}
L_b^{(d)} = - \frac{1}{4}
\left[
F_{\lambda \mu A}^{(d)} F^{\lambda \mu A}_{(d)}
\right],
 \end{equation}
where
\begin{equation} \label{8.11}
F^{\lambda \mu A}_{(d)} = \partial^{\lambda}
B^{ \mu A}_{(d)} - \partial^{\mu}
B^{ \lambda A}_{(d)}.
\end{equation}
Invoking the gauge condition (\ref{2.7}), which for the
mixed-index tensor subset corres\-ponding to bosons  yields
\begin{equation} \label{8.26}
i \left(
k_{A}^{(p)} - k_{A}^{(q)}
\right)
A^{\lambda A}_{(p\bar q)} = 0
\end{equation}
and
\begin{equation} \label{8.26a}
\partial_{\lambda} A^{\lambda A}_{(p\bar q)} = 0
\end{equation}
(that the application of the gauge condition to the
truncated set of fields is permissible will be verified
later),
the total boson
free-field Lagrangian may be written in the
alternative   form
\begin{eqnarray} \label{8.12}
\lefteqn{L_b^0 :=L_b^{(nd)}+L_b^{(d)} = } \nonumber \\
& - \frac{1}{2}
\left\{
\sum_{p \neq q} \left[
\partial_{\lambda} B_{ \mu A}^{(p \bar q)*}
\partial^{\lambda}B^{ \mu A}_{(p \bar q)}
+ \hat \omega_b^2  B_{ \lambda A}^{(p \bar q)*} B^{
\lambda
A}_{(p \bar q)}
\right]
+ \partial_{\lambda}B_{ \mu A}^{(d)}
\partial^{\lambda}B^{ \mu A}_{(d)}
\right\}.
\end{eqnarray}

Variation of the free-boson and fermion-boson interaction
Lagrangians with respect to the
non-diagonal fields $B^{\lambda A}_{(p\bar q)}$, $p
\neq q$, and the diagonal field $B_{(d)}^{\lambda A}$
yields
then for the boson field equations
\begin{equation} \label{8.28}
\left(\partial_{\mu}\partial^{\mu} - \hat \omega_b^2
\right)B_{\lambda A}^{(p\bar q)}
= -\frac{1}{2}
\left(
k_{A}^{(p)} + k_{A}^{(q)}
\right)
J_{\lambda}^{(\bar q p)}.
\end{equation}

Equation (\ref{8.28}) is applicable now for both
non-diagonal and diagonal bosons.  In the latter case this
follows by decomposing the field
equation for the net diagonal boson field $B_{(d)}^{\lambda
A}$ into field equations for the individual
constituents $B^{\lambda
A}_{(p\bar
p)}$, each constituent being generated by its specific
source
term in accordance with (\ref{8.28}).

The solution factorizes, as in the electromagnetic
case
(cf.eq.(\ref{3.7})),  into the product of a constant vector
in
harmonic space and a vector field in physical spacetime:
\begin{equation} \label{8.29}
B_{\lambda A}^{(p\bar q)} =:
\frac{1}{2}\left(
k_{A}^{(p)} + k_{A}^{(q)}
\right)
B_{\lambda}^{(p\bar q)},
\end{equation}
where the boson vector fields $B_{\lambda}^{(p\bar q)}$
satisfy  the field equations
\begin{equation} \label{8.32}
\left(\partial_{\mu}\partial^{\mu} - \hat \omega_b^2
\right)B_{\lambda}^{(p\bar q)}
= -J_{\lambda}^{(\bar q p)}.
\end{equation}

Applying eqs. (\ref{8.29}) and (\ref{8.32}), it can now be
readily
verified {\it a posteriori} that the generating currents
have zero divergence,
provided the generating fermions have the same harmonic
mass, which is the case. Thus the boson fields can indeed be
defined to
satisfy the truncated gauge conditions (\ref{8.26}) and
(\ref{8.26a}), as we had assumed.

Equations (\ref{8.29}) and (\ref{8.32}) are seen to
represent
the straightforward generalizations of the corresponding
electromagnetic relations (\ref{3.7}),(\ref{3.45}) -
(\ref{3.48}). However we have
chosen now for the general case a different normalization
than used in the
previous definition (\ref{3.7}) for the electromagnetic
field: by factoring out the harmonic-wavenumber dependence
in  the definition of $B_{\lambda}^{(p\bar q)}$ in
(\ref{8.29}), we have in effect removed the coupling
coefficients (e.g. the electric charge) in the source terms
of the boson field equations (\ref{8.32}). The coupling
coefficients will
be reintroduced later when the bosons $B_{\lambda}^{(p\bar
q)}$
are
renormalized in accordance with the gluon definitions of the
Standard
Model.

Substituting the factorized form (\ref{8.29}) into
(\ref{8.25}) and applying (\ref{8.26}), we obtain the
fermion field equations
\begin{equation} \label{8.33}
D_{\lambda}\psi^{(p)} =
\partial_{\lambda} \psi^{(p)}
- \frac{i  C_3}{2}
\left\{
 \hat B_{\lambda}^{(p)}  \psi^{(p)}
+  \sum_{q \neq p}
 B_{\lambda}^{(p\bar q)}
\psi^{(q)}
\right\}
= 0,
\end{equation}
where
\begin{equation} \label{8.33a}
\hat B_{\lambda}^{(p)} := A_1 B_{\lambda}^{(p\bar p)}
+ A_2 \sum_{q \neq p}  B_{\lambda}^{(q\bar q)}
\end{equation}
is the net diagonal boson acting on the fermion $(p)$
and the coefficients, invoking (\ref{8.2a}) and the
symmetrical-star symmetry
of the quark wavenumber vectors, are given by
\begin{eqnarray}
\label{8.30b}
C_3 & := & \frac{1}{2}
\left(
k_{A}^{(p)} + k_{A}^{(q)}
\right)
\left(
k^{A}_{(p)} + k^{A}_{(q)}
\right)
= \frac{1}{2} \hat\omega_f^2 \quad (p \neq q)
\\
\label{8.30}
A_1 & := & 2 k_{A}^{(p)}k^{A}_{(p)}/C_3 =
4  \\
\label{8.30a}
A_2 & := & 2 k_{A}^{(p)}k^{A}_{(q)} /C_3 =
- \; 2 \quad (p \neq q).
\end{eqnarray}

The symmetry of the quark configuration implies that
the three diagonal bosons are not independent.
Applying (\ref{8.2a}),(\ref{8.29}),(\ref{8.33a}),
(\ref{8.30}) and  (\ref{8.30a}) we
find
\begin{equation} \label{8.30c}
\sum_{p} \hat B_{\lambda}^{(p)} =
\sum_{p} B_{\lambda}^{(p \bar p)} = 0.
\end{equation}

As already mentioned, the empirical property that hadrons
are white implies for the
metron model that the wavenumber sum of all
quarks in a hadron vanishes.  Applying
eq.(\ref{8.30c}), it follows that the net integrated source
function generating the strong-interaction
far field of a hadron, consisting of the sum of all
diagonal-boson far fields, vanishes
(non-diagonal bosons can be ignored, as their finite
harmonic  mass yields an exponential  rather
than an $1/r$  fall off for large distances $r$ from the
particle core). We note that the mean-field cancellation
applies only
for  the  far fields, which depend only on the spatial
integrals of the source
functions, since the spatial distributions of the
currents
generating the individual diagonal boson fields can differ
for
different bosons within a hadron particle. The lack of far
fields represents an important distinction between
strong interactions and electroweak and gravity
interactions.

Substituting the factorized form (\ref{8.29}) of the boson
fields into the covariant derivative (\ref{8.25}) in the
fermion Lagrangian (\ref{8.6b}) and into the free-boson
Lagrangian (\ref{8.12}), we obtain finally as the metron
form of the total Lagrangian for the strongly
coupled fermion-boson system:
\begin{eqnarray} \label{8.34}
L^M_{st}&=&
-\frac{A_1 C_3}{4}\sum_p
\partial_{\lambda}B_{\mu}^{(p\bar p)}
\partial^{\lambda}B^{\mu}_{(p\bar p)}
-\frac{A_2 C_3}{4}\sum_{p \neq q}
\partial_{\lambda}B_{\mu}^{(p\bar p)}
\partial^{\lambda}B^{\mu}_{(q\bar q)}
\nonumber \\
&&
-\frac{C_3}{4}\sum_{p \neq q}
\left\{
\partial_{\lambda}B_{\mu}^{(p\bar q)*}
\partial^{\lambda}B^{\mu}_{(p\bar q)}
+ \hat \omega^2_b B_{\lambda}^{(p\bar q)*}
B^{\lambda}_{(p\bar q)}
\right\}
\nonumber \\
&& - \sum_p \bar \psi^{(p)}
\left (\gamma^{\lambda}
\partial_{\lambda} \psi^{(p)} + \hat
\omega_f \psi^{(p)}
\right)
\nonumber \\
&& + \frac{C_3}{2}
\left\{
\sum_p
\hat B_{\lambda}^{(p)}
J_{\lambda}^{(\bar p p)}
+ \sum_{q \neq p}
 B^{\lambda}_{(p \bar q)}
J_{\lambda}^{(\bar p q)}
\right\}.
\end{eqnarray}

We conclude this sub-section with some remarks on our
terminology for bosons.
Bosons are defined generally as the mixed-index metric
tensor fields generated by quadratic difference interactions
between fermions. We have distinguished between tensor
bosons
$B_{A \lambda}^{(b)}$ and vector bosons
$B_{\lambda}^{(b)}$, the latter being defined by the
factorization (\ref{8.29}) derived from the boson field
equations (\ref{8.28}).  Both tensor and vector bosons
$(b)$ have been characterized so far by an index pair $(p
\bar
q)$ identifying the pair of generating fermions
$(p),(q)$. These  determine the difference wavenumber
$k_{A}^{(p \bar q)} = k_{A}^{(p)} -
k_{A}^{(q)}$ characterizing the periodicity of the
boson and the sum
wavenumber $k_{A}^{(pq)} = k_{A}^{(p)}
+ k_{A}^{(q)}$ which defines the direction of the
vector boson in harmonic space.

To establish the connection to the Standard Model,   we will
need in the following to consider bosons defined more
generally as linear combinations of the above bosons. The
generalization is
required whenever different pairs of fermions $(p),(q)$
generate bosons with the same difference wavenumber
$k_{A}^{(p \bar q)}$. The
redefined boson fields, however, can then no longer be
attributed to a single pair of generating fermions.

Equation (\ref{8.34}) illustrates a case in point. Since the
harmonic sum vectors for different diagonal vector bosons
$(b),(b')$ are
not orthogonal, $k_{A}^{(b)} k^{A}_{(b')} \neq 0$,
the diagonal-boson sector of the free-boson Lagrangian,
expressed in terms of the
original bosons, is not diagonal -- in contrast to
the
free-gluon Lagrangian of the Standard Model. In the
following sub-section we shall diagonalize the free-boson
Lagrangian
through a suitable linear
transformation, the resultant bosons then being generated
by more than one current from more than one fermion (the
same holds also, of
course, for the bosons of the Standard Model).

\subsection*{Relation to chromodynamics}

The metron relations (\ref{8.32})-(\ref{8.34}) exhibit a
close structural similarity to the $ SU(3)$ chromodynamic
sector of the Standard Model. The set of metron bosons
$B_{\lambda}^{(p\bar q)}$ consists of eight independent real
fields:
three diagonal fields
 $B_{\lambda}^{(p\bar p)}$, or equivalently,
$\hat B_{\lambda}^{(p)}$ -- of which only two are
independent,
however -- and six real fields representing
the six non-diagonal components
$B_{\lambda}^{(p\bar
q)}$, $p \neq q$. The eight independent metron bosons  can
be related to
the
eight hypercharge gauge bosons $G^{(\rho)}_{\lambda}$
of the $SU(3)$ generators $\lambda_{\rho} \; (\rho =
1,\ldots,8)$.

Introducing vector notation ${\mbox{\boldmath$\psi$}}  =
\left(\psi^{(1)},\psi^{(2)},\psi^{(3)} \right)$, the
covariant
derivative for fermion fields is given in the
$ SU(3)$ model by
\begin{equation} \label{8.35}
D_{\lambda} {\mbox{\boldmath$\psi$}}  =
\left(
 \partial_{\lambda}
-\frac{i g_3}{2}\sum_{\rho} G_{\lambda}^{(\rho)}
\lambda_{\rho}
\right) {\mbox{\boldmath$\psi$}},
\end{equation}
where $g_3$ represents the strong-interaction coupling
coefficient and the  generators $\lambda_{\rho}$ of the
$SU(3)$ gauge group consist
of two traceless diagonal phase-shift generators, in the
standard  Gell-Mann notation \cite{ft36},
\begin{eqnarray} \label{8.37}
\lambda_3 &:=& \mbox{diag}(1,-1,0)\\
\label{8.38}
\lambda_8 &:=& \frac{1}{\sqrt{3}} \, \mbox{diag}(1,1,-2)
\end{eqnarray}
and six further non-diagonal generators. The latter can be
grouped into three pairs of generators, each generator pair
consisting of the first
two Pauli matrices $\sigma_1, \sigma_2$
acting on one of the
three different combinations of quark pairs.

Comparing (\ref{8.35}) with (\ref{8.33}), the
non-diagonal terms in the covariant
derivatives of the metron model and the
$ SU(3)$ model are seen to be
identical if the following assignments are made:
\begin{equation} \label{8.40}
g_3 =  C_3 /N_3
\end{equation}
and
\begin{equation} \label{8.41}
\mbox{non-diagonal} \; G^{(\rho)}_{\lambda} = \mbox{Re or
Im}
\;
\left \{
N_3 B_{\lambda}^{(p\bar
q)}
\right \}
\quad (p\neq q),
\end{equation}
specifically \cite{ft36}:
\begin{equation} \label{8.42}
\begin{array}{rclrcl}
G^{(1)}_{\lambda} &=& \mbox{Re }N_3 B_{\lambda}^{(1\bar 2)},
& G^{(2)}_{\lambda} &=& \mbox{Im }N_3 B_{\lambda}^{(1\bar
2)},
\nonumber \\
G^{(4)}_{\lambda} &=& \mbox{Re }N_3 B_{\lambda}^{(1\bar 3)},
& G^{(5)}_{\lambda} &=& \mbox{Im }N_3 B_{\lambda}^{(1\bar
3)},
\nonumber \\
G^{(6)}_{\lambda} &=& \mbox{Re }N_3 B_{\lambda}^{(2\bar 3)},
& G^{(7)}_{\lambda} &=& \mbox{Im }N_3 B_{\lambda}^{(2\bar
3)},
\end{array}
\end{equation}
where $N_3$ is a normalization factor. To reproduce the
non-diagonal sector of the Standard Model free-gluon
Lagrangian
\begin{equation} \label{8.42aa}
L_b^{SM} = - \frac{1}{2} \sum_{\rho} G^{(\rho)}_{\lambda}
G_{(\rho)}^{\lambda}
\end{equation}
we must set , according to (\ref{8.34}),
\begin{equation} \label{8.42a}
N_3 = \sqrt{C_3},
\end{equation}
so that
\begin{equation} \label{8.40a}
g_3 =  \sqrt {C_3}.
\end{equation}

The remaining terms containing the diagonal bosons
$\hat B_{\lambda}^{(p)}$ in the metron form (\ref{8.33}) of
the covariant derivative -- from which we
select, say, $\hat B_{\lambda}^{(1)}$ and $\hat
B_{\lambda}^{(2)}$
as the independent fields --  can
also be brought into agreement with the
corresponding terms
in the Standard Model covariant derivative (\ref{8.35}),
while yielding the correct form  for the diagonal sector of
the free-boson Lagrangian (\ref{8.42aa}), through the
linear transformation:
\begin{equation} \label{8.43}
N'_3  \left(
\begin{array}{c} \hat B_{\lambda}^{(1)} \\
                \hat B_{\lambda}^{(2)} \\
\end{array}
\right) =
\left(
\begin{array}{cc}
1 & \frac{1}{\sqrt{3}} \\
-1 & \frac{1}{\sqrt{3}}
\end{array}
\right)
\left(
\begin{array}{c} G_{\lambda}^{(3)} \\G_{\lambda}^{(8)}
\end{array}
\right),
\end{equation}
where the normalization
factor $N'_3 $ must be chosen in this case as
\begin{equation} \label{8.42b}
N'_3 = \left( \frac{C_3}{(\alpha_1 -\alpha_2)}
\right)^{\frac{1}{2}}.
\end{equation}

Expressed in terms of the gluon fields as defined in
(\ref{8.42}) - (\ref{8.43}), the metron
Lagrangian (\ref{8.34}) then becomes

\begin{eqnarray} \label{8.44}
L^M_{st} &=&
-\frac{1}{2}
\left(
\partial_{\lambda} G^{(3)}_{\mu}
\partial^{\lambda}G^{(3) \mu}
+ \partial_{\lambda} G^{(8)}_{\mu}
\partial^{\lambda}G^{(8) \mu}
\right)
 \nonumber \\
&&
-\frac{1}{2}  \sum_{non \; diag \; \rho}
 \left(
\partial_{\lambda} G^{(\rho) * }_{\mu}
\partial^{\lambda}G^{(\rho) \mu}
+  \hat \omega^2_b G_{\lambda}^{(\rho) *}
G^{(\rho) \lambda}
\right)
 \nonumber \\
&& - \sum_p \bar \psi^{(p)}
\left(\gamma^{\lambda}
\partial_{\lambda} \psi^{(p)} + \hat
\omega_f \psi^{(p)}
\right)
 \nonumber \\
&& + \frac{i g'_3}{2} \sum_{diag \, \rho}
G_{\lambda}^{(\rho)}
\left( \bar {{\mbox{\boldmath$\psi$}} } \gamma^{\lambda}
{\mbox{\boldmath$\psi$}}
\right)
+ \frac{i g_3}{2} \sum_{non-diag \, \rho}
G_{\lambda}^{(\rho)}
\left( \bar {{\mbox{\boldmath$\psi$}} } \gamma^{\lambda}
{\mbox{\boldmath$\psi$}}
\right),
\end{eqnarray}
where
\begin{equation} \label{8.44a}
g'_3 := (\alpha_1 -\alpha_2)
^{\frac{1}{2}}  g_3
= \sqrt 6 \, g_3.
\end{equation}

This is identical to the strong-interaction Lagrangian of
the
Standard Model, except for the difference in the coupling
coefficients for the diagonal and non-diagonal bosons (we
have excluded
the boson-boson coupling terms, which were not considered).
It should be recalled, however, that  the metron
interactions
considered here represent only a truncated subset of the
interactions of the full metron model: we have ignored the
quadratic   sum
interactions and all higher-order interactions, and the
analysis of the quadratic difference-interactions was also
restricted to the mixed-index tensor components.

A more fundamental difference is that
there exists no equivalent of the `mode-trapping' mean field
of the metron model in
the Standard
Model. In establishing the correspondence between the metron
and chromodynamic
models, we have accordingly not considered the mean
wave-guide
field or addressed the mechanism of
mode trapping in the metron model -- although, as
discussed in Section~\ref{The mode-trapping mechanism},
these are
 essential elements of the metron
model.
Similarly, we have not considered the
origin of the quark masses, which we also attribute to the
mode-trapping mechanism in the metron model (in the
following discussion of electroweak interactions, however,
we shall
discuss an alternative  metron equivalent of the Higgs
mechanism for the generation of the electroweak boson
masses).

These aspects were in effect
factored out of the above discussion by considering only the
coupled
quark-boson field equations as such,  without regard to
the mechanisms which determine the assumed form of the
metron
solutions.

For this reason, the metron
expressions (\ref{8.40a}), (\ref{8.44a}) for the
coupling coefficients $g_3,g'_3$ cannot, at this stage of
the analysis,  be compared quantitatively with the Standard
Model coupling coefficient, and the fact that the metron
coupling coefficients for the diagonal and non-diagonal
bosons are not the same is relatively immaterial. As in the
corresponding expression (\ref{3.46}) derived for the
elementary
electric charge in Section~\ref{The Maxwell-Dirac-Einstein
Lagrangian}, the
local coupling coefficient depends on the normalization of
the fields. This is different in the metron model, where we
are concerned with real fields, than in
quantum field theory, in which the normalization applies to
a
set of operators. In the case of the electric charge, the
normalized coupling coefficient could be expressed in
  Section~\ref{Particle interactions} in terms of an
integral property
$\beta_e$ of the electron (eq.(\ref{4.29a}))  by
considering the electrodynamic
far field generated by the net electron current, which was
determined by integrating the current
density over the electron core. Analagous computations need
to
be made for the metron representations of hadrons
to determine the metron
strong-interaction coupling coefficients quantitatively.

\section{Electroweak interactions}
\label{Electroweak interactions}

The metron interpretation of electroweak interactions can be
developed in a manner very similar to the chromodynamic
case, except
that
we must allow now also for  zero or almost zero
lepton masses. For this reason the minimal non-Euclidean
model (\ref{A.5}) is less suitable as reference model (cf.
Section~\ref{Lagrangians}), and we consider only the
minimal Euclidean model (\ref{8.3.14h}). To be consistent
with the chromodynamic
model, we extend also the minimal Euclidean model
through the addition of a fifth harmonic dimension, in this
case by simply adding
a last row
and column to the polarization tensor (\ref{8.3.14h}). If we
allow now in addition to the non-zero wavenumber component
$k_5$ of the minimal Euclidean model also an arbitrary
non-zero component $k_9$, requiring as before that all other
wavenumber components vanish (with the exception, discussed
below, of the  neutrino wavenumber component $k_6$), the
trace and divergence gauge conditions (\ref{3.13a}),
(\ref{3.13b})
require that the additional column and row must
be zero. Thus the extended form of the minimal-model
polarization tensor (\ref{8.3.14h})  becomes
\begin{equation} \label{8.3.15h}
\hat P^{a}_{AB}\psi_a = \frac{1}{\sqrt{2E}}
\left(
\begin{array}{ccccc}
0& 0& 0& 0& 0\\
0& 0 &\varphi_1^R & \varphi_2^R  &0 \\
0& \varphi_1^R & \varphi_1^L &  \varphi_2^L &0\\
0& \varphi_2^R & \varphi_2^L &  -\varphi_1^L&0 \\
0& 0& 0& 0& 0
\end{array}
\right).
\end{equation}

For the form (\ref{8.3.15h}), the fifth harmonic index
does not appear in the expression (\ref{3.31a}) for the
fermion
metric $M$. Thus it follows, as in the analagous discussion
of the polarization tensor (\ref{A.5}) for the minimal
non-Euclidean model $(+3,-1)$, that $\eta_{99}$ can have
either sign. The extension of the minimal Euclidean
model $(+4)$ to five dimensions therefore yields either an
Euclidean
model $(+5)$ or a non-Euclidean model $(+4,-1)$. The
non-Euclidean model has the formal advantage that it permits
the representation of massless fermions with non-zero
harmonic wavenumbers, but in the electroweak metron model
presented below, this feature is not implemented: in the
zero-mass limit, all wavenumber components tend to zero.

We recall that the motivation for introducing an additional
harmonic dimension was to describe colored quarks by a
polarization tensor which is independent of the quark
color. In the case of electroweak interactions, it will be
found that the original  minimal-model form (\ref{8.3.14h})
already yields a  polarization tensor which is independent
of flavor, so that from the viewpoint of electroweak
interactions there is no
need to extend harmonic space to five dimensions. As
example, we shall present results below only
for the model $(+4,-1)$, but all conclusions hold, with
minor modifications, also for
the models $(+5)$ or $(+4)$.

\subsection*{Metron representation of lepton-boson
interactions}

We consider first electroweak interactions in the lepton
sector. The lepton wavenumber vectors are
assumed to  consist of components $k_5,k_6$ in the
electroweak plane and an additional
component $k_9$. Specifically, for the lepton pair $\nu$ and
$e^-$   we set (cf. Fig \ref{Fig.8.1}b)
\begin{equation} \label{8.49}
\begin{array}{rcl}
(k_5^{(\nu)},k_6^{(\nu)},k_7^{(\nu)},k_8^{(\nu)},
k_9^{(\nu)})&
=& (0,k_6^{(\nu)},0,0,
k_9^{(\nu)})\\
(k_5^{(e)},k_6^{(e)},k_7^{(e)},k_8^{(e)},
k_9^{(e)})&=& (k_5^{(e)},0,0,0,
k_9^{(e)}).
\end{array}
\end{equation}

The wavenumber configuration (\ref{8.49})
applies also for the $(+5)$ Euclidean model. For the minimal
$(+4)$ Euclidean model, the wavenumber components $k_9$ are
simply suppressed. In this case $\kappa_{\nu e}$
vanishes in the expressions given below, and there is no
cross-coupling between the charged current and the neutral
boson $Z_{\lambda}$ \cite{ft37}.

The wavenumber components $k_5,k_6$ will be associated
with the electromagnetic and weak-interaction coupling
coefficients, respectively. Since the wavenumber component
$k_9$ occurs only in the electroweak interactions, we shall
refer to the harmonic wavenumber sub-space $k_5,k_6,k_9$
orthogonal to the chromodynamic plane $k_7,k_8$ as the {\it
extended} electroweak wavenumber space.

We assume that the metron solution contains only a
single left-handed neutrino field $\nu = \nu^L$, but both
left-handed and
right-handed electron components $e^L,e^R$.  Since the
neutrino has no
right-handed component, the polarization tensor
(\ref{8.3.15h}) is seen to be compatible with the
guage conditions (\ref{3.13a}), (\ref{3.13b}) even though,
in contrast to the general requirement for arbitrary fields,
the neutrino
 wavenumber component $k_6$
is non-zero.

The existence of
only a single left-handed neutrino field implies that the
neutrino harmonic mass $\hat \omega_{\nu}
= (k_{A}^{(\nu)}k^{A}_{(\nu)})^{1/2}$ must be
zero (cf.eqs. (\ref{8.3.15e}),(\ref{8.3.15ee})).
However, for formal reasons we retain a very small neutrino
mass,
neglecting nevertheless the small
right-handed field component with which this is accompanied.
To lowest order -- disregarding the lepton asymmetry which
we attribute in our case to the
mode-trapping mechanism -- we assume that the  mass of the
electron is the
same as that of the neutrino (i.e. very small but finite)
and that
\begin{equation} \label{8.49x}
k_5^{(e)} = - k_6^{(\nu)}, \quad k_9^{(e)} = - k_9^{(\nu)}.
\end{equation}
Thus the neutrino and electron wavenumber vectors are
identical up to a sign change in $k_9$ and a rotation in the
electroweak
plane (the sign change arises through the negative charge of
the
electron).

For the
left-handed components, the representations
(\ref{8.3.15h}),(\ref{8.49}}),(\ref{8.49x})
are invariant with respect to rotations in the electroweak
$k_5,k_6$-plane, in accordance with the
$SU(2)$ symmetry of the Standard Model. The symmetry does
not hold for the right-handed fields, for which we require
always $k_6 = 0$ in order to satisfy the gauge
condition. We note that  the same
polarization tensor has been assumed
for both the electron and the neutrino, which, as
has already been mentioned, simplifies the
analysis, particularly in
the treatment of non-diagonal boson interactions.

We consider now the boson fields
\begin{equation} \label{8.50}
g_{\lambda A}^{(l\bar m)} = \left[ g_{\lambda
A}^{(m\bar l)}\right ]^* = B_{\lambda A}^{(l\bar
m)} e^{iS^{l\bar m}}
\end{equation}
generated by quadratic
difference interactions between leptons $l,m$, where
\begin{equation} \label{8.50a}
S^{l \bar m} = S^l - S^m =
\left( k_A^{l)} -  k_A^{m)}  \right) x^A
= k_A^{l \bar m)}  x^A.
\end{equation}

 For the
lepton pair $\nu,e^-$, these consist of
two real diagonal bosons $g_{\lambda
A}^{(\nu\bar \nu)}$ and $g_{\lambda A}^{(e\bar
e)}$ (which, in constrast to the chromodynamic case,  are
linearly independent for the lepton wavenumber
configuration)
and the two real components of the complex
non-diagonal boson $g_{\lambda A}^{(\nu \bar e)}$.
The metron bosons can be related to
 the two real diagonal bosons and the two real components of
the
complex
non-diagonal boson of the electroweak $U(1)
\times SU(2)$
gauge group.

The details of the coupling can be determined  by
the
same transformation method as applied in the derivation
of the quark-boson chromodynamic interaction Lagrangian
(\ref{8.27}). Since $\left( \bar
\psi \gamma^{\lambda} \partial_{\lambda}
\psi
\right) = \left( \bar \psi^R
\gamma^{\lambda} \partial_{\lambda}
\psi^R
\right) + \left( \bar \psi^L
\gamma^{\lambda} \partial_{\lambda}
\psi^L
\right)$,   the
left- and right-handed leptons in the derivative terms
in
the fermion Lagrangian (\ref{8.51}), which generate the
electroweak bosons,
are not cross-coupled. Thus
the resulting
bosons $B_{\lambda A}^{(l\bar
m)}$  consist of a  single complex
non-diagonal boson
$B_{\lambda A}^{(\nu \bar e)}$ generated by the
left-handed electron and neutrino components, a diagonal
boson $B_{\lambda
A}^{(\nu \bar\nu)}$ generated by the left-handed
neutrino, and a second diagonal boson
$B_{\lambda A}^{(e \bar e)}$ generated by the electron
(which we  need not decompose into left- and
right-handed components).

 In analogy with the chromodynamic case, we obtain:
tensor boson field
equations of the form (\ref{8.28}), whose solutions
can be factorized in accordance with
(\ref{8.29})
 into a harmonic wavenumber term and a
vector boson field; a general
expression analagous to (\ref{8.33}) for the
covariant derivative;  and a coupled lepton-boson
interaction Lagrangian of the same form as (\ref{8.34}).
The only difference is in the geometry of the
interacting wavenumbers and in the distinction between
left- and right-handed fields.

Replacing the quark indices  in the chromodynamic
expressions
by the corresponding lepton indices, the covariant
derivatives become in the electroweak case
\begin{equation} \label{8.56}
 D_{\lambda} \left( \begin{array}{c}
\psi_L^{(\nu)}\\
\psi_L^{(e)}
\end{array} \right)
= \partial_{\lambda}
\left( \begin{array}{c}
\psi_L^{(\nu)}\\
\psi_L^{(e)}
\end{array} \right)
- \frac{iC_2}{2}
\left( \begin{array}{cc}
\hat B_{\lambda}^{(\nu)} &  B_{\lambda}^{(\nu \bar e)}\\
 B_{\lambda}^{(\nu \bar e)*} &
 \hat B_{\lambda}^{(e)}
\end{array} \right)
\left( \begin{array}{c}
\psi_L^{(\nu)}\\
\psi_L^{(e)}
\end{array} \right),
\end{equation}
\begin{equation} \label{8.57}
 D_{\lambda} \psi_R^{(e)}
= \partial_{\lambda}
\psi_R^{(e)}
- \frac{iC_2}{2}
 \hat B_{\lambda}^{(e)}
\psi_R^{(e)},
\end{equation}
where
\begin{eqnarray} \label{8.58}
C_2 & := & \frac{1}{2}
\left(
k_{A}^{(\nu)} + k_{A}^{(e)}
\right)
\left(
k^{A}_{(\nu)} + k^{A}_{(e)}
\right)
\nonumber \\
&=&  \frac{1}{2}
\left(
\hat \omega_e^2  + \hat \omega_{\nu}^2
+ 2 \kappa_{\nu e}^2
\right),
\end{eqnarray}
with \cite{ft38}
\begin{eqnarray} \label{8.59}
\hat \omega_e^2 &:=& k^{A}_{(e)} k_{A}^{(e)}
=  k_5^{(e)2} -   k_9^{(e)2}
\\
\label{8.59a}
\hat \omega_{\nu}^2 &:=& k^{A}_{(\nu)}
k_{A}^{(\nu)} =  k_6^{(\nu)2} -   k_9^{(\nu)2}
\\
\label{8.59b}
\kappa_{\nu e}^2 &:=&  k^{A}_{(\nu)}
k_{A}^{(e)} =   k_9^{(\nu)2} = k_9^{(e)2},
\end{eqnarray}
and where $\hat B_{\lambda}^{(\nu)}, \hat B_{\lambda}^{(e)}$
are
the net diagonal bosons acting on the leptons $\nu$ and
$e^-$, respectively,
\begin{equation} \label{8.60}
\hat B_{\lambda}^{(\nu)} := \frac{2}{C_2}
\left(
\hat \omega_{\nu}^2 B_{\lambda}^{(\nu \bar \nu)}
+ \kappa_{\nu e}^2 B_{\lambda}^{(e \bar e)}
\right)
\end{equation}
\begin{equation} \label{8.61}
\hat B_{\lambda}^{(e)} := \frac{2}{C_2}
\left(
\kappa_{\nu e}^2 B_{\lambda}^{(\nu \bar \nu)}
+ \hat \omega_{e}^2 B_{\lambda}^{(e \bar e)}
\right)
\end{equation}

In analogy with the chromodynamic result (\ref{8.34}), the
total lepton-boson
electroweak
Lagrangian consists then of the sum
\begin{equation} \label{8.62x}
L_{lb}^M = L_{b}^M + L_{l}^M +L_{lb}^M,
\end{equation}
of the free-boson Lagrangian
\begin{eqnarray} \label{8.75}
L_b^M &=& - \frac{1}{2}
\left\{
\hat \omega^2_{\nu}\partial_{\lambda} B_{\mu}^{(\nu \bar
\nu)}
\partial^{\lambda} B^{\mu}_{(\nu \bar \nu)}
+ \hat \omega^2_{e}\partial_{\lambda} B_{\mu}^{(e \bar e)}
\partial^{\lambda} B^{\mu}_{(e \bar e)}
+ 2 \kappa_{\nu e}^2 \partial_{\lambda} B_{\mu}^{(e \bar e)}
\partial^{\lambda} B^{\mu}_{(\nu \bar \nu)}
\right.
\nonumber \\
&& \left. + C_2
\left( \partial_{\lambda} B_{\mu}^{(\nu \bar e)*}
\partial^{\lambda} B^{\mu}_{(\nu \bar e)}
+ \hat \omega^2_{\nu \bar e}
B_{\lambda}^{(\nu \bar e)*}
 B^{\lambda}_{(\nu \bar e)}
\right)
\right\},
\end{eqnarray}
with
\begin{equation} \label{8.75a}
\hat \omega^2_{\nu \bar e} := \left( k_{A}^{(\nu)} -
k_{A}^{(e)} \right)
\left(
k^{A}_{(\nu)} - k^{A}_{(e)}
\right) = \hat \omega^2_{\nu} + \hat \omega^2_e  - 2 \hat
\kappa^2_{\nu e},
\end{equation}
the standard  free-fermion Lagrangian
\begin{equation} \label{8.75b}
L^M_l = -  \bar \psi^{(\nu)}
\gamma^{\lambda}
\partial_{\lambda} \psi^{(\nu)}
-  \bar \psi^{(e)}
\left (\gamma^{\lambda}
\partial_{\lambda} \psi^{(e)} + \hat
\omega_e \psi^{(e)}
\right)
\end{equation}
and the interaction Lagrangian
\begin{equation} \label{8.62}
L_{lb}^M = \frac{ C_2}{2}
\left\{
\hat B_{\lambda}^{(\nu)} J^{\lambda}_{(\bar \nu \nu)}
+
\hat B_{\lambda}^{(e)} J^{\lambda}_{(\bar e e)}
 +  B_{\lambda}^{(\nu \bar e)} J^{\lambda}_{(\bar \nu e )}
 +  B_{\lambda}^{(\nu \bar e)*} J^{\lambda *}_{(\bar \nu e)}
\right\},
\end{equation}
where
\begin{equation} \label{8.62a}
J^{\lambda}_{(\bar \nu \nu)} := i \left(
\bar \psi^{(\nu)}_L\gamma^{\lambda}\psi^{(\nu)}_L
\right),
\quad
J^{\lambda}_{(\bar e e)} :=
i \left(
\bar \psi^{(e)}\gamma^{\lambda}\psi^{(e)}
\right),
\quad
J^{\lambda}_{(\bar \nu  e)} :=
i \left(\bar \psi^{(\nu)}_L\gamma^{\lambda} \psi_L^{(e)}
\right).
\end{equation}
For positive square harmonic mass $\hat \omega^2_{\nu
\bar e}$ of the non-diagonal boson, the
wavenumber components must satisfy the inequality (cf.
eqs.(\ref{8.75a}),(\ref{8.49x}))
\begin{equation} \label{8.75aa}
k_5^{(e)2} = k_6^{(\nu)2} > 2 k_9^{(e)2} =  2 k_9^{(\nu)2}.
\end{equation}

We note that in the limit of vanishing lepton masses, the
coupling coefficients $\hat
\omega_{\nu}^2, \hat
\omega_{e}^2$ for the diagonal
bosons  $B_{\lambda}^{(\nu \bar \nu)},B_{\lambda}^{(e \bar
e)} $, eqs. (\ref{8.59}),(\ref{8.59a}),
(\ref{8.60}),
(\ref{8.61}), become zero: finite lepton masses $\hat
\omega_{\nu},\hat
\omega_e $ are
required formally to
generate diagonal bosons.
However, in the limit of zero mass, the  generating
trapped-mode fermion fields fall off as $1/r$
for large distances from the metron core, as opposed to the
exponential decrease
for a finite-mass trapped field
(cf. Section~\ref{The mode-trapping mechanism}). In this
limit, the
integral of the generating lepton current
$J^{\lambda}_{(\bar \nu \nu)}$ diverges. It should
thus be
possible --  within the
framework of a more complete mode-trapping analysis -- to
consider a
limiting transition to zero lepton mass such
that in the massless limit the product of the very small
coupling
coefficient and
the very large integral current yields  a
finite net (integrated) coupling coefficient. In fact, it
was shown already
Section \ref{The mode-trapping mechanism} that a
vanishing local coupling coefficient was a
prerequisite for
the existence of an asymptotically free wave-guide mode.
However, details of the trapped-mode solutions are not
considered in this paper, and it we shall accordingly assume
that the coupling coefficients $\hat
\omega_{\nu}^2, \hat
\omega_{e}^2$ are small but finite.

\subsection*{Relation to electroweak interactions in the
Standard Model}

The metron electroweak covariant derivatives (\ref{8.56}),
(\ref{8.57})
and interaction Lagrangian (\ref{8.62})
clearly exhibit a close resemblance to the
corresponding relations of the Standard
Model. In the Weinberg-Salam-Glashow model of electroweak
interactions,  the lepton covariant derivatives are given by
 \cite{ft36}
\begin{eqnarray} \label{8.63}
\lefteqn{
D_{\lambda} \left( \begin{array}{c}
\psi_L^{(\nu)}\\
\psi_L^{(e)}
\end{array} \right)
 =
\partial_{\lambda}
\left( \begin{array}{c}
\psi_L^{(\nu)}\\
\psi_L^{(e)}
\end{array} \right)
}
&& \\
&&
- \frac{i}{2}
\left( \begin{array}{cc}
-g_1 B_{\lambda} + g_2 W_{\lambda}^{(3)} &
 g_2 \left(W_{\lambda}^{(1)} - i W_{\lambda}^{(2)}\right)
  \\
g_2 \left(W_{\lambda}^{(1)} + i W_{\lambda}^{(2)}\right)
&
-g_1 B_{\lambda} - g_2 W_{\lambda}^{(3)}
\end{array} \right)
\left( \begin{array}{c}
\psi_L^{(\nu)}\\
\psi_L^{(e)}
\end{array} \right),
\nonumber
\end{eqnarray}
\begin{equation} \label{8.64}
\mbox{\hspace{-2cm}} D_{\lambda}
\psi_R^{(e)}
 = \partial_{\lambda}
\psi_R^{(e)}
+ ig_1
  B_{\lambda}
\psi_R^{(e)},
\end{equation}
where $B_{\lambda}$,
$W_{\lambda}^{(j)}, j=1,2,3$ represent the  hypercharge and
weak
isospin  bosons of the
 $U(1)$ and $SU(2)$ gauge
groups, respectively, with  associated coupling coefficients
$g_1$ and $g_2$.

Expressed in terms of the complex
non-diagonal bosons
\begin{equation} \label{8.65}
W_{\lambda}^{\pm} :=
\left(
W_{\lambda}^{(1)} \mp W_{\lambda}^{(2)}
\right) / \sqrt{2}
\end{equation}
and the rotated diagonal bosons
\begin{eqnarray} \label{8.66}
A_{\lambda} &:=& \cos \theta_w B_{\lambda}
+ \sin \theta_w W_{\lambda}^{(3)} \nonumber \\
Z_{\lambda} &:=& -\, \sin \theta_w B_{\lambda}
+ \cos \theta_w W_{\lambda}^{(3)},
\end{eqnarray}
where the electroweak mixing angle $\theta_w$ is defined by
\begin{equation} \label{8.67}
\sin \theta_w := g_1/\sqrt{(g_1^2 +g_2^2)}, \quad
\cos \theta_w := g_2/\sqrt{(g_1^2 +g_2^2)},
\end{equation}
eqs. (\ref{8.63}), (\ref{8.64}) may be written
\begin{eqnarray} \label{8.68}
\lefteqn{
D_{\lambda} \left( \begin{array}{c}
\psi_L^{(\nu)} \nopagebreak \\
\psi_L^{(e)}
\end{array} \right)
}
& \mbox{\hspace{0.8in}} = &
\partial_{\lambda}
\left( \begin{array}{c}
\psi_L^{(\nu)}\\
\psi_L^{(e)}
\end{array} \right)
\\
&& \mbox{\hspace{- 1.5in}} - i
\left( \begin{array}{cc}
 (\sqrt{(g_1^2 +g_2^2)}/2 ) Z_{\lambda}
& (g_2/\sqrt 2) W^+_{\lambda}  \\
(g_2/\sqrt 2) W^-_{\lambda} & - e  A_{\lambda}+
\left\{
( g_1^2 - g_2^2)/
\left(
2\sqrt{(g_1^2 +g_2^2)}
\right)
\right\}
Z_{\lambda}
\end{array} \right)
\left( \begin{array}{c}
\psi_L^{(\nu)}\\
\psi_L^{(e)}
\end{array} \right),
\nonumber
\end{eqnarray}
\begin{equation} \label{8.69}
 D_{\lambda}
\psi_R^{(e)}
 = \partial_{\lambda}
\psi_R^{(e)}
- i \left\{
- e A_{\lambda} +
\left(
g_1 ^2 / \sqrt{(g_1^2 +g_2^2)}
\right)
  Z_{\lambda}
\right\}
\psi_R^{(e)},
\end{equation}
where
\begin{equation} \label{8.69a}
e = g_1 g_2 /  \sqrt{(g_1^2 +g_2^2)}
\end{equation}
is the elementary charge.

The  lepton-boson interaction Lagrangian of
the Standard Model  becomes in this notation
\begin{eqnarray} \label{8.69b}
L_{lb}^{SM} &=& (\sqrt{(g_1^2 +g_2^2)}/2 ) Z_{\lambda}
J^{\lambda}_{(\bar \nu \nu)}
+ \left\{
( g_1^2 - g_2^2)/
\left(
2\sqrt{(g_1^2 +g_2^2)}
\right)
\right\}
Z_{\lambda}
 J^{\lambda}_{(\bar e e)}
\nonumber \\
&&
- e A_{\lambda} J^{\lambda}_{(\bar e e)}
+ (g_2/\sqrt 2)
\left(
W^+_{\lambda}
 J^{\lambda}_{(\bar \nu e )}
+  W^-_{\lambda}
 J^{\lambda}_{(\bar e \nu)}
\right).
 \end{eqnarray}

Comparing the metron and Standard Model covariant derivative
expressions (\ref{8.56}) and (\ref{8.68}), we can first
immediately identify --  in analogy
with the chromodyamic case -- the complex
non-diagonal metron bosons $B_{\lambda}^{(\nu \bar
e)}$  and $B_{\lambda}^{(e \bar \nu)} =
B_{\lambda }^{(\nu \bar
e)*}$   with the
charged weak-interaction
bosons $W^{\pm}_{ \lambda}$:
\begin{eqnarray} \label{8.70}
W^+_{\lambda} &=& N_2 B_{\lambda}^{(\nu \bar
e)}/ \sqrt{2} \nonumber \\
W^-_{\lambda} &=& N_2 B_{\lambda}^{(e \bar
\nu)} / \sqrt{2},
\end{eqnarray}
where the metron and $SU(2)$ coupling coefficients are
related through
\begin{equation} \label{8.71}
g_2 =  C_2/ N_2
\end{equation}
and $N_2$ is a scaling factor.
To yield the correct normalization of the non-diagonal
components in the free-boson
Lagrangian
\begin{equation} \label{8.74}
L_b^{SM} = - \frac{1}{2}
\left(
\partial_{\lambda} A_{\mu} \partial^{\lambda} A^{\mu}
+ \partial_{\lambda} Z_{\mu} \partial^{\lambda} Z^{\mu}
+ 2 \partial_{\lambda} W^+_{\mu} \partial^{\lambda} W_-
^{\mu}
\right),
\end{equation}
the scaling factor must be set to
\begin{equation} \label{8.72}
 N_2 = \sqrt {C_2},
\end{equation}
so that
\begin{equation} \label{8.73}
g_2 =  \sqrt { C_2}.
\end{equation}

The
diagonal-boson sector of the metron interaction Lagrangian
can also be brought to close
(but, as in the strong-interaction case, not perfect)
agreement  with the Standard Model electroweak Lagrangian
through a suitable linear transformation relating
the metron
bosons $B_{\lambda}^{(\nu \bar \nu)}, B_{\lambda}^{(e \bar
e)}$ to the corresponding diagonal bosons
$A_{\lambda},Z_{\lambda}$ of the Standard Model. The
transformation must reproduce the standard
form (\ref{8.74}) of the free-boson Lagrangian,
while yielding an interaction Lagrangian of
the general form (\ref{8.69b}). This is characterized,
in particular, by the vanishing cross-coupling
between the neutrino current $J^{\lambda}_{(\bar
\nu \nu)}$ and the
electromagnetic field $A_{\lambda}$ (the neutrino
carries no
electric charge).

The two conditions uniquely determine the transformation to
within the signs of the boson fields $A_{\lambda},
Z_{\lambda}$. These are determined by the  sign convention
chosen
for
the coupling coefficients.
The diagonalization of the
diagonal-boson sector of the free boson Lagrangian
(\ref{8.75}) in the standard isotropic form (\ref{8.74})
defines the
transformation to within an arbitrary rotation. The rotation
is then fixed by the second
condition that the coupling between $J^{\lambda}_{(\bar \nu
\nu)}$
and $A_{\lambda}$ is zero. One finds
\begin{eqnarray} \label{8.78a}
B_{\lambda}^{(\nu \bar \nu)} &=& \frac{1}{\hat \omega_{\nu}}
Z^{\lambda}
 + \frac{{\kappa_{\nu e}^2}}{ \hat \omega_{\nu} \Lambda}
A_{\lambda}
\nonumber \\
B_{\lambda}^{(e \bar e)} &=& - \frac{\hat
\omega_{\nu}}{\Lambda} A^{\lambda},
\end{eqnarray}
where
\begin{equation} \label{8.77a}
\Lambda^2 := \hat \omega_e^2  \hat \omega_{\nu}^2 -
\kappa^4_{\nu e}.
\end{equation}
A necessary condition in order that $L_b^M$ represents  a
negative
definite
form and can therefore be transformed into the normal
diagonal form (\ref{8.74}) is that $\Lambda^2 >0$. This is
already ensured, however, by
the inequality
(\ref{8.75aa}).

The transformation (\ref{8.78a}) corresponds to a
transformation of the wavenumber
base used in the factorization
of the diagonal tensor bosons into vector bosons from the
original non-orthogonal wavenumber vectors
$\mbox{k}^{(\nu)},\mbox{k}^{(e)}$ to the orthonormal
wavenumber
pair
\begin{eqnarray} \label{8.77ab}
\mbox{k}^{(Z)} =\frac{1}{\hat \omega_{\nu} }\mbox{k}^{(\nu)}
\\ \label{8.77ac}
\mbox{k}^{(A)} = \frac{\kappa^2_{\nu e}}{\hat \omega_{\nu}
\Lambda}\mbox{k}^{(\nu)}
 -\frac{\hat \omega_{\nu}}{\Lambda} \, \mbox{k}^{(e)} .
\end{eqnarray}
Thus the electroweak diagonal tensor boson may be
represented in the two alternative forms
\begin{eqnarray} \label{8.78ab}
B_{\lambda A}^{(d)} & = &
k_{A}^{(\nu)} B_{\lambda}^{(\nu \bar \nu)} +
k_{A}^{(e)} B_{\lambda}^{(e \bar e)} \quad \mbox{or}
 \\
\label{8.78aab}
B_{\lambda A}^{(d)} &=& k_{A}^{(Z)} Z_{\lambda} +
k_{A}^{(A)}
A_{\lambda}
\end{eqnarray}
The signs of $A_{\lambda}, Z_{\lambda}$ in the
transformation
(\ref{8.78a}) have been chosen such that  positive $k_5$ and
$k_6$ correspond to positive electrical and
`weak-interaction'
(not to be identified with isospin) charge, respectively,
the electron carrying negative electric charge
(cf. eq.(\ref{8.77ac})).

 In contrast to the original representation (\ref{8.78ab}),
the orthonormal representation (\ref{8.78aab}) no longer
specifies the individual fermion pairs which generate the
diagonal
bosons. However, it is in accord with the usual Standard
Model representation and is
more convenient for the extension of the analysis, in the
following
sub-section, to electroweak interactions between quarks. It
will be found that these
generate the
same set of electroweak bosons, but the diagonal bosons are
produced in different linear
combinations than in the lepton case. It is then useful to
have a common orthonormal representation for both sets of
interactions.

For the metron lepton-boson interaction Lagrangian we obtain
finally, in the boson notation of the Standard Model,
\begin{equation} \label{8.80}
L_{lb}^{M} = \hat \omega_{\nu} Z_{\lambda}
J^{\lambda}_{(\bar \nu \nu)}
+ \frac{\kappa_{\nu e}^2}{ \hat \omega_{\nu}}
Z_{\lambda}
 J^{\lambda}_{(\bar e e)}
- e_M A_{\lambda}
J^{\lambda}_{(\bar e e)}
+ \frac{g_2}{\sqrt 2}
\left(
W^+_{\lambda}
 J^{\lambda}_{(\bar \nu e )}
+  W^-_{\lambda}
 J^{\lambda}_{(\bar e \nu)}
\right),
\end{equation}
where
\begin{equation} \label{8.80a}
e_M := \frac{\Lambda}{ \hat \omega_{\nu}}.
\end{equation}

The metron form (\ref{8.80}) is seen to agree in general
structure with the
Standard Model lepton-boson electroweak interaction
Lagrangian (\ref{8.69b}) (the boson-boson interaction terms
were again not considered in either model). The
diagonal sectors of the two models can be matched more
closely through a
suitable choice of the wavenumbers determining the
square frequencies  $\hat \omega_{\nu }^2, \hat \omega_{
e}^2 $ and scalar product  $\kappa_{\nu e}^2$, which define
the metron coupling coefficients. However, as in the
chromodynamic case, a perfect agreement cannot be achieved
(in contrast to the exact match established for
the non-diagonal bosons). As before, this is not
particularly
disturbing in view of the basically different role of the
metron and Standard Model Lagrangians in the
description of particle states and interactions. For this
reason,  the metron
equivalent (\ref{8.80a}) of  the elementary charge carries
an index M as a reminder that the coupling coefficients of
the two models cannot
be compared quantitatively without considering the different
normalizations of the fields/operators in the two models. A
quantitative determination of the electric charge in terms
of integrated properties of the metron solution  was
presented  within the framework of the metron description of
the Maxwell-Dirac-Einstein system in Section~\ref{Particle
interactions}. A similar computation would need to be
carried out now for the extended electroweak system.   The
same
applies for the coupling
coefficients $g_2$, eq.(\ref{8.73}) and, as has already been
pointed out, for the strong-interaction coupling coefficient
$g_3$, eq.(\ref{8.40a}).

\subsection*{Electroweak interactions between quarks}

The metron picture of electroweak interactions between
leptons carries over with only minor modifications to
quarks. Consider the electroweak interactions between two
quarks, $u$, $d$, say, of the same color but different
flavor. The discussion is restricted again to a single
quark
family. To represent different quark flavors we modify the
original definitions (\ref{8.2a}), (\ref{8.2ab}) of the
quark wavenumber vectors, which were assumed to lie in the
strong-interaction color plane $k_7,k_8$, by including now
also wavenumber components $k_5,k_6,k_9$ in the extended
electroweak space.

The polarization tensor (\ref{8.2aa}) must then also be
modified such that the fermion matrix satisfies the relation
$M =(\hat \omega)^{-1} i
\gamma^4$, for the non-Euclidean model, or $M = E^{-1} i
\gamma^4$, for the Euclidean model, together
with the zero-trace and divergence gauge conditions. This
can be achieved,
for example, by rotating, or Lorentz transforming, the
harmonic sub-space such that the new wavenumber vectors lie
again in the  color plane. One can then apply the
original polarization relations   (\ref{8.2aa}) in the new
coordinate system and
transform back to the old coordinate system to obtain the
modified polarization tensors.

The transformation depends on
the additional electroweak wavenumber components
$k_5,k_6,k_9$ and therefore on the quark flavor.
 This
violates our simplifying assumption, made for both
strong and electroweak interactions, that the polarization
tensor is independent of the fermion color or flavor.
However, we assume that the
electroweak
 wavenumber components are small
compared with the
strong-interaction components. In this case the polarization
tensor can still be regarded as independent of the quark
flavor to lowest order. The flavor-dependent
modifications of the polarization tensor produce a weak
symmetry breaking in the strong interactions, but have no
impact to
lowest order on the electroweak interactions with which we
are concerned here.

Specifically, we assume that the harmonic wavenumber vectors
of the quarks are given by (cf. Fig \ref{Fig.8.1}b)
\begin{eqnarray} \label{8.96}
\mbox{k}^{(u)}&
=& - \frac{2}{3}
\mbox{k}^{(e)}
+\frac{1}{3}
\mbox{k}^{(\nu)}
+\mbox{k}^{(c)}
\\ \label{8.96aa}
\mbox{k}^{(d)}&
=& \quad
 \frac{1}{3}
\mbox{k}^{(e)}
-\frac{2}{3}\mbox{k}^{(\nu)}
+\mbox{k}^{(c)},
\end{eqnarray}
where
\begin{equation} \label{8.96a}
\mbox{k}^{(c)} =
(0,0,k_7^{(c)},k_8^{(c)},0)
\end{equation}
is the common  color wavenumber vector. The electromagnetic
wavenumber components $k_5^{(u)},k_5^{(d)}$ are determined
by the  standard assignment of charges to the up and down
quarks in
the
Standard Model.  The remaining extended weak-interaction
wavenumber
components
$k_6^{(u)},k_6^{(d)}$ and $k_9^{(u)}, k_9^{(d)}$ follow from
the requirement that the lepton and quark interactions $(\nu
\bar
e)$ and $(u \bar d)$ generate the same non-diagonal boson.
This yields two conditions:  the
 difference wavenumber vectors
must be identical, and the  directions of
the sum wavenumbers  in the extended
electroweak sub-space, which define the harmonic direction
of the non-diagonal electroweak tensor boson
(cf.(\ref{8.29})), must be the same. Indeed, eqs.
(\ref{8.96}), (\ref{8.96aa}) yield
\begin{equation} \label{8.97a}
\mbox{k}^{(u)}+\mbox{k}^{(d)}= -\frac{1}{3}
\left(
\mbox{k}^{(\nu)}+\mbox{k}^{(e)}
\right)
+ 2 \mbox{k}^{(c)}
\end{equation}

The wavenumber assignments (\ref{8.96}), (\ref{8.96aa}) also
ensure
that, apart from the color wavenumber vector
$\mbox{k}^{(c)}$,  the directions of the zero-wavenumber
diagonal tensor bosons for both quarks and leptons lie in
the same
plane in the extended electroweak sub-space spanned by
the vectors
$\mbox{k}^{(\nu)},\mbox{k}^{(e)}$ or, equivalently,
$\mbox{k}^{(Z)},\mbox{k}^{(A)}$. Thus with respect to the
extended electroweak sub-space orthogonal to the color
plane, quarks and
leptons
generate the same set of bosons.

However, in contrast to
their lepton counterparts, the quark-generated bosons also
have strong components in the color plane. The tensor-boson
factorization (\ref{8.29}) yields in the present case
\begin{eqnarray} \label{8.100}
B_{\lambda A}^{(u\bar u)} &=:&k_{A}^{(u)}
B_{\lambda}^{(u\bar u)} +
k_{A}^{(c)}
B_{\lambda}^{(q\bar q)}
\\
\label{8.101}
B_{\lambda A}^{(d\bar d)} &=:&
k_{A}^{(d)} B_{\lambda}^{(d\bar d)}+
k_{A}^{(c)}
B_{\lambda}^{(q\bar q)}
\\
\label{8.102}
B_{\lambda A}^{(u\bar d)} &=:&
\frac{1}{2}
\left(
k_{A}^{(u)} +k_{A}^{(d)}
\right)
B_{\lambda}^{(u\bar d)}+
k_{A}^{(c)}
B_{\lambda}^{(q\bar q)},
\end{eqnarray}
where we have divided the vector bosons defined on the right
hand side into components $B_{\lambda}^{(u\bar
u)},B_{\lambda}^{(s\bar d)}$ and $B_{\lambda}^{(u\bar d)}$
with harmonic directions lying in the extended electroweak
space and a vector boson  $B_{\lambda}^{(q \bar q)}$
associated with
the color wavenumber vector $\mbox{k}^c$. This is
common to all three tensor bosons and can be identified as
the
strong-interaction boson $B_{\lambda}^{(q \bar q)}$
considered in Section~\ref{Strong interactions}, where $(q)$
represents a quark of color $c$.  Since we are concerned
here only with the electroweak-interaction sector, we shall
discard this boson in the following.

The remaining quark-generated electroweak bosons can be
determined in the same way as the lepton-generated bosons
in the previous section. Using the boson representation
$Z_{\lambda},A_{\lambda}$ and $W^{\pm}_{\lambda}$ and
applying the
relations (\ref{8.96})-(\ref{8.97a}), (\ref{8.70}),
(\ref{8.72}) and (\ref{8.77ab}),(\ref{8.77ac}),
we obtain as generalization of (\ref{8.34}) and
(\ref{8.62x})-(\ref{8.62a}) for the metron form of the total
single-family electroweak Lagrangian
\begin{equation} \label{8.107}
L_{ew}^M = L_{b}^{M} + L_{f}^M + L_{ewint}^M,
\end{equation}
where  the free-boson and free-fermion Lagrangians
$L_{b}^{M}, L_{f}^M$ are
given
as before by eqs.(\ref{8.74}),(\ref{8.75b}), respectively
(the fermion
sum in (\ref{8.75b}) extending now over both leptons and
quarks) and the electroweak interaction Lagrangian is given
by
\begin{eqnarray} \label{8.110}
L_{ewint}^M &=&  Z_{\lambda}
\left\{
\hat \omega_{\nu} J_{\lambda}^{(\bar \nu \nu)}
+ \frac{\kappa_{\nu e}^2}{\hat \omega_{\nu}}
J_{\lambda}^{(\bar e e)}
+ \left(
- \frac{2 \kappa_{\nu e}^2}{3 \hat \omega_{\nu}}
+ \frac{1}{3} \hat \omega_{\nu}
\right)
J_{\lambda}^{(\bar u u)}
+ \left(
 \frac{ \kappa_{\nu e}^2}{3 \hat \omega_{\nu}}
- \frac{2}{3} \hat \omega_{\nu}
\right)
J_{\lambda}^{(\bar d d)}
\right\}
\nonumber \\
&& + A_{\lambda}
\left\{
  - e_M J_{\lambda}^{(\bar e e)}
+ \frac{2}{3} e_M J_{\lambda}^{(\bar u u)}
- \frac{1}{3} e_M J_{\lambda}^{(\bar d d)}
\right\}
\nonumber \\
&&
+\, \frac{g_2}{\sqrt{2}}
\left\{
W_{\lambda}^+
\left(
J^{\lambda}_{(\bar \nu e)} - \frac{1}{3}
J^{\lambda}_{(\bar u d)}
\right)
+
W_{\lambda}^-
\left(
J^{\lambda}_{(\bar e \nu )} - \frac{1}{3}
J^{\lambda}_{(\bar d u )}
\right)
\right\}.
\end{eqnarray}
The expression (\ref{8.110}) agrees in general
structure with the electroweak Lagrangian of the Standard
Model, but differs again in the details of the
weak-interaction coupling coefficients -- as to be expected.

An important  difference between the two models is that the
metron
model (as proposed here) preserves parity for the weakly
interacting quarks: there is no distinction in the
interactions between
left-handed and right-handed fermion fields when both fields
are present. Parity violation of the weak interactions
is attributed in the metron model entirely to the existence
of the massless (or
almost massless) left-handed neutrino,  which can interact
only
with the
left-handed electron component. Most of the classical
experiments on
the parity violation of the weak interactions involve
interactions with neutrinos. To test the metron picture,
it would be of interest to devise experiments to determine
the parity of weak
interactions involving only (up and down) quarks.

\subsection*{The Higgs mechanism}

In the Standard Model, the symmetry-breaking Higgs mechanism
is invoked to
generate the fermion masses and the masses
of the charged and neutral weak-interaction bosons. We shall
not resort to the Higgs mechanism to explain the fermion
masses, but attribute these simply to the mode-trapping
mechanism, which we assume produces a
non-$SU(2)$-symmetrical particle state.
However, an interaction analagous to the
Higgs mechanism  in the Standard Model is needed
to explain the boson masses in the
metron model, since the
lepton-boson interactions alone yield either zero boson
mass, for the diagonal bosons, or a small mass of the order
of the lepton mass, for the non-diagonal bosons.
In the following we therefore consider a simple
interaction which
generates boson masses in a manner  similar to the Higgs
mechanism.

As metron analogue of the Higgs field, consider a
periodic  perturbation
\begin{equation} \label{8.81}
g^{(h)}_{AB} = \hat g^{(h)}_{AB}(x)
\exp (ik^{(h)}_{A} x^{A})
+ c.c.
\end{equation}
of the harmonic components of the metric field. For
harmonic metric field components, the Lagrangian describing
the
interactions of the  field  with bosons can be
obtained, as shown generally in Section~\ref{Lagrangians}
and
implemented so far for fermions,
by replacing the
partial derivatives in the free-field Lagrangian
(\ref{3.30a}) by the appropriate covariant derivatives.
Applying (\ref{8.25}), (\ref{8.5d}) and (\ref{8.32}), the
Higgs Lagrangian, including both the free-field contribution
and  the  interaction of the Higgs field
with the electroweak bosons,  is accordingly given by
\begin{eqnarray} \label{8.82}
L_h & = & - \frac{1}{2}  \left
\{
\left[
\partial_{\lambda} g^{(h)}_{AB}
-i  k^{C}_{(h)}
\left(
B^{(\nu \bar e)}_{\lambda C} \exp i S_{\nu \bar e}
+
B^{(e \bar \nu)}_{\lambda C} \exp i S_{e \bar \nu}
+
 B^{(d)} _{ \lambda C}
\right)
g^{(h)}_{AB}
\right]^* \times
\right.
\nonumber \\
&&
\quad
\left.
\left[
\partial^{\lambda} g_{(h)}^{AB}
-i  k_{C}^{(h)}
\left(
B_{(\nu \bar e)}^{\lambda C} \exp i S_{\nu \bar e}
+
B_{(e \bar \nu )}^{\lambda C} \exp i S_{e \bar \nu}
+
 B_{(d)}^{ \lambda C}
\right)
g_{(h)}^{AB}
\right]
\right.
\nonumber \\
&&
\quad
\left.
+ \; \hat \omega^2_h g^{(h)*}_{AB}
 g_{(h)}^{AB}
\right\},
\end{eqnarray}
where
\begin{equation} \label{8.83}
B^{(\nu \bar e)}_{\lambda A}  =
B^{ (e \bar \nu)*}_{\lambda A}  =
 \frac{k^{(\nu)}_{A}+ k^{(e)}_{A}}{2}
B^{(\nu \bar e)}_{\lambda}
\end{equation}
\begin{equation} \label{8.84}
B^d_{\lambda A}  =
 k^{(\nu)}_{A} B^{(\nu \bar \nu)}_{\lambda}
+ k^{(e)}_{A}
B^{(e\bar e)}_{\lambda}.
\end{equation}
This may be written
\begin{equation} \label{8.85}
L_h  = - \frac{1}{2}
\left\{
\partial_{\lambda} g^{(h)*}_{AB}
\partial^{\lambda} g_{(h)}^{AB}
+ I + M_b
\right\},
\end{equation}
where $I$ represents the Higgs-boson interaction terms of
structure (ignoring derivatives) $g^{(h)*} B g^{(h)}$ and
$M_b$ is the boson mass matrix of the form $g^{(h)*} g^{(h)}
B^* B$ \cite{ft39}.

Previously, we had been concerned with the interaction
terms $I$ for the special case that the field
$g_{(h)}^{AB}$ corresponds to a fermion field.
Here we
are concerned with the higher-order terms $M_b$. We find
\begin{eqnarray} \label{8.86}
M_b & = &\frac{v^2}{2}
\left(
\kappa^2_{h\nu} + \kappa^2_{he}
\right)^2
B^{(\nu \bar e)*}_{\lambda}
B_{(\nu \bar e)}^{\lambda}
\nonumber \\
&& +\, v^2
\left(
\kappa^2_{h\nu} B^{(\nu \bar \nu)}_{\lambda}
+ \kappa^2_{he} B^{(e \bar e)}_{\lambda}
\right)
\left(
 \kappa^2_{h\nu} B_{(\nu \bar \nu)}^{\lambda}
+  \kappa^2_{he} B_{(e \bar e)}^{\lambda}
\right),
\end{eqnarray}
where
\begin{equation} \label{8.87}
v^2 =  g^{(h)*}_{AB}
 g_{(h)}^{AB}
\end{equation}
and
\begin{eqnarray} \label{8.88}
\kappa^2_{h\nu}&=& k_{A}^{(h)} k^{A}_{(\nu)}
\\
\label{8.88a}
\kappa^2_{he}&=& k_{A}^{(h)} k^{A}_{(e)}.
\end{eqnarray}

The first term on the right hand side of eq.(\ref{8.86})
yields the square mass $m^2_{W^{\pm}}$ of the charged boson
$W^{(\pm)}$. Substituting the relations (\ref{8.70}),
(\ref{8.72}) and (\ref{8.58}), we obtain
\begin{equation} \label{8.89}
m^2_{W^{\pm}} = v^2
\left(\kappa^2_{h\nu} + \kappa^2_{he} \right)^2
 \left(\hat \omega^2_{\nu} + \hat \omega^2_e
+ 2 \kappa^2_{\nu e}
\right)^{-1}.
\end{equation}

The second term  represents the mass matrix for the diagonal
bosons. The matrix is singular: mass is generated only for
the boson defined by the linear combination
$\left(
\kappa^2_{h\nu} B^{(\nu \bar \nu)}_{\lambda}
+ \kappa^2_{he} B^{(e \bar e)}_{\lambda}
\right)$; the orthogonal boson remains massless. Expressed
in terms of $A_{\lambda}, Z_{\lambda}$, using (\ref{8.78a}),
 the massive boson is given by
\begin{equation} \label{8.90}
\left(
\kappa^2_{h\nu} B^{(\nu \bar \nu)}_{\lambda}
+ \kappa^2_{he} B^{(e \bar e)}_{\lambda}
\right)
= \frac{\kappa^2_{h\nu}}{\hat \omega_{\nu}}
Z_{\lambda} + \frac{\kappa^2_{\nu e}}
{\hat \omega_{\nu } \Lambda}
\left(
-\kappa^2_{h\nu}\kappa^2_{\nu e}
+
\kappa^2_{he}\hat \omega^2_{\nu}
\right)
A_{\lambda}.
\end{equation}

To recover the Standard Model result that the massive
diagonal boson is identical to $Z_{\lambda}$, we require
\begin{equation} \label{8.91}
\frac{\kappa^2_{h\nu}}{\kappa^2_{h e}}
= \frac{\hat \omega^2_{\nu}}{\kappa^2_{\nu e}}.
\end{equation}
The square  mass of the $Z_{\lambda}$-boson is accordingly
\begin{equation} \label{8.92}
m^2_Z = v^2 \kappa^4_{h \nu} \hat \omega^{-2}_{\nu}.
\end{equation}

A simple solution of eq.(\ref{8.91}) is that  the Higgs and
neutrino
wavenumbers
are the same and that the Higgs and neutrino fields are in
fact  identical. The Higgs mechanism
represents in this case simply a higher-order neutrino-boson
interaction.

The condition (\ref{8.91}) implies, according to
(\ref{8.59a}), (\ref{8.59b}) and (\ref{8.88}),
(\ref{8.88a}), that the projection of the Higgs wavenumber
onto the plane spanned by the neutrino and electron
wavenumbers lies parallel to the neutrino wavenumber. This
is a non-symmetrical property.  In the Standard Model, the
symmetry breaking of the Higgs field is attributed to the
existence of non-symmetrical vacuum states for a
symmetrical potential (with an instability at the origin).
In the
metron model we argue similarly that the n-dimensional
gravitational equations, although symmetrical in harmonic
space, allow non-symmetrical trapped-mode solutions (as we
had in fact already assumed in allowing different masses for
the
electron and the neutrino).

Applying (\ref{8.91}) to (\ref{8.89}) and (\ref{8.92}), we
obtain for the
ratio of the charged and neutral
boson masses
\begin{equation} \label{8.93}
\frac{m_{W^{\pm}}}{m_Z} = \frac{\hat \omega^2_{\nu} +
\kappa^2_{\nu e}}
{
\hat \omega_{\nu}
\left( \hat \omega_{\nu}^2 + \hat \omega_e^2 +
\kappa_{\nu e}^2
\right)^{1/2}
}.
\end{equation}
The
neutrino and electron harmonic wavenumbers can be
chosen to reproduce the observed mass ratio
\begin{equation} \label{8.95}
\frac{m_{W^{\pm}}}{m_Z} = \cos \theta_w = 0.87.
\end{equation}
However, as pointed out above, there is little point in
tuning the metron model too closely to the Standard Model in
the present stage of the analysis. This must await detailed
computations
of the trapped-mode  metron solutions, which alone can
yield quantitative information on the
particle masses and other particle properties.

\section{Invariance properties}
\label{Invariance properties}

In the construction of the metron model, we have so far made
no use of general invariance considerations (apart from the
more technical application of invariance properties in the
determination of fermion-boson interactions). This is
in marked contrast to the Standard Model, which is founded
on the principles of gauge symmetry. However, it is in
keeping with
the general metron philosophy: specific symmetry properties
are attributed to the individual geometrical features of the
trapped-mode particle solutions, rather than to the
symmetries of the basic
Lagrangian, which
exhibits `only' the general gauge symmetry corresponding to
the invariance
with respect to coordinate transformations. We accordingly
assumed that the metron solutions exhibit the discrete
permutation symmetries associated with the $SU(2)$
and $SU(3)$ gauge groups of the Standard Model. The question
then arises: do the metron solutions exhibit also
continuous symmetry properties which can be related to the
continuous
symmetries of the $U(1)\times SU(2)\times SU(3)$ gauge group
of the Standard Model?

Since this question was already answered in the affirmative
for
the
special case of electromagnetic interactions in
Section~\ref{The Maxwell-Dirac-Einstein Lagrangian}, we may
anticipate
that the same holds also for the other interactions. In the
electromagnetic case, the
diffeomorphism
corresponding to the gauge group $U(1)$ was associated with
the transformation (\ref{3.37}),(\ref{3.38}), which was used
to locally remove
the electromagnetic field  as a technique for computing the
electromagnetic coupling terms.
The same approach can be applied also in
the general case. We consider a coordinate transformation
which does not change the basic geometrical symmetry of the
metron solution, i.e. does not affect the lowest order quark
configuration from which the boson fields are derived. The
Lagrangian for the set of transformed metron fields will
then be invariant under this transformation.

For a given set of fermions  $(p)$ and bosons $(p \bar q)$,
consider, as generalization of the local transformation
(\ref{8.13}), in analogy with the transformation
(\ref{3.49}), the
infinitesimal global coordinate
transformation
\begin{eqnarray} \label{8.111}
\check x^{A}  & =&    x^{A}    -
\xi^{A}
 \nonumber \\
\check x^{\lambda} &=& x^{\lambda},
\end{eqnarray}
in which the local relation (\ref{8.14}) is replaced
(after a sign change to conform with the notation of
Section~\ref{The Maxwell-Dirac-Einstein Lagrangian})
by the
global expression
\begin{equation} \label{8.112}
\xi^{A}= \sum_{p, q}
v_{ (p\bar
q) }^{A}
\epsilon_{p\bar q}
 \exp(iS^{p\bar q}),
\end{equation}
with constant complex vectors $v_{ (p\bar q) }^{A} =
v_{ (q\bar
p)}^{A*}$
and complex infinitesimal amplitudes $\epsilon_{ p\bar q } =
\epsilon_{ p\bar q }(x) =
\epsilon^* _{q\bar p}(x)$ which are
functions of physical spacetime.

In contrast to the translations considered in the
electromagnetic case, the periodic transformations
(\ref{8.111}),(\ref{8.112}) no longer represent a group when
applied to a finite set of fermion and boson
fields, since
the transformation generates higher-harmonic
Fourier components not contained in the original set of
fields. However, the group property is retained if one
considers the complete set of periodic fields consisting
of all possible higher-order products of the basic
fermion fields. The fermion-boson Lagrangians considered in
the previous sections represent
truncated versions  of the complete gravitational
Lagrangian, which is defined for the
infinite discrete set of Fourier components generated from
a given finite basic set of fermions. The
invariance considerations for the gravitational system
apply only for the complete Lagrangian, not for the
truncated Lagrangian. However,
relations between the invariance properties of the
gravitational system and the gauge symmetries of the
Standard Model can, of course, be established only for the
truncated gravitational system containing the fields
appearing in the Standard Model. Thus
the following invariance considerations indicate again -- as
in the case of the dynamical analysis --
that the
Standard Model can be regarded, from the metron viewpoint,
only as an approximation of the fully nonlinear
n-dimensional gravitational system.

To first order in $\epsilon^{p \bar q}$, the metric tensor
transforms under the coordinate transformation
(\ref{8.111}),(\ref{8.112}) as $g^{L M} \rightarrow
g^{L M} +
\delta g^{L M}$,
where
\begin{equation} \label{8.113}
\delta g^{L M} =  -\partial_{N}\xi^{L}
\eta^{N M} - \partial_{N}
\xi^{M}
\eta^{L N} +
 \xi^{N} \partial_{N}  g^{L M}.
\end{equation}
 If we consider only
the sub-set of periodic fields contained in the Standard
Model,
it can be shown that for fermions this
transformation has the same form as the  gauge
transformations of the Standard Model. We demonstrate the
identity for the gauge group $SU(3)$; the
derivation for the $U(1)\times SU(2)$ group (in the limit of
vanishing mass) is similar.

For fermions, the infinitesmal $SU(3)$ transformation is
given in the Standard Model  by
\begin{equation} \label{114}
\delta \mbox{\boldmath$\psi$} = \frac{i}{2}
 \left(
\sum_{\rho = 1,8} \epsilon_{\rho} \lambda^{\rho}
\right)
{\mbox{\boldmath$\psi$}},
\end{equation}
where $\epsilon_{\rho}$ are the infinitesmal Lie parameters
of the $SU(3)$ generators $ \lambda^{\rho}$.

Considering first the non-diagonal generators, $p \neq q$,
the corresponding
metron expression,
eqs.(\ref{8.112}),(\ref{8.113}), yields for the
harmonic-index metric field components corresponding to
fermions (noting that the
contributions from the
first two terms on the right hand side of (\ref{8.113})
 yield `bosonic' Fourier components with wavenumbers ${\bf
k}_{p \bar q}$, which are not included in the subset of
fields represented in the  Standard Model)
\begin{equation} \label{115}
\left(\delta \psi_{(p)} \right)_{nd} =  i
\sum_{q \neq p} k^{(q)}_{C} v_{(p \bar q)}^{C}
\epsilon_{p \bar q} \psi_{(q)}.
\end{equation}
Comparing (\ref{114}) and (\ref{115}), the non-diagonal
components of the $SU(3)$ and
diffeomorphism transformation relations are seen to be
identical if, for given vectors $v_{ (p\bar
q) }^{A}$, the Lie parameters $\epsilon_{\rho}$ and
$\epsilon_{p \bar q}$ are appropriately related. Setting,
for
example,
\begin{equation} \label{116}
v_{ (p\bar q) }^{A} = k_{(p)}^{A} -
k_{(q)}^{A},
\end{equation}
we find
\begin{equation} \label{117}
\mbox{Re or Im}\; \epsilon_{p \bar q} = C^{-1} \,
\epsilon_{\rho},
\end{equation}
where the constant (cf.eqs.(\ref{8.30}), \ref{8.30a}))
\begin{equation} \label{118}
C =2\left(
 k^{(p)}_{A} k_{(q)}^{A} - \omega^2_p
\right)
 = - \,\hat \omega_p^{2}
\end{equation}
and the cross-assignment of the indices $p \bar q$ and
$\rho$ is in accordance with the definitions (\ref{8.42}).

The diagonal components can be similarly related. The first
two terms on the right hand side of (\ref{8.113}) again
yield no contribution (in this case because
$\partial_{A}\xi^{B}$ vanishes) and one obtains,
equating the two transformation relations,
\begin{equation} \label{119}
\left(
\begin{array}{c}
k_{C}^{(1)} \\
k_{C}^{(2)} \\
k_{C}^{(3)}
\end{array}
\right)
\left(
v^{C}_{(1 \bar 1)}\epsilon_{1 \bar 1} +
v^{C}_{(2 \bar 2)}\epsilon_{2 \bar 2} +
v^{C}_{(3 \bar 3)}\epsilon_{3 \bar 3}
\right)
 =  \frac{1}{2}
\left(
\begin{array}{c}
\epsilon_3 + \frac{1}{\sqrt{3}}\epsilon_8 \\
-\epsilon_3 + \frac{1}{\sqrt{3}}\epsilon_8 \\
- \frac{2}{\sqrt{3}}\epsilon_8
\end{array}
\right).
\end{equation}
Equations (\ref{119}) represent three relations between the
three Lie parameters $\epsilon_{p \bar p}$ and the two  Lie
parameters
$\epsilon_3,\epsilon_8$. However, the equations are not
independent: their sum vanishes,
since the sum of the fermion wavenumbers vanishes, cf.
eq.(\ref{8.2a}). Thus for  given $v^{C}_{(p \bar p)}$,
eqs.(\ref{119}) uniquely determine the
three Lie parameters $\epsilon_{p \bar p}$ as linear
combinations of the two  Lie parameters
$\epsilon_3,\epsilon_8$, provided the vectors
$v^{C}_{(p\bar p)}$ are chosen such that their
projections onto the
color plane are not all parallel.

The correspondence between the
infinitesmal coordinate transformation
(\ref{8.111})-(\ref{8.113}) and the
$SU(3)$ gauge transformations can be demonstrated similarly
for boson fields. The  analysis of Section~\ref{The
Maxwell-Dirac-Einstein Lagrangian} for the electromagnetic
case can be generalized to periodic boson fields in the same
way
as for fermions. One finds, as in the electromagnetic case,
that the covariant derivative for the fermion fields and
thus the fermion-gluon
interaction Lagrangian are invariant with respect to both
transformations to lowest order in the boson fields. The
situation is a little more complicated for boson-boson
interactions, which were not considered
here. Although there exists a general structural symmetry
between the infinitesimal coordinate and $SU(3)$
transformations, the transformations differ again in detail.

In summary, the gauge transformations of the Standard Model
correspond to a particular class of
coordinate transformations in the metron model. The
transformations have the property that they map a given set
of Fourier components into the same set and thus do not
change the specific form of the gravitational Lagrangian
appropriate for this particular set of fields. However, the
invariance of the
Lagrangian applies strictly only for the
complete
Lagrangian, defined for the complete set of all periodic
fields
generated by interactions of arbitrary order between a given
basic set of fermion fields, rather than for the
Lagrangian of the truncated set considered in the metron
interpretation of the Standard Model. The correspondence
between the gauge symmetries
of the Standard Model and the diffeomorphic gauge symmetries
holds only if all gravitational fields not contained in the
Standard Model sub-set are discarded.  From the metron
viewpoint, the Standard Model appears therefore as an
approximation, the fundamental fermion fields (accepting
that
leptons and quarks are indeed the basic building blocks of
matter) generating not only the Standard Model bosons,
represented in the metron model by mixed-index,
quadratic difference-interaction fields, but also further
quadratic-interaction fields and a spectrum of
higher-order Fourier components.

\section{Summary and conclusions}
\label{Summary and conclusions}
\typeout{################################}
\typeout{################################}
\typeout{        START OF met4-9.tex}
\typeout{################################}
\typeout{################################}

In developing the metron model we followed the
deductive
method: starting from the basic Einstein vacuum field
equations (\ref{1.1}) in a higher-dimensional space, we
proceeded  to deduce the different properties of the
postulated nonlinear solutions of the equations in a natural
logical sequence. In retrospect, however, the
metron model can be seen to be a composite of a number of
rather independent concepts which were combined into
a unified theory. It is instructive to summarize these
different
concepts and their interrelationships
following the actual constructive development of the metron
picture. Historically, the metron
model evolved `accumulatively' from the following
considerations:
\begin{itemize}
\item
It is possible to resolve the wave-particle duality paradox
within the framework of a classical objective theory if it
is assumed that there exist quasi-point-like particles which
support, in addition to the classical electromagnetic and
gravitational fields, periodic fields with a frequency
proportional to the particle mass in accordance with de
Broglie's relation. The existence of
point-like particles explains the corpuscular nature of
matter,
while the periodic far fields of the particles give rise to
the interference phenomena observed in their interactions.
\item
A contradiction with Bell's theorem on the non-existence of
hidden-variable theories does not arise if the distant
interactions between such particles are assumed to be time
symmetric, following the general viewpoint of Tetrode,
Wheeler and
Feynman, and others. The periodic de Broglie fields then
also represent standing waves which do not decay through
radiation to infinity.
\item
Models for such point-like particles can be
constructed as trapped-mode solutions of  nonlinear wave
equations in n-dimensional space. The solutions
consist of a
superposition of a mean field and wave fields. With
respect to harmonic space, the mean
field is  uniform while the wave fields are periodic. All
fields are
highly localized in physical space. The wave fields are
trapped in physical space by the
mean field, which acts as a wave guide. The mean field is
generated, in turn, by the
radiation stresses (currents) of the wave fields.
\item
The simplest and most fundamental example of a nonlinear
system which can support such interacting wave and mean
fields are Einstein's equations in matter-free space.
\item
Particular periodic solutions in harmonic space of
Einstein's equations can be identified with the solutions of
the
Dirac equation and,  as pointed out by Kaluza and
Klein, of Maxwell's equations. Regarding
the fields as perturbations about a suitably chosen
 flat-space background metric, the
Maxwell-Dirac-Einstein Lagrangian can be recovered as the
lowest-order interaction Lagrangian of the nonlinear
Einstein Lagrangian. To establish this correspondence, the
dimension of harmonic
space must be at least four.
\item
Postulating the existence of further trapped-mode solutions,
the
general structure of weak and strong interactions, as
summarized in the Standard Model, can also be recovered,
thereby yielding a unified theory.
\item
Since all particle and field coupling phenomena are deduced
as properties of the nonlinear solutions of the
n-dimensional Einstein vacuum equations, which contain no
universal physical constants, it follows that the
theory must yield all physical constants.
\item
In deriving the formal expressions for the mass,
gravitational constant, charge, Planck's constant, etc. in
terms of the properties of the nonlinear metron
solutions, the gravitational coupling associated with the
particle masses was found to arise at  higher nonlinear
order than the electromagnetic coupling associated with the
particle charges. This explains the weakness of
gravitational forces
compared with other forces.
\end{itemize}

Not resolved is the question of the discreteness of
the particle spectrum. This appears as the most serious
conceptual uncertainty of the metron approach at this time.
It is speculated that discreteness can be
explained by stability considerations. Alternatively, if
this is not successful, it may be necessary to simply
postulate -- in analogy with string theory -- that our world
is periodic with respect to the harmonic space coordinates.

At present, the metron model is simply a hypothesis: no
computations of bound particle states have yet been carried
out for the real n-dimensional Einstein field
equations. However, it is hoped that the investigations
of the basic structure of the  theory presented in this
paper have revealed sufficient intriguing features to
motivate
more detailed quantitative investigations. The outcome of
such
efforts
should decide whether the full spectrum of elementary
particles and
all forces of nature can indeed be explained by a
deterministic theory
based on the simple generalizaton of Einstein's vacuum
equations to
higher dimensions - in fitting vindication of Einstein's
long held
conviction that ``God does not play dice''.

\subsection*{Acknowledgements}
The author is grateful for stimulating discussions  with
Rudolf Haag,
Gabriele Veneziano, Gerbrand Komen and a
number of other colleagues and, in particular, for the
detailed critical, constructive and
encouraging comments by Wolfgang Kundt.


\begin{thebibliography}{99}
\bibitem{ft1}
It has been pointed out by a reviewer and by Rudolf Haag,
personal communication, that Einstein also considered this
possibility. (Reviewer's comment: `Einstein likened the
source terms
in his field equations to "plywood" in contrast to the
"marble" of ... the curvature tensor'.)
\bibitem{ft3}
A complementary `accumulative' rather than deductive summary
of
the principal metron features is given in the concluding
Section~\ref{Summary
and conclusions}.
\bibitem{ft4}
We distinguish here between universal physical constants,
which
represent dimensionless ratios independent of the choice of
scales,
and scaling constants, which define units of measurement,
such as the velocity of light or the reference length scale
of
harmonic space.
\bibitem{ft5}
It remains to be explained, however, why left-handed metron
solutions are more prevalent in the universe than
the equally possible right-handed solutions - in analogy
with the unexplained higher abundance of particles
relative to anti-particles.
\bibitem{ft8}
For the real zero-wavenumber components, the two indices
represent the same field, $\varphi_{-p} = \varphi_{p}$,
but for simplicity of notation, both positive and negative
index components are retained also in this case, the net
real field $\varphi_{p}^{net}$ being represented as the sum
$\varphi_{p}^{net} = \varphi_{p} + \varphi_{-p}$ of two
equal constituents.
\end{thebibliography}
\begin{thebibliography}{99}
\bibitem{ft11}
This may be seen, for example, by Fourier decomposition.
For any given four-wavenumber $k$ satisfying the dispersion relation
$k_{\lambda} k^{\lambda} = - \hat \omega^2$, eq. (\ref{3.14})
has four independent eigensolutions, while each of the two
equations (\ref{3.18}), (\ref{3.19}) has two independent positive
and negative spin eigenstate solutions, $\psi^+_+,\psi^+_-$ and
$\psi^-_+ ,\psi^-_-$ , yielding again a total of four independent
solutions.
\bibitem{ft13}
A third term appearing in the full expression for $L_0$ can
be rewritten as a divergence term, invoking the gauge
condition (\ref{2.7}), and has been dropped in (\ref{3.28}).
The right hand side of eq.(\ref{3.27})  is assumed to be
symmetrized with respect to the first three and last two factors.
\bibitem{ft14}
Assuming, as stated above and discussed further
below, that the fermion parton fields contain only one of
the two Dirac brances.
\bibitem{ft16}
The form (\ref{A.5}), and the corresponding polarization tensor
(\ref{8.3.14h}) given later  for the Euclidean case,
can be readily shown to represent particular examples from
one-parameter families of solutions which yield the spinor metric
forms (\ref{A.3}) and (\ref{norm2}), respectively.
\bibitem{ft19}
Note that in contrast to the trajectory
and field equations (\ref{4.1}) - (\ref{4.5}), the line
integrals in eqs.  (\ref{4.9}) and (\ref{4.10})
are invariant with respect to the choice of the path
parameter $s$,
and the side condition $u^{\lambda}u_{\lambda} = -1$ is not
invoked in the variation of the action; the normalization
is introduced afterwards in the resultant trajectory and
field equations for convenience.
\bibitem{ft20}
This clearly applies not only for the quadratic form but
generally for the fully nonlinear spacetime gravitational
action.
\bibitem{ft21}
We consider here only mean far fields, ignoring the periodic
de Broglie far fields. These yield a mean source term only when
in resonance with the periodic interior fields in the core
region. This occurs in scattering problems, which will  be
discussed in Sections \ref{Bragg scattering} -
\ref{Atomic spectra}.
\bibitem{ft22}
The treatment of particle spin requires consideration  of the
first moments in the integration of the Lagrangians over the
tube cross-sections. We restrict the discussion here to the
zero'th moment, which characterizes the particle charge.
\end{thebibliography}
\begin{thebibliography}{99}
\bibitem{ft23}
For the coordinate notations for physical spacetime,
harmonic (extra) space and full space we refer to Table
\ref{ta2.1} in Part \ref{The Metron Concept}.
\bibitem{ft24}
`Effective' means here that the fields decrease exponentially
with distance from the test particle within the absorber; whether
or not there exist particles outside the effective absorbing
sphere is irrelevant, as they experience no field directly or
indirectly from the perturbed test particle.
\bibitem{ft25}
For consistency in the expansion order, we have not incorporated
the perturbation of the dispersion relation implied
by (\ref{5.a32}) into the right hand side of (\ref{5.a44}),
although this can be readily done by  replacing $ \omega^2_k$
by $ \omega'^2_k$ in the resonance denominator.
\bibitem{ft26}
In this context a comment is in order. It has been implicitly
assumed in the previous sections that the lowest-order far
fields represent free fields which fall off asymptotically as $1/r$.
However, as discussed in Sections~\ref{The mode-trapping mechanism}
and \ref{Particle interactions}, the periodic de Broglie fields
presumably represent trapped wave-guide modes with a (very weak)
exponential damping factor. This follows if the interactions
are of lowest (cubic) order   (cf. Section~\ref{The mode-trapping
mechanism}) and is furthermore required in order that the
particle charge, which is given by an integral over
the quadratic product of the de Broglie field (eqs.
(\ref{4.22}), (\ref{4.29a})), remains finite. Fortunately,
our analysis can absorb a very weak damping factor in the
lowest-order far fields, since  a damping factor arises anyway
at first order and is, in fact, explicitly invoked in the original
Wheeler-Feynman derivation.
\bibitem{ft28}
The lowest-order analysis of the metron trajectory and
far-field action integrals of Section~\ref{Particle interactions}
must be extended to next order to obtain the spin coupling
terms; however, this does not affect the symmetry structure
of the problem.
\bibitem{ft29}
The harmonic wavenumber components satisfy the  complementary
condition $k^{(i)}_{\alpha}k^{\alpha}_{(i)} = \, \omega_0^2$,
so that the field represents a free gravity wave in n-dimensional
space with $k^{(i)}_{L}k^{L}_{(i)} = 0$, cf. Parts~\ref{The
Metron Concept}, \ref{The Maxwell-Dirac-Einstein System}.
However, we shall not be concerned with the periodic harmonic-space
dependence here.
\bibitem{ft30}
We note that the concept of a resonant interaction between
the scattered particle and its scattered wave assumes that
the scattered far field generated by the de Broglie wave
of the incident particle is still present at the location
of the particle after the particle has interacted with the
lattice. This is indeed the case, as the scattered field
propagates at its group velocity, which is matched to the
particle propagation velocity.
\bibitem{ft31}
Note that although the intrinsic particle phase has been
regarded so far only along the particle trajectory, it is
in fact defined generally in space by the de Broglie wave
which the particle source field generates.
\end{thebibliography}
\begin{thebibliography}{}
\bibitem{ft37}
This can be remedied by including a non-zero component
$k_6^{(e)}$
in the electron wavenumber, but at the cost of introducing a
flavor
dependence into the polarization tensor.
\bibitem{ft38}
For better insight into the structure of these and the
following
relations, we retain separate notations for $\hat \omega_e^2
$ and $\hat \omega_{\nu}^2 $, although, according to
(\ref{8.49x}), $\hat \omega_e^2 =\hat \omega_{\nu}^2 $).
\bibitem{ft39}
We note as an aside that an additional reason for not
invoking the
Higgs mechanism to explain the electron mass is that the
identification
of the Higgs field with a field which is periodic in
harmonic
space -- which is necessary to generate the coupling terms
defining the mass matrix in eq.(\ref{8.86}) below -- rules
out a
normal Yukawa-type interaction of the form $\bar \psi^{(e)}
g^{(h)}
\psi^{(e)}$.
This would represent a periodic term which yields no
contribution to
the action integral.
\end{thebibliography}
\end{document}